\documentstyle [12pt,twoside,epsfig]{article}
\begin{document}
\thispagestyle{empty}
\fboxsep2ex


\noindent
\renewcommand{\thefootnote}{\fnsymbol{footnote}}
\setcounter{footnote}{1}
\begin{center}
{\LARGE \bf Homodyne Detection and Quantum\\[.5ex]
State Reconstruction\footnotemark}\\[2ex]
\end{center}
\footnotetext[2]{In \textsl{Progress in Optics}, Vol. XXXIX, ed. E. Wolf
(Elsevier, Amsterdam, 1999), p. 63--211.}
\renewcommand{\thefootnote}{\arabic{footnote}}
\setcounter{footnote}{0}


\noindent
\begin{center}
{\Large Dirk--Gunnar Welsch}\\[1ex]
Friedrich-Schiller-Universit\"{a}t Jena,  
Theoretisch-Physikalisches Institut\\
Max-Wien Platz 1, D-07743 Jena, Germany\\[2ex]
{\Large Werner Vogel}\\[1ex]
Universit\"{a}t Rostock,  
Fachbereich Physik\\  
Universit\"{a}tsplatz 3, D-18051 Rostock, Germany\\[2ex]
{\Large Tom\'{a}\v{s} Opatrn\'{y}}\\[1ex]
Palack\'{y} University,  
Faculty of Natural Sciences\\
Svobody 26, 77146 Olomouc, Czech Republic
\end{center}


\normalsize
\newpage

{
\tableofcontents

\newpage

\renewcommand{\thesection}{\arabic{section}.}
\renewcommand{\thesubsection}{\arabic{section}.\arabic{subsection}.}
\renewcommand{\thesubsubsection}{\arabic{section}.\arabic{subsection}.%
\arabic{subsubsection}.}

\section{Introduction}
\label{sec1}

Since the experimental demonstration of tomographic quantum-state
measurement on optical fields by Smithey, Beck, Raymer and Faridani
[1993] based on the theoretical work of Vogel, K., and Risken [1989]
a boom in studying quantum-state reconstruction problems has
been observed. Numerous workers have considered various schemes and
methods for extracting all knowable information on the state of a quantum
system from measurable data, and early work has been recovered.
In the history of quantum theory the concept of quantum state including
quantum-state measurement has been a matter of intense discussion.

The state of a quantum object is commonly
described by a normalized Hilbert-space vector $|\Psi\rangle$ or,
more generally, by a density operator $\hat{\varrho}$ $\!=$
$\!\sum_{i}$ $p_{i} |\Psi_{i}\rangle\langle\Psi_{i}|$ which is a
Hermitian and non-negative valued Hilbert-space operator of
unit-trace. The Hilbert space of the object is usually spanned
up by an orthonormalized set of basic vectors $|A\rangle$
representing the eigenvectors (eigenstates) of Hermitian operators
$\hat{A}$ associated with a complete set of simultaneously
measurable observables (physical quantities) of the
object.\footnote{For notational convenience we write $\hat{A}$,
   without further specifying the quantities belonging to the set.}
The eigenvalues $A$ of these operators are the values of the
observables that can be registered in a measurement. Here it
must be distinguished between an individual (single) and an ensemble
measurement (i.e., in principle, an infinitely large number of repeated
measurements on identically prepared objects). Performing
a single measurement on the object, a totally unpredictable
value $A$ is observed in general, and the state of the object
has {\em collapsed} to the state $|A\rangle$ according to
the von Neumann's projection definition of a measurement
(von Neumann [1932]). If the
same measurement is repeated immediately after the
first measurement (on the same object), the result is now well
predictable -- the same value $A$ as in the first measurement is
observed. Obviously, owing to the first measurement the object has been
prepared in the state $|A\rangle$. Repeating the measurement
many times on identically prepared object, the relative
rate at which the result $A$ is observed approaches the
diagonal density-matrix element $\langle A|\hat{\varrho}|A\rangle$
as the number of measurements tend to infinity. Measuring
$\langle A|\hat{\varrho}|A\rangle$ for all values of $A$
the statistics of $\hat{A}$ (and any function of $\hat{A}$)
are known. To completely describe the quantum state, i.e., to
determine all the quantum-statistical properties of the object,
knowledge of all density-matrix elements $\langle A|\hat{\varrho}|A'\rangle$
is needed. In particular, the off-diagonal elements essentially
determine the statistics of such sets of observables $\hat{B}$
that are not compatible with $\hat{A}$ ($[\hat{A},\hat{B}]$
$\!\neq$ $\!0$) and cannot be measured simultaneously with $\hat{A}$.
Obviously, the statistics of $\hat{B}$ can also be obtained directly
-- similar to the statistics of $\hat{A}$ -- from an ensemble
measurement yielding the diagonal density-matrix elements
$\langle B|\hat{\varrho}|B\rangle$ in the basis of the eigenvectors
$|B\rangle$ of $\hat{B}$. Now one can proceed to consider
other sets of observables that are not compatible with $\hat{A}$
and $\hat{B}$, and the rather old question arises of which and how
many incompatible observables must be measured in order to obtain all the
information contained in the density matrix of an arbitrary
quantum state. Already Pauli [1933] addressed the question whether
or not the wave function of a particle can be reconstructed from
the position and momentum probability distributions, i.e., the
diagonal density-matrix elements in the position and momentum
basis.

Roughly speaking, there have been two routes to collect
measurable data for reconstructing the quantum state of an object
under consideration. In the first, which closely follows the line
given above, a succession of (ensemble) measurements is made such that
a set of noncommutative object observables
is measured which carries the complete information about the quantum state.
A typical example is the tomographic quantum-state reconstruction
mentioned at the beginning. Here,
the probability distributions
$p(x,\varphi)$ $\!=$ $\!\langle x,\varphi|\hat{\varrho}
|x,\varphi\rangle$ of the quadrature components $\hat{x}(\varphi)$
of a radiation-field mode for all phases $\varphi$ within a $\pi$ interval
are measured, i.e., the expectation values of the
quadrature-component projectors
$|x,\varphi\rangle\langle x,\varphi|$.
In the second, the object is coupled to a reference system
(whose quantum state is well known) such that measurement of
observables of the composite system corresponds to ``simultaneous''
measurement of noncommutative observables of the object.
In this case the number of observables (of the composite system)
that must be measured in a succession of (ensemble) measurements
can be reduced drastically, but at the expense of the image
sharpness of the object.
As a result of the additional noise introduced by the
reference system only fuzzy measurements on the object
can be performed, which just makes a ``simultaneous'' measurement
of incompatible object observables feasible. A typical example
is the $Q$ function of a radiation-field mode, which
is given by the diagonal density-matrix elements
in the coherent-state basis, $Q(\alpha)$ $\!=$
$\!\pi^{-1}\langle\alpha|\hat{\varrho}|\alpha\rangle$. It can
already be obtained from one ensemble measurement of the complex
amplitude $\alpha$, which corresponds to a fuzzy measurement
of the ''joint'' probability distribution of two
canonically conjugated observables $\hat{x}(\varphi)$ and
\mbox{$\hat{x}(\varphi$ $\!+$ $\!\pi/2)$} of the object.

Since the sets of quantities measured via the one or the other
route (or an appropriate combination of them) carry the complete
information on the quantum state of
the object, they can be regarded, in a sense, as representations
of the quantum state, which can be more or less close
to (or far from) familiar quantum-state representations,
such as the density matrix in the Fock basis or the Wigner
function. In any case, the question arises of how to reconstruct
from the measured data specific quantum-state representations
(or specific quantum-statistical properties of the object
for which a direct measurement scheme is not available).
Again, there have been two typical concepts for solving the
problem. In the first, equations that relate the measured
quantities to the desired quantities are derived and tried to
be solved either analytically or numerically in order to obtain
the desired quantities in terms of the measured quantities.
In practice, the measured data are often incomplete, i.e.,
not all quantities needed for a precise reconstruction are
measured,\footnote{To compensate for incomplete data, some
   {\em a priori} knowledge of the quantum state is needed.}
and moreover, the measured data are inaccurate.
Obviously, any experiment can only run for a finite
time, which prevents one, on principle, from performing an
infinite number of repeated measurements in order to obtain
precise expectation values. These inadequacies give rise to systematic
and statistical errors of the reconstructed quantities, which
can be quantified in terms of confidence intervals.
In the second, statistical methods are used from the very
beginning in order to obtain the best {\em a posteriori} estimation
of the desired quantities on the basis of the available
(i.e., incomplete and/or inaccurate) data measured.
However the price to pay may be high. Whereas in the first
concept linear equations are typically to be handled and
estimates of the desired quantities (including statistical errors)
can often be directly sampled from the measured data, application of
purely statistical methods, such as the principle of {\em maximum entropy} or
{\em Bayesian inference}, require nonlinear equations to be
considered, and reconstruction in {\em real time} is impossible
in general.

The aim of this review is to familiarize physicists with the recent
progress in the field of quantum-state reconstruction and draw
attention to important contributions in the large body of work.
Although in much of what follows we consider the reconstruction of
quantum states of travelling optical fields, the basic-theoretical
concepts also apply to cavity fields and matter systems. Section
\ref{sec2} reviews the experimental schemes that have been considered
for quantum-state measurement on optical fields, with special emphasis
on optical homodyning (\S~\ref{sec2.1}). It begins by formulating
the basic ideas of four-port homodyne detection of the quadrature
components of single-mode fields (\S\S~\ref{sec2.1.1} and
\ref{sec2.1.2}) and then proceeds to extend the scheme to multimode
fields and multiport homodyne detectors (\S~\ref{sec2.1.3}), which
can also be used for ``simultaneous'' measurement of noncommuting signal
observables (\S\S~\ref{sec2.1.4} -- \ref{sec2.1.6}). Section
\ref{sec2.1.7} returns to the four-port scheme in \S~\ref{sec2.1.1}
and explains its use for measuring displaced Fock states. Homodyne
correlation measurements are shown to yield insight in phase-sensitive
field properties even in the case of low detection efficiencies
(\S~\ref{sec2.1.8}).  After addressing heterodyne detection
(\S~\ref{sec2.2}) and parametric amplification (\S~\ref{sec2.3}),
typical schemes for quantum-state measurement in high-$Q$ cavity QED
by test atoms are outlined (\S~\ref{sec2.4}).

Section~\ref{sec3} reviews typical methods for reconstruction
of quantum-state representations and specific
quantum-statistical properties of optical fields
from sets of measurable quantities carrying the complete information
on the quantum state. In \S~\ref{sec3.1} -- \ref{sec3.5} we focus
on the reconstruction of quantum-state representations
from quantities measurable in homodyne detection of
travelling optical fields. The next subsection (\S~\ref{sec3.6})
presents methods of reconstruction of quantum-state representations
of high-$Q$ cavity fields from measurable properties of test
atoms probing the cavity fields, and in \S~\ref{sec3.7} alternative
proposals are outlined. Section \ref{sec3.8} then addresses
the problem of direct reconstruction of specific quantities
from the measured data -- an important problem with respect to
quantities for which direct measurement schemes have not been
found so far. Whereas in \S\S~\ref{sec3.1} -- \ref{sec3.8}
it is assumed that all the data needed for a precise reconstruction
of the desired quantities can be precisely measured, at least
in tendency, in \S~\ref{sec3.9} statistical methods for processing
smeared and inaccurately measured incomplete data are outlined with the
aim to obtain optimum estimations of the quantum states.

Section \ref{sec4} summarizes methods for measurement and
reconstruction of quan\-tum-state representations of matter
system, with special emphasis on optical methods and
experimental demonstrations. In
\S~\ref{sec4.1} the problem of reconstruction of the
vibrational quantum state in an excited electronic state of
a two-atom molecule is addressed. Typical schemes for
determining the motional quantum state of trapped atoms
are outlined in \S~\ref{sec4.2}, and in \S~\ref{sec4.4}
the problem of quantum-state measurement on Bose--Einstein
condensates is considered. Sections \ref{sec4.5} and
\ref{sec4.3}, respectively, present schemes suitable for
determining the motional quantum state of atom- and
electron-wave packets. In the debate on quantum-state
measurement, which is nearly as old as quantum mechanics,
spin systems has been attracted much attention, because
of low-dimensional Hilbert space. Section \ref{sec4.6}
outlines the main ideas in this field. Finally, \S~\ref{sec4.7}
explains the basic ideas of Compton and $X$-ray scattering for
measuring the single-particle quantum state of crystal lattices.


\section{Phase-sensitive measurements of light}
\label{sec2}

The statistical properties of a classical radiation field are known
when the amplitude statistics and the phase statistics are known.
Whereas the amplitude statistics can be obtained from intensity
measurements, determination of the phase statistics needs
phase-sensitive measurement. Obviously the same is true in quantum optics.
In order to obtain the complete information about the quantum
state a radiation field is prepared in interferometric measurements
that respond to amplitude and phase variables must be performed.


\subsection{Optical homodyning}
\label{sec2.1}

{\em Homodyne} detection has been a powerful method for measuring
phase sensitive properties of travelling optical fields which are
suitable for quantum-state reconstruction, and a number of
sophisticated detection schemes have been studied. In the four-port
basic scheme, a signal field is combined, through a lossless beam
splitter, with a highly stable reference field which has the same
mid-frequency as the signal field. The reference field, also called
local oscillator, is usually prepared in a coherent state of large
photon number. The superimposed fields impinge on photodetectors, the
numbers of the emitted (and electronically processed) photoelectrons
being the homodyne detection output (for the basic ideas, see Yuen and
Shapiro, J.H.  [1978a,b, 1980]; Shapiro, J.H., and Yuen [1979];
Shapiro, J.H., Yuen and Machado Mata [1979]).  The observed
interference fringes, which vary with the difference phase between the
two fields, reflect the quantum statistics of the signal field and can
be used -- under certain circumstances -- to obtain the quantum state
of the signal field.
The homodyne output can be fully given in terms of the joint-event
probability distribution of the detectors in the output channels.
In balanced homodyning, difference-event distributions are measured.
In particular, the difference-event statistics measurable by a perfect
four-port homodyne detector directly yields the
quadrature-com\-ponent statistics of the signal field, which
has offered novel possibilities of quan\-tum-state measurement.


\subsubsection{Basic scheme}
\label{sec2.1.1}

Let us start the analysis with the four-port scheme and first restrict
attention to single-mode fields mode-matched to the local-oscillator
(Fig.~\ref{fig2.1}).\footnote{Here
   and in the following spatial-temporal modes are considered. They are
   nonmonochromatic in general and defined with a particular spatial,
   temporal and spectral form (Appendix \protect\ref{app1}).}
\begin{figure}[htb]
 \unitlength=1cm
 \begin{center}
 \begin{picture}(6,3)
 \put(0,0){
 \includegraphics{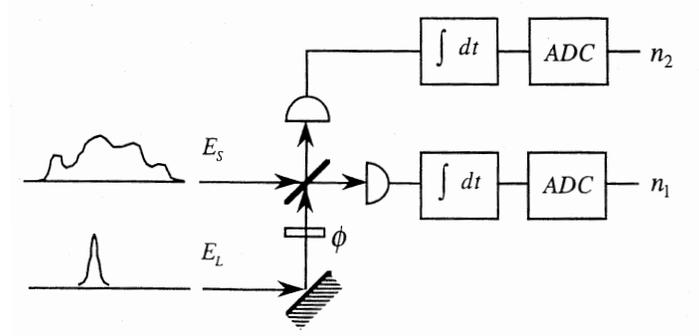}}
 \end{picture}
 \end{center}
\begin{center}
\protect\parbox{.9\textwidth}{
\caption{
A signal pulse field $E_{S}$ interferes with a shorter
local-oscillator pulse $E_{L}$ at a $50/50$ beam splitter. The
local-oscillator phase $\phi$ determines which quadrature amplitude of the
signal is detected. The superposed fields are detected with
high-efficiency photodiodes having response times much longer than the pulse
durations. The photocurrents are integrated and sampled by analog-to-digital
converters, to yield pulse photoelectron numbers $n_{1}$ and $n_{2}$.
(After Raymer, Cooper, Carmichael, Beck and Smithey [1995].)
\label{fig2.1}
}
}
\end{center}
\end{figure}%
The action of a lossless beam splitter (as an example of a linear four-port
coupler) on quantum fields has widely been studied (see, e.g.,
Richter, G., Brunner and Paul, H., [1964];
Brunner, Paul, H., and Richter, G., [1965]; Paul, H., Brunner and
Richter, G., [1966]; Yurke, McCall and Klauder [1986];
Prasad, Scully and Martienssen [1987]; Ou, Hong and Mandel [1987a];
Fearn and Loudon [1987]; Campos, Saleh and Teich [1989]; Leonhardt
[1993]; Vogel, W., and Welsch [1994]; Luis and S\'{a}nchez-Soto [1995]),
and it is well known that the input--output relations can be
characterized by the SU(2) Lie algebra.

In the Heisenberg picture, the photon destruction operators of the
outgoing modes, $\hat{a}'_{k}$ ($k$ $\!=$ $\!1,2$), can be obtained
from those of the incoming modes, $\hat{a}_{k}$, by a unitary
transformation
\begin{equation}
\hat{a}'_{k} = \sum_{k'=1}^2 U_{kk'} \,
\hat{a}_{k'}\,,
\label{2.1}
\end{equation}
where $U_{11}$ and $U_{21}$ are the transmittance and reflectance of
the beam splitter from one side, and $U_{22}$ and $U_{12}$ are those
from the other side. The unitarity of the scattering matrix
$U_{kk'}$ $\!=$ $\!|U_{kk'}|e^{i\varphi_{kk'}}$ implies that the
following restrictions are placed on it:
\begin{equation}
|U_{11}|^2 + |U_{21}|^2 =1, \quad
|U_{22}| = |U_{11}|,\ |U_{12}| = |U_{21}|,
\label{2.2}
\end{equation}
\begin{equation}
\varphi_{11} - \varphi_{21} + \varphi_{22} -\varphi_{12} = \pm \pi.
\label{2.3}
\end{equation}
The relations (\ref{2.2}) and (\ref{2.3}) ensure that the
bosonic commutation relations for the photon creation and destruction
operators are preserved.
Using the angular momentum representation, the unitary transformation
(\ref{2.1}) can be given by
\begin{equation}
\label{2.5}
\hat{a}'_{k} = \hat{V} \hat{a}_{k} \hat{V}^\dagger,
\end{equation}
where the operator $\hat{V}$ reads as\footnote{The
   parameters in eq.~(\ref{2.6}) are related
   to the parameters in eqs.~(\ref{2.2}) and (\ref{2.3}) as
   $\alpha$ $\!=$ $\!{\textstyle\frac{1}{2}}
   (\varphi_{11}$ $\!-$ $\!\varphi_{22}$ $\!+$ $\!\varphi_{21}$
   $\!-$ $\!\varphi_{12}$ $\!\pm$ $\! \pi)$,
   $\gamma$ $\!=$ $\!{\textstyle\frac{1}{2}}
   (\varphi_{11}$ $\!-$ $\!\varphi_{22}$ $\!-$ $\!\varphi_{21}$ $\!+$
   $\!\varphi_{12}$ $\!\mp$ $\!\pi)$,
   $\cos^{2}(\beta/2)$ $\!=$ $\!|U_{11}|^{2}$,
   $\sin^{2}(\beta/2)$ $\!=$ $\!|U_{21}|^{2}$,
   $\delta$ $\!=$ $\!\textstyle\frac{1}{2}
   (\varphi_{11}$ $\!+$ $\!\varphi_{22})$.}
\begin{equation}
\label{2.6}
\hat{V} = e^{-i\alpha \hat L_{3}} e^{-i\beta \hat{L}_{2}}
e^{-i\gamma \hat{L}_{3}} e^{-i\delta \hat{N}}.
\end{equation}
Here, $\hat{N}$ $\!=$ $\!\hat{n}_{1}$ $\!+$ $\!\hat{n}_{2}$
is the operator of the total photon number ($\hat{n}_{k}$
$\!=$ $\!\hat{a}^{\dagger}_{k}\hat{a}_{k}$), and the operators
\begin{equation}
\label{2.7}
\hat{L}_{1} = \textstyle\frac{1}{2} \left( \hat{a}^{\dagger}_{1}
\hat{a}_{2} + \hat{a}^{\dagger}_{2} \hat{a}_{1} \right) , \
\hat{L}_{2} = \textstyle\frac{1}{2i} \left( \hat{a}^{\dagger}_{1}
\hat{a}_{2} - \hat{a}^{\dagger}_{2}\hat{a}_{1} \right) , \
\hat{L}_{3} = \textstyle\frac{1}{2} \left( \hat{a}^{\dagger}_{1}
\hat{a}_{1} - \hat{a}^{\dagger}_{2}\hat{a}_{2} \right)
\end{equation}
obey the familiar commutation relations of angular-momentum
operators.

In the Schr\"{o}dinger picture the photonic operators are left unchanged,
but the state is transformed. Let $\hat{\varrho}$ and $\hat{\varrho}'$
be the input- and output-state density operator of the two modes,
respectively. It is easily seen that if they are related to each other as
\begin{eqnarray}
\label{2.11}
\hat{\varrho}' = \hat{V}^{\dagger} \hat{\varrho} \hat{V},
\end{eqnarray}
then the two pictures lead to identical expectation values of
arbitrary field quantities.

Let us now assume that the operation of the photodetectors in the
two output channels of the beam splitter can be described by means of
the standard photodetection theory as outlined in Appendix \ref{app3};
i.e., the numbers of photoelectric events counted in a given time interval
(integrated photocurrents) are proportional to the numbers of
photons falling on the detectors. From eq.~(\ref{A3.7}) the joint-event
distribution of measuring $m_1$ and $m_2$ events in the two output
channels 1 and 2, respectively, is then given (in the Heisenberg picture) by
\begin{equation}
\label{2.12}
P_{m_{1},m_{2}} = \left\langle \! :
\prod_{k=1}^2 \frac{1}{m_{k}!}\,(\eta_{k} \hat{n}'_{k})^{m_k}
e^{-\eta_{k} \hat{n}'_k} : \! \right\rangle ,
\end{equation}
where
\begin{equation}
\label{2.13}
\hat{n}'_{k} = \hat{a}'^{\dagger}_{k} \hat{a}'_{k} \, ,
\end{equation}
with $\hat{a}'_{k}$ being given by eq.~(\ref{2.1}) [or eq.~(\ref{2.5})].
Note that for $\eta$ $\!=$ $\!1$ (i.e., perfect detection) $P_{m_{1},m_{2}}$
is nothing more than the joint-photon-number distribution of the two
outgoing modes. In particular when the reference mode is prepared in
a coherent state $|\alpha_{L}\rangle$, then in the expectation values
of normally ordered operator functions, $\hat{a}'_{k}$ can be
replaced with ($\hat{a}$ $\!\equiv$ $\!\hat{a}_{1}$)
\begin{equation}
\label{2.14}
\hat{a}'_{k} = U_{k1} \, \hat{a} + U_{k2} \, \alpha_{L} \, .
\end{equation}


\subsubsection{Quadrature-component statistics}
\label{sec2.1.2}

Let us now turn to the question of what information on the
quantum statistics of the signal mode can be obtained
from the homodyne output. From
eq.~(\ref{2.13}) together with eqs.~(\ref{2.14}) and (\ref{2.3})
we easily find that the output photon number $\hat{n}'_{k}$ can be
rewritten as
\begin{equation}
\label{2.15}
\hat{n}'_{k} = |U_{k2}|^{2} |\alpha_{L}|^{2}
+ |U_{k1}|^{2} \hat{a}^{\dagger} \hat{a}
+ (-1)^{k+1} |U_{k2}| |U_{k1}| \hat{F}(\varphi) ,
\end{equation}
where
\begin{equation}
\label{2.16}
\hat{F}(\varphi)
= 2^{1/2} |\alpha_{L}| \, \hat{x}(\varphi) ,
\end{equation}
with
\begin{equation}
\label{2.17}
\hat{x}(\varphi)
= 2^{-1/2} \left(\hat{a} \, e^{-i\varphi}
+ \hat{a}^{\dagger} e^{i\varphi}\right)
\end{equation}
being the quadrature-component operator, and
\begin{equation}
\label{2.18}
\varphi
= \varphi_{L} + \varphi_{12} - \varphi_{11}.
\end{equation}
We see that $\hat{n}'_{k}$ consists of
three terms. The first and the second terms represent the photon numbers
of the local oscillator and the signal, respectively. More interestingly,
the third term that arises from the interference of the two modes
is proportional to a quadrature component of the signal mode,
$\hat{x}(\varphi)$, the rapidly varying optical phase $\omega t$ that
usually occurs in a field strength (quadrature component) is replaced with
the phase parameter $\varphi$. In particular, when the signal and the
local oscillator come from the
same source, then the phase parameter $\varphi$ can be controlled easily,
so that the dependence on $\varphi$ of $\hat{x}(\varphi)$ can
be monitored; e.g., by shifting the phase difference between the signal
and the local oscillator in the input ports of the beam splitter
in Fig.~\ref{fig2.1}.

If a balanced (i.e., $50$\%:$50$\%) beam splitter is used, the relation
$|U_{k2}|$ $\!=$ $\!|U_{k1}|$ $\!=1/\sqrt{2}$ is valid, and hence the
difference photon number is a signal-mode field strength $\hat{F}(\varphi)$,
\begin{equation}
\label{2.19}
\hat{n}'_{1} - \hat{n}'_{2} = \hat{F}(\varphi);
\end{equation}
i.e., it is proportional to a signal-mode quadrature component.
This result suggests that it is advantageous to use a balanced scheme
and to measure the difference events or the corresponding
difference of the photocurrents of the detectors in the two output channels
in order to eliminate the intensities of the two input fields
from the measured output. The method
is also called {\em balanced} homodyning and can be used advantageously
in order to suppress perturbing effects
due to classical excess noise of the local oscillator
(Mandel [1982]; Yuen and Chan [1983]; Abbas, Chan and Yee [1983];
Schumaker [1984]; Shapiro, J.H., [1985]; Yurke [1985];
Loudon [1986]; Collett, Loudon and Gardiner [1987]; Yurke and Stoler [1987];
Yurke, Grangier, Slusher and Potasek [1987]; Drummond [1989];
Blow, Loudon, Phoenix and Shepherd [1990]; Huttner, Baumberg, Ryan and
Barnett [1992]). In particular, when the local
oscillator is much stronger than the signal field, then
even small intensity fluctuations of the local oscillator may significantly
disturb the signal-mode quadrature components that are desired to be observed.
For example, if the quadrature-component variance is intended to
be derived from the variance of events measured by a single detector, the
classical noise of the local oscillator and the quantum noise of the signal
would contribute to the measured data in the same manner, so that the two
effects are hardly distinguishable.
Since in the two output channels identical classical-noise effects
are observed, in the balanced scheme they eventually cancel in the
measured signal due to the subtraction procedure.

Strictly speaking, from the arguments given above it is only established
that for chosen phase parameters the mean value of the measured difference
events or photocurrents is proportional to the expectation value of a
quadrature component of the signal mode. From a more careful
(quantum-mechanical) analysis it can be shown that in perfect balanced
homodyning the quadrature-component
statistics of the signal mode are indeed
measured, provided that the local oscillator is sufficiently strong; i.e.,
$|\alpha_{L}|^{2}$ is large compared with the mean number of signal
photons (Carmichael [1987]; Braunstein [1990]; Vogel, W., and Grabow [1993];
Vogel, W., and Welsch [1994]; Raymer, Cooper, Carmichael, Beck and Smithey
[1995]; Munroe, Boggavarapu, Anderson, Leonhardt and Raymer [1996]).
The probability distribution of the difference events,
\begin{equation}
\Delta m = m_{1} - m_{2}\,,
\label{2.20}
\end{equation}
can be derived from the joint-event distribution (\ref{2.12}) as
\begin{equation}
P_{\Delta m} =  \sum_{m_2} P_{\Delta m\!+\!m_2,m_2}.
\label{2.21}
\end{equation}
Examples of the dependence of the difference-event distribution
$P_{\Delta m}$ on the local-oscillator strength and the
detection efficiency are shown in Fig.~\ref{fig2.2}.
In the limit of the local oscillator being sufficiently strong, so that
$|\alpha_{L}|^2$ is large compared with the average number of
photons in the signal mode, the difference-event distribution
$P_{\Delta m}$ can be given by (assuming that the two detectors
have the same quantum efficiency $\eta$)
\begin{equation}
P_{\Delta m} = \frac{1}{\eta\sqrt{2}|\alpha_{L}|}\;
p\!\left(x=\!\frac{\Delta m}{\eta\sqrt{2}|\alpha_{L}|}\,,\varphi
\,;\eta\right)\!,
\label{2.22}
\end{equation}
with the phase $\varphi$ being determined by the phase parameters of
the apparatus and of the local oscillator according to
eq.~(\ref{2.18}).  In eq.~(\ref{2.22}), $p(x,\varphi;\eta)$ is a
convolution of the quadrature-component distribution of the signal
mode, $p(x,\varphi)$ $\!=$ $\!\langle
x,\varphi|\hat{\varrho}|x,\varphi\rangle$ [cf. Appendix \ref{app2}], with
a Gaussian:\footnote{This result is obtained directly (in the limit
  of a strong local oscillator) from the basic equations (\ref{2.12}),
  (\ref{2.21}) of photocounting theory (Vogel, W., and Grabow [1993]),
  which include the effects of imperfect detection.}
\begin{equation}
p(x,\varphi;\eta)
= \int {\rm d}x' \, p(x',\varphi) \,
p(x\!-\!x';\eta),
\label{2.23}
\end{equation}
\begin{equation}
p(x;\eta) =
\sqrt{\frac{\eta}{\pi (1\!-\!\eta)}} \,
\exp\!\left(-\frac{\eta x^{2}}{1\!-\!\eta}\right)\!.
\label{2.24}
\end{equation}
\begin{figure}[htb]
 \unitlength=1cm
 \begin{center}
 \begin{picture}(10,6.1)
 \put(0,0){
 \includegraphics{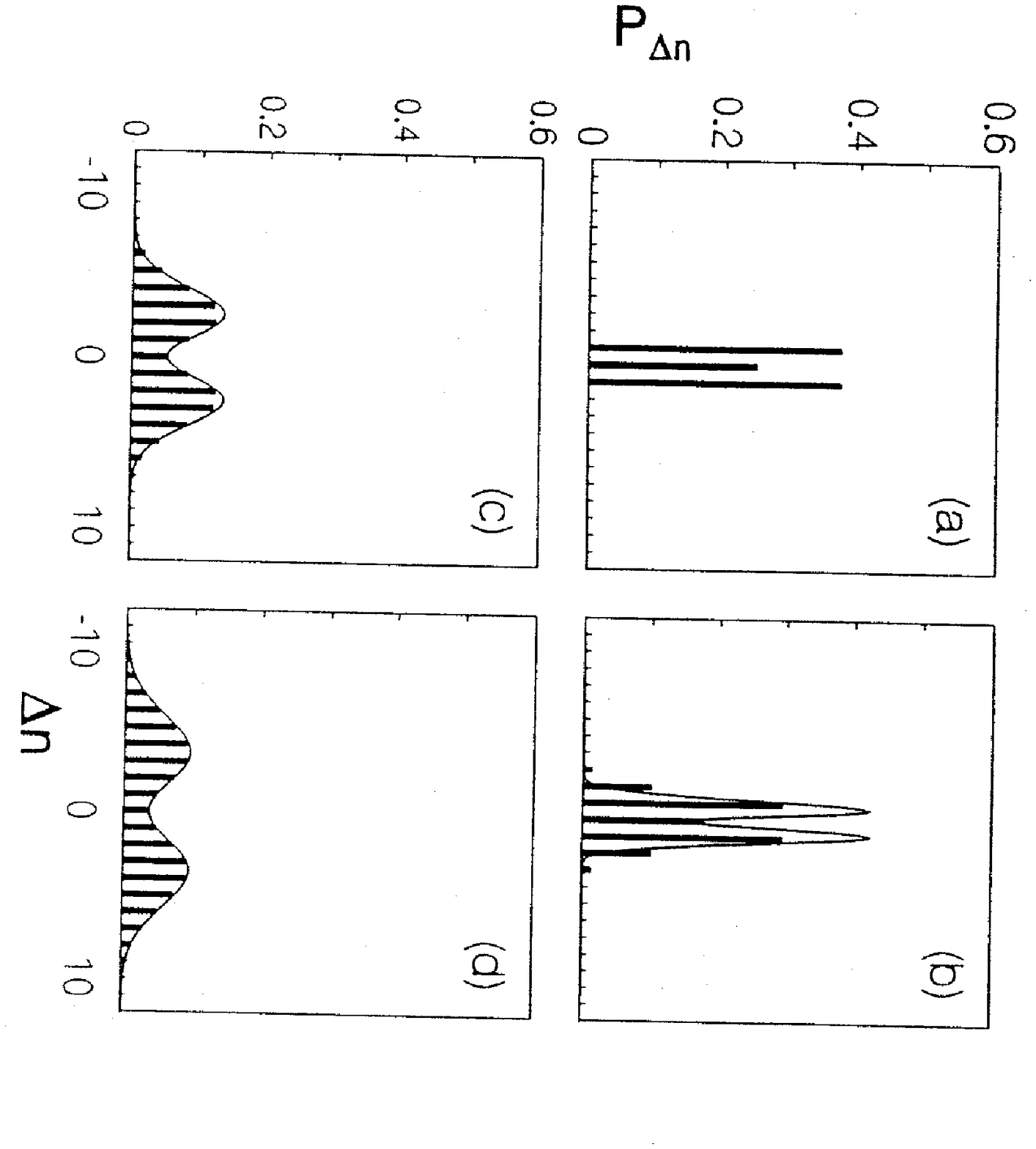}}
 \end{picture}
 \end{center}
\begin{center}
\protect\parbox{.9\textwidth}{
\caption{
  Difference statistics $P_{\Delta n}$ for a single-photon state,
  quantum efficiency $\eta$ $\!=$ $\!0.75$, and various photon numbers
  of the local oscillator: $|\alpha_{L}|^{2}$ $\!=$ $\!0$ (a), $0.5$
  (b), $5$ (c), $10$ (d). The solid curves represent the smeared
  quadrature-component distribution $p(x,\varphi;\eta)$,
  eq.~(\protect\ref{2.23}), which is a convolution of $p(x,\varphi)$
  with $p(x;\eta)$.
(After Vogel, W., and Grabow [1993].)
\label{fig2.2}
}
}
\end{center}
\end{figure}%
In particular, for perfect detection, $\eta$ $\!=$ $\!1$, the
measured difference-event distribution $P_{\Delta m}$ is identical
with the difference-photon-number distribution. In this case
$p(x;\eta)$ reduces to a $\delta$ function; i.e.,
$p(x,\varphi;\eta$ $\!=$ $\!1)$ $\!=$ $\!p(x,\varphi)$, and hence
$P_{\Delta m}$ is (apart from a scaling factor) exactly a
quadrature-component distribution $p(x,\varphi)$ of the signal mode.
Note that when the local oscillator is sufficiently
strong, then $\Delta m/(\sqrt{2}|\alpha_{L}|)$ in eq.~(\ref{2.22})
is effectively continuous. In other words, single-photon resolution is
not needed in order to measure the (continuous) quadrature-component
statistics with high accuracy. In this case, highly efficient linear
response photodiodes can be used, which do not discriminate between single
photons, but nearly reach $100$\% quantum efficiency (Polzik, Carri and
Kimble [1992]).

The Gaussian $p(x;\eta)$ [with dispersion $\sigma^{2}$ $\!=$ $(1$ $\!-$
$\!\eta)/(2\eta)$] obviously reflects the noise associated with
nonperfect detection ($\eta$ $\!<$ $\!1$). In particular, for
$50$\% quantum efficiency (i.e., $\eta$ $\!=$ $\!1/2$) the function
$p(x;\eta)$ is the (phase-independent) quadrature-component distribution
of the vacuum ($\sigma^{2}$ $\!=$ $1/2$).
Obviously, the measured quadrature-com\-po\-nent distribution
$p(x,\varphi;\eta)$ does not correspond to the quadrature component
$\hat{x}(\varphi)$ of the signal mode, but it corresponds to
the quadrature component $\hat{x}(\varphi;\eta)$ of a
superposition of the quadrature component of the
signal, $\hat{x}(\varphi)$, and that of an additional (Gaussian) noise
source, $\hat{x}_{N}(\varphi)$,
\begin{equation}
\label{2.24a}
\hat{x}(\varphi;\eta) = \sqrt{\eta} \, \hat{x}(\varphi)
+ \sqrt{1\!-\!\eta} \, \hat{x}_{N}(\varphi),
\end{equation}
so that $\hat{x}(\varphi;\eta)$ is effectively a combined two-mode
quadrature component. In particular, eq.~(\ref{2.24a}) reveals that the
effect of nonperfect detection can be modelled, assuming a (virtual) beam
splitter is placed in front of a perfect detector, since
eq.~(\ref{2.24a}) exactly corresponds to a beam-splitter transformation
(\S~\ref{sec2.1.1}) (Yurke and Stoler [1987]; Leonhardt and
Paul, H., [1993b, 1994a,b]). In this case the
fields are only partly detected, together with some fraction of
vacuum noise introduced through the ``unused'' input ports of the beam
splitters.\footnote{For a discussion
   of eqs.~(\ref{2.23}) and (\ref{2.24}) in terms of moments of the
   quadrature components $\hat{x}(\varphi;\eta)$ and $\hat{x}(\varphi)$,
   see Banaszek and W\'{o}dkiewicz [1997a].}

If the signal field is not perfectly mode-matched to the
local-oscillator mode, the non-mode-matched part can give rise to
additional noise in the measured quadrature-component distributions
of the mode-matched signal (Raymer, Cooper, Carmichael,
Beck and Smithey [1995]). It can be shown that when the local oscillator
is strong enough to dominate the mode-matched signal field but not
necessarily the nonmatched signal, then the detection efficiency reduces to
\begin{equation}
\label{2.25}
    \tilde{\eta} = \eta \left(1
    + \frac{\overline{N}_{B}}{|\alpha_{L}|^{2}} \right)^{-1} ,
\end{equation}
where $\overline{N}_B$ is the mean number of photons in the nonmatched
signal; i.e., $\eta$ is replaced with $\tilde{\eta}$ in eq.~(\ref{2.22}).
As expected, $\tilde{\eta}$ approaches $\eta$ only in the limit
$|\alpha_{L}|^{2}$ $\!\gg$ $\!\overline{N}_{B}$.

In terms of the characteristic functions of the distributions involved,
the content of eqs.~(\ref{2.22}) -- (\ref{2.24}) can be given by
\begin{equation}
\label{2.26}
\Omega_{\Delta}(y) \ = \ \Psi(\sqrt{2}y,\varphi;\eta)
= \exp\!\left[-\frac{(s-1)\eta+1}{2\eta}\,y^{2}\right]
\Phi\!\left(i y e^{i\varphi};s\right)\!.
\end{equation}
Here,
\begin{equation}
\label{2.27}
\Omega_{\Delta}(y) = \sum_{\Delta m} P_{\Delta m} \,
\exp\!\left(i\frac{\Delta m}{\eta|\alpha_{L}|}\,y\right)
\end{equation}
is the characteristic function of the measured probability distribution
of the scaled difference events $\Delta m/(\eta|\alpha_{L}|)$, and
$\Psi(z,\varphi;\eta)$ and $\Phi(\beta;s)$, respectively,
are the characteristic functions, i.e., the Fourier transforms, of the
(owing to nonperfect detection) smeared quadra\-ture-com\-po\-nent
distribution $p(x,\varphi;\eta)$ and the $s$-parametrized phase-space
function $P(\alpha;s)$ [see eqs.~(\ref{A2.21})\footnote{The
   characteristic function $\Psi(z,\varphi;\eta)$ of $p(x,\varphi;\eta)$
   is defined according to eq.~(\ref{A2.21}) with $p(x,\varphi;\eta)$
   in place of $p(x,\varphi)$. It is seen to be related to the characteristic
   function $\Psi(z,\varphi)$ of $p(x,\varphi)$ as $\Psi(z,\varphi;\eta)$
   $\!=$ $\!\exp[-\frac{1}{4}(\eta^{-1}$ $\!-$ $\!1)z^{2}]$
   $\!\Psi(z,\varphi)$.\label{fn2}}
and (\ref{A2.18})]. Equation~(\ref{2.26})
reveals that in the case of perfect detection the characteristic function
of the (scaled) difference-event distribution is equal, along a line in the
complex plane, to the characteristic function of the Wigner function,
which can be regarded as a suitable representation
of the quantum state of the signal mode.
Hence, when the difference-event statistics are measured on a sufficiently
dense grid of phases $\varphi$ within a $\pi$ interval, then all knowable
information on the signal-mode quantum state can be obtained.
The result is obviously a consequence of the fact that knowledge of the
quadrature-component distribution for all phases within a $\pi$ interval
is equivalent to knowledge of the quantum state; see Appendix \ref{app2.5}
(Vogel, K., and Risken [1989]).


\subsubsection{Multimode detection}
\label{sec2.1.3}

Relations of the type given in eq.~(\ref{2.26}) are also valid for
(higher than four-port) multiport homodyning (Walker [1987]; K\"{u}hn,
Vogel, W., and Welsch [1995]; Ou and Kimble [1995]). Let us consider a linear,
lossless $2(N$ $\!+$ $\!1)$-port coupler (Fig.~\ref{fig2.3})
and assume, e.g., that
$N$ spatially separated signal modes (channels $1,$ $\!\ldots,$ $\!N$)
and a strong local-oscillator mode (channel $N\!+\!1$) of the same
frequency are mixed to obtain $(N$ $\!+$ $\!1)$ output modes impinging
on photodetectors.\footnote{All the modes are assumed
   to be matched to each other.}
The input and output photon operators
$\hat{a}_{k}$ and $\hat{a}'_{k}$, respectively, are then related
to each other by a SU($N$ $\!+$ $\!1$) transformation, extending
eq.~(\ref{2.1}) to
\begin{equation}
\label{2.28}
\hat{a}'_{k} = \sum_{k'=1}^{N+1} U_{kk'} \, \hat{a}_{k'}
\end{equation}
($k$ $\!=$ $\!1,\ldots,N$ $\!+$ $\!1$).
Note that any discrete finite-dimensional unitary matrix can be constructed
in the laboratory using devices, such as beam splitters, phase shifters
and mirrors (Reck, Zeilinger, Bernstein and Bertani [1994]; for further
readings, see Jex, Stenholm and Zeilinger [1995]; Mattle, Michler,
Weinfurter, Zeilinger and Zukowski [1995]; Stenholm [1995];
T\"{o}rm\"{a}, Stenholm and Jex [1995]; T\"{o}rm\"{a}, Jex and
Stenholm [1996]).
\begin{figure}[t]
 \unitlength=1cm
 \begin{center}
 \begin{picture}(10,7)
 \put(0,0){
 \includegraphics{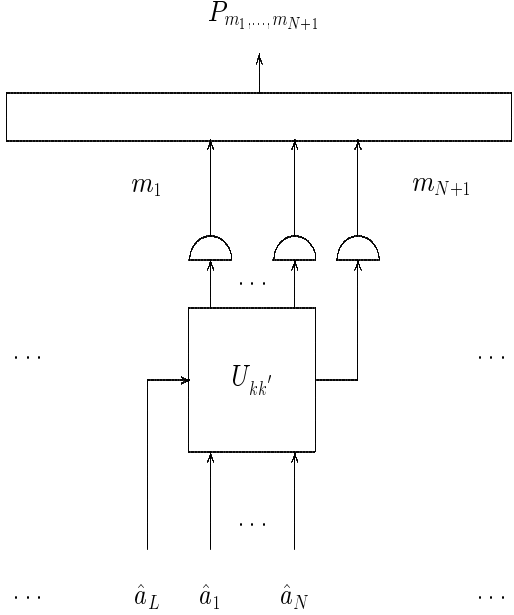}}
 \end{picture}
 \end{center}
\begin{center}
\protect\parbox{.9\textwidth}{
\caption{
Scheme of homodyne $2(N$ $\!+$ $\!1)$-port detection.
$N$ signal modes $\hat{a}_{k}$ ($k$ $\!=$ $\!1,\ldots,N$) of frequency
$\omega$ and a strong local oscillator $\hat{a}_{L}$ of
the same frequency are combined by a lossless
$2(N$ $\!+$ $\!1)$-port device to give $N$ $\!+$ $\!1$ output
modes ($U_{kk'}$: unitary transformation matrix of the
device). Simultaneous detection of the $N$ $\!+$ $\!1$ output
modes yields the $(N$ $\!+$ $\!1)$-fold joint count distribution
$P_{m_{1},\ldots,m_{N}}$.
(After Vogel and Welsch [1995].)
\label{fig2.3}
}
}
\end{center}
\end{figure}%
Simultaneous detection of the output modes yields an
$(N\!+\!1)$-fold joint event distribution $P_{m_{1},\ldots,m_{N+1}}$.
With regard to difference-event measurements, let us consider the scaled
difference events
\begin{equation}
\label{2.29}
 \Delta\tilde{m}_{l} = \frac{m_{l}}{\eta_{l}|\alpha_{L}|}
  - \frac{  \big|U_{ l\,N+1} \big|^2 }
                     {  \big|U_{ \bar{k}\,N+1 }\big|^2 } \,
  \frac{m_{\bar{k}}}{\eta_{\bar{k}}|\alpha_{L}|}\,.
\end{equation}
Here, $l$ $\!=$ $\!1,$ $\!\ldots,$ $\!\bar{k}$ $\!-$ $\!1,$ $\!\bar{k}$ $\!+$
$\!1,$ $\!\ldots,$ $\!N$ $\!+$ $\!1$, the reference channel being
denoted by $\bar{k}$. It can be shown
that when the photon number $|\alpha_{L}|^{2}$ of the coherent-state
local oscillator
is much greater than unity and dominates all the signal modes at the
detectors,\footnote{Apart from this condition, the analysis also allows
   the local oscillator to be in other than coherent states (Walker [1987];
   Leonhardt [1995]; Kim, M.S., and Sanders [1996]; for measuring the degree
   of {\em squeezing} in a signal field by using squeezed local oscillators,
   see Kim, C., and Kumar [1994]).
   This can be done by writing the local-oscillator state $\hat{\varrho}_{L}$
   in the form $\hat{\varrho}_{L}$ $\!=$ $\!\hat{D}(\alpha_{L})$
   $\!\hat{\varrho}'_{L}$ $\!\hat{D}^{\dagger}(\alpha_{L})$,
   where $\hat{D}(\alpha_{L})$ is the coherent displacement operator
   and $\hat{\varrho}'_{L}$ is a state that remains finite as the
   complex amplitude $|\alpha_{L}|$ becomes large.}
then the characteristic
function of the joint probability distribution of the scaled difference
events, $\Omega_{\Delta}(\{ y_l\} )$, can be related to the characteristic
function of the $s$-parametrized multimode phase-space
function of the signal field, $\Phi(\{ \beta_{n}\} ;s)$,
$n$ $\!=$ $\!1,\ldots,N$, as\footnote{Note that $\Phi(\{ \beta_{j} \} ;s)$
   $\!=$ $\!\exp$ $\![\sum_{n}$ $\!\frac{1}{2}$ $\!s$ $\!|\beta_{n}|^2]$
   $\!\langle$ $\!\hat{D}(\{ \beta_{n}\} )$ $\!\rangle$,
   where $\hat{D}(\{ \beta_{n} \} )$ is the $N$-mode coherent displacement
   operator, and the Fourier transform of $\Phi(\{ \beta_{n} \} ;s)$
   is the $s$-parametrized $N$-mode phase-space function.}
\begin{equation}
\label{2.30}
\Omega_{\Delta}(\{ y_{l}\} ) =
 \exp\!\bigg[-\frac{(s\!-\!1)\eta\!+\!1}{2\eta}
   \sum_{n=1}^{N} \left|\beta_{n}\right|^2 \bigg] \,\Phi(\{ \beta_{n} \};s),
\end{equation}
where equal detection efficiencies have been
assumed ($\eta_{k}$ $\!=$ $\eta_{k'}$ $\!\equiv$ $\!\eta$), and
\begin{equation}
\label{2.31}
 \beta_{n} = i e^{i\varphi_{L}}
    \sum_{k=1}^{N+1} U_{k\,N+1} U_{kn}^{\ast} \, y_{k} \, ,
\quad
{\rm with} \quad
\sum_{k=1}^{N+1} |U_{k\,N+1}|^{2} y_{k} = 0
\end{equation}
(K\"{u}hn, Vogel, W., and Welsch [1995]).
Note that when $N$ $\!=$ $\!1$ (four-port apparatus) and
$|U_{k2}|$ $\!=$ $|U_{k1}|$ $\!=$ $\!1/\sqrt{2}$,
then eq.~(\ref{2.30}) [together with eq.~(\ref{2.31})] reduces
to eq.~(\ref{2.26}). It should be pointed out that joint measurements on
combinations of multiports of the above described type can also be used
in order to detect (groups of) correlated modes of different frequencies.
In particular, the combination of two balanced four-port schemes
can be used to measure the joint quadrature-component distribution
$p(x_{1},x_{2},\varphi_{1},\varphi_{2})$ of a (correlated) two-mode
signal field for all values of the two phases $\varphi_{1}$ and
$\varphi_{1}$ within $\pi$ intervals in order to determine the two-mode
quantum state (Raymer, Smithey, Beck, Anderson and McAlister [1993]).

   From eqs.~(\ref{2.30}) and (\ref{2.31}) it can be seen that for perfect
detection the characteristic function of the $N$-fold joint scaled
difference-event distribution is nothing other than the characteristic
function of the Wigner function of the $N$-mode input radiation field
at certain values of the $N$ complex arguments $\beta_{n}$, and it
corresponds to the characteristic function of a joint quadrature-component
distribution. To obtain all knowable information on the
$N$-mode quantum state, i.e., the joint quadrature-component
distributions for all relevant phases or the complete Wigner function,
the arguments $\beta_{n}$ must be allowed to attain arbitrary complex values.
This may be achieved by means of an appropriate succession of
(ensemble) measurements or one (ensemble) measurement, including
in the apparatus appropriately chosen reference modes
whose quantum states are known (or a combination of the two
methods) (Vogel, W., and Welsch [1995]). In the first case, phase shifters
in the apparatus may be used to appropriately vary (similar to
the four-port scheme) the scattering matrix $U_{kk'}$ from
measurement to measurement. In the second case, the scattering
matrix is left unchanged, but additional reference inputs
are used, so that a joint quadrature-component distribution of signal
and reference modes is effectively measured (see \S~\ref{sec2.1.4}).

Instead of measuring the $N$-fold joint difference-event statistics
in a $2(N$ $\!+$ $\!1)$-port scheme, the information on the $N$-mode signal
field can also be obtained from the difference-event statistics measured
in a standard four-port scheme with controlled signal-mode
superposition (Opatrn\'{y}, Welsch and Vogel, W., [1996, 1997a,b];
Raymer, McAlister and Leonhardt [1996]; McAlister and Raymer [1997a,b]).
In the scheme, which applies to both spatially separable and
nonseparable modes, the signal input is formed as a superposition of all
the signal modes. Controlling the expansion coefficients in the
superposition, from the measured difference-event distributions
the same information on the quantum statistics of the $N$-mode signal-field
can be obtained as from the $N$-fold joint difference-event
distribution measured in a \mbox{$2(N$ $\!+$ $\!1)$}-port scheme.

If the signal modes are separated spatially, they can be superimposed
using an optical multiport interferometer such that each of the signal
modes is fed into a separate input channel of the interferometer and
the superimposed mode in one of the output channels is used as
the (signal) input of a four-port homodyne detector.
If the signal field is a pulse-like mode,
the signal pulse can be combined, through the beam splitter
in the four-port scheme, with a sequence of short local-oscillator
pulses in order to analyse the signal pulse in terms of
the modes associated with the local-oscillator pulses.
In this case the superposition of the component modes to be measured
is achieved by a proper formation of the local oscillator pulses.

\begin{figure}[htb]
 \unitlength=1cm
 \begin{center}
 \begin{picture}(10,4)
 \put(0,0){
 \includegraphics{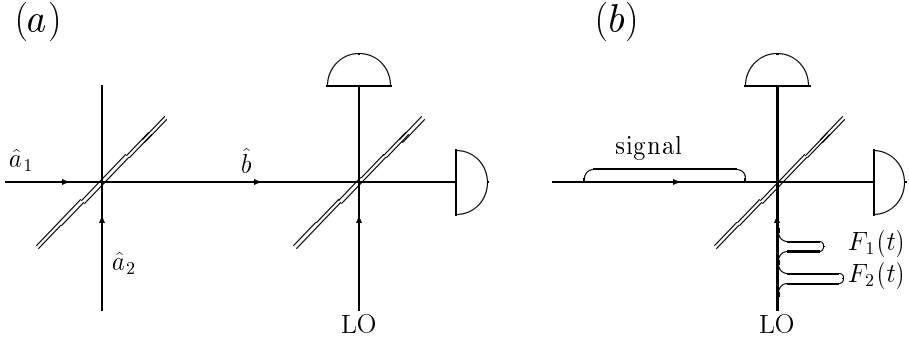}}
 \end{picture}
 \end{center}
\begin{center}
\protect\parbox{.9\textwidth}{
\caption{
Two possible schemes for reconstructing two-mode density
matrices via measurement of combined quadrature components.
(a) Two modes ($\hat{a}_{1}$ and $\hat{a}_{2}$) are mixed by
a beam splitter and one of the interfering output modes ($\hat{b}$)
is used as signal mode in balanced homodyning in order to measure the sum
quadratures of the two modes (LO is the strong local oscillator).
(b) A signal pulse and a sequence of two short (strong) local-oscillator
pulses with envelopes $F_{1}(t)$ and  $F_{2}(t)$ are superimposed
in balanced homodyne detection.
(After Opatrn\'{y}, Welsch and Vogel, W., [1997b].)
\label{fig2.4}
}
}
\end{center}
\end{figure}%
To illustrate the method, let us consider the simplest case when a
(correlated) two-mode field is perfectly detected (Fig.~\ref{fig2.4}).
Then the measured quantitity is a weighted sum of the
quadrature components $\hat{x}_{1}(\varphi_{1})$ and
$\hat{x}_{2}(\varphi_{2})$ of the two modes,
\begin{equation}
\label{2.32}
\hat{x} =
\hat{x}_{1}(\varphi_{1}) \cos\alpha +
\hat{x}_{2}(\varphi_{2}) \sin\alpha,
\end{equation}
where the phases $\varphi_{1}$ and $\varphi_{2}$ and the
relative weight [superposition parameter $\alpha$,
$\alpha$ $\!\in$ $\!(0,\pi/2)$] are assumed to be controlled in the
measurement. For spatially separated modes [Fig.~\ref{fig2.4}(a)] the
phases can be controlled by phase shifters placed in front of a beam
splitter used for mixing the modes, and the parameter $\alpha$
can be controlled by the transmittance (reflectance) of the beam splitter.
If the quadrature statistics of a signal pulse are measured in a
set of short local-oscillator pulses [Fig.~\ref{fig2.4}(b)], then
the parameter $\alpha$ can be controlled by changing the mutual
intensities of the two pulses. Again, the phases can be controlled
interferometrically.\footnote{If signal and local oscillator come
   from different sources, only the phase difference $\delta\varphi$
   $\!=$ $\varphi_{1}$ $\!-$ $\!\varphi_{2}$ can be controlled.}
Let $p_{J}(x_{1},x_{2},\varphi_{1},\varphi_{2})$ and
$p_{S}(x,\alpha,\varphi_{1},\varphi_{2})$ be respectively,
the joint quadrature-component distribution
and the (measured) sum quadrature-component distribution of the two
modes, and $\Psi_{J}(z_{1},z_{2},\varphi_{1},\varphi_{2})$
and $\Psi_{S}(z,\alpha,\varphi_{1},\varphi_{2})$ the corresponding
characteristic functions. It can be shown that
\begin{equation}
\label{2.33}
\Psi_{J}(z\cos\alpha,z\sin\alpha,\varphi_{1},\varphi_{2})
= \Psi_{S}(z,\alpha,\varphi_{1},\varphi_{2}).
\end{equation}
In other words, the joint-quadrature component distribution for
all phase values (within $\pi$ intervals) can be obtained from the
sum quadrature component distribution for all phase values and all
values of the weighting factor, which shows the possibility
to obtain the two-mode quantum state from the measured
statistics of the summed quadratures. The effect of nonperfect
detection can be taken into account according to eq.~(\ref{2.26}).


\subsubsection{$Q$ function}
\label{sec2.1.4}

As mentioned in \S~\ref{sec2.1.3}, the $2N$-fold manifold of data
necessary for a reconstruction of the quantum state of an $N$-mode
signal field can be obtained by including additional reference modes
in the detection scheme in Fig.~\ref{2.3} whose quantum states are
known, such as the vacuum inputs in ``unused'' input channels. Let us
suppose that the $2(N\!+\!1)$-port apparatus is extended to a
$2(2N\!+\!1)$-port device, $N$ input channels being ``unused''. Since
the vacuum inputs are not correlated to each other and to the signal
input, the characteristic function of the quantum state of the
$(2N)$-mode input field can be factorized,
\begin{equation}
\label{2.34}
\Phi(\{ \beta_{n} \};s)
= \Phi(\{ \beta_{i}\} ;s)\,\Phi(\{ \beta_{j} \} ;s),
\quad {\rm with} \quad
\Phi(\{ \beta_{j} \} ;s) = \prod_{j=N+1}^{2N}
e^{(s-1)|\beta_{j}|^2/2}
\end{equation}
($i$ $\!=$ $\!1,$ $\!\ldots,$ $\!N$, signal channels;
$j$ $\!=$ $\!N$ $\!+$ $\!1,$ $\!\ldots,$ $\!2N$, vacuum
channels). Equation~(\ref{2.31}) (with $2N$ in place of $N$) can
then be inverted easily in order to obtain $2N$ real arguments $y_{l}$
of the characteristic function $\Omega_{\Delta}(\{ y_{l}\} )$
of the measured distribution for any
(freely chosen) $N$ complex arguments $\beta_{i}$ of the
characteristic function $\Phi(\{ \beta_{i}\} ;s)$ of the $N$-mode
signal-quantum state. In other words, in the extended detection scheme
there is a one-to-one correspondence between the measured $(2N)$-fold
joint difference-event distribution and the quantum state of the
$N$-mode signal field (Vogel, W., and Welsch [1995]).

Obviously, one local oscillator per group of
(equal-frequency) signal modes and one additional reference input per
signal mode are sufficient for measuring a joint difference-event
distribution that can be regarded as a measure of the signal-mode quantum
state. Needless to say, both the number of local oscillators and/or the
number of additional reference inputs may be increased, and the quantum
state of the signal may of course be expressed in terms of an appropriately
chosen joint difference-event distribution recorded in such an extended
measurement scheme.
In the case of a single-mode signal field a six-port
scheme is the minimum in order to determine the quantum state of
the signal (Zucchetti, Vogel, W., and Welsch [1996]; Paris, Chizhov and
Steuernagel [1997]). The situation is quite similar to classical optics,
in which the six-port scheme is the minimum in order to determine
the complex amplitude (Walker [1987]).
\begin{figure}[htb]
 \unitlength=1cm
 \begin{center}
 \begin{picture}(10,6)
 \put(0,0){
 \includegraphics{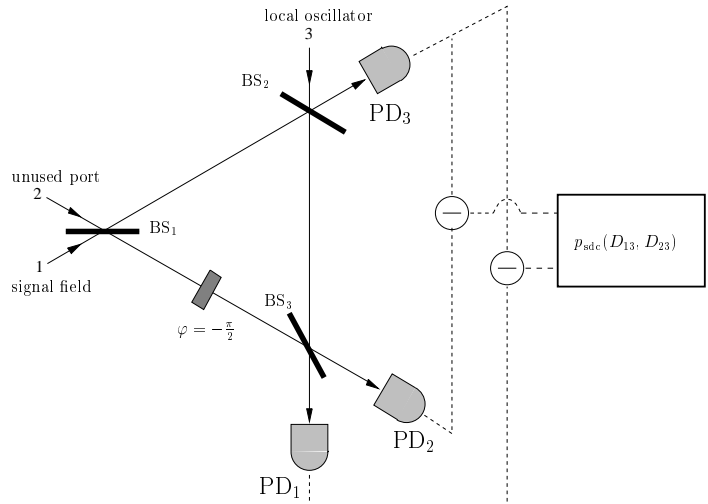}}
 \end{picture}
 \end{center}
\begin{center}
\protect\parbox{.9\textwidth}{
\caption{
Scheme of balanced six-port homodyning for the detection of
the $Q$ function. Three input fields (signal, local oscillator, vacuum)
are combined by three symmetric beam splitters BS$_{i}$ ($i$ $\!=$
$\!1,2,3$) and a ($-\pi/2$) phase shifter, where two ($1/2:1/2$)
beam splitters (BS$_{1}$ and BS$_{3}$) and one ($2/3:1/3$)
beam splitter (BS$_{2}$) are used. The joint difference
statistics is recorded by the detectors PD$_{i}$ in the output
channels [$p_{\rm sdc}(D_{13},D_{23})$ corresponds to
$P_{\Delta m_{1},\Delta m_{2}}$ in eq.~(\protect\ref{2.37})].
(After Zucchetti, Vogel, W., and Welsch [1996].)
\label{fig2.5}
}
}
\end{center}
\end{figure}%
Let us consider a $3$ $\!\times$ $\!3$ coupler and assume that the
reference mode is in the vacuum state (Fig.~\ref{fig2.5}). In this
case, it can be shown that application of eqs.~(\ref{2.30}),
(\ref{2.31}) and (\ref{2.34}) yields
\begin{equation}
\label{2.35}
\Phi(\beta;s) =
\exp\!\left( \textstyle{\frac{1}{2}} s |\beta|^2 \right) \,
\Phi(\beta)   =
\exp\!\left\{ {\textstyle{\frac{1}{2}}}|\beta|^2
\left[ s + (2\!-\!\eta)\eta^{-1} \right] \right\}
\Omega_{\Delta}(y_{1},y_{2}),
\end{equation}
where $\beta$ is relatated to $y_{1}$ and $y_{2}$ as
\begin{equation}
\label{2.36}
\beta  =
\textstyle{\frac{1}{3}}
i [(\phi^{\ast}-1)y_{1}+(\phi-1)y_{2}]
\,e^{i\varphi_{{L}}},
\end{equation}
with $\phi$ $\!=$ $\!\exp(-i2\pi/3)$ [$\Phi(\beta)$
$\!\equiv$ $\!\Phi(\beta;0)$]. Equation~(\ref{2.36}) reveals that
\mbox{$\Phi(\beta;s$ $\!=$ $\!1$ $\!-$ $\!2\eta^{-1})$} $\!=$
$\!\Omega_{\Delta}(y_1,y_2)$; i.e., measurement of the joint
difference-event distribution $P_{\Delta m_{1},\Delta m_{2}}$ in balanced
six-port homodyning is equivalent to measurement of a
phase-space function $P(\alpha;s)$ of the signal mode
(in non-orthogonal coordinates in the phase space):
\begin{equation}
\label{2.37}
P_{\Delta m_{1},\Delta m_{2}}
= \frac{\sqrt{3}}{2\eta^2|\alpha_{L}|^2} \, P(\alpha;s),
\end{equation}
with
\begin{equation}
\label{2.37a}
\alpha =
- \frac{\Delta m_{1}}{\eta\alpha_{L}^\ast}\sqrt{\phi}
- \frac{\Delta m_{2}}{\eta\alpha_{L}^\ast}\sqrt{\phi^\ast}
\end{equation}
and $s$ $\!=$ $\!1$ $\!-$ $\!2\eta^{-1}$
[$\sqrt{\phi}$ $\!=$ $\!\exp(-i\pi /3)$].
The function $P(\alpha;s$ $\!=$ $\!1$ $\!-$ $\!2\eta^{-1})$ can
be regarded as a smoothed $Q$ function of the signal mode, which
approaches the $Q$ function as the quantum efficiency $\eta$ goes
to unity. In particular for perfect
detection ($\eta$ $\!=$ $\!1$) the $Q$ function of the signal mode
is measured (in non-orthogonal phase-space coordinates).

Identifying in eq.~(\ref{A2.20}) $s'$ with $\!1$ $\!-$ $\!2\eta^{-1}$, this
equation can be regarded as a prescription for obtaining other
phase-space functions $P(\alpha;s)$ from the measured one. When
$s$ $\!<$ $\!1$ $\!-$ $\!2\eta^{-1}$ is valid, then the $\beta$ integration
in eq.~(\ref{A2.20}) can be performed separately to obtain $P(\alpha;s)$
as a convolution of $P(\alpha;s$ $\!=$ $\!1$ $\!-$ $\!2\eta^{-1})$
with a Gaussian, which reveals that all the phase-space functions which
are typically broader than the measured one can be obtained simply by
convolving the measured distribution with a Gaussian.
In the opposite case, when $s$ $\!>$ $\!1$ $\!-$ $\!2\eta^{-1}$, then the
integration over $\gamma$ must be done first in eq.~(\ref{A2.20}). However,
experimental inaccuracies may be exploding via the inverse Gaussian
and prevent a stable reconstruction of $P(\alpha;s)$ with
reasonable precision. Since the maximum value of $s$ for an always
stable reconstruction, i.e., \mbox{$s$ $\!=$ $\!1$ $\!-$ $\!2\eta^{-1}$},
tends to minus unity as $\eta$ goes to unity, the
upper boundary of $s$ corresponds to the $Q$ function.

As mentioned above, higher than six-port homodyne detectors can also
be used in order to measure the signal-mode quantum state in the phase
space. Among them, eight-port homodyne detection has widely been studied
(Walker and Carroll [1984]; Walker [1987]; Lai and Haus [1989];
Noh, Foug\`{e}res and  Mandel [1991, 1992a,b];
Hradil [1992]; Freyberger and Schleich [1993]; Freyberger, Vogel, K., and
Schleich [1993a,b]; Leonhardt and Paul, H., [1993a]; Luis and
Pe\v{r}ina [1996a]). Let us consider two
modes that are superimposed, in accordance with eq.~(\ref{2.1}),
by a $50$\%:$50$\% beam splitter, and assume that one of the incoming
modes, say, the second is in the vacuum. It can then be shown that
the joint probability distribution of the two outgoing quadrature components
$\hat{x}'_{1}(\varphi)$ and $\hat{x}'_{2}(\varphi\!+\!\pi/2)$
is the (scaled) $Q$ function of the first incoming
mode,\footnote{Equation~(\ref{2.38}) can be proved correct,
   applying the beam splitter transformation (\ref{2.1}) and expressing
   the characteristic function of the two-mode (outgoing) quadrature-component
   distribution $p\left(x'_{1},x_{2}',\varphi,\varphi\!+\!\pi/2\right)$
   in terms of the characteristic function of the single-mode (incoming)
   $Q$ function $Q(\alpha_{1})$.}
\begin{equation}
\label{2.38}
p\left(x'_{1},x_{2}',\varphi,\varphi\!+\!\pi/2\right)
\sim
Q\!\left[\alpha_{1}\!=\!2^{-1/2}(x'_{1}\!+\!ix'_{2})\right]
\end{equation}
(Lai and Haus [1989]; Leonhardt and Paul, H., [1993a]; for details, also see
Leonhardt and Paul, H., [1995]; Leonhardt [1997c]).
Hence, using the two outgoing modes of the beam splitter as incoming
signal modes in two separate balanced four-port homodyne detectors
and measuring (for a $\pi/2$ phase difference) the joint difference-event
statistics of the two homodyne detectors, the $Q$ function of the
original signal mode is obtained, provided that perfect detection
is accomplished.\footnote{Note
   that $\hat{x}(\varphi)$ and $\hat{x}(\varphi\!+\!\pi/2)$ play the same role
   as position and momentum of a harmonic oscillator in quantum mechanics,
   because of $[ \hat{x}(\varphi), \hat{x}(\varphi\!+\!\pi/2)]$ $\! =$ $\! i$.
   Therefore also the notations \mbox{$\hat{q}$ $\!\equiv$ $\!\hat{x}(\varphi)$}
   and $\hat{p}$ $\!\equiv$ $\!\hat{x}(\varphi\!+\!\pi/2)$ are frequently used.
   Equation~(\protect\ref{2.38}) is an example of ``simultaneous'' measurement
   of a pair of conjugate quantities. Actually, the quantities
   $\hat{q}'_{1}$ $\!\equiv$ $\!\hat{x}'_{1}(\varphi)$
   and \mbox{$\hat{p}'_{2}$ $\!\equiv$ $\!\hat{x}'_{2}(\varphi\!+\!\pi/2)$}
   are measured, which can be regarded as the conjugate
   quantities $\hat{q}_{1}$ $\!\equiv$ $\!\hat{x}_{1}(\varphi)$ and
   \mbox{$\hat{p}_{1}$ $\!\equiv$ $\!\hat{x}_{1}(\varphi\!+\!\pi/2)$},
   respectively, ``smoothed'' by the introduction of additional (vacuum) noise
   necessary for a simultaneous (but approximate) measurement of $\hat{q}_{1}$
   and $\hat{p}_{1}$. Accordingly, an additional uncertainty is introduced
   in the measurement, and it can be shown that
   the uncertainty product for $\hat{q}'_{1}$ and
   $\hat{p}'_{2}$ is twice that of the measurements of $\hat{q}_{1}$ and
   $\hat{p}_{1}$ made individually, $\Delta q'_{1}\Delta p'_{2}$
   $\!\ge$ $\!1$ (Arthurs and Kelly [1965]; for a review, see
   Stenholm [1992]).}
Altogether the setup forms an balanced eight-port
homodyne detector, with a signal input, a vacuum input and two
local-oscillator inputs. Alternatively, the beam splitter and the
two four-port homodyne detectors can be combined into an eight-port
apparatus with two vacuum inputs and one local-oscillator input
(Fig.~\ref{fig2.6}).
\begin{figure}[htb]
 \unitlength=1cm
 \begin{center}
 \begin{picture}(10,5.2)
 \put(0,0){
 \includegraphics{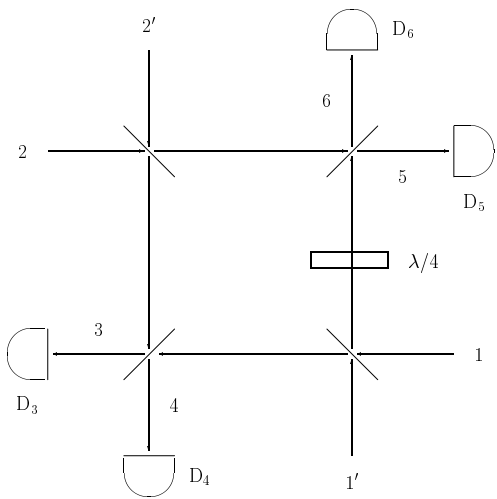}}
 \end{picture}
 \end{center}
\begin{center}
\protect\parbox{.9\textwidth}{
\caption{
Balanced homodyne eight-port detection scheme for measuring the
$Q$ function, using $50$\%:$50$\% beam splitters and
a $\lambda/4$ phase shifter. The signal is fed into port $1$ and the
strong local oscillator is fed into port $2$, and
vacuum is fed into the ports $1'$ and $2'$. The
joint difference-event probabbility distribution
$P_{\Delta m_{1},\Delta m_{2}}$, eq.~(\protect\ref{2.39}),
is measured in the channels
$3$ and $4$ ($\Delta m_{1}$) and $5$ and $6$ ($\Delta m_{2}$).
(After Vogel, W., and Welsch [1994].)
\label{fig2.6}
}
}
\end{center}
\end{figure}%
A straightforward calculation shows (similar to the
six-port homodyne detector) that measurement of the joint difference-event
distribution $P_{\Delta m_{1},\Delta m_{2}}$ again yields the phase-space function
$P(\alpha;s$ $\!=$ $\!1$ $\!-$ $\!2\eta^{-1})$,
\begin{equation}
\label{2.39}
 P_{\Delta m_{1},\Delta m_{2}} = \frac{1}{\eta^2|\alpha_{L}|^2}\,
   P\!\left(\alpha\!= -
   \frac{\Delta m_{1}\!-\!i\Delta m_{2}}{ \eta \alpha_{L}^\ast };\,
   s = 1 - 2\eta^{-1} \right)
\end{equation}
(Freyberger and Schleich [1993]; Freyberger, Vogel, K., and
Schleich [1993a,b]; Leonhardt and Paul, H., [1993b];
Vogel, W., and Welsch [1994]; D'Ariano, Macchiavello and Paris [1995];
Kocha\'{n}ski and W\'{o}dkiewicz [1997]).
Compared with the six-port scheme, $P_{\Delta m_{1},\Delta m_{2}}$ is the
(scaled) function $P(\alpha;s$ $\!=$ $\!1$ $\!-$ $\!2\eta^{-1})$ in
an orthogonal basis [cf. eqs.~(\ref{2.37}) and (\ref{2.39})].
The balanced eight-port homodyne detector was first used by Walker
and Carroll in order to demonstrate the feasibility of measuring
the components of the complex amplitude of optical signals,
extending earlier microwave techniques (Walker and Carroll [1984]).

Finally, it was proposed to measure the $Q$ function by
{\em projection synthesis}, mixing the signal mode with a reference
mode that is prepared in a quantum state such that, for appropriately
chosen parameters, the joint-photon-number probabilities in the two output
channels of the beam splitter realize the coherent-state projector
$\hat{\Pi}(\alpha)$ $\!=$ $\!\pi^{-1}|\alpha\rangle\langle\alpha|$
for truncated signal states (Baseia, Moussa and Bagnato [1997]).
The method was first introduced to
synthesize (for truncated states) the phase-state projector
$\hat{\Pi}(\phi)$ $\!=$ $\!|\phi\rangle\langle\phi|$ (Barnett and Pegg
[1996]; cf. \S~\ref{sec3.8.3}).


\subsubsection{Probability operator measures}
\label{sec2.1.5}

As mentioned, the reference mode with which a signal mode is mixed
has not necessarily to be in the vacuum state in order to obtain,
in principle, all knowable information on the quantum state
of the signal mode. If the reference mode is allowed to be prepared
in a quantum state $\hat{\varrho}_{R}$, then a joint measurement
of the ($\pi/2$-shifted) quadrature-components of the interfering
fields can be regarded as a realization of a complex amplitude
measurement (Walker [1987]). Each
$\hat{\varrho}_{R}$ implies a {\em probability operator measure}
(POM) over the complex amplitude which can be characterized by
a positive valued Hermitian operator
\begin{equation}
\label{2.40}
\hat{\Pi}(\alpha) = \pi^{-1}
\hat{D}(\alpha)\,\hat{\varrho}_{R}\,\hat{D}^{\dagger}(\alpha),
\end{equation}
with
\begin{equation}
\label{2.42}
\int {\rm d}^{2}\alpha \, \hat{\Pi}(\alpha) = \hat{I}.
\end{equation}
The joint probability density ${\rm prob}(\alpha)$ of obtaining a
result $\alpha$ $\!=$ $\!{\rm Re}\,\alpha$ $\!+$ $\!i{\rm Im}\,\alpha$
from a measurement described by this POM is\footnote{The concept
   of POM was introduced in quantum theory in order to
   generalize the familiar quantum probability theory based on orthogonal
   projectors (Davies, E.B., and Lewis [1970]; Holevo [1973];
   Davies, E.B., [1976]; Helstrom [1976]). In particular, it can be used for
   describing ``simultaneous'' measurement of noncommuting observables
   of an object, the corresponding POM being related to a quantum
   measurement of commuting observables in some extended Hilbert space.
   Note that for nonperfect detection ($\eta$ $\!<$ $\!1$) the
   photocounting formula (\ref{A3.7}) is also a POM.}
\begin{equation}
\label{2.43}
{\rm prob}(\alpha) =
\left\langle\hat{\Pi}(\alpha)\right\rangle.
\end{equation}
Note that from the properties of $\hat{\Pi}(\alpha)$ it follows
that ${\rm prob}(\alpha)$ $\!\ge$ $\!0$ and
$\int {\rm d}^{2}\alpha$ ${\rm prob}(\alpha)$ $\!=1$. The operational
probability density distribution ${\rm prob}(\alpha)$ in eq.~(\ref{2.43}),
which is also called {\em propensity} (Popper [1982]),
can be given by a convolution of the Wigner function $W(\alpha)$ of the
signal prepared in a state $\hat{\varrho}$ with the Wigner function
$W_{R}(\alpha)$ of the reference mode prepared in a state
$\hat{\varrho}_{R}$,
\begin{equation}
\label{2.44}
{\rm prob}(\alpha) =
\int {\rm d}^{2}\beta \, W_{R}(\beta\!-\!\alpha)\,W(\beta)
\end{equation}
(Husimi [1940]; Arthurs and Kelly, Jr., [1965]; Kano [1965];
Wootters and Zurek [1979];
O'Connell and Rajagopal [1982]; Rajagopal, A.K., [1983];
W\'{o}dkiewicz [1984, 1986, 1988]; Takahashi and Sait\^{o} [1985];
Walker [1987]; Lai and Haus [1989];
Hradil [1992]; Lalovi\'{c}, Davidovi\'{c} and
Bijedi\'{c} [1992]; Stenholm [1992]; Davidovi\'{c}, Lalovi\'{c} and
Bijedi\'{c} [1993]; Leonhardt [1993]; Leonhardt and Paul, H., [1993b, 1995];
Chaturvedi, Agarwal and Srinivasan [1994]; Raymer [1994];
Bu\v{z}ek, Keitel and Knight [1995a,b];
Paris, Chizhov and Steuernagel [1997]; W\"{u}nsche and  Bu\v{z}ek [1997]).
Obviously, the reference mode acts as a filter and smoothes the Wigner
function of the signal. A filter system of this type is also called
{\em quantum ruler} (Aharonov, Albert and Au [1981]). It is needed in
order to resolve the noncommutative quadrature components of the signal.

The particular realization of the filter strongly
influences the outcome of the measurement. The class of phase-space
functions that can be obtained includes the $s$-parametrized
functions with $s$ $\!\leq$ $\!-1$. In particular, detection
of the $Q$-function implies that $\hat{\varrho}_{R}$ $\!=$
$\!|0\rangle\langle 0|$, so that $\hat{\Pi}(\alpha)$ $\!=$
$\!\pi^{-1}|\alpha\rangle\langle\alpha|$, and hence ${\rm prob}(\alpha)$
$\!=$ $\!\pi^{-1}\langle\alpha|\hat{\varrho}|\alpha\rangle$ $\!=$
$\!Q(\alpha)$. This is the case when vacuum is fed into the reference
channels in the homodyne detection schemes in
Figs.~\ref{fig2.5} and \ref{fig2.6},
and the signal is mixed with it. If the signal is mixed with a
squeezed vacuum, then the POMs of the form given in eq.~(\ref{2.40}) with
$\hat{\varrho}_{R}$ $\!=$ $\!\hat{S}(\xi)|0\rangle\langle 0|
\hat{S}^{\dagger}(\xi)$ can be realized (Walker [1987]; Lai and Haus [1989];
Leonhardt [1993]; Kim, M.S., and Sanders [1996]; Kocha\'{n}ski and
W\'{o}dkiewicz [1997]).\footnote{It should be noted
   that these POMs can also be realized in heterodyne detection
   (\S~\ref{sec2.2}) (Yuen and Shapiro, J.H., [1980]; Yuen [1982]).
   Further, they can be realized in an unbalanced homodyne scheme with vacuum
   input but not equal-part signal-beam splitting (Leonhardt [1993, 1997c]).}

Probability operator measures over the complex amplitude and the
photon number can also be defined such that the joint probability
density for an outcome $\alpha$ $\!=$ $\!{\rm Re}\,\alpha$ $\!+$
$\!i{\rm Im}\,\alpha$ and $n$ is given by
\begin{equation}
\label{2.44a}
{\rm prob}(\alpha,n) =
\left\langle\hat{\Pi}(\alpha,n)\right\rangle,
\end{equation}
where $\hat{\Pi}(\alpha,n)$ must be a positive valued Hermitian operator
that can be used to resolve the idendity,
\begin{equation}
\label{2.44b}
\sum_{n}\int {\rm d}^{2}\alpha \, \hat{\Pi}(\alpha,n) = \hat{I}.
\end{equation}
A probability operator measure of this form can be realized by ten-port
homodyne detection (Luis and Pe\v{r}ina [1996b]).\footnote{The ten-port
   scheme can be regarded as a reduced twelve-port scheme, the latter
   being suited for measuring simultaneously the Stokes
   parameters of a two-mode field.}
In the scheme, a signal mode is fed into one of the input channels of an
unbalanced beam splitter whose second input channel is ``unused''.
The mode in one of the output channels
of the beam splitter (say, the transmitted signal)
is used as the input mode in an eight-port homodyne detector, and
a photodetector is placed in the other output channel in order to measure
the photon-number statistics of the mode in this channel,\footnote{For a
   direct measurement of the photon-number statistics,
   see \S~\ref{sec2.1.7}.}
together with the eight-port complex-amplitude measurement.
The realized POM is related closely to photon-added coherent states,
\begin{equation}
\label{2.44c}
\hat{\Pi}(\alpha,n) = g_{n}(\alpha) \,
\hat{a}^{\dagger n}|\alpha\rangle\langle\alpha|\,\hat{a}^{n},
\end{equation}
where $g_{n}(\alpha)$ $\!=$ $\!\pi^{-1}t^{-2}$
$\!\exp[|\alpha|^{2}(1$ $\!-$ $\!t^{-2})]$ $\!(1$ $\!-$ $\!t^{2})^{n}/n!\,$,
with $t$ being the (absolute value of the) transmittance of the
beam splitter ($0$ $\!<$ $\!t$ $\!<$ $\!1$). Note that when
$t$ $\!\to$ $\!1$ then $\hat{\Pi}(\alpha,n)$ $\!\to$
$\!\pi^{-1}|\alpha\rangle\langle\alpha|$ $\!\delta_{n0}\,$; i.e.,
the familiar coherent-state POM is realized.


\subsubsection{Positive $P$ function}
\label{sec2.1.6}

It is well known that for $s$ $\!>$ $\!-1$
the $s$-parametrized phase-space functions
$P(\alpha;s)$ (see Appendix \ref{app2.4}) do not necessarily exist as
positive functions, and for $s$ $\!>$ $\!0$
they are not necessarily well-behaved.
The latter also applies to the $P$ function,
which is widely used to calculate averages of normally ordered
(i.e., measurable) quantities.
In order to avoid using singular functions, generalized $P$
representations may be used (Drummond and Gardiner [1980]; see also
Gardiner [1983, 1991]), such as the positive $P$ function,\footnote{Note
   that $\hat{\varrho}$ $\!=$ $\int {\rm d}^2\alpha$
   $\!\int {\rm d}^2 \alpha'$ $\!|\alpha\rangle\langle{\alpha'}^\ast|$
   $\!(\langle{\alpha'}^\ast|\alpha\rangle)^{-1}$ $\!P(\alpha,\alpha')$
   in this representation.}
\begin{equation}
\label{2.44f}
P(\alpha,\alpha') = \frac{1}{4\pi^2}\,
\exp\!\left(-{\textstyle\frac{1}{4}}|\alpha-{\alpha'}^\ast|^2\right)
\left\langle{\textstyle\frac{1}{2}}(\alpha+{\alpha'}^\ast)|\,\hat{\varrho}\,
|{\textstyle\frac{1}{2}}(\alpha+{\alpha'}^\ast)\right\rangle\!.
\end{equation}

The possibility of measuring the single-mode $Q$ function in perfect
(six-port or eight-port) homodyne detection offers also the possibility
of measuring the quantum state of the mode in terms of the positive
$P$ distribution using more involved multiport homodyne detection
schemes (Agarwal and Chaturvedi [1994]).
Let us again consider a signal mode and a reference mode that are
superimposed by a 50\%:50\% beam splitter and assume that the
reference mode is in the vacuum. It can then be shown that
the joint $Q$ function of the two outgoing modes,
$Q$ $\!(\alpha_1,$ $\!\alpha_2)$, is related
to the $Q$ function of the signal mode, $Q(\alpha)$,
as\footnote{Equation~(\ref{2.45}) can be proved correct,
   applying the beam splitter transformation (\ref{2.1}) and expressing
   the characteristic function of the two-mode (outgoing) $Q$ function
   $Q(\alpha_{1},\alpha_{2})$
   in terms of the characteristic function of the signal-mode (incoming)
   $Q$ function $Q(\alpha)$. Note that $Q(\alpha_{1},\alpha_{2})$
   is the two-mode $s$-parametrized phase-space function
   $P(\alpha_{1},\alpha_{2};s$ $\!=$ $\!-1)$.}
\begin{equation}
Q(\alpha_{1},\alpha_{2}) =
\label{2.45}
\frac{1}{\pi^2}\,\exp\!\left(-{\textstyle\frac{1}{2}}
|\alpha_{1}\!-\!\alpha_{2}|^2\right)\,
Q\!\left(\alpha = \frac{\alpha_{1}\!+\!\alpha_{2}}{\sqrt{2}}\right)\!,
\end{equation}
which reveals that $Q(\alpha_{1},\alpha_{2})$ is nothing but the
(scaled) positive $P$ function of the signal mode,
\begin{equation}
\label{2.46}
Q(\alpha_{1},\alpha_{2})=
4 P\!\left(\alpha = \sqrt{2} \alpha_{1},\,
\beta = \sqrt{2} \alpha_{2}^\ast\right)\!.
\end{equation}
Hence, if each of the two output modes of the beam splitter is used
as an input mode of a multiport apparatus (such as the six- or eight-port
homodyne detector outlined in \S~\ref{sec2.1.4}) that measures the
$Q$ function, then measurement of the joint $Q$ function of the two
modes yields the positive $P$ function of the signal-mode under study.
Needless to say that for imperfect detection a smeared positive
$P$ function is measured.

The positive $P$ function is an example of a measurable phase-space
function\footnote{For a method suggested to measure the
   positive $P$ function of a quantum-mechanical particle, see
   Braunstein, Caves and Milburn [1991].}
that is defined
as a function of two complex amplitudes $\alpha$ and $\beta$ (per mode).
The concept can also be extended -- similar to \S~\ref{sec2.1.5} --
to other than vacuum reference inputs (in the beam splitter and/or
the multiport homodyne detectors) in order to obtain generalized phase-space
functions defined in a four-dimensional phase space. A simple example
is the smeared positive $P$ function mentioned.


\subsubsection{Displaced-photon-number statistics}
\label{sec2.1.7}

Let us return to the four-port homodyne detection scheme (Fig.~\ref{fig2.1})
and answer the question of which quantity is measured when
a signal mode is mixed with a local-oscillator mode
by an unbalanced beam splitter and only a single-channel homodyne output
is measured. We rewrite eq.~(\ref{2.14}) as
\begin{equation}
\label{2.47}
\hat{a}'_{k} = U_{k1}(\hat{a} - \alpha)
             = U_{k1}\hat{D}(\alpha)\hat{a}\hat{D}^{\dagger}(\alpha),
\end{equation}
where
\begin{equation}
\label{2.48}
\alpha = - \frac{U_{k2}}{U_{k1}} \, \alpha_{L} \, ,
\end{equation}
and apply the photocounting formula (\ref{A3.4}). The probability of
detecting $m$ events in the $k$th output channel can then be given by
\begin{equation}
\label{2.49}
P_{m} = \left\langle :
\frac{1}{m!}\,[\eta \hat{n}(\alpha)]^{m}
e^{-\eta \hat{n}(\alpha)}  : \right\rangle,
\end{equation}
where $\hat{n}(\alpha)$ $\!=$
$\hat{D}(\alpha)\hat{n}\hat{D}^{\dagger}(\alpha)$ is the displaced
photon-number operator of the signal mode,
and $\eta$ $\!=$ $\!|U_{k1}|^{2}\eta_{D}$
(with $\eta_{D}$ being the quantum efficieny of the detector used).
Equation~(\ref{2.49}) reveals that the observed probability distribution
$P_{m}$ is nothing but the displaced photon-number distribution of the
signal field measured with quantum efficiency $\eta$. In particular,
when a signal and a strong local oscillator, $|\alpha_{L}|$ $\!\to$
$\!\infty$, are mixed by a beam splitter with high transmittance,
\mbox{$|U_{11}|$ $\!=$ $\!|U_{22}|$ $\!\to$ $\!1$}, and low reflectance,
\mbox{$|U_{21}|$ $\!=$ $\!|U_{12}|$ $\!\to$ $\!0$}, such that the product
$|U_{12}||\alpha_{L}|$ is finite, then for high quantum efficiency
\mbox{($\eta$ $\!\to$ $\!1$)} the displaced
photon-number probability distribution of the signal is measured
(Wallentowitz and Vogel, W., [1996a];
Banaszek and W\'{o}dkiewicz [1996]; Paris [1996a]),
\begin{equation}
\label{2.50}
P_{m} \to  p_{m}(\alpha) = \langle m,\alpha|\hat{\varrho}|m,\alpha\rangle,
\end{equation}
where $|m,\alpha\rangle$ $\!=$ $\hat{D}(\alpha)|m\rangle$ are the
displaced photon-number states. It should be noted
that for chosen $m$ the quantity $p_{m}(\alpha)$ as a function of
$\alpha$ can be regarded (apart from the factor $\pi^{-1}$) as a
propensity for the signal-mode complex amplitude, which can also be
measured in multiport homodyning (see \S~\ref{sec2.1.5}). For chosen
$\alpha$ it is an ordinary probability for the displaced signal-mode
photon number, which can already be obtained from the four-port
detector outlined here. In order to obtain in this scheme
$p_{m}(\alpha)$ as a function of $\alpha$, a succession of (ensemble)
measurements must be performed. Hence, measurement of the
displaced-signal-mode photon-number statistics as a function of the
complex parameter $\alpha$ is equivalent to measurement of the
signal-mode quantum state, and it is expected that it yields more data
than the minimum necessary for reconstructing it (\S~\ref{sec3}).

In contrast to balanced homodyning, measurement of the displaced
pho\-ton-num\-ber statistics in unbalanced homodyning requires
highly efficient photodetectors that can discriminate between $n$
and $n$ $\!+$ $\!1$ photons in order to resolve the discrete nature of
the photon number. Presently, such detectors are not available.
Photomultipliers and streak cameras can discriminate between
single photons, provided that the field does not contain more
than about $10$ photons, but the quantum efficiency of about
$10$\% -- $20$\% is extremely low. Currently available avalanche
photodiodes operating in the Geiger regime may reach about
$80$\% quantum efficiency (Kwiat, Steinberg, Chiao, Eberhard and
Petroff [1993]), but they do not discriminate between single photons.
They can only indicate the presence of photons, because of saturation.

The problem may be overcome using multichannel coincidence-event
measurement techniques also called {\em photon chopping}.
In particular it was proposed to use highly efficient avalanche
photodiodes and a beam splitter array to divide the number of readout
photons among the photodiodes, so that none is likely to receive more than
one photon (Ho, Lane, La Porta, Slusher and Yurke [1990];
Song, Caves and Yurke [1990]; Paul, H., T\"{o}rm\"{a}, Kiss and Jex
[1996a]).\footnote{For detection of squeezing via coincidence-event
  measurement, see Janszky, Adam, P., and Yushin [1992]; for Fock state
  detection and preparation, see Paul, H., T\"{o}rm\"{a}, Kiss and Jex
  [1996b].}
Alternatively, it was proposed to directly
defocuse the field to be measured onto an array of photodiodes
(Wallentowitz and Vogel, W., [1996a]).

Let us assume that the mode to be detected enters one of
the input ports of a balanced linear $2N$-port apparatus (the other
input ports being ``unused''), and multiple coincidences are measured
at the output, placing avalanche photodiodes in the $N$ output
channels. If the signal mode contains
less than $N$ $\!+$ $\!1$ photons, then there is a one-to-one
correspondence between the measured coincidence-event distribution
$\tilde{p}_{n}(N)$ and the photon-number distribution $p_{n}$ of the
signal. To be more specific, it can be shown that\footnote{Here it is
  assumed that the balanced $2N$-port realizes a unitary
  transformation ${\bf U}_{N}$ $\!=$ ${\bf U}_{2}\otimes{\bf
    U}_{N/2}$.}
\begin{equation}
\label{2.51}
\tilde p_{m}(N) = \sum_{n=m}^{N} \tilde{P}_{m|n}(N) \, p_{n},
\end{equation}
where
\begin{equation}
\label{2.52}
\tilde{P}_{m|n}(N) = \frac{1}{N^{n}} {N \choose m}
\sum_{i=0}^{m} (-1)^{i} {m \choose i} (m-i)^{n}
\quad {\rm if} \quad
m \le n \le N ,
\end{equation}
and $\tilde{P}_{m|n}(N)$ $\!=$ $\!0$ if $n$ $\!<$ $\!m$
(Paul, H., T\"{o}rm\"{a}, Kiss and Jex [1996a]). Here,
$\tilde{P}_{m|n}(N)$ is the probability of registration
of $m$ clicks under the condition that $n$ photons are present.
The probabilities $\tilde{P}_{m|n}(N)$ form an $N$ $\!\times$ $\!N$
upper triangular matrix $A_{mn}$ $\!=$ $\tilde{P}_{m|n}(N)$
which can be inverted in order to
calculate $p_{n}$ from $\tilde{p}_{n}(N)$,
\begin{equation}
\label{2.51a}
p_{n} = \sum_{m=n}^{N} (A^{-1})_{nm} \, \tilde{p}_{m}.
\end{equation}


\subsubsection{Homodyne correlation measurements}
\label{sec2.1.8}

As already mentioned, the accuracy with which a (quantum) field can be
measured by a homodyne detector is limited by the overall quantum efficiency
of the device [see, e.g., Eqs~(\ref{2.23}) and (\ref{2.24})].
To overcome this limitation, one may perform homodyne correlation
measurements (Ou, Hong and Mandel [1987b]).
In contrast to ordinary balanced homodyning, the (time-delayed)
intensity correlation between the two outgoing fields is measured.
In particular, the
information on squeezing is obtained from the time dependence of the
measured correlation function. Since the measured coincidence events
are proportional to the product of the detection efficiencies of the
two detectors, small quantum efficiencies may reduce the measured
signal (which could be compensated by longer measurement times) but do
not smooth out the desired information.
However, a drawback of the method is that the classical noise
of the local oscillator is not balanced out, i.e., even small
relative fluctuations of the (strong) local oscillator may prevent
the quantum noise effects of a weak signal from being measured.

It was therefore proposed to use a weak local oscillator whose
intensity is comparable to that of the signal (Vogel, W., [1991, 1995]).
In this case, the classical noise of the (highly stablized)
local oscillator may be reduced below the level of the quantum
fluctuations of the signal. Moreover, simultaneous measurement of
different kinds of correlation functions of the signal field is possible.
To illustrate this, let us consider the difference between the
measured second-order intensity correlation function
$G^{(2)}(t,t$ $\!+$ $\!\tau)$ for short and long delay
times $\tau$,
\begin{equation}
\Delta G^{(2)}(t) = G^{(2)}(t,t) - \lim_{\tau \to \infty}
G^{(2)}(t,t+\tau),
\label{2.1.8-1}
\end{equation}
and restrict attention to stationary fields, so that the time
argument $t$ can be omitted. Decomposing $\Delta G^{(2)}$
with respect to powers of the local-oscillator amplitude $E_{L}$,
one may observe the following effects. The zeroth-order term yields the
normally ordered intensity ($I$) fluctuation of the signal,
\begin{equation}
\Delta G_0^{(2)} \sim
\Big\langle :\left(\Delta \hat{I} \right)^2: \Big\rangle.
\label{2.1.8-2}
\end{equation}
The second-order term
is related to the normally ordered electric-field variance of the signal,
\begin{equation}
\Delta G_2^{(2)} \sim E_{L}^2
\, \Big\langle : \left[\Delta \hat{E} (\varphi)\right]^2
                :\Big\rangle,
\label{2.1.8-3}
\end{equation}
where $\hat{E}(\varphi)$ corresponds to the quadrature-component
operator $\hat{x}(\varphi)$,
$\varphi$ being the phase difference between signal and local
oscillator.
Eventually, the first-order term represents the correlation between
the two signal-field observables,
\begin{equation}
\Delta G_1^{(2)} \sim {E}_{L}
\, \Big\langle :\Delta \hat{E}(\varphi)
                \Delta \hat{I} : \Big\rangle.
\label{37}
\end{equation}
Note that all these quantum-statistical moments
can be separated from each other by using their dependences
on the phase shift $\varphi$
(for the measurement of the corresponding spectral properties,
see Vogel, W., [1995]).


\subsection{Heterodyne detection}
\label{sec2.2}

It is worth noting that multiport homodyning for measurement
of the complex amplitude is equivalent to (four-port) {\em heterodyne}
detection (Yuen and Shapiro, J.H., [1980]; Yuen [1982]; Yuen and Chan [1983];
Shapiro, J.H., [1985]; Shapiro, J.H., and Wagner [1984];
Caves and Drummond [1994]).
In the scheme, an optical field is combined, through a
beam splitter, on the surface of a photodetector with a strong local
oscillator whose frequency $\omega_{0}$ is offset by an amount
$\Delta\omega$ from that of the signal mode in the optical field
($\Delta\omega$ $\!\ll$ $\!\omega_{0}$). The measured photocurrent is
electrically filtered
in order to select the complex valued component at frequency
$\Delta\omega$. The classical statistics of this component correspond,
under certain circumstances, to the quantum statistics of the two-mode
operator
\begin{equation}
\label{2.52c}
\hat{\alpha}
=\hat{a}_{S} + \hat{a}^{\dagger}_{I},
\end{equation}
where the subscripts $S$ and $I$, respectively, are used to denote the
signal mode at frequency $\omega_{0}$ $\!+$ $\!\Delta\omega$ and the imaging
mode at frequency $\omega_{0}$ $\!-$ $\!\Delta\omega$. Obviously,
the imaging mode can be used to probe the signal mode. To show this,
we first note that the measured quantity $\hat{\alpha}$ can always
be brought into the form
\begin{equation}
\label{2.52d}
\hat{\alpha} = \hat{\alpha}_{1} + i\hat{\alpha}_{2},
\end{equation}
where $\hat{\alpha}_{1}$ and $\hat{\alpha}_{2}$ are commuting Hermitian
operators having a quantum-me\-cha\-ni\-cal joint probability
density $p(\alpha)$ $\!\equiv$ $\!p(\alpha_{1},\alpha_{2})$,
which can be found easily through its characteristic function
\begin{equation}
\label{2.52e}
\phi(\beta) = \left\langle
e^{\beta\hat{\alpha}^{\dagger}-\beta^{\ast}\hat{\alpha}}
\right\rangle
= \int {\rm d}^{2}\alpha \,
e^{\beta\alpha^{\ast}-\beta^{\ast}\alpha} \, p(\alpha) .
\end{equation}
In particular, if the imaging mode is in the vacuum, then measurement
of $p(\alpha)$ can be seen to yield the $Q$ function of the signal mode,
because of
\begin{equation}
\label{2.52f}
\phi(\beta) =
\left\langle
e^{-\beta^{\ast}\hat{a}_{S}} e^{\beta\hat{a}_{S}^{\dagger}}
\right\rangle
=\Phi(\beta;s\!=\!-1)
\end{equation}
(cf. Appendix \ref{app2.4}).


\subsection{Parametric amplification}
\label{sec2.3}

Finally it should be mentioned that linear amplification of a signal
mode and measurement of the complex amplitude of the amplified
signal may be regarded as an equivalent of the homodyne
measurement of the complex amplitude of the original signal
(Bandilla and Paul, H., [1969, 1970]; Paul, H., [1974]). If the signal mode is
strongly enhanced such that it behaves classically, the conjugated
quadrature components can be measured simultaneously. The scheme can
be realized by (nondegenerate) parametric amplification, in which the
photon destruction operators $\hat{a}'_{S}$ and $\hat{a}'_{I}$, respectively,
of the amplified signal and idler modes are obtained from
the SU(1,1) input--output relations
\begin{equation}
\label{2.52a}
\hat{a}'_{S}
= \sqrt{g}\,\hat{a}_{S} + \sqrt{g-1}\,\hat{a}^{\dagger}_{I}
\end{equation}
\begin{equation}
\label{2.52b}
\hat{a}'_{I}
= \sqrt{g-1}\,\hat{a}_{S} + \sqrt{g}\,\hat{a}^{\dagger}_{I},
\end{equation}
where $g$ is the (linear) signal-gain factor (for details, see, e.g.,
Mollow and Glauber [1967a,b]; Caves and Schumaker [1985];
Yurke, McCall and Klauder [1986]).

It can be shown that when the signal mode is mixed with an idler mode that
is initially in the vacuum state and the strong-amplification
limit is realized, i.e., $g$ $\!\to$ $\!\infty$, then the $Q$ function
of the original signal mode can be inferred from the measured distribution
function for the complex amplitude of the amplified signal mode by
appropriately rescaling this distribution (Leonhardt [1994]; see
also Leonhardt and Paul, H., [1995]). It is worth noting that
a degenerate parametric amplifier that operates as a squeezer
[$\hat{a}_{I}^{\dagger}$ $\!\to$ $\!e^{i\phi}\hat{a}_{S}^{\dagger}$
in eq.~(\ref{2.52a})] may be used to (partially) compensate for
the detection losses in balanced four-port homodyne detection
for measuring quadrature components (Leonhardt and Paul, H., [1994a];
\S~\ref{sec3.9.1}), since it makes it possible to amplify a
quadrature component with no increase of noise (Caves [1982]).

Further, parametric amplification can be used to determine the
quantum state of the signal mode by direct photodetection of the
amplified signal if the idler mode is
initially prepared in, e.g., a coherent state, without restriction
to the $Q$ function and without restriction to the strong-amplification
limit (Kim, M.S., [1997a]; \S~\ref{sec3.5}). In particular,
when the idler mode is prepared in a strong coherent state such
that it can be treated classically, $\hat{a}_{I}$ $\!\to$ $\!\alpha_{I}$,
then -- with regard to the photocounting formula (\ref{A3.4}) for
detecting the amplified signal --
eq.~(\ref{2.52a}) takes the form of eq.~(\ref{2.47}),
with $\alpha$ $\!=$ $\!-[(g$ $\!-$ $\!1)/g]^{1/2}\alpha_{I}^{\ast}$
in place of eq.~(\ref{2.48}). In this case, eq.~(\ref{2.49}) also applies
to the detection of the amplified signal, where now $\hat{n}$ $\!=$
$\!\hat{n}_{S}$ and $\eta$ $\!=$ $\!g\,\eta_{D}$.\footnote{Note that
   eq.~(\ref{2.52a}) implies that the signal and idler operators in
   the photocounting formula (\ref{A3.4}) are subject to different orderings,
   so that eq.~(\ref{2.47}) does not apply in general.}


\subsection{Measurement of cavity fields}
\label{sec2.4}

Let us now consider the possibilities of phase-sensitive measuements
of high-$Q$ cavity fields. The realization of such fields, for
example in a micromaser (Meschede, Walther and M\"uller [1985]; Rempe,
Walther and Klein [1987]; Brune, Raimond, Goy, Davidovich and Haroche
[1987]), necessitates a strong suppression of cavity losses.
That is, only a very small fraction of the
field escapes from the cavity and could be used, e.g., in homodyne
detection. Since the quantum efficiency of such a scheme is very
low, it is impossible, in general, to extract the (complete) quantum
statistics of the field inside the cavity from the homodyne data
measured outside the cavity. To overcome the problem, methods
have been developed which use atoms that travel through the cavity
and probe the field inside. The system that is directly
observed in the experiment is not the cavity field itself but
the atoms after their interaction with the field. Under certain
circumstances, the measured atomic occupation probabilities can
then be related uniquely (in a more or less direct way) to
the quantum state of the cavity field (\S~\ref{sec3.6}).
\begin{figure}[htb]
 \unitlength=1cm
 \begin{center}
 \begin{picture}(6,2.5)
 \put(0,0){
 \includegraphics{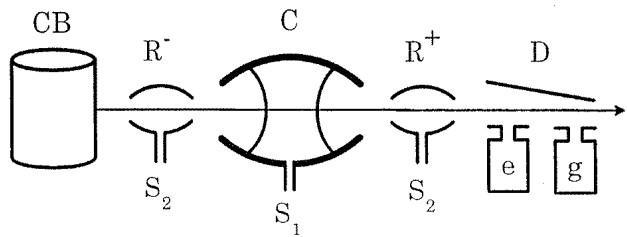}}
 \end{picture}
 \end{center}
\begin{center}
\protect\parbox{.9\textwidth}{
\caption{
Set-up for the detection of a high-$Q$ cavity mode. The
Rydberg atoms are prepared in the state $|e\rangle$
in the box CB and cross a microwave zone R$^{-}$ (source S$_{2}$)
between CB and C which mixes the (excited) state $|e\rangle$ and
the (ground) state $|g\rangle$ before the atoms enter the cavity C.
A second microwave pulse (source S$_{2}$) applied between C and D in
the zone R$^{+}$ again mixes the states $|e\rangle$ and
$|g\rangle$ before the atoms reach the detector D, which
renders it possible to detect linear superpositions of
these two states. The microwave source S$_{1}$ can be used
to feed into the cavity a (classical) field.
(After Brune and Haroche [1994].)
\label{fig2.7}
}
}
\end{center}
\end{figure}%
Let us consider a measurement scheme typically used in Rydberg-atom
superconducting cavity QED (see, e.g., Brune and Haroche [1994] and
references therein).\footnote{Here circular Rydberg atoms
   are used in which one electron is placed on a highly excited
   energy level with large value of principal quantum number $n$, and
   the angular momentum takes its maximum value ($l$ $\!=$ $\!m$
   $\!=$ $\!n$ $\!-$ $\!1$). For $n$ $\!\approx$ $\!50$ the atomic-dipole
   decay time is of the order of magnitude of $10^{-2}\,$s. The cavity
   damping time is of the order of magnitude of $10^{-3}\ldots 10^{-1}\,$s
   for superconducting cavities at subKelvin
   temperatures. The $Q$ factor defined as the ratio of the cavity
   frequency to the cavity damping rate (inverse damping time) then
   reaches values of about $10^{8}\ldots 10^{10}$. The Rabi frequency,
   which is typically of the order of magnitude of $10\ldots 100\,$kHz,
   is large compared with the atomic and the cavity decay rates, so that
   relaxation can be disregarded in first approximation (strong coupling
   regime).}
The atoms that cross the cavity are prepared,
after leaving an oven and before entering the cavity, into a
superposition of Rydberg states $|e\rangle$ and $|g\rangle$
of different energies (Fig.~\ref{fig2.7}). After preparation
of the state $|e\rangle$, which involves laser and radiofrequency
excitation, a microwave (Ramsey) zone can be used to apply a resonant pulse
to the atoms, mixing the state $|e\rangle$ with another state $|g\rangle$
of different energy. In this way an initial atomic state of the form
$c_{e}|e\rangle$ $\!+$ $\!c_{g}|g\rangle$ can be injected into the
cavity. The atoms then interact with the cavity mode while crossing it.
After leaving the cavity, the atoms can again cross a microwave
zone in order to realize state mixing before they enter a detector
for measuring their energies by field ionization. This detection can be
made energy-sensitive and one can thus count the atoms
in $|e\rangle$ and $|g\rangle$. The dependence of the measured data
on the various parameters characterizing the experimental conditions
(such as state mixing and interaction time) can then be
used to gain insight into the quantum state of the cavity mode
(Vogel, W., Welsch and Leine [1987]; Bardroff, Mayr and Schleich [1995];
Bardroff, Mayr, Schleich, Domokos, Brune, Raimond and Haroche [1996]).

The cavity and the atomic transition can be tuned to coincide
(resonant regime) or slightly detuned (dispersive
regime), with a frequency difference
$\delta\omega$ such that any exchange of energy between atom and field
is made impossible. In the resonance regime, the system
can be described (in its simplest version) by the Jaynes--Cummings model
(Jaynes and Cummings [1963]; Paul, H., [1963]; for a review, see, e.g.,
Shore and Knight [1993]).
The Hamiltonian consists of three terms describing the
free two-level (circular Rydberg) atom, the free cavity mode and
the atom--field coupling, the latter being given by
\begin{equation}
\label{2.53}
\hat{H}' = \hbar \kappa
\left( \hat{\sigma}_{+}\hat{a} + \hat{a}^{\dagger} \hat{\sigma}_{-} \right) .
\end{equation}
Here, $\hat{\sigma}_{+}$ $\!=$ $\!|e\rangle\langle g| $ and
$\hat{\sigma}_{-}$ $\!=$ $\!|g\rangle\langle e| $, respectively, are
the atomic raising and lowering operators, and $\kappa$ is the
coupling constant between the atom and the cavity mode
(\mbox{$\Omega$ $\!=$ $2\kappa$} is the vacuum Rabi frequency,
which is the rate at which the atom and the empty cavity exchange a
photon at exact resonance.)

In the dispersive regime, the atom--field interaction produces a
dephasing of the field and also dephases the atom's wave function
by an angle depending upon the number of photons in the cavity
and on the quantum state of the atom. The nonresonant
interaction can be described by an effective Hamiltonian,
\begin{equation}
\label{2.54}
\hat{H}'' = \hbar \, \frac{\kappa^{2}}{\delta \omega} \,
\left( \hat{\sigma}_{-} \hat{\sigma}_{+} \hat{a}^{\dagger}\hat{a}
- \hat{a}\hat{a}^{\dagger} \hat{\sigma}_{+} \hat{\sigma}_{-} \right).
\end{equation}
The phase difference induced by the interaction of the two parts of
the system can be measured when the two microwave zones in
Fig.~\ref{fig2.7} are active. It can be shown that the probability
that a probe atom which is initially prepared in the state $|e\rangle$
is finally detected in the state $|g\rangle$ (i.e., after the combined
action of the two microwave zones on it) oscillates as a function of
the microwave frequency, the position of the intereference (Ramsey)
fringes being dependent on the additional dephasing produced by the
cavity mode. It is worth noting that the Ramsey interferometer
outlined here can serve as an apparatus for a {\em quantum
  non-demolition measurement} (QND) (Braginsky, Vorontsov and Khalili
[1977]; also see Braginsky and Khalili [1992]) of the photon number in
the cavity (Brune, Haroche, Lefevre, Raimond and Zagury [1990]; Brune,
Haroche, Raimond, Davidovich and Zagury [1992]).

When in the dispersive regime an additional microwave generator
is used in order to resonantly couple a classical (strong) oscillator
to the cavity mode, then the Wigner function of the cavity mode
can, in principle, be measured directly (Lutterbach and Davidovich [1997]).
Owing to the coupling of the classical oscillator to cavity mode, a
displacement in phase space of the (initial) quantum state of the cavity mode
is performed such that the density operator $\hat{\varrho}$ is replaced
with $\hat{D}^{\dagger}(\alpha)\hat{\varrho}\hat{D}(\alpha)$.\footnote{Note
   that this coupling acts in a similar way as the mode mixing at an
   unbalanced beam splitter outlined in \S~\ref{sec2.1.7}.}
The atoms initially prepared in state $|e\rangle$ cross
the first Ramsey zone where they see a $\pi/2$ pulse, so that
$|e\rangle$ $\!\to$ $2^{-1/2}(|e\rangle$ $\!+$ $\!e^{i\psi}|g\rangle)$.
The atoms then interact dispersively with the cavity field, and after
that they undergoe a $\pi/2$ pulse in the second Ramsey zone before
detection. Let $P_{e}$ and $P_{g}$ be the probabilities of detecting a
probe atom in the states $|e\rangle$ and $|g\rangle$, respectively.
It can be shown that the difference between these probabilities is
\begin{equation}
\label{2.55}
\Delta P = P_{e}-P_{g} = - {\rm Re} \left\{ e^{i(\psi+\phi)}
\left\langle
\hat{D}(\alpha) \,
e^{ 2i \phi \hat{a}^{\dagger} \hat{a} } \hat{D}^{\dagger}(\alpha)
\right\rangle
\right\} ,
\end{equation}
where $\phi$ $\!=$ $\!(\Omega^{2}/\delta)\tau$ is the additional
dephasing of the atoms in the cavity, with $\tau$ being the interaction
time. In particular, if $\phi$ $\!=$ $\!\psi$ $\!=$ $\!\pi/2$, then
\begin{equation}
\label{2.56}
\Delta P =
{\textstyle\frac{1}{2}} \pi W(\alpha),
\end{equation}
which reveals that indeed the Wigner function is detected, provided
that during the (ensemble) measurement the phases are controlled
and each probe atom meets the cavity field in exactly the same state.

The method of displaced states\footnote{An analogous method can be used
   in order to determine the quantum state of the center-of-mass motion
   of trapped ions (cf. \S~\ref{sec4.2.2}).}
can also be used
for quantum-state measurement in the resonance regime, without
phase-sensitive state mixing (Wilkens and Meystre [1991];
Bodendorf, Antesberger, Kim, M.S., and Walther [1998]). Here, the probe
atoms are prepared in one of the two considered states $|e\rangle$ and
$|g\rangle$ or in a statistical mixture of them.
The measured two-level occupation statistics of the atoms as a function
of the interaction time and/or the displacement parameter
contain all knowable information on the quantum state of the
cavity mode.

The method of probing a cavity field using two-level atoms that travel
through the cavity can also be extended to other than purely two-level
probe atoms. So it was proposed to completely
transfer, by adiabatic passage, the quantum state of the field onto
an internal Zeeman submanifold
of an atom (Walser, Cirac and Zoller [1996]). Utilizing a tomography
of atomic angular momentum states by Stern--Gerlach measurements
(Newton and Young [1968]), this angular momentum state can then be
determined uniquely, by a finite number of magnetic dipole measurements.

The spatial variation of the cavity mode implies a dependence on
the atomic position of the Rabi frequency, which offers the possibility
of probing the quantum state of the mode via atomic deflection
(Akulin, Fam and Schleich [1991]; Herkommer, Akulin and Schleich [1992];
Freyberger and Herkommer [1994]). In the scheme in Fig.~\ref{fig2.8}
a narrow beam of resonant atoms is proposed to pass through a node
region of the mode, where the interaction Hamiltonian can be assumed to
be a linear function of the atomic position.
The cavity photons repulse the atoms so that the
transverse momentum distribution of the atoms is changed
during their passage through the cavity.
When the cavity mode is (initially) in a pure state, then the
measurable change provides (for appropriately prepared atoms)
all the information needed for reconstructing it.
\begin{figure}[htb]
 \unitlength=1cm
 \begin{center}
 \begin{picture}(6,3.5)
 \put(0,0){
 \includegraphics{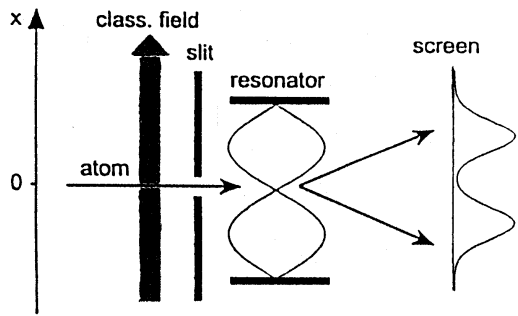}}
 \end{picture}
 \end{center}
\begin{center}
\protect\parbox{.9\textwidth}{
\caption{
Atomic-deflection measurement scheme. A classical field prepares
the two-level atoms in a superposition $|g\rangle$ $\!+$
$\!e^{i\varphi}|e\rangle$ of the ground state $|g\rangle$
and the excited state $|e\rangle$. The so-prepared atoms pass
a narrow slit which confines them to a region \mbox{$\Delta x$
$\!\ll$ $\!\lambda$} centred around the node of the standing
light field at $x$ $\!=$ $\!0$. After the resonant interaction
with the field the deflected atoms travel to a screen put up far
away from the resonator. The spatial distribution on this
screen reflects the momentum distribution of the atoms right
after the interaction.
(After Freyberger and Herkommer [1994].)
\label{fig2.8}
}
}
\end{center}
\end{figure}%

To extend the method to the reconstruction
of mixed states, the effects of the classical field and the cavity
field on the atoms can be combined such that the classical field
travels orthogonally with respect to the cavity mode and both fields
overlap in the region where the atoms cross them
(Schneider, Herkommer, Leonhardt and Schleich [1997]).
In particular, when the classical field is sufficiently strong,
then it can play a similar role as the local oscillator in balanced
four-port optical homodyning. It is worth noting that under certain
circumstances the transverse momentum distribution of the atoms
which is observed in the dispersive regime is
nothing but a smeared quadrature-component distribution of
the cavity mode (\S~\ref{sec3.6.2}).


\section{Quantum-state reconstruction}
\label{sec3}

When in an experiment a sequence of quantities $\hat{A}_{i}$
($\hat{A}_{i}$ $\!=$ $\!\hat{A}_{i}^{\dagger}$) can be measured,
such that the density operator of the sytem under study can
be given by
\begin{equation}
\label{3.0}
\hat{\varrho} = \sum_{i} \hat{B}_{i} \langle\hat{A}_{i}\rangle,
\end{equation}
then all knowable information on the quantum state of the system can be
obtained in principle (Fano [1957]).\footnote{In connection with
   the study of the Pauli problem (Pauli [1933]) the sufficient
   set of observables that has to be mesured in order to obtain the
   complete information on the quantum state was also called {\em quorum} of
   a quantum-state measurement (Park and Band [1971]). It was shown
   that for the one-dimensional motion of a particle a quorum can by
   given by all symmetrized products of powers of position and momentum
   (cf. \S~\ref{sec3.8.1}) or alternatively, all time derivatives
   of all powers of the position (Band and Park [1979]; Park, Band and
   Yourgrau [1980]).}
In practice it must be ensured that
enough data can be measured and that they can be decoded with
reasonable accuracy and acceptable effort -- a task which is not
trivial in general.
In particular if the measured data are directly related to
a smeared quantum-state representation in the phase space, it may be
an effort to derive from it the quantum state in another basis
for studying specific quantum-statistical properties.
Typical examples of complete and overcomplete basis sets of observables
for which measurement schemes have been designed are the
generalized projectors
$\hat{A}_{i}$ $\!\to$  $\hat{\Pi}(x,\varphi)$ $\!=$
$\pi^{-1}|x,\varphi\rangle\langle x,\varphi|$ [see eq.~(\ref{A2.26})]
and $\hat{A}_{i}$ $\!\to$
$\hat{\Pi}(\alpha)$ $\!=$ $\pi^{-1}|\alpha\rangle\langle\alpha|$
[see eq.~(\ref{A2.16}) with $s$ $\!=$ $\!-1$],
respectively, for the quadrature-component states and the coherent
states. If a scheme is not designed for measuring at least a
[in the sense of eq.~(\ref{3.0})] complete set of observables,
then some {\em a priori} knowledge of
the quantum state is required in order to compensate for the
lack of data.


\subsection{Optical homodyne tomography}
\label{sec3.1}

   From \S~\ref{sec2.1.2} we know that in a succession of (phase-shifted)
balanced four-port homodyne measurements the quadrature-component
distribution $p(x,\varphi)$ of a signal mode can be obtained for
various values of the phase $\varphi$, provided that $100$\%
quantum efficiency is realized. The quantum state is then known
when $p(x,\varphi)$ is known for all values of $\varphi$ within
a $\pi$ interval (Vogel, K., and Risken [1989]). That is to say,
all quantum-statistical properties can be obtained from the
quadrature-component distributions measured in a $\pi$ interval.

In particular, $p(x,\varphi)$ can be used to reconstruct the
Wigner function. From eq.~(\ref{A2.23}) ($s$ $\!=$ $\!0$) it can be
derived that
\begin{equation}
\label{3.1}
p(x,\varphi) = \int {\rm d}y \,
W(x \cos\varphi \! - \! y \sin\varphi,
        x \sin\varphi \! + \! y  \cos\varphi) ,
\end{equation}
where the definition $W(q,p)$ $\!\equiv$ $\!2^{-1}$ $\!W[\alpha$ $\!=$
$\!2^{-1/2}(q$ $\!+$ $\!ip)]$ is used. Equation~(\ref{3.1}) reveals
that the quadrature-component distributions can be regarded as marginals
of the Wigner function. An integral relation of the form given in
eq.~(\ref{3.1}) is also called {\em Radon transformation}
(Radon [1917]). Inversion of the Radon transformation (\ref{3.1})
yields the Wigner function in terms of the quadrature-component
distribution.\footnote{Inverse Radon transformation techniques are
   well known from {\em tomographic imaging} (for mathematical details,
   see, e.g., Natterer [1986]). In quantum mechanics, the problem of
   tomographic reconstruction of the Wigner function of a particle
   moving in one dimension was first addressed by
   Bertrand and Bertrand [1987].} From  eq.~(\ref{A2.24})
($s$ $\!=$ $\!0$) the Wigner function is
derived to be\footnote{For phase-randomized
   states, see Leonhardt and Jex [1994].}
\begin{equation}
\label{3.2}
W(q,p) = \frac{1}{4\pi^{2}}
\int_{0}^{\pi} {\rm d}\varphi \int {\rm d}z \int {\rm d}x \,
|z| \exp\!\left[iz(q\cos\varphi \! + \! p\sin\varphi \! - \! x)\right]
p(x,\varphi).
\end{equation}
Performing the $z$ integral first would lead to an integral kernel that
is not well-behaved. To overcome this difficulty in the numerical
calculation, regularization techniques can be used. In the
{\em filtered back projection algorithm}, the $z$ integral is
truncated such that
\begin{equation}
\label{3.3}
W(q,p) \simeq W_{c}(q,p) = \frac{1}{2\pi^{2}}
\int_{0}^{\pi} {\rm d}\varphi \int {\rm d}x \,
K_{c}(q\cos\varphi \! + \! p\sin\varphi \! - \! x) \, p(x,\varphi),
\end{equation}
with
\begin{equation}
\label{3.4}
K_{c}(y) = {\textstyle\frac{1}{2}}
\int_{-z_{c}}^{+z_{c}} {\rm d}z \, |z| \, e^{iyz}
\end{equation}
($z_{c}$ $\!>$ $\!0$). Here, rapid variations of the
quadrature-component distributions which correspond to frequencies
higher than the cut-off frequency $z_{c}$ are suppressed, i.e.,
the exact distributions are effectively replaced with
somewhat smeared ones.

In the pioneering experimental demonstration
of the method also called {\em optical homodyne tomography}
(Smithey, Beck, Raymer and Faridani [1993]; Smithey, Beck, Cooper,
Raymer and Faridani [1993]), a pulsed signal field is superimposed
with a pulsed coherent-state field much stronger than the signal.
The quadrature-component distributions are
measured for a squeezed signal field and a vacuum signal field.
The squeezed field is generated by using a walk-off compensated,
travelling-wave optical parametric amplifier. The generated
down-conversion signal centred at $1064$~nm consists of two
orthogonally polarized fields, the signal and idler, and
has a bandwidth estimated to be $10^{4}$ times that of the
laser pump ($532$~nm, $300$~ps, $420$ pulses per second).
The laser pump and the local-oscillator field ($1064$~nm, $400$~ps)
are obtained from a common laser source, and each local-oscillator pulse
contains a mean number of photons of about $4\times 10^{6}$. The
interfering fields are detected with high-quantum-efficiency ($\sim 85\%$)
photodiodes, and the resulting current pulses are temporally integrated and
subtracted. The measurements and reconstructions are performed for a
squeezed signal field and for a vacuum signal field,
Figs.~\ref{fig3.1} and \ref{fig3.2}, the
field mode detected being selected by the spatial-temporal mode of the
local-oscillator field. The method of
optical homodyne tomography was also extended subsequently to
the continuous wave-regime, including squeezed vacuum with a high
degree of quantum noise reduction (Breitenbach, M\"{u}ller,
Pereira, Poizat, Schiller and Mlynek [1995]; Schiller, Breitenbach,
Pereira, M\"{u}ller and Mlynek [1996]) and bright squeezed light
(Breitenbach and Schiller [1997]; Breitenbach, Schiller and Mlynek [1997]).

\begin{figure}[htb]
 \unitlength=1cm
 \begin{center}
 \begin{picture}(6,9)
 \put(0,0){
 \includegraphics{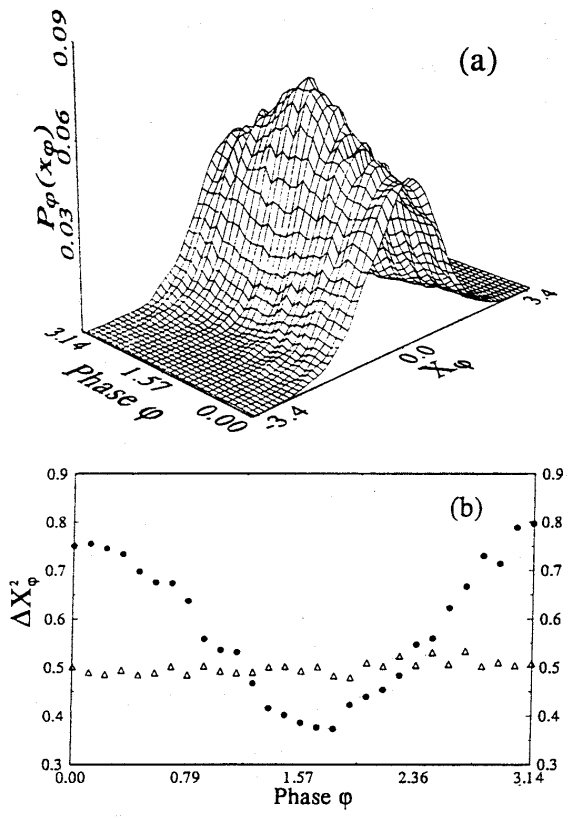}}
 \end{picture}
 \end{center}
\vspace*{5ex}
\begin{center}
\protect\parbox{.9\textwidth}{
\caption{
(a) In balanced four-port homodyne detection measured
qua\-dra\-ture-com\-po\-nent
distributions at various values of the local-oscillator phase
[$P_{\varphi}(x_{\varphi})$ corresponds to $p(x,\varphi)$].
(b) Variances of quadrature components vs local-oscillator
phase: circles, squeezed state; triangle, vacuum state.
In the experiment $4000$ repeated measurements of the
photoelectron difference number at $27$ values of the
relative phase $\varphi$ are made.
(After Smithey, Beck, Raymer and Faridani [1993].)
\label{fig3.1}
}
}
\end{center}
\end{figure}%

\pagebreak
As mentioned in \S~\ref{sec2.1.2}, the measured quadrature-component
distributions do not correspond, in general, to the true signal mode, but
they must be regarded as the distributions of a superposition of the
signal and an additional noise source [eq.~(\ref{2.24a})],
because of non-perfect detection.
Substituting in eq.~(\ref{3.1}) for $p(x,\varphi)$ the measured
distributions $p(x,\varphi;\eta)$ with $\eta$ $\!<$ $\!1$
[eq.~(\ref{2.23})] and performing the
inverse Radon transform on them, the Wigner function of a noise-assisted
signal field is effectively reconstructed.
Equivalently, the reconstructed Wigner function can be regarded
as an $s$-parametrized phase-space function of the true signal field,
however with $s$ $\!<$ $\!0$.
The characteristic function
$\Psi(z,\varphi;\eta)$ of $p(x,\varphi;\eta)$ typically measured when
\mbox{$\eta$ $\!<$ $\!1$} and the characteristic function $\Phi(\beta;s)$
of the phase-space function $P(\alpha;s)$ are related to each other
according to eq.~(\ref{A2.22}), with $\Psi(z,\varphi;\eta)$ in place of
$\Psi(z,\varphi)$ and \mbox{$s$ $\!-$ $\!1$ $\!+$ $\!\eta^{-1}$}
in place of $s$ in the exponential [cf. eq.~(\ref{2.26}) and footnote
\ref{fn2}]. Hence, making in the exponentials in eqs.~(\ref{A2.23})
and (\ref{A2.24}) for $s$ the substitution \mbox{$s$ $\!-$ $\!1$ $\!+$
$\!\eta^{-1}$} yields the relations between $P(\alpha;s)$ and
$p(x,\varphi;\eta)$. In particular, eq.~(\ref{A2.24}) can be used
to obtain any phase-space function $P(q,p;s)$
$\equiv$ $\!2^{-1}$ $\!P[\alpha$ $\!=$ $\!2^{-1/2}(q$ $\!+$ $\!ip;s)]$
from the measured quadrature-component distributions
in principle:\footnote{When $s$ $\!>$ $\!1$ $\!-$ $\!\eta^{-1}$,
   then in eq.~(\ref{3.5}) an inverse Gaussian occurs which may lead to an
   artifical enhancement of the inaccuracies of the measured data, so that a
   stable deconvolution might be impossible and the noise dominates
   the reconstruction of $P(q,p;s)$. This effect is not observed
   when $s$ $\!\leq$ $\!1$ $\!-$ $\!\eta^{-1}$, and hence a
   stable reconstruction with reasonable precision
   of $P(\alpha;s)$ for $s\!\leq$ $\!1$ $\!-$
   $\!\eta^{-1}$ may therefore be expected to
   be feasible for any quantum state (see also \S~\ref{sec3.9}).
   In particular, reconstruction of the $Q$ function is always possible
   if $\eta$ $\!>$ $\!1/2$.
   When \mbox{$s\!<$ $\!1$ $\!-$ $\!\eta^{-1}$},
   then in eq.~(\ref{3.5})
   the $z$ integral can be performed first to obtain $P(q,p;s)$ in a form
   suited for application of sampling techniques (\S~\ref{sec3.3.1}):
   $P(q,p;s)$ $\!=$ $\!\int_{0}^{\pi} {\rm d}\varphi$
   $\!\int {\rm d}x$ $\!K(q,p,x,\varphi;s;\eta)$ $\!p(x,\varphi;\eta)$,
   with $\!K(q,p,x,\varphi;s;\eta)$ being a well-behaved integral kernel.}
\begin{eqnarray}
\label{3.5}
\lefteqn{
P(q,p;s) = \frac{1}{4\pi^{2}}
\int_{0}^{\pi} {\rm d}\varphi \int {\rm d}z \int {\rm d}x \,
\bigg\{\exp\!\left[{\textstyle\frac{1}{4}} (s\!-\!1\!+\!\eta^{-1})z^{2}\right]
}
\nonumber \\ && \hspace{15ex} \times \
|z| \exp\!\left[iz(q\cos\varphi \! + \! p\sin\varphi \! - \! x)\right]
p(x,\varphi;\eta)
\bigg\}.
\end{eqnarray}
Obviously, when $s$ $\!=$ $\!1$ $\!-$ $\!\eta^{-1}$, then eq.~(\ref{3.5})
takes the form of eq.~(\ref{3.2}); i.e., replacing in eq.~(\ref{3.1})
$p(x,\varphi)$ with $p(x,\varphi;\eta)$, inverse Radon transformation
yields the signal-mode phase-space function $P(q,p;s$ $\!=$ $\!1$
$\!-$ $\!\eta^{-1})$ in place of the Wigner function,
\begin{equation}
\label{3.5z}
p(x,\varphi;\eta) = \int {\rm d}y \,
P(x \cos\varphi \! - \! y \sin\varphi,
        x \sin\varphi \! + \! y  \cos\varphi;s = 1 \!-\! \eta^{-1}).
\end{equation}
\begin{figure}[t]
 \unitlength=1cm
 \begin{center}
 \begin{picture}(6,7.9)
 \put(0,0){
 \includegraphics{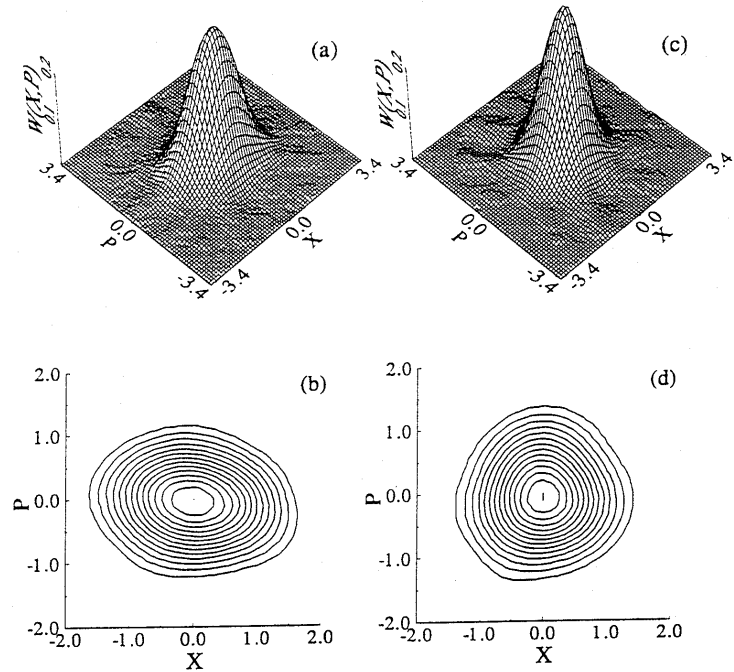}}
 \end{picture}
 \end{center}
\begin{center}
\protect\parbox{.9\textwidth}{
\caption{
   From the measured quadrature-component distributions
reconstructed Wigner distributions for (a),(b) a squeezed
state and (c),(d) a vacuum state, viewed in 3D and as contour
plots, with equal numbers of constant-height contours [$W(X,P)$
corresponds to $W(q,p)$]. The reconstruction is performed
by using inverse Radon transformation according to eq.~(\protect\ref{3.3}).
(After Smithey, Beck, Raymer and Faridani [1993].)
\label{fig3.2}
}
}
\end{center}
\end{figure}%

The proposal was made to combine squeezing and balanced homodyning
such that generalized quadrature-components
\begin{equation}
\label{3.5a}
\hat{x}(\mu,\nu) = \mu\,\hat{q} + \nu \, \hat{p}
\end{equation}
for all real parameters $\mu$ and $\nu$ can be measured, and to
reconstruct the quantum state from the corresponding distributions
$p(X,\mu,\nu)$ (Mancini, Man'ko, V.I., and Tombesi [1995]; D'Ariano,
Mancini, Man'ko, V.I., and Tombesi [1996]; Man'ko, O.V., [1996];
Mancini, Man'ko, V.I., and Tombesi [1997]).\footnote{For an application
  of the method (also called {\em symplectic tomography}) to
  trapped-ion quantum state reconstruction, see Mancini, Man'ko, V.I.
  and Tombesi [1996]; Man'ko, O.V.  [1997].}
In fact, $\hat{x}(\mu,\nu)$ can be related to $\hat{x}(\varphi)$ since
it represents a scaled quadrature-component,
\begin{equation}
\label{3.5b}
\hat{x}(\mu,\nu)
= |F|\left(\hat{a}e^{-i\varphi}+\hat{a}^{\dagger}e^{i\varphi}\right)
= 2^{1/2}|F|\,\hat{x}(\varphi),
\end{equation}
with
\begin{equation}
\label{3.5c}
2^{1/2}|F| = \sqrt{ \mu^{2}+\nu^{2} }\,,
\qquad
\tan\varphi = \frac{\nu}{\mu}
\end{equation}
(cf. \S~\ref{sec2.1.2} and Appendix \ref{app2}), which imples that
\begin{equation}
\label{3.5d}
p(x,\mu,\nu)
=  \frac{1}{\sqrt{\mu^{2}+\nu^{2}}}\;
p\!\left(\frac{x}{\sqrt{\mu^{2}+\nu^{2}}},
\varphi\!=\!\arctan\frac{\nu}{\mu}\right)\!.
\end{equation}
Hence, measurement of a particularly scaled quadrature-component
distribution by means of an ordinary
homodyne detector already yields all scaled quadrature components.
Performing the analysis with a variable scaling parameter,
it can be shown that eqs.~(\ref{3.1}) and (\ref{3.2}),
respectively, can be given in terms of $p(x,\mu,\nu)$ by\footnote{For
   a detailed discussion of the transformation properties, see
   also W\"{u}nsche [1997].}
\begin{equation}
\label{3.5e}
p(x,\mu,\nu)
= \frac{1}{2\pi} \int {\rm d}k \int {\rm d}q \int {\rm d}p \,
e^{-ik(x-q\mu-p\nu)} W(q,p)
\end{equation}
and
\begin{equation}
\label{3.5f}
W(q,p)
=\frac{z^{2}}{(2\pi)^{2}}\int {\rm d}x \int {\rm d}\mu \int {\rm d}\nu \,
e^{iz(x-q\mu-p\nu)} p(x,\mu,\nu).
\end{equation}
Comparing eq.~(\ref{3.5f}) with eq.~(\ref{3.2}), we see that in any
case a three-fold Fourier transformation is required in order
to obtain the Wigner function from the homodyne data. Note
that $z$ in eq.~(\ref{3.5f}) can be chosen arbitrarily, which reflects
the above mentioned fact of overcomplete data.


\subsection{Density matrix in quadrature-component bases}
\label{sec3.2}
\label{sec4b}

The reconstructed Wigner function can be used in order to calculate
the density matrix in a quadrature-component basis
(Smithey, Beck, Raymer and Faridani [1993]; Smithey, Beck, Cooper,
Raymer and Faridani [1993]). The definition
of the Wigner function given in Appendix \ref{app2.4} can be rewritten, on
expanding the density operator in a quadrature-component basis, as
\begin{equation}
\label{3.6}
W(q,p) =
\frac{1}{\pi}
\int {\rm d}x' \, e^{2iyx'}
\langle x\!-\!x',\varphi |\hat{\varrho}| x\!+\!x',\varphi\rangle,
\end{equation}
with
\begin{equation}
\label{3.7}
x = q \cos\varphi + p \sin\varphi,
\quad
y = - q \sin\varphi + p \cos\varphi,
\end{equation}
which for $\varphi$ $\!=$ $\!0$ is nothing but the well-known Wigner
formula (Wigner [1932]). Hence the density matrix in a
quadrature-component basis can be obtained by Fourier transforming the
Wigner function:
\begin{equation}
\label{3.8}
\langle x\!-\!x',\varphi |\hat{\varrho}| x\!+\!x',\varphi\rangle
= \int {\rm d}y \, e^{-2ix'y}
W(x\cos\varphi\!-\!y\sin\varphi,x\sin\varphi\!+\!y\cos\varphi) .
\end{equation}
Note that eq.~(\ref{3.8}) reduces to eq.~(\ref{3.1}) for $x'$ $\!=$ $\!0$.

It is worth noting that the reconstruction of the density matrix
from the homodyne data can be accomplished with two Fourier integrals,
avoiding the detour via the Wigner function (K\"{u}hn, Welsch and
Vogel, W., [1994]; Vogel, W., and Welsch [1994]).
Writing the density-matrix elements as
\begin{equation}
\langle x\!-\!x',\varphi | \hat{\varrho} | x\!+\!x',\varphi \rangle
= \frac{1}{2\pi} \int {\rm d}z \, e^{-ixz} \Psi(z,x',\varphi),
\label{3.9}
\end{equation}
the characteristic function $\Psi(z,x',\varphi)$ can be shown to
be the characteristic function of a quadrature-component distribution,
\begin{equation}
\label{3.9a}
\Psi(z,x',\varphi) = \Psi(\tilde{z},\tilde{\varphi}) ,
\end{equation}
with
\begin{equation}
\label{3.10}
\tilde{z} = \tilde{z}(z,x') =\left[z^{2}\! + \! (2x')^{2}\right]^{1/2}
\quad {\rm and} \quad
\tilde{\varphi} = \varphi - {\textstyle\frac{1}{2}\pi}
+ \arg\!\left(2x'\! +\! i z\right).
\end{equation}
Hence, the density-matrix elements can be obtained from the
quadrature-component distributions by means of a simple two-fold Fourier
transformation:\footnote{For the two-fold Fourier transformation
   that relates the density-matrix elements to $p(x,\mu,\nu)$,
   eq.~(\ref{3.5e}), see D'Ariano, Mancini, Man'ko and Tombesi [1996].}
\begin{equation}
\label{3.11}
\langle x\!-\!x',\varphi | \hat{\varrho} | x\!+\!x',\varphi \rangle
= \frac{1}{2\pi} \int {\rm d}z \, e^{-ixz}
\int {\rm d}\tilde{x} \,
e^{i\tilde{z}\tilde{x}} p(\tilde{x},\tilde{\varphi}).
\end{equation}
It is worth noting that eq.~(\ref{3.11}) can be used to obtain the
density matrix in different representations,
by varying the phase of the quadrature component defining
the basis. Further, eq.~(\ref{3.11}) can also be extended, in principle, to
imperfect detection, expressing $\Psi(z,\varphi)$ in eq.~(\ref{3.9})
in terms of $\Psi(z,\varphi;\eta)$ (cf. footnote \ref{fn2};
for details, see K\"{u}hn, Welsch and Vogel, W., [1994];
Vogel, W., and Welsch [1994]).

For the numerical implementation of the reconstruction based on
eq.~(\ref{3.11}) spline-expansion techniques can be used
(Zucchetti, Vogel, W., Tasche and Welsch [1996]),\footnote{Choosing a
   finite set of nodes $\{x_{n}\}$, an approximate spline function $f_{s}(x)$
   of $f(x)$ is given by $f_{s}(x)$ $\!=$ $\!\sum_{n}$ $\!f(x_{n+1})
   B_{\Delta x_{n},n}(x)$, where $B_{\Delta x_{n},n}(x)$ $\!=$
   $\!(x\!-\!x_{n})/\Delta x_{n}$ if $x_{n}$ $\!\leq$ $\!x$ $\!\leq$
   $\!x_{n+1}$, $B_{\Delta x_{n},n}(x)$ $\!=$
   $\!(x_{n+2}\!-\!x)/\Delta x_{n+1}$ if $x_{n+1}$ $\!\leq$ $\!x$ $\!\leq$
   $\!x_{n+2}$, and $B_{\Delta x_{n},n}(x)$ $\!=$ $\!0$ elsewhere,
   and $\Delta x_{n}$ $\!=$ $x_{n+1}$ $\!-$ $\!x_{n}$; for mathematical details
   of spline expansion, see de Boor [1987].}
\begin{equation}
\label{3.11a}
\langle x\!-\!x',\varphi | \hat{\varrho} | x\!+\!x',\varphi \rangle
\simeq \sum_{m,n} K_{mn}(x,x',\varphi)\,
p(\tilde{x}_{m+1},\tilde{\varphi}_{n+1}),
\end{equation}
where
\begin{equation}
\label{3.11b}
K_{mn}(x,x',\varphi)
= (2\pi)^{-1} \sum_{k}
\underline{B}_{\Delta\tilde{x}_{m},m}[\tilde{z}(\xi_{k+1},x')]
B_{\Delta z_{n},n}(\xi_{k+1})
\underline{B}_{\Delta\xi_{k},k}(-x),
\end{equation}
with $z_{n}$ $\!=$ $-2x'\cot(\tilde{\varphi}_{n}$ $\!-$ $\!\varphi)$
[$\underline{B}_{(\dots)}(k)$, Fourier transform of $B_{(\dots)}(x)$].
As an illustration of this method, in
Figs.~\ref{fig3.2A}(a) and \ref{fig3.2A}(b)
the reconstructed density matrices in the
``position'' basis, $\varphi$ $\!=$ $\!0$, and the ``momentum'' basis,
$\varphi$ $\!=$ $\!\pi/2$, of a squeezed vacuum state are shown.
The homodyne data were recorded by T. Coudreau, A.Z. Khoury
and E. Giacobino. In the underlying experimental scheme
the squeezing effect is obtained in a probe beam that interacts with
cold atoms in a nearly single-ended cavity (Lambrecht, Coudreau,
Steinberg and Giacobino [1996]).
The phases $\varphi$ $\!=$ $\!0$ and
$\varphi$ $\!=$ $\!\pi/2$ that define the quadrature-component
bases in Fig.~\ref{fig3.2A} coincide with the phases
of minimal and maximal field noise, respectively.
\begin{figure}[htb]
 \unitlength=1cm
 \begin{center}
 \begin{picture}(6,3.5)
 \put(0,0){
 \includegraphics{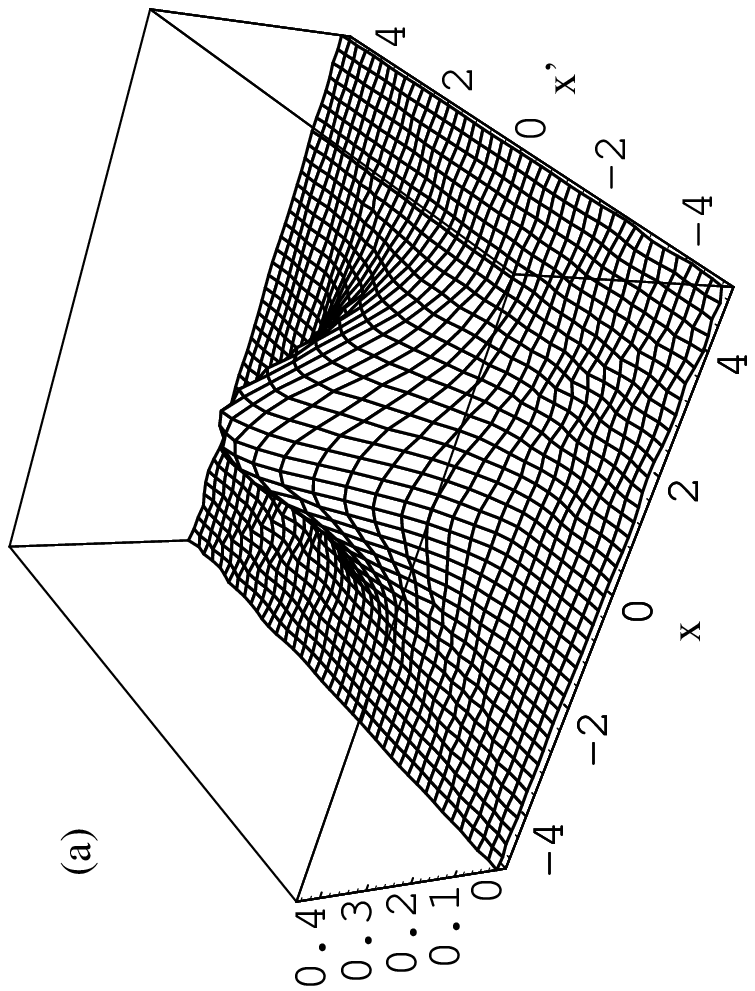}}
 \put(0,0){
 \includegraphics{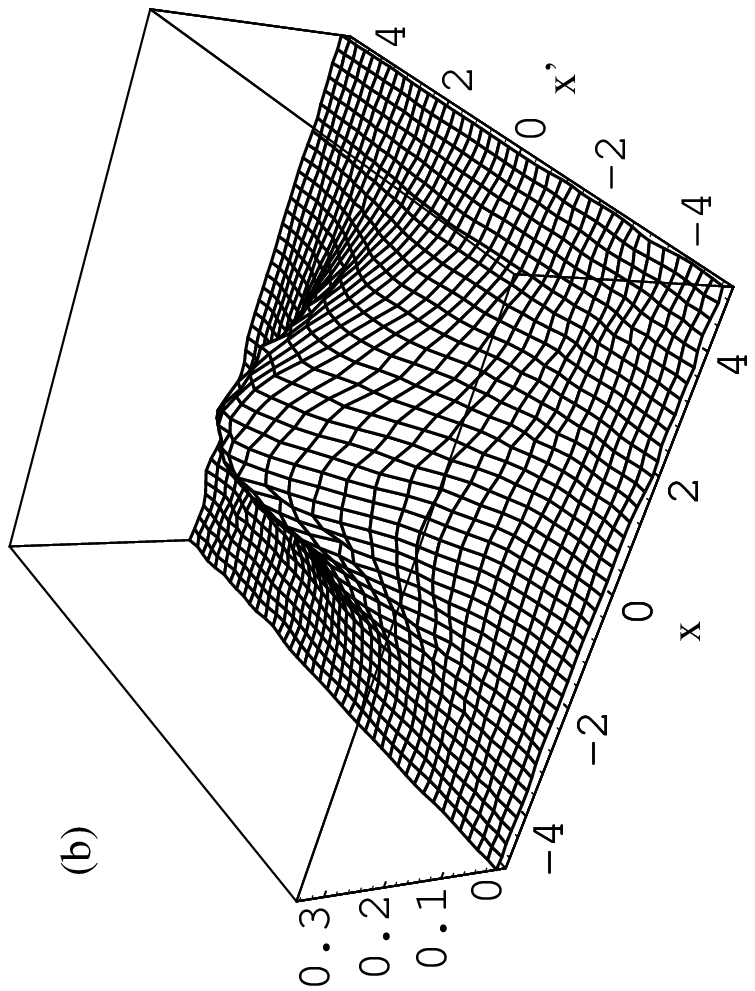}}
 \end{picture}
 \end{center}
\begin{center}
\protect\parbox{.9\textwidth}{
\caption{
   From measured quadrature-component distributions
reconstructed (real) density matrix
$\langle x,\varphi | \hat{\varrho} | x',\varphi \rangle$ of a squeezed
state in (a) the ``position'' basis ($\varphi$ $\!=$ $\!0$) and (b)
the ``momentum'' basis ($\varphi$ $\!=$ $\!\pi/2$). The
homodyne data were obtained by T. Coudreau, A.Z. Khoury and E.
Giacobino, using the experimental setup reported by
Lambrecht, Coudreau, Steinberg and Giacobino [1996].
In the experiment, the quadrature-components were measured
at $48$ phases, and at each phase
$7812$ measurements were performed.
\label{fig3.2A}
}
}
\end{center}
\end{figure}%


\subsection{Density matrix in the Fock basis}
\label{sec3.3}

Stimulated by the tomographic reconstruction of the Wigner function
(\S~\ref{sec3.1}) much effort has been made to obtain the density
matrix in the Fock basis from measurable data as direct as possible.
Let us again start with analysing balanced four-port homodyning.


\subsubsection{Sampling of quadrature-components}
\label{sec3.3.1}

The problem of reconstruction of the density matrix in the
Fock basis from the quadrature-component
distributions can be solved, in principle, by
relating the den\-sity-matrix elements to derivatives of the
$Q$ function (cf. \S~\ref{sec3.3.3}) and expressing the $Q$
function in terms of the quadrature-component
distributions, using eq.~(\ref{A2.24}), with $s$ $\!=$ $\!-1$
(D'Ariano, Macchiavello and Paris [1994a,b]).
An equivalent formalism, which is suited for practice and which
can also be applied to the reconstruction of the density matrix in
other than the photon-number basis, is based on the expansion of
the density operator as given in eq.~(\ref{A2.26}). In the
photon-number basis this equation reads as\footnote{For the relation
   between the density-matrix elements and $p(x,\mu,\nu)$,
   eq.~(\ref{3.5e}), see D'Ariano, Mancini, Man'ko and Tombesi [1996]).}
\begin{equation}
\label{3.13}
\varrho_{mn} =
\langle m| \hat{\varrho} |n\rangle =
\int_0^\pi {\rm d}\varphi \int {\rm d}x \,
K_{mn}(x,\varphi)
\, p(x,\varphi).
\end{equation}
The integral kernel (also called {\em pattern function})
\begin{equation}
\label{3.14}
K_{mn}(x,\varphi)
= \langle m | \hat{K}(x,\varphi) | n \rangle
= f_{mn}(x) \, e^{i(m-n)\varphi},
\end{equation}
with
\begin{equation}
\label{3.14a}
\hat{K}(x,\varphi)
= \frac{1}{2\pi} \int {\rm d}z \, |z|\,
\exp\!\left\{iz\left[\hat{x}(\varphi)\!-\!x\right]\right\}\!,
\end{equation}
is studied in detail in a number of papers
(D'Ariano [1995]; D'Ariano, Leonhardt
and Paul, H., [1995]; Leonhardt, Paul, H., and D'Ariano [1995]; Leonhardt,
Munroe, Kiss, Richter, Th., and Raymer [1996]; Richter, Th., [1996a];
W\"{u}nsche [1997]).
The function $f_{mn}(x)$ (Fig.~\ref{fig3.2a}) is well behaved, and it is
worth noting that it can be given by (Richter, Th., [1996a]; Leonhardt,
Munroe, Kiss, Richter, Th., and Raymer [1996])
\begin{equation}
\label{3.15}
f_{mn}(x) = \frac{d}{dx}\left[\psi_{m}
(x)\varphi_{n}(x)\right],
\end{equation}
where $\psi_{m}(x)$ and and $\varphi_{m}(x)$ are the regular
and irregular
(real)
solutions of the harmonic-oscillator Schr\"{o}dinger
equation for a chosen energy value $E_{m}$.\footnote{Strictly speaking,
   the irregular (i.e., not normalizable) function $\varphi_{n}(x)$
   has to be chosen such that $\psi_{n}\varphi_{n}'$ $\!-$
   $\psi_{n}'\varphi_{n}$ $\!=$ $\!2/\pi$ (for details, see Leonhardt [1997c]).
   Note that $f_{mn}(x)$ is not determined uniquely.}
\begin{figure}[t]
 \unitlength=1cm
 \begin{center}
 \begin{picture}(6,7.5)
 \put(0,0){
 \includegraphics{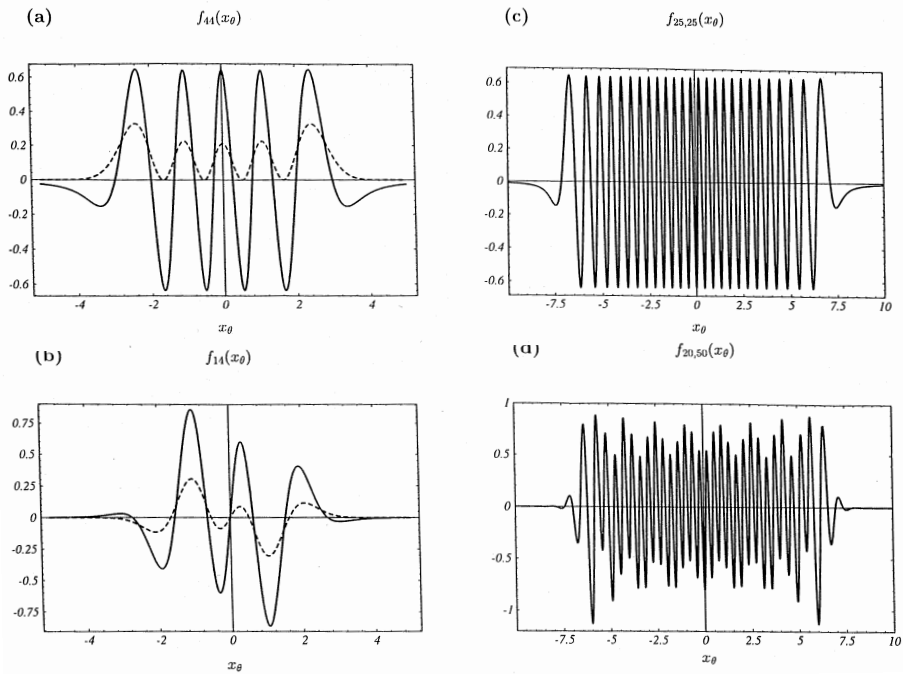}}
 \end{picture}
 \end{center}
\begin{center}
\protect\parbox{.9\textwidth}{
\caption{
Plots of some typical pattern functions $f_{mn}(x)$ (lines) along
with the products of regular wave functions $\psi_{n}(x)\psi_{m}(x)$
(dashed lines). In (a) the diagonal pattern function with $n$ $\!=$
$\!m$ $\!=$ $\!4$ is depicted. Oscillations between $-2/\pi$ and
$+2/\pi$ are clearly visible in the classically allowed region.
Then the function swings over and decays like $x^{-2}$. In (b)
the off-diagonal pattern function with $n$ $\!=$ $\!1$, $m$ $\!=$ $\!4$
is depicted. It is less oscillating than the diagonal pattern function
and it decays faster in the forbidden zone. In (c) and (d) highly
oscillating pattern functions are shown for (c) $n$ $\!=$ $\!m$
$\!=$ $\!25$ and (d) $n$ $\!=$ $\!20$, $m$ $\!=$ $\!50$.
(After Leonhardt, Munroe, Kiss, Raymer and Richter, Th., [1996].)
\label{fig3.2a}
}
}
\end{center}
\end{figure}%

Equation (\ref{3.13}) reveals that the density matrix in the
Fock basis can be sampled directly from the measured quadrature-component
statistics, since $\varrho_{mn}$ can be regarded as a statistical
average of the (bounded) sampling function $K_{mn}(x,\varphi)$
(D'Ariano, Leonhardt and Paul, H., [1995]; Leonhardt,
Paul, H., and D'Ariano [1995];
Leonhardt, Munroe, Kiss, Richter, Th., and Raymer [1996]; Leonhardt [1997c];
D'Ariano [1997a]). In an experiment each outcome $x$ of $\hat{x}(\varphi)$,
with $\varphi$ $\!\in$ $\![0,\pi)$, contributes individually to
$\varrho_{mn}$, so that $\varrho_{mn}$ is gradually building up
during the data collection. That is to say, $\varrho_{mn}$
can be sampled from a sufficiently large set of homodyne data
in {\em real time}, and the mean value obtained from different
experiments can be expected to be normal-Gaussian distributed
around the true value, because of the central-limit theorem.
Moreover, the sampling method can also be used to estimate
the statistical error (see also \S~\ref{sec3.9.1}).

Experimentally, the method was successfully applied
to the determination of the density matrix of squeezed light
generated by a continuous-wave optical parametric amplifier
(Schiller, Breitenbach, Pereira, M\"{u}ller and
Mlynek [1996]; Breitenbach and Schiller [1997]; Breitenbach,
Schiller and Mlynek [1997]).
In the experiment the spectral component of the photocurrent in a
small band around a radiofrequency $\Omega$ is measured (overall
quantum efficiency $\sim$ $\!$ $\!82$\%).
In this case the measurement is on a two-mode quadrature-component
$\hat{x}(\varphi)$ $\!=$ $\!e^{i\phi_{R}}$
$\![\hat{a}(\omega\!+\!\Omega)e^{-i\varphi}$ $\!+$
$\!\hat{a}(\omega\!-\!\Omega)e^{i\varphi}]$, $\phi_{R}$ being the
phase of the radio-frequency local oscillator, so that the scheme is
basically a heterodyne detector. Examples of reconstructed diagonal
density-matrix elements are shown in Fig.~\ref{fig3.3} (for
reconstructed off-diagonal density-matrix elements, see, e.g.,
Schiller, Breitenbach, Pereira, M\"{u}ller and Mlynek [1996];
Breitenbach and Schiller [1997]).
\begin{figure}[htb]
 \unitlength=1cm
 \begin{center}
 \begin{picture}(6,6)
 \put(0,0){
 \includegraphics{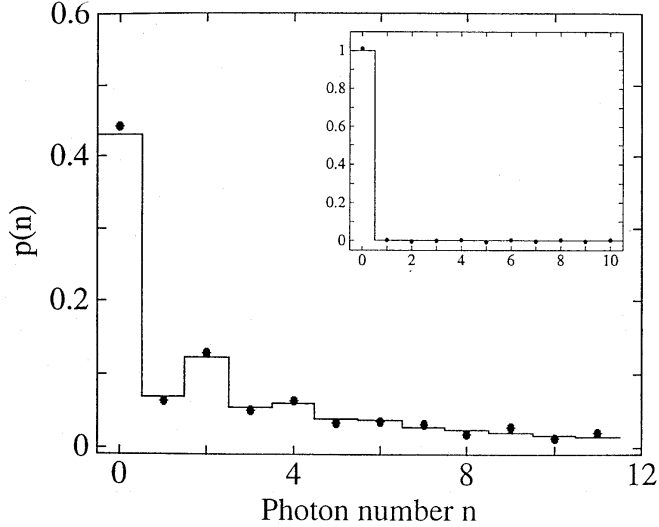}}
 \end{picture}
 \end{center}
\begin{center}
\protect\parbox{.9\textwidth}{
\caption{
Photon-number distribution of a squeezed vacuum and the vacuum state
(inset) reconstructed from the quadrature-component distributions
according to eq. (\protect\ref{3.13}).
Solid points refer to experimental data, histograms to theory.
The experimentally determined statistical error is $0.03$.
(After Schiller, Breitenbach, Pereira, M\"{u}ller and Mlynek [1996].)
\label{fig3.3}
}
}
\end{center}
\end{figure}%

Further, the method was used successfully to measure
the time-resolved photon-number statistics of a $5$~ns
pulsed field with a sampling time of $330$~fs, set by the
duration of the local-oscillator pulse (Munroe, Boggavarapu,
Anderson and Raymer [1995]). From eqs.~(\ref{3.13}) and (\ref{3.14})
it is easily seen that the photon-number probability distribution
$p_{n}$ $\!=$ $\!\varrho_{nn}$ can be
given by
\begin{equation}
\label{3.15a}
p_{n} = \pi
\int {\rm d}x \, f_{nn}(x) \, \bar{p}(x) ,
\end{equation}
where $\bar{p}(x)$ $\!=$ $\!(2\pi)^{-1}\int {\rm d}\varphi$ $\!p(x,\varphi)$
is the phase-averaged quadrature-component distribution.
Equation (\ref{3.15a}) reveals that for sampling the photon-number
statistics the phase need not be controlled -- a situation that
is typically realized when the signal and the local oscillator come
from different sources.
In the experiment, an argon-laser-pumped Ti:saphire laser is used
in combination with a chirped-pulse regenerative amplifier to
generate ultrashort, transform limited local-oscillator pulses ($330$~fs)
at a wavelength of $830$~nm and a repitition rate of $4$~kHz with
approximately $10^{6}$ photons per pulse. The signal is from a
laser diode of wavelength $830$~nm and pulse width $5$~ns.
Figure \ref{fig3.3a} shows examples of the (with $\sim$ $\!65$\% overall
quantum efficiency) measured quadrature-component
probability distributions and the resulting photon-number
probability distributions at two times in the signal pulse.
In particular, the photon-number statistics are seen to change from
nearly Poissonian statistics (laser above threshold) to
thermal-like statistics (laser below threshold)
(for ultrafast homodyne detection of two-time photon-number
correlations, see \S~\ref{sec3.8.1}).
\begin{figure}[htb]
 \unitlength=1cm
 \begin{center}
 \begin{picture}(6,6)
 \put(0,0){
 \includegraphics{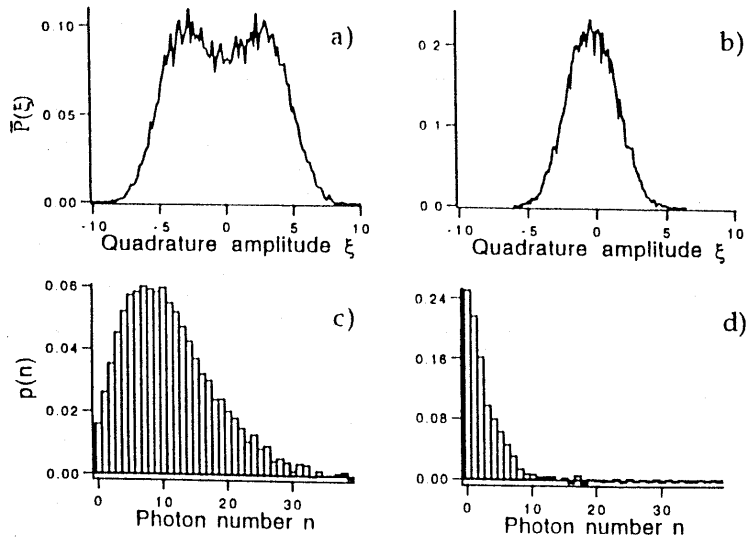}}
 \end{picture}
 \end{center}
\begin{center}
\protect\parbox{.9\textwidth}{
\caption{
Measured phase-averaged quadrature-component distributions for
(a) \mbox{$t$ $\!=$ $\!4.0$~ns} and (b) $t$ $\!=$ $\!6.0$~ns, and
the the resulting time-resolved photon-number distributions for
(c) $t$ $\!=$ $\!4.0$~ns and (d) $t$ $\!=$ $\!6.0$~ns,
obtained from (a) and (b) by using eq.~(\protect\ref{3.15a}).
$\bar{P}(\xi)$ and $p(n)$, respectively, correspond to
$\bar{p}(x)$ and $p_{n}$.
(After Munroe, Boggavarapu, Anderson and Raymer [1995].)
\label{fig3.3a}
}
}
\end{center}
\end{figure}%

So far in the formulas, perfect detection has been assumed.
The problem of extending eq.~(\ref{3.13}) to imperfect detection such
that $p(x,\varphi)$ and $K_{mn}(x,\varphi)$, respectively, are replaced
with $p(x,\varphi;\eta)$ and a sampling function $K_{mn}(x,\varphi;\eta)$
that compensates for the losses has also been
considered,\footnote{Equation (\ref{3.16}) follows
   from eqs.~(\ref{A2.26}) and (\ref{A2.27}), expressing $\Psi(z,\varphi)$
   in terms of $\Psi(z,\varphi;\eta)$; see footnote \ref{fn2}.}
\begin{equation}
\label{3.16}
K_{mn}(x,\varphi;\eta)
= \langle m| \hat{K}(x,\varphi;\eta) |n\rangle,
\end{equation}
where
\begin{equation}
\label{3.16a}
\hat{K}(x,\varphi;\eta)
= \frac{1}{2\pi} \int {\rm d}z \, |z|\,
\exp\!\left\{iz\left[\hat{x}(\varphi)\!-\!x\right]
+{\textstyle\frac{1}{4}}(\eta^{-1}-1)z^{2} \right\}\!
\end{equation}
(D'Ariano [1995]; D'Ariano, Leonhardt and Paul, H., [1995];
Leonhardt, Paul, H., and D'Ari\-a\-no [1995]; D'Ariano [1997a];
D'Ariano and Paris [1997a]).
It has been shown that $K_{mn}(x,\varphi;\eta)$ is a well-behaved
bounded function provided that $\eta\!>$ $\!1/2$.

It is worth noting that the reconstruction formula (\ref{3.13}) also
applies to other than harmonic-oscillator systems (Leonhardt and Raymer
[1996]; Richter, Th., and W\"{u}nsche [1996a,b];
Kr\"{a}hmer and Leonhardt [1997b, 1997c]; Leonhardt [1997a];
Leonhardt and Schneider [1997]). To be more specific, $p(x,\varphi)$ in
terms of $\varrho_{mn}$ reads as\footnote{Note that substituting in
   eq.~(\ref{3.13}) for $p(x,\varphi)$ the expression on the right-hand side
   of eq.~(\ref{3.17}), carrying out the $\varphi$ integral and using
   the orthonormalization relation (\ref{3.20}) just yields an identity.}
\begin{equation}
\label{3.17}
p(x,\varphi)
= \sum_{m,n}
g_{mn}(x)
e^{-i(m-n)\varphi} \, \varrho_{mn},
\end{equation}
where
\begin{equation}
\label{3.18}
g_{mn}(x) =
\psi_{m}
(x)\,\psi_{n}(x),
\end{equation}
and from eqs.~(\ref{3.13}) and (\ref{3.17}) together with
eqs.~(\ref{3.14}), (\ref{3.15}) and
(\ref{3.18}) it follows that the functions
$\pi f_{mn}(x)$, eq.~(\ref{3.15}), are orthonormal to products
of energy eigenfunctions $g_{mn}(x)$, eq.~(\ref{3.18}),
\begin{equation}
\label{3.20}
\pi \int {\rm d}x \, f_{mn}(x) \,
g_{m'n'}(x)
= \delta_{mm'}\delta_{nn'}
\quad {\rm for} \quad
E_{m}\!-\!E_{n}=E_{m'}\!-\!E_{n'}.
\end{equation}
It can be shown that $\psi_{m}(x)$ and
$\varphi_{m}(x)$ can be the regular and irregular
solutions, respectively, which solve a Schr\"{o}dinger equation
\begin{equation}
\label{3.21}
\left[-\frac{1}{2} \frac{d^{2}}{dx^{2}} + U(x)\right]\phi_{m}(x)
= E_{m}\phi_{m}(x)
\end{equation}
for chosen energy $E_{m}$, with $U(x)$ being an arbitrary
potential.\footnote{For a proof, see, e.g., Leonhardt and
   Schneider [1997]; Leonhardt [1997c].}
This offers the possibility of reconstruction of the density matrix
(in the energy representation) of a particle in an arbitrary
one-dimensional potential $U(x)$ from the time-dependent position
distribution of the particle. The quadrature-component distribution
in eq.~(\ref{3.13}) has to be regarded as a time-dependent position
distribution $p(x,t)$ $\!=$ $\!\langle x|\hat{\varrho}(t)|x\rangle$
and the phase integral converts into a time integral,
\begin{equation}
\label{eq4.5a}
\varrho_{mn} = \lim_{T\to\infty}\frac{\pi}{T}
\int_{-T/2}^{+T/2} {\rm d}t \int {\rm d}x \,
e^{i\nu_{mn} t} f_{mn}(x) \, p(x,t),
\end{equation}
[$\varrho_{mn}$ $\!=$ $\varrho_{mn}(t)|_{t=0}$; $\nu_{mn}$, transition
frequencies]. Obviously, the position distribution $p(x,t)$ can
also be used for reconstruction of other quantum-state
representations, such as tomographic reconstruction of the
Wigner function. Note that the
time interval $T$, in which the position distribution can
be measured is limited in general (for a data analysis scheme
for determining the quantum state of a freely evolving
one-dimensional wave packet, see Raymer [1997]; see also
\S~\ref{sec4.5.1}).

In the numerical implementation of eq.~(\ref{3.13}) [or eq.~(\ref{eq4.5a})]
the sampling function can be calculated on the basis of the analytical result
given in eqs.~(\ref{3.14}) and (\ref{3.15}), using appropriate
numerical routines (Leonhardt, Munroe, Kiss, Richter, Th., and Raymer
[1996]). An alternative way is the direct inversion of the
underlying basic equation (\ref{3.17})
that expresses the quadrature-component
distributions $p(x,\varphi)$ in terms of the density-matrix
elements $\varrho_{mn}$ (Tan [1997]).\footnote{The underlying basic
   equation for eq.~(\ref{eq4.5a}) is given by eq.~(\ref{3.17}) with
   $p(x,t)$ and $e^{-i\nu_{mn}t}$ in place of $p(x,\varphi)$ and
   $e^{-i(m-n)\varphi}$, respectively. Inverting it numerically,
   the time need not be infinitely large as it might be suggested
   from the analytical result given in eq.~(\ref{eq4.5a}) (Opatrn\'{y},
Welsch and Vogel, W., [1997c]).}
For any physical quantum state the
density-matrix elements $\varrho_{mn}$ must eventually decrease
to zero with increasing $m$($n$). Therefore it follows that
the expression on the right-hand side of eq.~(\ref{3.17}) can
always be approximated to any desired degree of accuracy by
setting $\varrho_{mn}$ $\!\approx$ $\!0$ for \mbox{$m$($n$) $\!>$
$\!n_{\rm max}$}. To obtain the (finite) number of density-matrix
elements from the measured quadrature-component distributions,
the resulting equation can always be inverted numerically,
using standard methods, such as least-squares inversion
(Appendix \ref{app4}; see also \S~\ref{sec3.9.2}). It should
be pointed out that the method also applies when the detection
efficiency is less than unity. Recalling eqs.~(\ref{2.23}) and
(\ref{2.24}), it is easily seen that $p(x,\varphi;\eta)$ is
related to $\varrho_{mn}$ according to eq.~(\ref{3.17}), with
$g_{mn}(x)$ being replaced with
\begin{equation}
\label{3.19}
g_{mn}(x;\eta) = \int {\rm d}x' \,
\psi_{m}
(x') \psi_{n}(x') \,
p(x\!-\!x';\eta).
\end{equation}

When the characteristic function $\Psi(z,\varphi)$ of the
quadrature-component distribution $p(x,\varphi)$ can be measured
directly (\S~\ref{sec4.2.1}), then the density matrix in the Fock
basis can be obtained from $\Psi(z,\varphi)$ by replacing in
eqs.~(\ref{3.13}) and (\ref{3.14}) the $x$ integral with the $z$ integral
over $\underline{f}_{mn}(z)\Psi(z,\varphi)$, where
\begin{equation}
\label{3.19b}
\underline{f}_{mn}(z) = \int {\rm d}x \, e^{izx} f_{mn}(x).
\end{equation}
More explicitly, the result can be given by (Wallentowitz and
Vogel, W., [1996b])
\begin{equation}
\label{eq4.13}
\varrho_{n\,n+k}
=
\int_{0}^{\pi} {\rm d}\varphi \,
e^{-i k \varphi}
\int_{0}^{\infty}
{\rm d}z \,
S_{n}^{(k)}(z)
\left\{
 \begin{array}{ll}
    {\rm Re}\,\Psi(z,\varphi) & \ {\rm if} \ $k$ \ {\rm even}, \\[1ex]
    {\rm Im}\,\Psi(z,\varphi) & \ {\rm if} \ $k$ \ {\rm odd}
  \end{array} \right.
\end{equation}
($k$ $\!\ge$ $\!0$),
with
\begin{equation}
\label{eq4.14}
S_{n}^{(k)}(z)
=
\frac{\sqrt{2}}{\pi}
\sqrt{\frac{2 \, n!}{(n\!+\!k)!}}
\left( \frac{z}{2} \right)^{k+1} {\rm L}_{n}^{(k)}(z^{2}/2) \, e^{-z^{2}/4}
\left\{
  \begin{array}{ll}
  (-2)^{k/2} & \ {\rm if} \ $k$ \ {\rm even},\\[1ex]
  (-2)^{(k-1)/2} & \ {\rm if} \ $k$ \ {\rm odd}
  \end{array}
\right.
\end{equation}
[${\rm L}_{n}^{(k)}(x)$, Laguerre polynomial].

So far, reconstruction of arbitrary quantum states has
been considered, which can require measurement of the quadrature
components at a large number of phases (see also \S\ref{sec3.9.1}),
which reveals that the Pauli problem [i.e., reconstructing a
quantum state from $p(x,\varphi)$ and $p(x,\varphi$ $\!+$ $\!\pi/2)$]
cannot be uniquely solved in general. However,
if there is some {\em a priori} information on the quantum state to be
reconstructed, then $p(x,\varphi)$ need not be known for all phases
within a $\pi$ interval. In particular when the state is known to be
a pure state that is a finite superposition of Fock states,
\begin{equation}
\label{3.19a}
|\Psi\rangle = \sum_{n=0}^{n_{\rm max}} c_{n} |n\rangle,
\end{equation}
then it can be reconstructed from two quadrature-component
distributions $p(x,\varphi)$ and $p(x,\varphi$
$\!+$ $\!\pi/2)$.\footnote{The Pauli problem of determining
   the quantum state of a particle from the position distribution and
   the momentum distribution has been studied widely, and
   it has turned out that it cannot be solved uniquely even
   if the particle moves in a one-dimensional potential and
   is prepared in a pure but arbitrary state
   (for the problem including finite dimensional spin systems,
   see Pauli [1933]; Feenberg [1933]; Kemble [1937];
   Reichenbach [1946]; Gale, Guth and Trammell [1968];
   Lamb [1969]; Trammell [1969]; Band and Park [1970, 1971];
   Park and Band [1971];  d'Espagnat [1976]; Kreinovitch [1977];
   Corbett and Hurst [1978]; Prugove\v{c}ki [1977];
   Corbett and Hurst [1978]; Vogt [1978]; Band and Park [1979];
   Park, Band and Yourgrau [1980]; Ivanovi\'{c} [1981, 1983]
   Moroz [1983, 1984]; Royer [1985, 1989]; Friedman [1987];
   Pavi\v{c}i\'{c} [1987]; Wiesbrock [1987]; Busch and Lahti [1989];
   Wootters and Fields [1989]; Stulpe and Singer [1990];
   Bohn [1991]; Weigert [1991]).\label{general}}
In this case the problem reduces to solving blocks of linear
equations for the unknown coefficients
in the Fock-state expansion of the state (Or{\l}owski and Paul, H.
[1994]).\footnote{These coefficients can also be obtained from
   $p(x,\varphi)$ and $\partial / \partial \varphi$ $\!
   p(x,\varphi)|_{\varphi = 0}$ (Richter, Th., [1996c]). More generally,
   it can be shown that a pure state can always be
   determined from \mbox{$p(x,\varphi)$ and $\partial / \partial \varphi$
   $\! p(x,\varphi)$}, i.e., from the position distribution and its
   time derivative in the case of a particle that moves in
   a one-dimensional potenial and is prepared in an arbitrary pure
   state (Feenberg [1933], for the problem, also see Gale, Guth and
   Trammell [1968]; Royer [1989]).}


\subsubsection{Sampling of the displaced Fock-states on a circle}
\label{sec3.3.2}

   From \S~\ref{sec2.1.7} we know that in unbalanced homodyning
the photon-number distribution of the transmitted signal
mode is, under certain conditions, the displaced photon-number
distribution of the signal mode, $p_{m}(\alpha)$,
the displacement parameter \mbox{$\alpha$ $\!=$ $\!|\alpha|e^{i\varphi}$}
being controlled by the
local-oscillator complex amplitude
[see eq.~(\ref{2.50})]. Expanding the density operator in the Fock
basis, $p_{m}(\alpha)$ can be related to the density matrix of the
signal mode as
\begin{equation}
\label{3.22}
p_{m}(\alpha)
= \langle m,\alpha|\hat{\varrho}|m,\alpha \rangle
= \sum_{k,n}
\langle n|m,\alpha\rangle
\langle k|m,\alpha\rangle^{\ast} \varrho_{kn} ,
\end{equation}
where the expansion coefficients $\langle n|m,\alpha\rangle$ can be
taken from eq.~(\ref{A2.12c}) or eq.~(\ref{A2.12c-t}).
Equation (\ref{3.22}) can always be inverted in order to obtain
$\varrho_{mn}$ in terms of $p_{k}(\alpha)$ (Mancini, Tombesi and
Man'ko, V.I.
[1997]; Mancini, Man'ko, V.I., and Tombesi [1997]):\footnote{Equation
   (\ref{3.22a}) directly follows from eqs.~(\ref{A2.16}) and (\ref{3.37}).}
\begin{equation}
\label{3.22a}
\varrho_{mn} = \sum_{k=0}^{\infty}
\int {\rm d}^{2}\alpha \, K_{mn}^{k}(\alpha) \, p_{k}(\alpha),
\end{equation}
where
\begin{equation}
\label{3.22b}
K_{mn}^{k}(\alpha)
= \langle m|\hat{K}^{k}(\alpha)|n\rangle
= \frac{2}{(1\!-\!s)}\left(\frac{s\!+\!1}{s\!-\!1}\right)^{k}
\langle m|\hat{\delta}(\alpha\!-\!\hat{a};-s)|n\rangle
\end{equation}
is bounded for $s$ $\!\in$ $\!(-1,0]$, the $\hat{\delta}$ operator
being given by eq.~(\ref{A2.15}).
This offers the possibility
of direct sampling of $\varrho_{mn}$ from the displaced
Fock-state probability distribution $p_{m}(\alpha)$.

Since $p_{m}(\alpha)$ as a function of
$\alpha$ for chosen $m$ already determines the quantum state
(\S\S~\ref{sec2.1.5} and \ref{sec3.3.3}), it
is clear that when $m$ is allowed to be varying, then
-- in contrast to eq.~(\ref{3.22a}) -- $p_{m}(\alpha)$ need not
be known for all complex values of $\alpha$ in order to reconstruct
the density-matrix elements $\varrho_{kn}$ from $p_{m}(\alpha)$. In
particular, it is sufficient to know $p_{m}(\alpha)$ for all values of $m$
and all phases $\varphi$, $|\alpha|$ being fixed (Leibfried, Meekhof,
King, Monroe, Itano and Wineland [1996]\footnote{Here the method was
   first used for reconstructing experimentally the density matrix of the
   center-of-mass motion of a trapped ion (also see \S~\ref{sec4.2.2}).};
Opatrn\'{y} and Welsch [1997]; Opatrn\'{y}, Welsch, Wallentowitz and
Vogel, W., [1997]). For chosen $|\alpha|$ we regard $p_{m}(\alpha)$ as a
function of $\varphi$ and introduce the Fourier coefficients
\begin{equation}
\label{3.23}
p_{m}^{s}(|\alpha|) = \frac{1}{2\pi}
\int_{0}^{2\pi} {\rm d}\varphi \, e^{is\varphi} p_{m}(\alpha)
\end{equation}
($s$ $\!=$ $\!0,1,2,\ldots$), which are related to the density-matrix
elements whose row and column indices differ by $s$ as
\begin{equation}
\label{3.24}
p_{m}^{s}(|\alpha|) = \sum_{n=0}^{\infty} G_{mn}^{s}(|\alpha|)\,\varrho_{n+s\,n} ,
\end{equation}
where
\begin{eqnarray}
\label{3.25}
\lefteqn{
G_{mn}^{s}(|\alpha|) = e^{-|\alpha|^{2}}m!\sqrt{n!(n\!+\!s)!}
}
\nonumber \\[.5ex] && \hspace{2ex} \times
\sum_{j=0}^{J} \sum_{l=0}^{L}
\frac{(-1)^{j+l}|\alpha|^{2(m+n-j-l)+s}}
{j!(m\!-\!j)!(n\!+\!s\!-\!j)!l!(m\!-\!l)!(n\!-\!l)!}\,,
\end{eqnarray}
with $J$ $\!=$ $\!{\rm min}(m,n$ $\!+$ $\!s)$ and $L$ $\!=$ $\!{\rm min}(m,n)$.
Inverting eq.~(\ref{3.24}) for each value of $s$ yields the sought
density matrix. Since there has not been an analytical solution,
eq.~(\ref{3.24}) has been inverted numerically,\footnote{For examples,
   see, e.g., Opatrn\'{y}, Welsch, Wallentowitz and Vogel, W., [1997].}
setting \mbox{$\varrho_{mn}$ $\!\approx$ $\!0$}
for $m$($n$) $\!>$ $\!n_{\rm max}$
(cf. the last paragraph but two of \S~\ref{sec3.3.1}) and using
least-squares inversion (Appendix \ref{app4}; \S~\ref{sec3.9.2}).
In this way $\varrho_{m+s\,m}$ can be given by
\begin{equation}
\label{3.25a}
\varrho_{m+s\,m} = \sum_{n=0}^{N} F_{mn}^{s}(|\alpha|)
\, p_{n}^{s}(|\alpha|) ,
\end{equation}
where it is assumed that $p_{n}$ is measured for $n$ $\!=$ $\!0,1,2,\ldots,N$,
with $N$ $\!\le$ $\!n_{\rm max}$, and the matrix $F_{mn}^{s}(|\alpha|)$ is
calculated numerically.
Combining eqs.~(\ref{3.25a}) and (\ref{3.23}),
$\varrho_{m+s\,m}$ can be given in a form suitable for statistical
sampling.

An extension of eq.~(\ref{3.24}) to nonperfect detection is
straightforward. In this case eq.~(\ref{2.49}) applies, and the
measured probability distribution $P_{m}(\alpha;\eta)$ can be related
to $p_{m}(\alpha)$ as shown in Appendix \ref{app3}. From eqs.~(\ref{3.23})
and (\ref{A3.3}) it can be seen that when in eq.~(\ref{3.24}) the Fourier
component $p_{m}^{s}(|\alpha|)$ is replaced with the actually measured
one, then the matrix $G_{mn}^{s}(|\alpha|)$ has to be replaced with
the matrix
\begin{equation}
\label{3.26}
G_{mn}^{s}(|\alpha|,\eta)
= \sum_{l} P_{m|l}(\eta)\,G_{ln}^{s}(|\alpha|).
\end{equation}
Similarly, multichannel detection of the photon-number distribution
can be taken into account. In particular when a photon-chopping scheme as
outlined in \S~\ref{sec2.1.7} is used, then the Fourier components
of the measured $N$-coincidence-event probability distribution
can again be related to the density-matrix elements according to
eq.~(\ref{3.24}), but now with
\begin{equation}
\label{3.27}
G_{mn}^{s}(|\alpha|,\eta,N)
= \sum_{k,l} \tilde{P}_{m|k}(N) P_{k|l}(\eta)\,G_{ln}^{s}(|\alpha|)
\end{equation}
in place of $G_{mn}^{s}(|\alpha|,\eta)$, where $\tilde{P}_{m|k}(N)$
can be taken from eq.~(\ref{2.52}).


\subsubsection{Reconstruction from propensities}
\label{sec3.3.3}

As already mentioned, the displaced-Fock-state probability
$p_{m}(\alpha)$, eq.~(\ref{3.22}),
as a function of $\alpha$ defines for each $m$ a propensity
${\rm prob}(\alpha)$ $\!=$ $\!\pi^{-1}$ $\!p_{m}(\alpha)$
of the type given in eq.~(\ref{2.43}). The simplest
example is the $Q$ function ($m$ $\!=$ $\!0$), which can be measured in
perfect eight-port balanced homodyne detection, with the vacuum as quantum
filter (\S~\ref{sec2.1.4}). Propensities contain all knowable information
on the quantum state, and therefore all relevant properties of it
can be obtained from them in principle. However, in order to obtain
the ``unfiltered'' quantum state, the additional noise introduced
by the filter must be ``removed'', which may be an effort in practise.

  When the $Q$ function is known, then the density matrix in the
photon-number basis can be calculated straightfowardly, using the
well-known relation\footnote{Writing
   $\pi$ $\!e^{|\alpha|^2}$ $\!Q(\alpha)$ $\!=$ $\!e^{|\alpha|^2}$
   $\!\langle \alpha | \hat{\varrho} | \alpha \rangle$ $\!=$ $\!\sum_{m,n}$
   $\!e^{|\alpha|^2}$ $\!\langle \alpha | m \rangle$
   $\!\langle m | \hat{\varrho} | n \rangle$ $\!\langle n| \alpha \rangle$
   and recalling that $\langle n|\alpha \rangle$ $\!=$ $\!(n!)^{-1/2}$
   $\!\alpha^{n}$ $\!e^{-|\alpha|^{2}/2}$,
   eq.~(\ref{3.12}) can be derived easily.}
\begin{equation}
\label{3.12}
\varrho_{mn} =
\langle m | \hat{\varrho} | n \rangle =
\frac{\pi}{\sqrt{m!n!}}\,
\left. \frac{\partial^{m+n}}
{\partial\alpha^{\ast m}\partial\alpha^{n}}
 \, e^{|\alpha|^2}Q(\alpha) \right|_{\alpha=\alpha^\ast=0},
\end{equation}
which corresponds to
\begin{equation}
\label{3.28}
\hat{\varrho} = \pi \,
\left.
\exp\!\left(\hat{a}^{\dagger}\frac{\partial}{\partial\alpha^{\ast}}\right)
\exp\!\left( \hat{a} \frac{\partial}{\partial\alpha}\right)
Q(\alpha)\right|_{\alpha=\alpha^\ast=0}
\end{equation}
(W\"{u}nsche [1991, 1996a]). Equation (\ref{3.28}) can also be extended to
other than vacuum filters, in order to obtain the density operator
in terms of derivatives of more general propensities of the type given in
eq.~(\ref{2.43}) (W\"{u}nsche and  Bu\v{z}ek [1997]).
In order to obtain the density matrix from the homodyne data measured,
derivatives on them must be carried out, which probably could be done
with sufficient accuracy only for states that contain very few
photons.

In practice it may be more convenient to handle integrals rather
than derivatives. It can be shown that\footnote{Equation (\ref{3.29})
   can be derived, applying eq.~(\ref{A2.17b}) to $\hat{F}$ $\!=$ $\!|n\rangle
   \langle m|$ and calculating the $c$-number function $F(\alpha;s)$
   associated with $\hat{F}$ in chosen order. Note that for $s$ $\!=$
   $\!-1$ the integral form (\ref{3.29}) corresponds to the
   differential form (\ref{3.12}).}
\begin{equation}
\label{3.29}
\varrho_{mn} = \left(\frac{s\!-\!1}{2}\right)^{n} \sqrt{\frac{n!}{m!}}
\,\sum_{p=0}^{\infty} \int {\rm d}^{2}\alpha P(\alpha;s) \, \alpha^{m-n}
{p\!+\!n \choose p}\left(\frac{1\!-\!s}{2}\right)^{p}
L_{p+n}^{m-n}\!\left(\frac{2|\alpha|^{2}}{1\!-\!s}\right),
\end{equation}
where $m$ $\!\ge$ $\!n$ and $s$ $\!\leq$ $\!-1$,
$L_{n}^{m}(z)$ being the Laguerre polynomial
(Paris [1996b]). In particular when $s$ $\!=$ $\!1$
$\!-$ $\!2\eta^{-1}$, then $P(\alpha;s)$ is just the smoothed
$Q$ function measured in nonperfect detection (cf. \S~\ref{sec2.1.4}).
Unfortunately, eq.~(\ref{3.29}) is not suitable for statistical
sampling, since the integral must be performed first and after
that the summation can be carried out. Moreover, the inaccuracies
of the measured $P(\alpha;s)$ together with the Laguerre
polynomials can give rise to an error explosion in the reconstructed
density matrix, so that an exact reconstruction of the density matrix
from measured data may be expected to be possible only
for states that contain finite (and not too large) numbers of
photons. In this case the
$p$ sum in eq.~(\ref{3.29}) can be truncated at $p$ $\!=$ $\!N$ $\!-$ $\!m$,
where the value $N$ has to be chosen large enough to ensure that
$\langle\hat{a}^{\dagger\,i} \hat{a}^{j}\rangle$ $\!=$ $\!0$ for
\mbox{$i,j$ $\!\ge$ $\!N$}. Now the
$p$ sum can be performed first, and a (state-dependent) integral kernel
for statistical sampling can be calculated (Paris [1996b, 1996c]).


\subsection{Multimode density matrices}
\label{sec3.4}

The extension of the methods outlined in \S\S~\ref{sec3.2} and \ref{sec3.3}
to the reconstruction of multimode density matrices from the corresponding
multimode joint quadrature-component distributions or multimode joint
propensities is straightforward. The situation is somewhat different
when combined distributions, i.e., distributions that are related to
linear combinations of the modes, are measured. Let us consider
the two-mode detection schemes shown in Fig.~\ref{fig2.4}
in \S~\ref{sec2.1.3}. When the sum quadrature-component
distribution of two modes, $p_{S}(x,\alpha,\varphi_{1},\varphi_{2})$,
is known for all phases $\varphi_{1}$ and $\varphi_{2}$ within
$\pi$ intervalls and all superposition parameters $\alpha$ $\!\in$
$\!(0,\pi/2)$, then it can be shown, on recalling eq.~(\ref{2.33}),
that the reconstruction of the two-mode density matrix in a
quadrature-component basis can be accomplished with a three-fold
Fourier integral (Opatrn\'{y}, Welsch and Vogel, W., [1996, 1997b]): \\
\parbox{\textwidth}{
\begin{eqnarray}
\label{3.30}
\lefteqn{
\left\langle x_{1}\!-\!x_{1}',x_{2}\!-\!x'_{2},\varphi
|\hat \varrho |
x_{1}\!+\!x'_{1}, x_{2}\!+\!x'_{2},\varphi\right\rangle
}
\nonumber \\[.5ex] && \hspace{2ex}
= \left( \frac{1}{2\pi} \right)^{2}
\int {\rm d}z_{1}  \int {\rm d}z_{2} \,
e^{-i(x_{1}z_{1}+x_{2}z_{2})}
\int {\rm d}x \, e^{iyx}
p_{S}( x, \alpha,\tilde{\varphi}_{1},\tilde{\varphi}_{2}),
\end{eqnarray}
}
where
\begin{equation}
\label{3.31}
y = \sqrt{z_{1}^{2}\!+\!z_{2}^{2}\!+\!(2x'_{1})^{2}\!+\!(2x'_{2})^{2} },
\qquad
\tan\alpha = \sqrt{
\frac{z_{2}^{2}\!+\!(2x_{2}')^{2}}{z_{1}^{2}\!+\!(2x'_{1})^{2}}} \, ,
\end{equation}
and
\begin{equation}
\label{3.32}
\tilde{\varphi}_{k} = \varphi - {\textstyle\frac{1}{2}\pi}
+ \arg\!\left(2x_{k}'\! +\! i z_{k}\right)
\end{equation}
($k$ $\!=$ $\!1,2$).\footnote{For an extension to imperfect detection,
   see Opatrn\'{y}, Welsch and Vogel, W., [1997b].}

The generalization to the $N$-mode case is straightforward. Suppose that
we can measure the probability distribution of a weighted sum of
quadratures $\hat x$ $\!=$ $\!\sum_{k=1}^{N}$ $\! f_{k}(\{ \alpha _{l} \} )
\hat x_{k}$, where $\alpha _{l}$ are
$N$ $\!-$ $\!1$ parameters that can be controlled in the experiments,
$l$ $\!=$ $\!1,2,\dots N\!-\!1$, and
$f_{k}$ are real functions which satisfy $\sum_{k=1}^{N}$
$\! f_{k}^{2}(\{ \alpha_{l} \} )$ $\!=1$ identically for each set of
parameters $\{ \alpha _{l} \}$. Let $g_{l}$ be the inversions of $f_{k}$,
$g_{l}[\{ f_{k}(\{ \alpha _{m} \} ) \} ]$ $\!=$ $\!\alpha _{l}$
(i.e., the set $\{ \alpha _{l} \}$ parameterizes the surface
of an $N$-di\-men\-sion\-al sphere). From the measured sum
quadrature-component probability distribution
$p_{S}(x,\{ \alpha _{l} \}, \{ \varphi_{k} \})$ its characteristic
function $\Psi _{s} (z, \{ \alpha _{l} \}, \{ \varphi_{k} \} )$ can be
calculated as a Fourier transform. The characteristic function of the joint
quadrature-component probability distribution can then be calculated as
\begin{eqnarray}
\label{3.33}
\Psi _{j}(\{ z_{k} \}, \{ \varphi_{k} \} )
= \Psi _{s} \left(z, \left\{ g_{l} (\{ z_{k}/z \}, \{ \varphi_{k} \} )
\right\}
\right) ,
\end{eqnarray}
[$z$ $\!\equiv$ $\!( \sum_{k=1}^{N} z_{k}^{2} )^{1/2}$], from which
the $N$-mode density matrix can be obtained by an $N$-fold Fourier
transform; the whole $N$-mode density matrix reconstruction is
thus accomplished by an $N\!+\!1$-fold integration of the measured data.

In the Fock basis the reconstruction of a two-mode density matrix
from the combined quadrature-component distribution
$p_{S}(x,\alpha,\varphi_{1},\varphi_{2})$ can be accomplished
with a four-fold integration (Raymer, McAlister and Leonhardt [1996];
McAlister and Raymer [1997b]; Richter, Th., [1997a]):
\begin{eqnarray}
\label{3.34}
\langle m_{1},m_{2}|\hat{\varrho} | n_{1},n_{2} \rangle
=
\!\!\int \! {\rm d}x
\!\int_{0}^{\pi} \!\!\! {\rm d}{\alpha}
\!\!\int_{0}^{\pi} \!\! {\rm d}\varphi_{1}
\!\!\int_{0}^{\pi} \!\! {\rm d}\varphi_{2} \,
R^{m_{2}n_{2}}_{m_{1}n_{1}}(x,\alpha,\varphi_{1}\varphi_{2})
p_{s}(x,\alpha,\varphi_{1},\varphi_{2}),
\quad
\end{eqnarray}
where the integral kernel
$R^{m_{2}n_{2}}_{m_{1}n_{1}}(x,\alpha,\varphi_{1}\varphi_{2})$ is
suitable for application of statistical sampling. It can be given by
\begin{eqnarray}
\label{3.35}
R^{m_{2}n_{2}}_{m_{1}n_{1}}(x,\alpha,\varphi_{1}\varphi_{2})
= r^{m_{2}n_{2}}_{m_{1}n_{1}}(x,\alpha)
\,e^{i(m_{1}-n_{1})\varphi_{1}}
e^{i(m_{2}-n_{2})\varphi_{2}},
\end{eqnarray}
where
\begin{eqnarray}
\label{3.36}
r^{m_{2}n_{2}}_{m_{1}n_{1}}(x,\alpha) =
\frac{1}{2\pi}
\int {\rm d}y \,
f_{m_{2}n_{2}}(y)
\,\frac{\partial^{2}}{\partial x^{2}}
\,\psi_{m_{1}}\!\!\left( \frac{x\!-\!y\cos \alpha}{\sin \alpha} \right)
\psi_{n_{1}}\!\!\left( \frac{x\!-\!y\cos \alpha}{\sin \alpha} \right).
\end{eqnarray}
Here, $\psi_{m_{1}}(y)$ and $\psi_{n_{1}}(y)$ are
harmonic-oscillator energy eigenfunctions, and
$f_{m_{2}n_{2}}(y)$ is given by eq.~(\ref{3.15}). An alternative
integral expression for $r_{m_1n_1}^{m_2n_2}(x,\alpha)$ reads
\begin{equation}
\label{3.36a}
r_{m_1n_1}^{m_2n_2}(x, \alpha) =
\frac{1}{(2\pi)^{2}}\int {\rm d}z \,
|z| e^{-izx} \underline{f}_{m_1n_1}(z\cos\alpha)
\,\underline{f}_{m_2n_2}(z\sin\alpha),
\end{equation}
with $\underline{f}_{mn}(z)$ being given by eq.~(\ref{3.19b}).
Since $\underline{f}_{mn}(z)$ is the kernel function for reconstruction
of the single-mode density matrix in the Fock basis from the
quadrature-component characteristic function $\Psi(z,\phi)$, from
eqs.~(\ref{eq4.13}) and (\ref{eq4.14}) it is seen that it can be
expressed in terms of the associated Laguerre polynomial. Using this in
eq.~(\ref{3.36a}), then the integral can be performed to obtain
a representation of $r_{m_1n_1}^{m_2n_2}(x,\alpha)$ as a finite
sum over confluent hypergeometric functions (Richter, Th., [1997a]).

Finally, the problem of reconstruction of the quantum state
of unpolarized light was studied, by considering the
(two-mode) density operator
\begin{equation}
\label{3.35a}
\hat{\varrho} = \sum_{n=0}^{\infty} p_{n}
\left(
\frac{1}{n\!+\!1}\sum_{k=0}^{\infty}
|k,n\!-\!k\rangle\langle k,n\!-\!k|
\right)
\end{equation}
(Lehner, Leonhardt and Paul, H., [1996]). Obviously, the probability
distribution $p_n$ of finding $n$ completely unpolarized photons
in the signal field characterizes uniquely the quantum state
(\ref{3.35a}). Two schemes for determining $p_n$ were studied.
First, expressions for the sampling function were derived for the case
when the two orthogonal polarization modes are delivered to two balanced
homodyne detectors and the joint quadrature-component distributions
are measured (Kr\"{a}hmer and Leonhardt [1997a]. Second,
it was shown that measurement of the quadrature-component distributions
of any linearly polarized component of the signal field is sufficient
to determine $p_n$, the corresponding sampling function being
closely related to the single-mode function (Richter, Th., [1997b]).


\subsection{Local reconstruction of $P(\alpha;s)$}
\label{sec3.5}

The displaced photon-number statistics that is measurable in unbalanced
homodyning (\S~\ref{sec2.1.7}) can be used for a
pointwise reconstruction of $s$-parametrized phase-space functions
(Wallentowitz and Vogel, W., [1996a]; Banaszek and W\'{o}dkiewicz
[1996]).\footnote{For a squeezed coherent local oscillator,
   see Banaszek and W\'{o}dkiewicz [1998].}
  From eq.~(\ref{A2.17}) together with eqs.~(\ref{A2.15}) and
(\ref{A2.12a}) it is easily seen (cf. also Moya-Cessa and Knight [1993])
\begin{equation}
\label{3.37}
P(\alpha;s) = \frac{2}{\pi(1\!-\!s)} \sum_{m=0}^{\infty}
\left(\frac{s\!+\!1}{s\!-\!1}\right)^{m} p_{m}(\alpha).
\end{equation}
Hence, all the phase-space functions $P(\alpha;s)$, with
$s$ $\!<$ $1$, can be obtained from $p_{m}(\alpha)$ for each
phase-space point $\alpha$ in a very direct way, without
integral transformations. In particular
when $s$ $\!=$ $\!-1$, then eq.~(\ref{3.37}) reduces to the
well-known result that $Q(\alpha)$ $\!=$ $\!\pi^{-1}p_{0}(\alpha)$,
with $p_{0}(\alpha)$ $\!=$ $\!\langle\alpha|\hat{\varrho}|\alpha\rangle$.
Further, choosing $s$ $\!=$ $\!0$ in eq.~(\ref{3.37}), we arrive at
the Wigner function,
\begin{eqnarray}
\label{3.38}
W(\alpha) = \frac{2}{\pi}
\sum_{m=0}^{\infty} (-1)^{m} p_{m}(\alpha).
\end{eqnarray}
Equation (\ref{3.38}) reflects nothing but the well-known fact that the
Wigner function is proportional to the expectation value of the displaced
parity operator (Royer [1977]).\footnote{For proposals of measuring the
   Wigner function of a particle using this fact, see (Royer [1985, 1989]).}
Experimentally, the method was first applied to the reconstruction of
the Wigner function of the center-of-mass motion of a trapped ion
(Leibfried, Meekhof, King, Monroe, Itano and Wineland [1996];
\S~\ref{sec4.2.2}).

   From \S~\ref{sec2.1.7} we know that in unbalanced homodyning
the quantum efficiency \mbox{$\eta$ $\!=$ $|U_{k1}|^{2}\eta_{D}$}
is always less than unity even if $\eta_{D}$ $\!=$ $\!1$, because of
$|U_{k1}|$ $\!<$ $\!1$. Equation (\ref{3.37}) can be extended to
imperfect detection in order to obtain
(Wallentowitz and Vogel, W., [1996a]; Banaszek and W\'{o}dkiewicz
[1997b])\footnote{This can be easily proved correct, applying
   eq.~(\ref{A3.3a}) and expressing $p_{m}(\alpha)$ in eq.~(\ref{3.37})
   in terms of $P_{m}$, eq.~(\ref{2.49}).}
\begin{equation}
\label{3.37a}
P(\alpha;s) = \frac{2}{\pi(1\!-\!s)} \sum_{m=0}^{\infty}
\left[\frac{\eta(s\!-\!1)\!+\!2}{\eta(s\!-\!1)}\right]^{m} P_{m}
\end{equation}
[$\alpha$ $\!=$ $\!-(U_{k2}/U_{k1})\alpha_{L}$, eq.~(\ref{2.48})],
where the measured photon-number distribution $P_{m}$ can be regarded
as a smoothed displaced photon-number distribution
$p_{m}(\alpha;\eta)$ of the signal mode.  The method works very well
and is well suited for statistical sampling if $s$ $\!<$ $\!1$ $\!-$
$\!\eta^{-1}$; i.e., when the weighting factors improve the
convergence of the series.\footnote{
  The feasibility of
  reconstruction of the Wigner function
  of truncated states
  was also demonstrated
  for $\eta$ $\!<$ $\!1$ (Wallentowitz and Vogel, W., [1996a]).}
  Hence reconstruction of the
  $Q$ function with reasonable precision is always possible if $\eta$
  $\!>$ $\!1/2$ (for computer simulations of measurements, see
  Wallentowitz and Vogel, W., [1996a]; Banaszek and W\'{o}dkiewicz
  [1997b]).

As already mentioned in \S~\ref{sec2.3},
measurement of the photon-number distribution
of a linearly amplified signal can also be used -- similarly to measurement
of the displaced photon-number statistics in unbalanced homodyning --
for reconstructing the quantum state of the signal mode. It can be shown
that when the idler mode is prepared in a coherent state, then
the phase-space funcion $P(\alpha;s)$ of the signal mode can be related
to the measured photon-number distribution as\footnote{From eq.~(\ref{A3.4})
   it can be found that for $\alpha$ $\!=$ $\!0$, eq.~(\ref{3.37}) relates
   the measured distribution $P_{m}$ to the phase-space function of the
   detected mode at the origin of the phase space, $P_{\rm det}(0,s)$. Equation
   (\ref{3.37b}) can then be proved correct, using the unitary tranformation
   (\ref{2.52a}) and expressing $P_{\rm det}(\alpha,s)$ in terms of phase-space
   functions of the signal and idler modes by convolution (see also
   Leonhardt [1994]; Kim, M.S., and Imoto [1995]).}
\begin{equation}
\label{3.37b}
P(\alpha;s) = \frac{2g}{\pi(2\!-\!g\!-\!s)} \sum_{m=0}^{\infty}
\left[\frac{\eta_{D}(s\!+\!g\!-\!2)\!+\!2}{\eta_{D}(s\!+\!g\!-\!2)}\right]^{m}
P_{m}
\end{equation}
(Kim, M.S., [1997a,b]), where $\alpha$ $\!=$
$\!-[(g$ $\!-$ $\!1)/g]^{1/2}\alpha_{I}^{\ast}$. Note that the gain
factor $g$ and the quantum efficiency $\eta_{D}$ of the detector enter
separately into eq.~(\ref{3.37b}) [in contrast
to $\eta$ $\!=$ $|U_{k1}|^{2}\eta_{D}$ in eq.~(\ref{3.37a})].


\subsection{Reconstruction from test atoms in cavity QED}
\label{sec3.6}

Let us now turn to the problem of reconstruction of the quantum state
of a high-$Q$ cavity field from measurable properties of test atoms
in detection schemes outlined in \S~\ref{sec2.4}. Though at a first
glance the schemes look quite different from the homodyne detection
schemes, there are a number of remarkable similarities between them.


\subsubsection{Quantum state endoscopy and related methods}
\label{sec3.6.1}

Let us first consider a two-level (test) atom that resonantly interacts with a
single-mode cavity field according to a $k$-photon Jaynes--Cummings model,
the atom--field interaction Hamiltonian being given by
\begin{equation}
\label{3.39}
\hat H^{(k)} = \hbar \kappa^{(k)}
\left(
\sigma_{+} \hat{a}^{k}  + \hat{a}^{\dagger k} \hat{\sigma}_{-}
\right) ,
\end{equation}
which for $k$ $\!=$ $\!1$ reduces to eq.~(\ref{2.53}).
When the atoms are initially prepared in superposition states
$|\pm \rangle$ $\!=$ $\!2^{-1/2}$ $\!(|g\rangle$ $\!\pm$ $e^{-i\psi}$
$\!|e\rangle )$ and the excited-state occupation probabilities
$P_{e}^{\pm}(t)$ are measured as functions of time, then the cavity-mode
density-matrix elements $\varrho_{n\,n+k}$ can be determined (Vogel, W.,
Welsch and Leine [1987]). To be more specific, it can be shown
that the difference $P_{e}^{-}(t)$ $\!-$ $\!P_{e}^{+}(t)$ reads as
\begin{equation}
\label{3.40}
P_{e}^{-}(t) - P_{e}^{+}(t) = 2 \sum_{n=0}^{\infty} a_{n}^{(k)}
\sin\!\left(
\Omega_{n\,n+k} t
\right) ,
\end{equation}
where
$\Omega_{n\,n+k}$ $\!=$ $\!2 \kappa^{(k)}$ $\!\{(n\!+\!1)$
$\!\cdots$ $\!(n\!+\!k)\}^{1/2}$,
and
\begin{equation}
\label{3.41}
a_{n}^{(k)}
= \frac{e^{i\psi}}{2i}\,\varrho_{n\,n+k} + {\rm c.c.};
\end{equation}
i.e., the off-diagonal density-matrix elements $\varrho_{n\,n+k}$ can
be obtained directly from the coefficients $a_{n}^{(k)}$ for two phases
$\psi$, such as $\psi$ $\!=$ $\!0$ and $\psi$ $\!=$ $\!\pi/2$.
Provided that the interaction time $t$ can be varied in a sufficiently
large interval $(0,T)$, the Fourier transform of $P_{e}^{-}(t)$ $\!-$
$\!P_{e}^{+}(t)$ consits of sharp peaks, whose values yield the
sought coefficients $a_{n}^{(k)}$ as\footnote{If $T$ is not large enough,
   then the peaks in the Fourier integral contain non-negligible contributions
   of the tails of the corresponding sinc functions. In this case, the
   coefficients $a_{n}^{(k)}$ can be calculated from a set of linear
   equations obtained from eq.~(\ref{3.40}) for different times.}
\begin{equation}
\label{3.42}
a_{n}^{(k)} = \frac{2}{T} \int_{0}^{T} {\rm d}t \,
\sin\!\left(
\Omega_{n\,n+k} t
\right)
\left[P_{e}^{-}(t) - P_{e}^{+}(t)\right]
\end{equation}
($T$ $\!\to$ $\!\infty$). To measure the diagonal density-matrix elements
$\varrho_{nn}$, it is sufficient to prepare the atom in the
excited state, $P_{e}(t)|_{t=0}$ $\!=$ $\!1$,
and observe (for arbitrary $k$) the atomic-state inversion
$\Delta P$ $\!=$
$\!P_{e}$ $\!-$ $\!P_{g}$ $\!=$ $\!2P_{e}$ $\!-$ $\!1$,
\begin{equation}
\label{3.44}
\Delta P(t) = \sum_{n=0}^{\infty}
\varrho_{nn} \cos\!\left(
\Omega_{n\,n+k} t
\right) ,
\end{equation}
from which $\varrho_{nn}$ can be obtained by Fourier
transformation.\footnote{For $k$ $\!=$ $\!1$ very precise measurements
   of the Rabi oscillations have been performed recently (Brune,
   Schmidt--Kaler, Maali, Dreyer, Hagley, Raimond and Haroche [1996]),
   the peaked structure of the Fourier-transformed data being interpreted
   as a direct experimental verification of field quantization
   in a cavity.}

In cavity QED the $1$-photon Jaynes--Cummings model is typically
realized, so that the method -- also called {\em quantum state
endoscopy} -- does not yield the off-diagonal
density-matrix elements $\varrho_{n\,n+k}$ with \mbox{$k$ $\!>$ $\!1$}.
When the quantum state is {\em a priori} known to be a pure
state such that $\varrho_{mn}$ is given by $\varrho_{mn}$
$\!=$ $c_{_{m}}c^{\ast}_{n}$ with $c_{_{m}}c^{\ast}_{m+1}$
$\!\not\equiv$ $\!0$ $\!\forall$ $\!m$,\footnote{This condition is
   not satisfied, e.g., for even and odd coherent states as typical
   examples of {\em Schr\"{o}dinger-cat\/}-like states. For even and odd
   coherent states, see Dodonov, Malkin and Man'ko, V.I., [1974].
   For a review of Schr\"{o}dinger cats, see Bu\v{z}ek and Knight [1995].}
then eq.~(\ref{3.40}) [together with eq.~(\ref{3.41})] can be taken
at a sufficiently large number of time points (and at least at
two phases) in order to obtain [after truncating the state at a sufficiently
large photon number $n_{\rm max}$ according to eq.~(\ref{3.19a})]
a system of conditional equations for the expansion coefficients $c_{m}$,
which can be solved numerically (Bardroff, Mayr and Schleich [1995];
Bardroff, Mayr, Schleich, Domokos, Brune, Raimond and Haroche
[1996]).

The reconstruction problem for arbitrary quantum states
can be solved by performing a displacement of the
initial state of the cavity field such that $\hat{\varrho}$ is
replaced with $\hat{D}^{\dagger}(\alpha)\hat{\varrho}\hat{D}(\alpha)$,
and hence ($k$ $\!=$ $\!1$)
\begin{equation}
\label{3.45}
\Delta P(t) = \sum_{n=0}^{\infty}
\varrho_{nn}(\alpha)
\cos\!\left(\Omega\sqrt{n\!+\!1}\,t \right)
\end{equation}
($\Omega$ $\!=$ $\!2\kappa$, $\kappa$ $\!\equiv$ $\!\kappa^{(1)}$),
where $\varrho_{nn}(\alpha)$ $\!=$
$\!\langle n|\hat{D}^{\dagger}(\alpha)\hat{\varrho}\hat{D}(\alpha) |n\rangle$
$\!=$ $\langle n,\alpha|\hat{\varrho}|n,\alpha\rangle$.
Now $\varrho_{nn}(\alpha)$ can again be obtained from $\Delta P(t)$
by Fourier transformation, and from $\varrho_{nn}(\alpha)$ the quantum
state of the cavity mode can be obtained, applying, e.g., the methods
outlined in \S\S~\ref{sec3.3.2} and \ref{sec3.5}. Alternatively,
the quantum state can also be reconstructed when the interaction time
is left fixed and only $\alpha$ $\!=$ $\!|\alpha|e^{i\varphi}$
is varied (Bodendorf, Antesberger, Kim, M.S., and Walther [1998]). Regarding
the measured atomic occupation probabilities as functions of
$\alpha$ $\!=$ $\!|\alpha|e^{i\varphi}$, $\Delta P(t)$ $\!=$
$\!\Delta P(t,\alpha)$, and introducing for chosen $t$ and
$|\alpha|$ the Fourier coefficients
\begin{equation}
\label{3.46}
\Delta P^{s}(t,|\alpha|) = \frac{1}{2\pi}
\int_{0}^{2\pi} {\rm d}\varphi \, e^{is\varphi} \Delta P(t,\alpha)
\end{equation}
($s$ $\!=$ $\!0,1,2,\ldots$), it can be shown that they are related
to the density-matrix elements by equations of the form of
\begin{equation}
\label{3.47}
\Delta P^{s}(t,|\alpha|)
= \sum_{n}^{\infty} Y_{n}^{s}(t,|\alpha|)\,\varrho_{n+s\,n}
\end{equation}
(for explicit expressions for $Y_{n}^{s}(t,|\alpha|)$, see
Bodendorf, Antesberger, Kim, M.S., and Walther [1998]). Inverting
eq.~(\ref{3.47}), which resembles, in a sense, eq.~(\ref{3.24}) in
\S~\ref{sec3.3.2}, for each value of $s$ then yields the density matrix
of the cavity mode. Similarly to eq.~(\ref{3.24}), the inversion can be
carried out numerically; e.g., by means of least-squares inversion
(Appendix \ref{app4}), choosing an appropriate set of values of $|\alpha|$.

In the two-mode {\em nonlinear atomic homodyne detection} scheme
(Wilkens and Mey\-stre [1991]) it is assumed
that the signal cavity mode is mixed with a local-oscillator cavity
mode according to the interaction Hamiltonian
\begin{equation}
\label{3.47a}
\hat{H}' = \hbar\kappa
\left[\hat{\sigma}_{+}(\hat{a} + \hat{a}_{L})
+ (\hat{a}^{\dagger} + \hat{a}_{L}^{\dagger})
\hat{\sigma}_{-}\right].
\end{equation}
Equation (\ref{3.45}) can then be found treating the effect of
the local oscillator semiclassically (i.e., replacing the operator
$\hat{a}_{L}$ with a $c$ number
$\alpha_{L}$, $\hat{a}_{L}$ $\!\to$ $\!\alpha_{L}$).
In this case the scheme is obviously equivalent to an initial displacement
$\alpha$ $\!=$ $\!-\alpha_{L}$
of the density operator of the cavity mode,
so that the atomic-state inversion
$\Delta P(t)$ is exactly given by eq.~(\ref{3.45}).
In particular, when
$|\alpha_{L}|$
is sufficiently large, then
$\Delta P(t)$
can be rewritten as
\begin{equation}
\label{3.48}
\Delta P(t) =
{\textstyle \frac{1}{2}}
\left[ e^{i2\kappa t |\alpha_{L}|}
\,\Phi\!\left (ie^{i\varphi_{L}}\kappa t \right)
+ {\rm c.c.}
\right].
\end{equation}
In other words, for
$|\alpha_{L}|$
$\!\to$ $\!\infty$ the atomic occupation probabilities $P_{e(g)}(t)$
can be related directly to the characteristic function $\Phi(\beta)$
of the Wigner function $W(\beta)$ of the cavity mode. Varying the
interaction time and the phase
$\varphi_{L}$ of $\alpha_{L}$,
the whole function
$\Phi(\beta)$ can be scanned in principle. Knowing
$\Phi(\beta)$, the Wigner function can then be obtained by Fourier
transformation.\footnote{Recall
   that in the dispersive regime the Wigner function can be measured
   directly (Lutterbach and Davidovich [1997]; see \S~\ref{sec2.4}),
   without any reconstruction algorithm.}
Since for appropriately
chosen arguments $\Phi(\beta)$ is nothing but the characteristic
function of the quadrature-component distribution $p(x,\vartheta)$
for all values of $\vartheta$ within a $\pi$ interval (see
Appendix \ref{app2.5}), the density matrix in both a quadrature-component basis
and the photon number basis can be reconstructed straightforwardly
from eq.~(\ref{3.48}) (\S\S~\ref{sec3.2} and \ref{sec3.3.1}).
Later it was found that the semiclassical treatment
of the local oscillator restricts the time scale to times less than a
vacuum Rabi period, because of the quantum fluctuations in the
local-oscillator cavity mode
(Zaugg, Wilkens and Meystre [1993]; Dutra, Knight and Moya-Cessa [1993]),
and it was shown that this difficulty can be
overcome when the atoms are weakly coupled to the local oscillator
but strongly coupled to the signal (Dutra and Knight [1994]).

{\em A priori} knowledge on the quantum state to be measured
is required in {\em magnetic tomography} (Walser, Cirac and Zoller
[1996]). Here, the idea of quantum state mapping between multilevel atoms
and cavity modes prepared in truncated states (Parkins, Marte, Zoller,
Carnal and Kimble [1995]) is
combined with a tomography of atomic angular momentum states by Stern-Gerlach
measurements (Newton and Young [1968]).
It is assumed that an angular-momentum degenerate two-level atom
passes adiabatically through the spatial
profile of a classical laser beam [Rabi frequency: $\Omega(t)$] and,
with a spatiotemporal displacement $\tau$ $\!>$ $\!0$, through
the profile of a quantized cavity mode [atom--cavity coupling:
$g(t$ $\!-$ $\!\tau)$] such that the coupled atom--cavity system
evolves according to the time-dependent Hamiltonian
\begin{eqnarray}
\label{3.52}
\lefteqn{
\hat{H}(t) = \hbar\omega\hat{a}^{\dagger}\hat{a}
+\hbar\omega_{eg}\sum_{m_{e}=-J_{e}}^{J_{e}}
|J_{e},m_{e}\rangle\langle J_{e},m_{e}|
}
\nonumber \\ && \hspace{2ex}
-i\hbar\Omega(t)\left(e^{i\omega_{L}t}\hat{A}_{1}
                  -\hat{A}_{1}^{\dagger}e^{-i\omega_{L}t}\right)
+i\hbar g(t\!-\!\tau)\left(\hat{a}^{\dagger}\hat{A}_{0}
                  -\hat{A}_{0}^{\dagger}\,\hat{a}\right),
\end{eqnarray}
where the atomic deexcitation operators $\hat{A}_{0},\hat{A}_{\pm 1}$
are defined by
\begin{equation}
\label{3.53}
\hat{A}_{\sigma} = \sum_{|m_{g}|\leq J_{g},\,|m_{e}|\leq J_{e}}
|J_{g},m_{g}\rangle\langle J_{e},m_{e}|
\,C_{\sigma,m_{g},m_{e}}^{1,J_{g},J_{e}},
\end{equation}
$C_{\sigma,m_{g},m_{e}}^{1,J_{g},J_{e}}$ being Clebsch--Gordan
coefficients.

It can be shown that
if the time-de\-pen\-dent change of the Hamiltonian during the total
interaction time is much less than the characteristic transition
energies and if the delay and shape of the pulse sequences are
chosen such that
\begin{equation}
\label{3.54}
0 \stackrel{-\infty\leftarrow t}{\longleftarrow}
g(t\!-\!\tau)/\Omega(t)
\stackrel{t\to +\infty}{\longrightarrow} \infty ,
\end{equation}
then a coupled atom--cavity-field density operator
$\hat{\varrho}_{AF}$ that can be factorized initially into a pure
atomic state and a field state containing less than $2J_{g}$ photons
will be mapped to a product of atomic ground state superpositions
and the cavity vacuum
\begin{equation}
\label{3.55}
|J_{g},J_{g}\!-\!1\rangle \langle J_{g},J_{g}\!-\!1|
\otimes \hat{\varrho}_{F}
\stackrel{-\infty\leftarrow t}{\longleftarrow}
\hat{\varrho}_{AF}(t)
\stackrel{t\to +\infty}{\longrightarrow}
\hat{\varrho}_{A} \otimes |0\rangle\langle 0|,
\end{equation}
with
\begin{equation}
\label{3.56}
\hat{\varrho}_{F} = \sum_{m,n=0}^{2J_{g}-1}
(\varrho_{F})_{mn}|m\rangle\langle n|
\end{equation}
and\footnote{Note that with reverse adiabatic passage, an internal
   atomic state is preapred uniquely by reading out the cavity state.}
\begin{equation}
\label{3.57}
\hat{\varrho}_{A} = \sum_{m,n=0}^{2J_{g}-1}
s_{m}s_{n}(\varrho_{F})_{mn}
|J_{g},J_{g}\!-\!m\!-\!1\rangle\langle J_{g},J_{g}\!-\!n\!-\!1|,
\end{equation}
$s_{m},s_{n}$ being possible sign changes. Hence, the original
cavity-field density matrix $\langle m|\hat{\varrho}_{F}|n\rangle$ is
known if the final atomic density matrix
\mbox{$\langle J_{g},J_{g}$ $\!-$ $\!m$ $\!-$ $\!1|\hat{\varrho}_{A}
|J_{g},J_{g}$ $\!-$ $\!n$ $\!-$ $\!1\rangle$} is known. Atomic states
of this type can be determined from magnetic dipole measurements
using conventional Stern--Gerlach techniques (see \S~4.5).


\subsubsection{Atomic beam deflection}
\label{sec3.6.2}

When a cavity mode is known to be in a pure state, then
the expansion coefficients of the state in the photon-number
basis can be inferred from the measured deflection of two-level
probe atoms during their passage through the cavity
(Freyberger and Herkommer [1994]; for an application of related
schemes to the reconstruction of the transverse motional quantum state of
two-level atoms, see \S~\ref{sec4.5.1}).
A narrow slit put in front
of one node of the standing wave (see Fig.~\ref{fig2.8})
transmits the atoms only in a small region $\Delta x$ $\!\ll$
$\!\lambda$ centered around \mbox{$x$ $\!=$ $\!0$}
($x$, atomic center-of-mass position; $\lambda$, wavelength of the mode).
The dependence on $\hat{x}$ of the coupling strength $\kappa$ $\!=$
$\kappa(\hat{x})$ $\!=$ $\!\kappa'\sin(k\hat{x})$ in eq.~(\ref{2.53})
can then be approximated by $\kappa$ $\!=$ $\!\kappa'k \hat{x}$,
$k$ $\!=$ $\!2\pi /\lambda$, so that the interaction Hamiltonian
of the combined system reads as
\begin{equation}
\label{3.49}
\hat{H}' = \hbar \kappa' k \hat{x}
\left( \hat{\sigma}_{+}\hat{a} + \hat{a}^{\dagger} \hat{\sigma}_{-} \right) .
\end{equation}
It is further assumed that the system is initially prepared in a state
$|\Psi\rangle$ $\!=$ $\!|\Psi_{F}\rangle|\Psi_{A}\rangle$, where
\begin{equation}
\label{3.50}
|\Psi_{F}\rangle = \sum_{n=0}^{\infty} c_{n}|n\rangle,
\quad
|\Psi_{A}\rangle = 2^{-1/2}\int {\rm d}x \, \psi(x) \, |x\rangle \left(
|g\rangle + c e^{i\varphi} |e\rangle \right).
\end{equation}
The subscripts $F$ and $A$ label the states of the cavity field and the atom,
respectively, $\psi(x)$ being the initial wave function of the atomic
center-of-mass motion. Assuming $c$ $\!=$ $\!1$ and
calculating the temporal evolution of the state vector in the
Raman--Nath regime,\footnote{In this approximation it is assumed that
   the transverse displacement of the atoms during the interaction time
   $t$ is small compared to the wavelength of the field, so that the
   kinetic energy of the atomic center-of-mass motion can be disregarded
   in the Hamiltonian of the combined system.
   For times longer than the interaction time the now deflected atoms
   move freely, and the spatial distribution of the atoms on a screen put up
   far away from the resonator is a picture of their momentum distribution
   at the moment when they leave the interaction zone.}
it can be shown
that the atomic momentum probability density for the interaction time $t$
is given by
\begin{eqnarray}
\label{3.51}
\lefteqn{
P(p,t) = {\textstyle\frac{1}{4}}\hbar k \sum_{n=1}^{\infty}
|\underline{\psi}(p\!+\!\sqrt{n}\hbar k\kappa't)|^{2}
|c_{n}+e^{i\varphi}c_{n\!-\!1}|^{2}
} \nonumber \\ \hspace{5ex} &&
+ {\textstyle\frac{1}{4}}\hbar k  \sum_{n=1}^{\infty}
|\underline{\psi}(p\!-\!\sqrt{n}\hbar k\kappa't)|^{2}
|c_{n}-e^{i\varphi}c_{n\!-\!1}|^{2}
+ \, {\textstyle\frac{1}{2}} \hbar k |\underline{\psi}(p)|^{2} |c_{0}|^{2} ,
\hspace*{10ex}
\end{eqnarray}
where $\underline\psi(p)$ is the initial wave function of the atomic
center-of-mass motion in the momentum basis. For an initial
position distribution $|\psi(x)|^{2}$, such as a Gaussian, with
$\Delta x$ $\!\ll$ $\!\lambda$, and sufficiently strong interaction,
i.e., $\kappa't$ $\!\to$ $\!\infty$, the corresponding shifted momentum
distribution $|\underline{\psi}(p$ $\!\pm$ $\!\sqrt{n}\hbar k\kappa't)|^{2}$
becomes sharply peaked at $p$ $\!=$ $\!\mp$ $\!\sqrt{n}\hbar k\kappa't$,
$n$ $\!=$ $1,2,\ldots$. In this case, $|\underline{\psi}(p)|^{2}$ and
$|\underline{\psi}(p$ $\!\pm$ $\!\sqrt{n}\hbar k\kappa't)|^{2}$
obviously select in the measured momentum distribution $P(p,t)$ a
sequence of peaks whose heights are proportional to $|c_{0}|^{2}$ and
$|c_{n}$ $\!\pm$ $\!e^{i\varphi}c_{n\!-\!1}|^{2}$, respectively, for
$n$ $\!=$ $\!0$ and $n$ $\!>$ $\!0$. In this way $|c_{0}|^2$ and
-- if the measurement is performed at two phases $\varphi$ -- the
sequence of products $c_{0}^{\ast}c_{1}$, $c_{1}^{\ast}c_{2}$, \ldots
can be obtained, from which (after truncating the state at
a sufficiently large photon number $n_{\rm max}$) the expansion
coefficients $c_{n}$ can be calculated easily (up to an unimportant
overall phase factor).\footnote{For an analysis of a diffuse peak
   structure, see Freyberger and Herkommer [1994].}

To allow the cavity mode to be prepared in a mixed state, a
running reference field aligned along the $y$-axis (i.e., perpendicular
with respect to the standig-wave cavity field aligned along the $x$-axis
and the $z$-direction of the moving atoms) can be included in the scheme
such that the atoms initially prepared in the electronic ground state
simultaneously interact with the two fields
(Schneider, Herkommer, Leonhardt and Schleich [1997]). In the
dispersive regime the effective interaction Hamiltonian for the system
can then [on extending eq.~(\ref{2.54}) to two modes] be given by
\begin{eqnarray}
\label{3.51a}
\lefteqn{
\hat{H}'' = \hbar \, \frac{\kappa^{2}(\hat{x})}{\delta \omega} \,
\left( \hat{\sigma}_{-} \hat{\sigma}_{+} \hat{a}^{\dagger}\hat{a}
- \hat{a}\hat{a}^{\dagger} \hat{\sigma}_{+} \hat{\sigma}_{-} \right)
+
\hbar \, \frac{\kappa_{L}^{2}}{\delta \omega} \,
\left( \hat{\sigma}_{-} \hat{\sigma}_{+} \hat{a}_{L}^{\dagger}\hat{a}_{L}
- \hat{a}_{L}\hat{a}_{L}^{\dagger} \hat{\sigma}_{+} \hat{\sigma}_{-} \right)
}
\nonumber \\ && \hspace{15ex}
+
\hbar \, \frac{\kappa(\hat{x})\kappa_{L}}{\delta \omega} \,
\left[ \left(\hat{\sigma}_{-} \hat{\sigma}_{+}
-\hat{\sigma}_{+} \hat{\sigma}_{-}\right)
\left(\hat{a}^{\dagger}\hat{a}_{L}e^{i\hat{y}}
+ e^{-i\hat{y}}\hat{a}_{L}^{\dagger}\hat{a}\right)\right].
\hspace*{5ex}
\end{eqnarray}
Here the index $L$ refers to the running reference field, which has the
same frequency as the cavity field and can play a similar role
as the local oscillator in a homodyne detector.
In the calculations it is assumed that the atoms
are initially prepared in a state $|\Psi_{A}\rangle$ of the form given
in eq.~(\ref{3.50}) with $c$ $\!=$ $\!0$,
and the reference field is prepared in a strong coherent state
and can be treated classically ($\hat{a}_{L}$
$\!\to$ $\alpha_{L}$ $\!=$ $\!|\alpha_{L}|e^{i\varphi_{L}}$,
\mbox{$|\alpha_{L}|$ $\!\to$ $\!\infty$)}.\footnote{Note that an atom
   that is initially in the electronic ground state ($c$ $\!=$ $\!0$)
   stays in the electronic ground state, because of the
   effective Hamiltonian (\ref{3.51a}).}
The cavity field can be
prepared in an arbitrary mixed state described by a density
operator $\hat{\varrho}$.
Again making the approximation
$\kappa(\hat{x})$ $\!\approx$ $\!\kappa'k\hat{x}$ and asuming
that the initial atomic wave function $\psi(x,y)$ factorizes
into a very narrow $y$-wave function $\cong\delta(y$ $\!-$ $\!y_{0})$
and a Gaussian in the $x$-direction,
it can be shown that the transverse momentum distribution $P(p,t)$ of the
atoms for an interaction time
$t$ is a (scaled) smeared quadrature-component
distribution $p(x,\varphi;\eta)$ of the cavity mode,
\begin{equation}
\label{3.51b}
P(p,t) = \frac{\eta}{\sqrt{2}|\tilde{\kappa}|} \,
p\!\left(x\!=\frac{p}{\sqrt{2}|\tilde{\kappa}|},\varphi;\eta\right),
\end{equation}
$\tilde{\kappa}$ $\!=$ $\!\Omega_{L}\kappa'\hbar k|\alpha_{L}|t/\delta$.
Here, the quantum efficiency $\eta$ with which the quadrature-component
distribution is measured is given by
\begin{equation}
\label{3.51c}
\eta = \left(1+
\frac{(\hbar k)^{2}}{2(\Delta x)^{2}\tilde{\kappa}^{2}}\right)^{-1},
\end{equation}
and the phase $\varphi$ is determined by the reference-field phase
and the position $y_{0}$, $\varphi$ $\!=$ $-\varphi_{L}$
$\!-$ $\!ky_{0}$. The result proves that the atomic probe carries the
(smeared) quadrature-component information on the cavity field that
corresponds to the reference angle $\varphi$. Hence,
after having measured the transverse momentum probability distribution
$P(p,t)$ of the deflected atoms for a sequence of phases $\varphi_{L}$
of the (local-oscillator) reference field, the phase-space function
$P(q,p;s$ $\!=$ $\!1$ $\!-$ $\!\eta^{-1})$ of the cavity
mode can be reconstructed from eq.~(\ref{3.5z}) by means of
tomographic imaging (\S~\ref{sec3.1}). Note that the larger
the effective coupling
constant $\tilde{\kappa}$ is, the closer is
$P(p,q;s)$ to the Wigner function. However the initial
position uncertainty must not exceed a fraction of the
wavelength of the cavity mode, because of the approximations
made. Needless to say, the interaction time $t$ must be
small compared to the decay time of the cavity mode.


\subsection{Alternative proposals}
\label{sec3.7}

Since the quantum state of a radiation field mode is known
when its density-matrix elements $\varrho_{mn}$ in the photon-number
basis are known, it was proposed to measure them directly (Steuernagel
and Vaccaro [1995]). The diagonal elements can be measured by direct
photodetection in principle. In order to measure the off-diagonal elements,
it was proposed to prepare a probe field in a superposition of
two photon-number states,
\begin{equation}
\label{3.58}
|a_{mn}\rangle = N_{a} ( |m\rangle + a\,|n\rangle),
\end{equation}
$N_{a}$ $\!=$ $\!(1$ $\!+$ $\!|a|^{2})^{-1/2}$,
and combine it with the signal mode at a beam splitter. The observed
joint photon-number probabiltity in the two output channels of the
beam splitter then reads as ($m$ $\!>$ $\!n$)
\begin{equation}
\label{3.59}
p_{q,k+m-q} \sim \langle b_{kl}|\hat{\varrho}|b_{kl}\rangle,
\end{equation}
$l$ $\!-$ $\!k$ $\!=$ $\!m$ $\!-$ $\!n$,
where $|b_{kl}\rangle$ is again a state of the type given in
eq.~(\ref{3.58}), so that
\begin{equation}
\label{3.60}
\langle b_{kl}|\hat{\varrho}|b_{kl}\rangle
= N_{b}^{2}\left(\varrho_{kk} + |b|^{2}\varrho_{ll}
+ b \varrho_{kl} + b^{\ast} \varrho_{kl}^{\ast} \right).
\end{equation}
When for chosen difference $k$ $\!-$ $\!l$ $\!=$ $\!n$ $\!-$ $\!m$ the
diagonal elements $\varrho_{ll}$ and $\varrho_{kk}$ are known from a
direct photon-number measurement, then the off-diagonal elements
$\varrho_{k\,k+m-n}$, $k$ $\!=0,1,2,\ldots$, can be obtained from two
(ensemble) measurements for different superposition parameters.  The
whole density matrix can then be obtained by means of a succession of
(ensemble) measurements, varying the difference \mbox{$m$ $\!-$ $\!n$}
of photon numbers in the reference state (\ref{3.58}) from measurement
to measurement. The presently unresolved problem, however, consists in
building an apparatus that prepares travelling waves in superpositions of
two photon-number states $|m\rangle$ and $|n\rangle$ for arbitrary
difference \mbox{$m$ $\!-$ $\!n$} in a controllable way.

For a radiation-field mode that is known to be prepared in a pure state
$|\Psi\rangle$ and that contains only a finite number of photons,
eq.~(\ref{3.19a}), it was proposed to extend the method of photon
chopping outlined in \S~\ref{sec2.1.7} to a complete determination of
the expansion coefficients $c_{n}$ (Paul, H., T\"{o}rm\"{a}, Kiss and
Jex [1996a, 1997a]).\footnote{For a proposal to extend
   the method to a two-mode signal field, see Paul, H., T\"{o}rm\"{a},
   Kiss and Jex [1997b].}
First the photon-number distribution
$p_{n}$ $\!=$ $\!|c_{n}|^2$ is determined according to eqs.~(\ref{2.51})
-- (\ref{2.51a}). Then the balanced $2N$-port apparatus
($N$ $\!\ge$ $\!n_{\rm max}$) is used for mixing the signal mode
with $N$ $\!-$ $\!2$ reference modes in the vacuum state and one reference
mode prepared in a coherent state $|\alpha\rangle$ (in place of the $N$
vacuum reference inputs in \S~\ref{sec2.1.7}).
Using the selected statistics when only photons in the first
$N/2$ output channels are observed, it can be shown that the probability
${p}_{n}(N)$ of detecting $n$ photons with at most one photon
in each channel is given by (Paul, H., T\"{o}rm\"{a}, Kiss and Jex [1996a])
\begin{equation}
\label{3.61}
p_{n}(N) = {N/2 \choose n} \frac{e^{-|\alpha|^{2}}}{N^{n}}
\left|\sum_{m=0}^{n} {n \choose m}  \sqrt{(n\!-\!m)!}
\, \alpha^{m} c_{n-m} \right|^{2}.
\end{equation}
Determining the coincidence probability distribution $p_{n}(N)$
for two coherent states with different phases, the phases $\varphi_{n}$ of
the expansion coefficients $c_{n}$ $\!=$ $\!|c_{n}|e^{i\varphi_{n}}$ can
be calculated step by step (for known amplitudes $|c_{n}|$ and up to an
overall unimportant phase) from the two measured coincidence-event
distributions. More involved formulas are obtained
in the case when the complete coincidence statistics is included in
the analysis; i.e., photons are allowed to be observed in any output
channel and realistic photodetection is considered (Paul, H., T\"{o}rm\"{a},
Kiss and Jex [1997a]). It turns out that the reconstruction scheme with
two coherent states requires photodetectors
that can discriminate between $0,1,2,\ldots$ photons, i.e., detectors that
have rather low quantum efficiency. Using avalanche photodiodes, which
can have high quantum efficiency, the reconstruction might become rather
involved, because additional reference beams with different phase
properties must be used. Similarly, when a mixed quantum state
is tried to be reconstructed, then extra reference beams (reference
phases) must be used, the maximum number of reference beams being limited
by the cutoff in the photon number.

It can also be shown (Bia{\l}ynicka-Birula and Bia{\l}ynicki-Birula [1994];
Vaccaro and Barnett [1995]) that
when a radiation-field mode is prepared in a pure state $|\Psi\rangle$
that is a finite superposition of photon-number states, eq.~(\ref{3.19a}),
then it can be reconstructed
from the photon-number distribution $p_{n}$ $\!=$ $|c_{n}|^{2}$ and the
(Pegg--Barnett)\footnote{See Pegg
   and Barnett [1988].}
truncated canonical phase distribution
$P_{PB}(\phi)$ $\!=$ $\!|\psi(\phi)|^{2}$, with\footnote{Note that
   $\psi(\phi)$ $\!=$ $\!\langle\Phi_{PB}(\phi)|\Psi\rangle$,
   where $|\Psi\rangle$ is given by eq.~(\ref{3.19a}), and the
   truncated phase state $|\Phi_{PB}(\phi)\rangle$ reads as
   $|\Phi_{PB}(\phi)\rangle$ $\!=$
   $\!(n_{\rm max}\!+\!1)^{-1/2}$ $\!\sum_{n=0}^{n_{\rm max}}$
   $\!e^{-in\phi} |n\rangle$ (cf. \S~\ref{sec3.8.3}).}
\begin{equation}
\label{3.62}
\psi(\phi)
= |\psi(\phi)|e^{i\lambda(\phi)}
= \frac{1}{\sqrt{n_{\rm max}\!+\!1}} \sum_{n=0}^{n_{\rm max}}
|c_{n}| e^{i\varphi_{n}} \, e^{in\phi}.
\end{equation}
Taking eq.~(\ref{3.62}) at $2(n_{\rm max}$ $\!+$ $\!1)$ values of
$\phi$, such as $\phi_{l}$ $\!=$ $l\pi/($ $\!n_{\rm max}$ $\!+$ $\!1)$,
the resulting equations can be regarded as conditional equations
for the unknown phases $\varphi_{n}$, provided that all absolute
values $|\psi(\phi_{l})|$ and $|c_{n}|$ are known. Apart from the
fact that the quantum state must be known to be a pure state, the
question remains of how to obtain the phase statistics (see also
\S~\ref{sec3.8.3}).


\subsection{Reconstruction of specific quantities}
\label{sec3.8}

Since the density matrix in any basis contains the full
information about the quantum state of the system under consideration,
all quantum-statistical properties can be inferred from it.
Let $\hat{F}$ be an operator whose expectation value,
\begin{equation}
\label{3.62a}
\langle\hat{F}\rangle = \sum_{mn} F_{nm}\,\varrho_{mn}\,,
\end{equation}
is desired to be determined. One may be tempted to calculate it from the
reconstructed density matrix (or another measurable quantum-state
representation). However, an experimentally determined density matrix
always suffers from various inaccuracies which can propagate (and increase)
in the calculation process (cf. \S~\ref{sec3.9.1}).
Therefore it may be advantageous to determine directly the quantities
of interest from the measured data, without reconstructing
the whole quantum state. In particular, substituting in eq.~(\ref{3.62a})
for $\varrho_{mn}$ the integral representation (\ref{3.13}) one
can try to obtain an integral representation
\begin{equation}
\label{3.62b}
\langle\hat{F}\rangle
=\int_{0}^{\pi} {\rm d}\varphi \int {\rm d}x
\, K_{F}(x,\varphi)\,p(x,\varphi),
\end{equation}
with
\begin{equation}
\label{3.62c}
K_{F}(x,\varphi)
= \sum_{mn} F_{nm}\,f_{mn}(x)\,e^{i(m-n)\varphi},
\end{equation}
suited for direct sampling of $\langle\hat{F}\rangle$ from the
quadrature component distributions $p(x,\varphi)$ [provided
that both the sampling function $K_{F}(x,\varphi)$ and the integral in
eq.~(\ref{3.62b}) exist].


\subsubsection{Normally ordered photonic moments}
\label{sec3.8.1}

It is often sufficient to know some moments of the
photon creation and destruction operators of a radiation-field mode
rather than its overall quantum state.
Let us again consider the measurement of the quadrature-component
probability distributions $p(x,\varphi)$ in balanced homodyning
(\S~\ref{sec2.1.2}).
Then the normally ordered moments of the creation and
destruction operators, $\langle \hat{a}^{\dagger n} \hat{a}^{m} \rangle$
can be related to $p(x,\varphi)$ as\footnote{Equation (\ref{3.63})
   can be proven correct if both sides are expressed in terms
   of the density matrix in the photon-number basis and standard summation
   rules for the Hermite polynomials are used.}
\begin{equation}
\label{3.63}
\langle \hat{a}^{\dagger n} \hat{a}^{m} \rangle =
\frac{n! m!}{\pi 2^{(n+m)/2}(n\! + \! m)!}
\int_{0}^{\pi} {\rm d}\varphi \int {\rm d}x \,
p(x,\varphi) \, {\rm H}_{n+m}(x) e^{i(m-n)\varphi} ,
\end{equation}
where ${\rm H}_{n}(x)$ is the Hermite polynomial
(Richter, Th., [1996b]).\footnote{A more involved transformation was
   suggested by Bia{\l}ynicka-Birula and Bia{\l}ynicki-Birula [1995];
   see footnote \ref{Bial}.}
Equation (\ref{3.63}) is just of the
type given in eq.~(\ref{3.62b}) and offers the possibility
of direct sampling of normally ordered moments
$\langle \hat{a}^{\dagger n} \hat{a}^{m} \rangle$ from the
homodyne data. It is worth noting that knowledge of $p(x,\varphi)$
at all phases within a $\pi$ interval is not necessary to
reconstruct $\langle \hat{a}^{\dagger n} \hat{a}^{m} \rangle$ exactly,
and therefore the $\varphi$ integral in eq.~(\ref{3.63}) can be replaced
with a sum. It was shown that any normally ordered moment
$\langle \hat{a}^{\dagger n} \hat{a}^{m} \rangle$ can already be
obtained from $p(x,\varphi)$ at \mbox{$N$ $\!=$ $\!n$ $\!+$ $\!m$
$\!+$ $\!1$} discrete different phases $\varphi_{k}$ (W\"{u}nsche
[1996b, 1997]).

The method is especially useful, e.g., for a
determination of the moments of photon number. Note that for finding
the mean number of photons $\langle \hat n \rangle$ from the
Fock-basis density matrix a relatively large number of measured
diagonal elements $\varrho_{nn}$ must be included into the calculation,
in general, each of which being determined with some error. After
calculation of the sum $\langle\hat{n}\rangle$ $\!=$
$\!\sum_{n}n\varrho_{nn}$ the error of the result
can be too large to be acceptable, and for higher-order moments severe
error amplification may be expected.
Equation (\ref{3.63}) can easily be extended to nonperfect detection
\mbox{($\eta$ $\!<$ $\!1$)}, since replacing $p(x,\varphi)$ with
$p(x,\varphi;\eta)$ simply yields $\eta^{(n+m)/2}
\langle \hat a^{\dagger n} \hat a^{m} \rangle$.\footnote{Recall
   that an imperfect detector can be regarded
   as a perfect detector with a beam splitter in front of it, so that
   the destruction operator $\hat{a}$ of the mode that is originally
   desired to be detected is transformed according to a beam-splitter
   transformation (\S~\ref{sec2.1.1}), \mbox{$\hat{a}(\eta)$ $\!=$
   $\!\sqrt{\eta}$ $\!\hat{a}$ $\!+$ $\!\sqrt{1-\eta}$ $\!\hat{b}$},
   where $\hat{b}$ is the photon destruction operator of a reference
   mode prepared in the vacuum state.}

Provided that the whole manifold of moments $\langle \hat a^{\dagger n}
\hat a^{m} \rangle$ has been determined, then the quantum state is
known in principle (W\"{u}nsche [1990, 1996b]; Lee [1992];
Herzog [1996b]).\footnote{This is also true for other
   than normally ordered moments, such as symmetrically ordered
   moments (Band and Park [1979]; Park, Band and Yourgrau [1980]).}
To be more specific, the density operator
can be expanded as [cf. eq.~(\ref{3.0})]
\begin{equation}
\label{3.64}
\hat \varrho = \sum_{k,l=0}^{\infty} \hat{a}_{k,l} \,
\langle \hat{a}^{\dagger k} \hat{a}^{l} \rangle,
\end{equation}
where
\begin{equation}
\label{3.65}
\hat{a}_{k,l} = \sum_{j=0}^{ \{ k,l \} }
\frac{(-1)^{j}  |l\!-\!j \rangle \langle k\!-\! j | }
{j! \sqrt{(k\!-\!j)! (l\!-\!j)!}}
\end{equation}
[$\{k,l\}$ $\!=$ $\!{\rm min}\,(k,l)$],
provided that $\langle \hat{a}^{\dagger k} \hat{a}^{l} \rangle$ exists
for all values of $k$ and $l$ and the series (\ref{3.64})
(in chosen basis) exists as well. Since all the moments
do not necessarily exist for any quantum state, and the series
need not necessarily converge for existing moments, the quantum-state
description in terms of density matrices is more universal than that
in terms of normally ordered moments.\footnote{An example of
   nonexisting normally ordered moments is realized by a quantum
   state whose photon-number distribution behaves like
   $p_{n}$ $\!\sim$ $\!n^{-3}$ for $n$ $\!\to$ $\!\infty$. Even
   though it is a normalizable state with finite energy, its moments
   $\langle \hat{n}^{k} \rangle$ do not exist for $k$ $\!\ge$ $\!2$.
   For a thermal state the relation $\langle \hat a^{\dagger n}
   \hat a^{m} \rangle$ $\!=$ $\!\delta_{nm}n!\bar{n}^{n}$ is valid
   ($\bar{n}$, mean photon number), which implies that the
   series (\ref{3.64}) for $\varrho_{mn}$ does not exist. Note that
   the problem of nonconvergence of the series (\ref{3.64}) may be
   overcome by analytic continuation.}

The extension of the method to the reconstruction of normally-ordered
moments of multimode fields from joint quadrature-component distributions
is straightforward. Moreover, normally-ordered moments of multimode fields
can also be reconstructed from combined distributions considered
in \S~\ref{sec2.1.3} (Opatrn\'{y}, Welsch and Vogel, W., [1996, 1997a];
McAlister and Raymer [1997a,b]; Richter, Th., [1997a]). For simplicity,
let us restrict
attention here to two-mode moments $\langle \hat{a}_{1}^{\dagger n_{1}}
\hat{a}_{2}^{\dagger n_{2}} \hat{a}_{1}^{m_{1}} \hat{a}_{2}^{m_{2}} \rangle$
(the subscripts $1$ and $2$ refer to the two modes) and assume that
the weighted sum $\hat{x}$ $\!=$ $\!\hat{x}_{1}(\varphi_{1})$
$\!\cos\alpha$ $\!+$ $\!\hat{x}_{2}(\varphi_{2})$ $\!\sin\alpha$,
eq.~(\ref{2.32}), is measured in a homodyne detection scheme
outlined in \S~\ref{sec2.1.3} (see Fig.~\ref{fig2.4}).
Measuring the moment $\langle\hat{x}^{n}\rangle$ for $n$ $\!+$ $\!1$
values of $\alpha$, a set of $n$ $\!+$ $\!1$ linear
algebraic equations can be obtained whose solution yields
the two-mode quadrature-component moments
$\langle\hat{x}_{1}^{n-k}(\varphi_{1})
\hat{x}_{2}^{k}(\varphi_{2})\rangle$ ($k$ $\!=$ $\!0,1,2,\ldots,n$).
Varying $\varphi_{1}$ and $\varphi_{2}$, the procedure can be repeated
to obtain the dependence on $\varphi_{1}$ and $\varphi_{2}$ of
the two-mode quadrature-component moments. These can then be used
to determine the normally ordered moments of the photon creation and
destruction operators of the two modes step by step. In a closed
form, $\langle \hat{a}_{1}^{\dagger n_{1}} \hat{a}_{2}^{\dagger n_{2}}
\hat{a}_{1}^{m_{1}} \hat{a}_{2}^{m_{2}} \rangle$ can be
obtained from the combined quadrature-component distribution
$p_{S}(x,\alpha,\varphi_{1},\varphi_{2})$ as
\begin{eqnarray}
\label{3.66}
\lefteqn{
\left\langle \hat a_{1}^{\dagger
n_{1}} \hat a_{2}^{\dagger n_{2}} \hat a_{1}^{m_{1}} \hat
a_{2}^{m_{2}} \right\rangle =
C_{n_{1}m_{1}} C_{n_{2}m_{2}}
\int {\rm d}x \int_{\Omega} \! {\rm d}\alpha \int_{0}^{\pi}
\! {\rm d}\varphi_{1} \int_{0}^{\pi} \! {\rm d}\varphi_{2} \,
\Big[ p_{\rm s}(x,\alpha ,\varphi_{1},\varphi_{2})
}
\nonumber \\[.5ex] && \hspace{15ex}
\times \,
H_{n_{1}\!+\!n_{2}\!+\!m_{1}\!+\!m_{2}}(x)
\, F_{n_{2}\!+\!m_{2}}^{(n_{1}\!+\!n_{2}\!+\!m_{1}\!+\!m_{2})} (\alpha)
\, e^{i(n_{1}-m_{1})} e^{i(n_{2}-m_{2})}
\Big]
\hspace*{5ex}
\end{eqnarray}
[$C_{nm}$ $\!=$ $\!n!m!/(\pi 2^{(n+m)/2}$ $\!(n$ $\!+$ $\!m)!)$].
The functions $F_{k}^{(l)}(\alpha)$ form a biorthonormal system to the
functions $G_{k}^{(l)}(\alpha)$ = ${l \choose k}$ cos$^{l-k}\alpha$
sin$^{k}\alpha$ in some $\Omega$ interval, so that $\int_{\Omega}$
$\!\rm{d}\alpha$ $\!F_{k}^{(l)}(\alpha)$ $\!G_{k'}^{(l)}(\alpha)$
$\!=$ $\!\delta_{kk'}$. Note that since for given $l$ there is a
finite number of functions $G_{k}^{(l)}(\alpha)$, the system of
functions $F_{k}^{(l)}(\alpha)$ is not uniquely determined and
can be chosen in different ways.
In particular when only the phase difference
$\Delta \varphi$ $\!=$ $\!\varphi_{2}$ $\!-$ $\!\varphi_{1}$
is controlled and the overall phase $\varphi_{1}$ is
averaged out (this is the case when the signal and the local oscillator
stem from different sources), then the moments
$\langle \hat{a}_{1}^{\dagger n_{1}} \hat{a}_{2}^{\dagger n_{2}}
\hat{a}_{1}^{m_{1}} \hat{a}_{2}^{m_{2}} \rangle$ can be reconstructed
for $n_{1}$ $\!-$ $\!m_{1}$ $\!=$ $\!m_{2}$ $\!-$ $\!n_{2}$.
If both phases $\varphi_{1}$ and $\varphi_{2}$ are averaged out, then
those moments $\langle \hat{a}_{1}^{\dagger n_{1}} \hat{a}_{2}^{\dagger n_{2}}
\hat{a}_{1}^{n_{1}} \hat{a}_{2}^{n_{2}} \rangle$ can still be obtained,
which carry the information about the photon-number correlation
in the two modes.

\begin{figure}[htb]
 \unitlength=1cm
 \begin{center}
 \begin{picture}(6,5)
 \put(0,0){
 \includegraphics{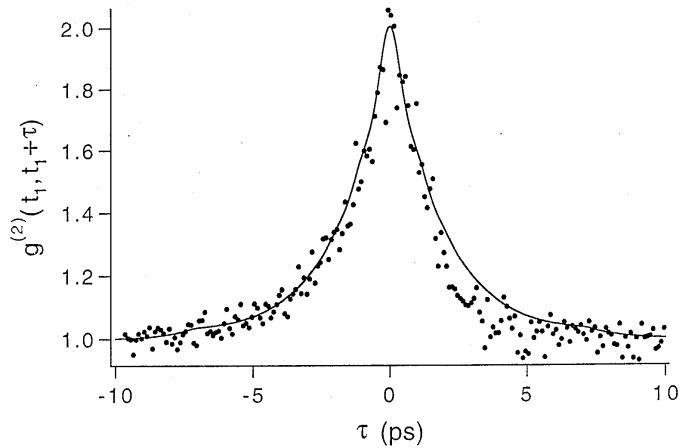}}
 \end{picture}
 \end{center}
\begin{center}
\protect\parbox{.9\textwidth}{
\caption{
The second-order coherence, eq.~(\protect\ref{3.67}), experimentally
determined via balanced four-port homodyne detection (dots) and
from the measured optical spectrum (solid line). The value
of $t_{1}$ is set to occur near the maximum of the signal pulse.
(After McAlister and Raymer [1997a].)
\label{fig3.4}
}
}
\end{center}
\end{figure}%
The method was used to experimentally demonstrate the determination of the
ultrafast two-time photon number correlation of a $4$~ns optical pulse
(McAlister and Raymer [1997a]). In the experiment, the local-oscillator
pulses are derived from a Ti:saphire-based laser system which
generates ultrashort, near transform-limited pulses ($150$~fs) at a
wavelength of $830$~nm and a repetition rate of $1$~kHz. The signal is
from a single-spatial-mode superluminescent diode. The broadband
emission at $830$~nm is spectrally filtered to produce a $4$~ns
pulse having a $0.22$~nm spectral width. For each value of
the relative delay $\Delta t$ $\!=$ $\!t_{1}$ $\!-$ $\!t_{2}$
between two local-oscillator
pulses the phase-averaged combined quadrature-component distribution
$\bar{p}_{S}(x,\alpha)$ $\!=$ $\!(2\pi)^{-2}\int{\rm d}\varphi_{1}$
$\!\int{\rm d}\varphi_{2}$ $\!p_{S}(x,\alpha,\varphi_{1},\varphi_{2})$
is measured for three different values of $\alpha$: $0$, $\pi/2$ and $\pi/4$.
   From these the normalized second-order coherence function
\begin{equation}
\label{3.67}
g^{(2)}(t_{1},t_{2}) =
\frac{ \langle \hat{a}_{1}^{\dagger} \hat{a}_{2}^{\dagger}
\hat{a}_{1} \hat{a}_{2} \rangle }
{ \langle \hat{a}_{1}^{\dagger} \hat{a}_{1} \rangle
\langle \hat{a}_{2}^{\dagger} \hat{a}_{2} \rangle}  \, ,
\end{equation}
is computed ($\hat{a}_{1}$ and $\hat{a}_{2}$ being the photon
destruction operators of the spatial-temporal modes defined by
the local-oscillator pulses centered at times
$t_{1}$ and $t_{2}$).
This experiment represents an extension of
measurements of the Hanbury-Brown--Twiss
correlations\footnote{See Hanbury Brown and Twiss [1956a,b,1957a,b];
   for a more discussion of the Hanbury Brown and Twiss
   experiments, see Pe\v{r}ina [1985], Mandel and Wolf [1995].}
to a sub-picosecond region.
Results are shown in Fig.~\ref{fig3.4}.


\subsubsection{Quantities admitting normal-order expansion}
\label{sec3.8.2}

The basic relation (\ref{3.63}) can also be used [similar to
eqs.~(\ref{3.62a}) -- (\ref{3.62c})] to find
sampling formulas for the mean values of quantities that can be given
in terms of normally ordered moments of photon creation and destruction
operators (D'Ariano [1997b]). Let us consider an operator $\hat{F}$
and assume that it can be given by a normal-order expansion as
\begin{equation}
\label{3.68}
\hat{F} = \sum_{n,m=0}^{\infty} f_{nm}
\, \hat{a}^{\dagger n} \hat{a}^{m}.
\end{equation}
The mean value $\langle \hat{F} \rangle$ can then be obtained
from $p(x,\varphi;\eta)$ according to\footnote{Recall that when
   the $c$-number function $F(\alpha;s)$
   that is associated with an operator $\hat{F}$ in $s$ order
   exists for $s$ $\!\leq$ $\!-1$, then $\langle\hat{F}\rangle$
   can be obtained from the phase-space function $P(\alpha;s)$
   (measurable, e.g., in balanced eight-port homodyning; see
   \S~\ref{sec2.1.4}) according to eq.~(\ref{A2.17b}), provided
   that the integral exists.}
\begin{equation}
\label{3.69}
\langle \hat{F} \rangle
= \int_{0}^{\pi} {\rm d}\varphi \int {\rm d}x
\, K_{F}(x,\varphi;\eta) \, p(x,\varphi;\eta) ,
\end{equation}
where the integral kernel
\begin{equation}
\label{3.69a}
K_{F}(x,\varphi;\eta)
= {\rm Tr}\left[\hat{F}\hat{K}(x,\varphi;\eta)\right]
\end{equation}
[$\hat{K}(x,\varphi;\eta)$ being defined in eq.~(\ref{3.16a})]
can be given by
\begin{equation}
\label{3.70}
K_{F}(x,\varphi;\eta) = \frac{1}{\pi\sqrt{2\pi\eta^{-1}}}
\int {\rm d}y \,e^{-\eta y^{2}/2}
G_{F}\!\left( y \!+\! i\sqrt{2\eta^{-1}} \, x ,\varphi \right) ,
\end{equation}
with
\begin{equation}
\label{3.71}
G_{F}(z,\varphi) = \sum_{n,m=0}^{\infty}
f_{nm} {n\!+\!m \choose m}^{-1}
(-iz)^{n+m} e^{i(m-n)\varphi} .
\end{equation}
Here it is assumed that the integral kernel $K_{F}(x,\varphi;\eta)$
exists and the $x$ integral in eq.~(\ref{3.69}) exists as
well.\footnote{Equations (\ref{3.69})
   -- (\ref{3.71}) can be obtained by taking the average of eq.~(\ref{3.68}),
   substituting for $\langle\hat{a}^{\dagger n} \hat{a}^{m}\rangle$
   the right-hand side (multiplied by $\eta^{-(n+m)/2}$) of eq.~(\ref{3.63}),
   and using the generating function of the Hermite polynomials with the
   argument $(i/\sqrt{2\eta}) ({\rm d}/{\rm d}z)$.}.
The integral kernel (\ref{3.70}) exists if for $z$ $\!\to$ $\!\infty$
the function $G_{F}(z,\varphi)$ grows slower than $e^{\eta z^{2}/2}$, so
that the integral in eq.~(\ref{3.70}) converges (for examples,
see D'Ariano [1997b]; D'Ariano and Paris [1997b]).\footnote{Since
   the integral kernel $K_{F}(x,\varphi;\eta)$ does not depend on $\varphi$
   if $\hat{F}$ is a function of the photon-number operator, this case
   can be regarded as the first realization of Helstrom's {\em quantum
   roulette wheel} (D'Ariano and Paris [1997b]).}
The integral kernel (\ref{3.70})
is therefore applicable to such quantum states whose
(smeared) quadrature-component distributions $p(x,\varphi;\eta)$
tend to zero sufficiently fast as $|x|$ goes to infinity such
that eq.~(\ref{3.69}) converges even if $K_{F}(x,\varphi;\eta)$
increases with $|x|$. At this point it should be noted that
$K_{F}(x,\varphi;\eta)$ is determined only up to a function
$\Theta(x,\varphi)$; i.e., $K_{F}(x,\varphi;\eta)$ in eq.~(\ref{3.69})
can by replaced with $K'_{F}(x,\varphi;\eta)$ $\!=$ $\!K_{F}(x,\varphi;\eta)$
$\!+$ $\!\Theta(x,\varphi)$ such that
\begin{equation}
\label{3.71a}
\int_{0}^{\pi} {\rm d}\varphi \int {\rm d}x \,
\Theta(x,\varphi) \, p(x,\varphi;\eta) = 0
\end{equation}
for any normalizable quantum state. Hence, if the integral kernel
$K_{F}(x,\varphi;\eta)$ that is obtained from eq.~(\ref{3.70})
is unbounded for $|x|$ $\!\to$ $\!\infty$ such that the $x$ integral in
eq.~(\ref{3.69}) does not exist for any normalizable quantum
state, it cannot be concluded that $\hat{F}$ cannot be
sampled from the quadrature-component distributions of any
normalizable quantum state, since a different, bounded kernel may exist.


\subsubsection{Canonical phase statistics}
\label{sec3.8.3}

The quantum-mechanical description of the phase and its measurement
has turned out to be troublesome and is still a matter of discussion.
Many papers have dealt with the problem and an extensive
literature is available (for reviews, see Luk\v{s} and
Pe\v{r}inov\'{a} [1994], Lynch [1995], Royer [1996], Pegg and Barnett
[1997]). Here we confine ourselves to the canonical phase that
is obtained in the attempt -- in close analogy to the classical description --
to decompose the photon destruction operator into
amplitude and phase such that $\hat{a}$ $\!=$ $\!\hat{E}\sqrt{\hat{n}}$.
The (nonorthogonal and unnormalizable) phase states,
\begin{equation}
\label{3.72}
|\phi \rangle = \sum_{n=0}^{\infty} e^{in\phi} |n\rangle
\end{equation}
(London [1926, 1927]),
which are the right-hand eigenstates of the one-sided unitary operator
$\hat{E}$, \mbox{$\hat{E}|\phi\rangle$ $\!=$ $\!e^{i\phi}|\phi\rangle$},
can then be used to define -- in the sense of a POM -- the canonical
phase distribution $P(\phi)$ $\!=$ $\langle\hat{\Pi}(\phi)\rangle$,
with $\hat{\Pi}(\phi)$ $\!=$ $\!|\phi\rangle\langle\phi|$.\footnote{For
   a two-mode orthogonal projector realization in the
   relative-photon-number basis, see Ban [1991a--d,
   1992, 1993]; Hradil [1993].}
Note that the Hermitian operators $\hat{C}$ $\!=$ $\!(\hat{E}$ $\!+$
$\!\hat{E}^{\dagger})/2$ and \mbox{$\hat{S}$ $\!=$ $\!(\hat{E}$ $\!-$
$\!\hat{E}^{\dagger})/(2i)$} are the Susskind--Glogower sine and cosine
operators (Susskind and Glogower [1964]).\footnote{For constructing
   the operators corresponding to classical phase-dependent
   quantities, see Bergou and Englert [1991].}

Various proposals have been made to measure $P(\phi)$
using homodyne detection or related schemes.\footnote{The proposal was
   made to regard the quadrature-component distribution at $x$ $\!=$
   $\!0$ as (unnormalized) phase distribution,
   $P(\phi)$ $\!\sim$ $\!p(x$ $\!=$ $\!0,\phi)$
   (Vogel, W., and Schleich [1991]), which yields a phase statistics that is
   quite different from the canonical phase statistics in general
   (for an improvement, see Bu\v{z}ek and Hillery [1996]).}
So the Wigner function
or the density matrix reconstructed from the homodyne data could
be used to infer the phase statistics in principle
(Beck, Smithey and Raymer [1993]; Smithey, Beck, Cooper and
Raymer [1993]; Breitenbach and Schiller [1997]).
Further it was proposed to obtain $P(\phi)$ $\!=$
$\langle\hat{\Pi}(\phi)\rangle$ from a homodyne measurement of (all) normally
ordered moments $\langle\hat{a}^{\dagger m+n}\hat{a}^{n}\rangle$,
because  $\hat{\Pi}(\phi)$ has a series expansion of the type given
in eq.~(\ref{3.68}),
\begin{equation}
\label{3.73}
\hat{\Pi}(\phi) = (2\pi)^{-1} \left[ 1 +
\sum_{m=1}^{\infty}\sum_{n=0}^{\infty}\left(
e^{im\phi}\hat{a}^{\dagger m+n}\hat{a}^{n}
+ e^{-im\phi}\hat{a}^{\dagger n}\hat{a}^{m+n}
\right)\right],
\end{equation}
(Bia{\l}ynicka-Birula and Bia{\l}ynicki-Birula [1995]).\footnote{Here
   it was proposed
   to obtain $\langle\hat{a}^{\dagger m+n}\hat{a}^{n}\rangle$ from
   the homodyne data via the $(2n$ $\!+$ $\!m)$th derivative of
   $\int_{0}^{2\pi}{\rm d}\varphi$ $\!e^{-im\varphi}
   e^{-\lambda^{2}/2}\langle\exp[i\lambda\hat{x}(\varphi$ $\!-$
   $\!\pi/2)]\rangle$.\label{Bial}}
Unfortunately $\langle\hat{\Pi}(\phi)\rangle$
cannot be given (even for $\eta$ $\!=$ $\!1$) by an integral
relation of the form (\ref{3.69}) suited to statistical sampling,
because the integral kernel does not exist (see footnote \ref{nonexist}).
It was therefore proposed to introduce parametrized phase distributions
$P(\phi,\epsilon)$ such that for any \mbox{$\epsilon$ $\!>$ $\!0$}
$P(\phi,\epsilon)$ can be obtained by direct sampling from
$p(x,\varphi)$, and $P(\phi,\epsilon)$ $\!\to$ $\!P(\phi)$
for $\epsilon$ $\!\to$ $\!0$ (Dakna, Kn\"{o}ll and Welsch
[1997a,b]). The sampling functions can then be obtained as
convergent sums of the sampling functions for the
density-matrix elements in the Fock-basis (see \S~\ref{sec3.3}).
Since the value of the parameter $\epsilon$ for which
$P(\phi,\epsilon)$ $\!\approx$ $\!P(\phi)$ is determined by
the number of photons at which the quantum state under study
can effectively be truncated, the method is state-dependent.
Further, it was proposed to measure the canonical phase by
projection synthesis, mixing the signal mode with a reference
mode that is prepared in a quantum state such that, for appropriately
chosen parameters, the joint-photon-number probabilities in the two
output channels of the beam splitter directly correspond to the
canonical phase statistics of the signal mode (Barnett and Pegg [1996];
Pegg, Barnett and Phillips [1997]).
Apart from the direct photon-number measurement needed and the
fact that the quantum state under study must again be truncated,
so that the method is state-dependent, the
difficult problem remains to design an apparatus that
produces the reciprocal binomial states needed.

The problem of homodyne measurement of the canonical phase
can be solved when the exponential phase moments $\Psi_{k}$
[i.e., the Fourier components of $P(\phi)$] are considered
and not the phase distribution itself,
\begin{equation}
\label{3.74}
P(\phi) = (2\pi)^{-1} \sum_{k=-\infty}^{+\infty} e^{-ik\phi} \Psi_{k},
\end{equation}
where $\Psi_{k}$ $\!=$ $\!\langle \hat{E}^{k} \rangle$ if $k$ $\!>$ $\!0$,
and $\Psi_{k}$ $\!=$ $\Psi_{-k}^{\ast}$ if $k$ $\!<$ $\!0$
($\Psi_{0}$ $\!=$ $1$). Then it can be shown that \mbox{($k$ $\!>$ $\!0$)}
\begin{equation}
\label{3.75}
\Psi_k = \int_{0}^{2\pi} {\rm d}\varphi \int {\rm d}x
\,e^{ik\varphi}\, K_{k}(x)\,p(x,\varphi)
\end{equation}
(Opatrn\'{y}, Dakna and Welsch [1997, 1998];
Dakna, Opatrn\'{y} and Welsch [1998]). The integral kernel $K_{k}(x)$
(Fig.~\ref{fig3.5}) can be used for sampling the exponential phase
moments from the homodyne data for any normalizable
state.\footnote{Note that when $K_{k}(x)$ is calculated according to
   eq.~(\ref{3.70}), then it does not apply to all normalizable states
   (for analytical expressions and properties of $K_{k}(x)$,
   including nonperfect detection, see Dakna, Opatrn\'{y} and Welsch [1998]).}
\begin{figure}[htb]
 \unitlength=1cm
 \begin{center}
 \begin{picture}(6,3.9)
 \put(0,0){
 \includegraphics{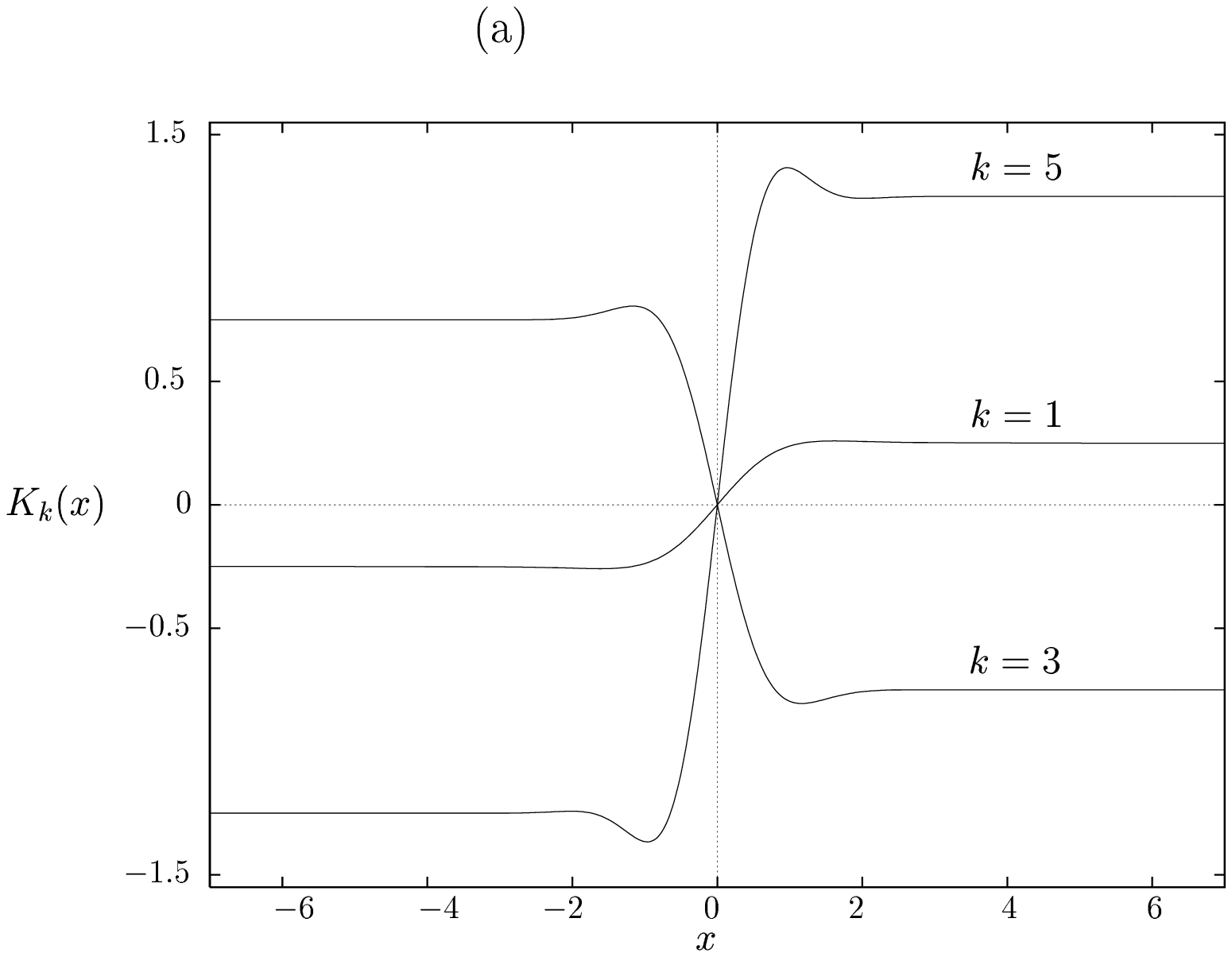}}
 \put(0,0){
 \includegraphics{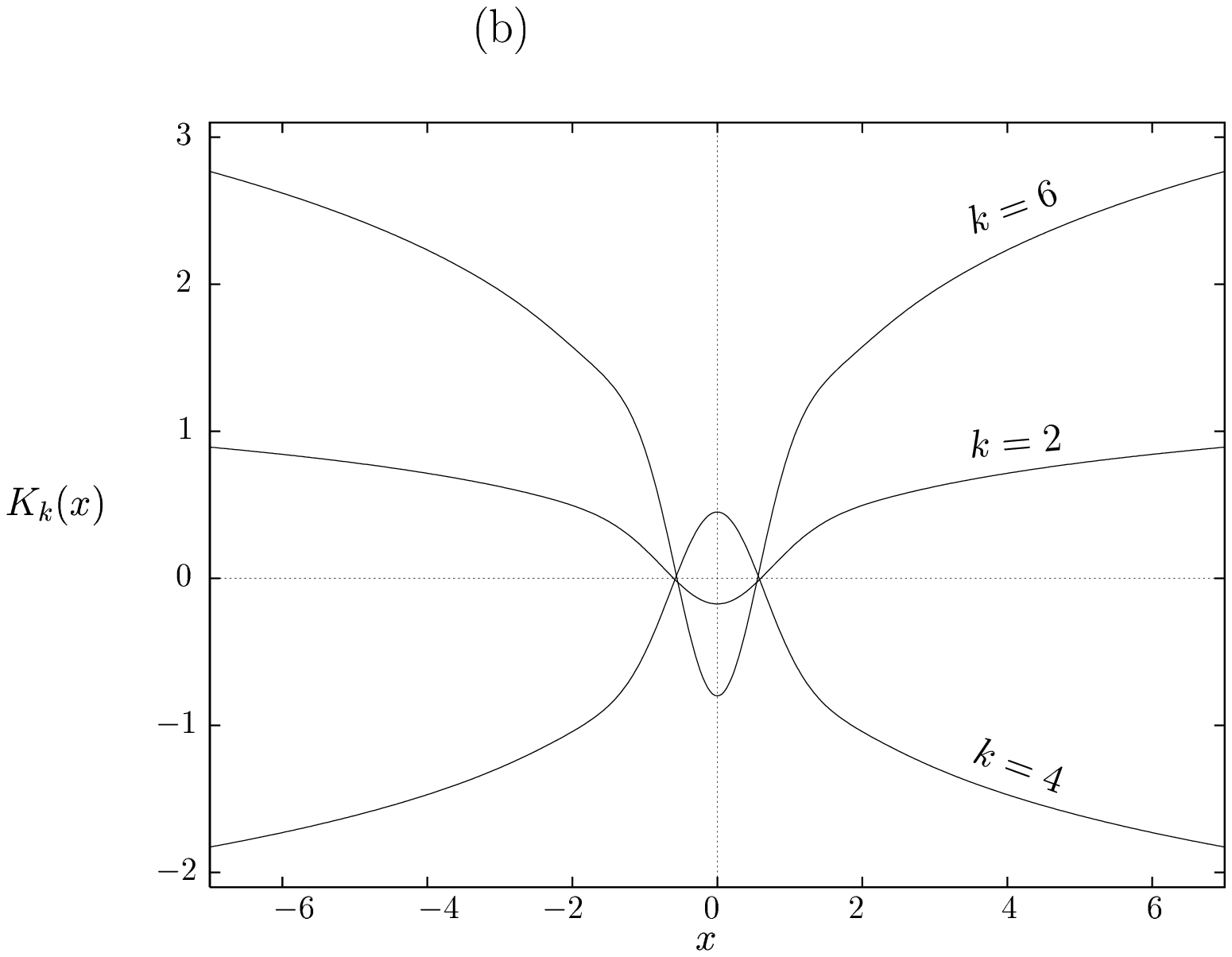}}
 \end{picture}
 \end{center}
\begin{center}
\protect\parbox{.9\textwidth}{
\caption{
The $x$-dependent part $K_{k}(x)$ of the sampling function
$K_{k}(x,\varphi)$ $\!=$ $e^{ik\varphi}K_{k}(x)$ for the
determination of the exponential moments $\Psi_{k}$
$\!=$ $\!\langle\hat{E}^{k}\rangle$ of the canonical phase
from the quadrature-component distributions $p(x,\varphi)$
according to eq.~(\protect\ref{3.75})
for various (a) odd and (b) even $k$.
(After Dakna, Opatrn\'{y} and Welsch [1998].)
\label{fig3.5}
}
}
\end{center}
\end{figure}%
In particular, $K_{k}(x)$ rapidly approaches
the classical limit\footnote{Already from the classical kernel
   (\ref{3.76}) it is seen that $\sum_{k=-\infty}^{+\infty}
   e^{ik(\varphi-\phi)}K_{k}(x)$
   does not exist, and hence $P(\phi)$ cannot be obtained from
   $p(x,\varphi)$ by means of an integral transformation of the form
   given in eq.~(\ref{3.75}).\label{nonexist}}
\begin{equation}
\label{3.76}
K_{k}^{(cl)}(x) =
\left\{
\begin{array}{ll}
{\textstyle\frac{1}{4}}(-1)^{(k-1)/2} k \,{\rm sign}\,x
& {\rm if} \ k \ {\rm odd}, \\[1ex]
(2\pi)^{-1} (-1)^{(k+2)/2} k \ln x
& {\rm if} \ k \ {\rm even}
\end{array}
\right.
\end{equation}
as $|x|$ increases, and it differs from the classical limit only
in a small interval around the origin.
It is worth noting that the  method applies
to quantum and classical systems in a unified way and bridges the
gap between quantum and classical phase. In particular,
the integral kernel $K_{k}^{(cl)}(x)$ as given in eq.~(\ref{3.76}) is
nothing but the integral kernel for determining the radially
integrated Wigner function, which reveals that in the classical
limit the canonical phase distribution is simply the radially
integrated probability density for the complex
amplitude $\alpha$. Note that any radially integrated propensity
${\rm prob}\,(\alpha)$, such as the radially integrated $Q$ function
measurable, e.g., in balanced eight-port homodyne detection, can be used
to define an operational phase probability density
(Noh, Foug\`{e}res and Mandel [1991, 1992a,b, 1993a,b,c]),
which in the quantum regime differs from the canonical
phase distribution in general (for further readings, see Turski [1972];
Paul, H., [1974]; Shapiro, J.H., and Wagner [1984];
Hradil [1992, 1993]; Vogel, W., and Welsch [1994]; Leonhardt,
Vaccaro, B\"{o}hmer and Paul, H., [1995]; Leonhardt [1997c]).


\subsubsection{Hamiltonian and Liouvillian}
\label{sec3.8.4}

So far, measurement and reconstruction of quantum-state representations
and averages of particular quantities at certain time have been
considered. The quantum state of an object at chosen time $t$ is of
course a result of state evolution from an initial time $t_{0}$,
the state evolution being governed by the Hamiltonian $\hat{H}$ of the
system or a Liouvillian $\hat{L}$ in the more general case of the system
being open. Since the Liouvillian of an object expresses in most
concentrated form the dynamics of the object, knowledge of the
Liouvillian is essential for understanding the behaviour of the object,
and the question may arise of how to experimentally determine its form.
To answer the question, it was proposed to appropriately apply
quantum-state reconstruction routines, such as direct sampling
of the density matrix in balanced homodyning (D'Ariano and Maccone [1997]).

Let us consider a system that is initially prepared in some known
state $\hat{\varrho}_{\rm in}$ $\!\equiv$ $\hat{\varrho}(t_{0})$
and assume that in the further course of time it evolves to a
state $\hat{\varrho}_{\rm out}$ $\!\equiv$ $\hat{\varrho}(t)$
at time~$t$,
\begin{equation}
\label{3.77}
\hat{\varrho}_{\rm out} = \hat{G} \,
\hat{\varrho}_{\rm in},
\end{equation}
where for a system that is homogeneous in time the superoperator
$\hat{G}$ $\!\equiv$ $\!\hat{G}(t,t_{0})$ can be given by the exponential
of the (time-independent) Liouvillian $\hat{L}$ of the system,
\begin{equation}
\label{3.78}
\hat{G} =
e^{\hat{L}(t-t_{0})}.
\end{equation}
The superoperators $\hat{G}$ and $\hat{L}$ can then be obtained by
measuring (reconstructing) $\hat{\varrho}_{\rm out}$ for various probe
inputs $\hat{\varrho}_{\rm in}$.
A typical example may be a radiation-field mode which (at time $t_{0}$)
is fed into a nonlinear resonator-like equipment giving rise to
amplification and damping. The quantum state of the outgoing
mode (at time $t$) can then be related to that of the incoming
mode according to eq.~(\ref{3.77}). In the Fock basis,
eq.~(\ref{3.77}) reads as
\begin{equation}
\label{3.79}
\langle m|\hat{\varrho}_{\rm out}|n\rangle
= \sum_{l,k=0}^{\infty} G_{mn}^{lk}
\langle l|\hat{\varrho}_{\rm in}|k\rangle.
\end{equation}
Now let us assume that in a succession of measurements
the quantum states of the outgoing mode
are reconstructed for an appropriately chosen (overcomplete) set of
input quantum states $\hat{\varrho}_{\rm in}^{\mu}$,
$\mu$ $\!=$ $\!1,2,\ldots$, such that for chosen $m$ and $n$ the
block of equations
\begin{equation}
\label{3.79a}
\langle m|\hat{\varrho}_{\rm out}^{\mu}|n\rangle
= \sum_{l,k=0}^{\infty} G_{mn}^{lk}
\langle l|\hat{\varrho}_{\rm in}^{\mu}|k\rangle
\end{equation}
can be used as conditional equations for the matrix
elements $G_{mn}^{lk}$. Repeating the procedure for all
values of $m$ and $n$, the $G$ matrix and [by using
eq.~(\ref{3.78})] the $L$ matrix can be calculated in principle.
Apart from a truncation of the Hilbert space in numerical
implementations, the main problem that remains in practice is
of course the (controlled) initial-state preparation such
that all (relevant) matrix elements of the Liouvillian
can be probed.

In particular, when the Liouvillian is phase-insensitive, then
mixtures of Fock states evolves into mixtures of Fock states.
For this case a setup was proposed which uses correlated twin beams
produced by a nondegenerate parametric amplifier in combination
with conditional photon-number measurement on one beam in order to
prepare the other beam in (random) photon-number states. These
states can then be used as inital states for probing the Liouvillian.
To realize the scheme with available techniques, it should
be noted that the reconstruction algorithm can be extented such
that the quantum efficiency of the photon-number measurement need not
be unity (D'Ariano [1997c]).


\subsection{Processing of smeared and incomplete data}
\label{sec3.9}

In practice, there are always experimental inaccuracies that limit
the precision with which the quantum state in a chosen representation
can be determined. Typical examples of inaccuracies\footnote{For
   limits upon the measurement of state vectors expressed
   in terms of channel capacities for the transmission of information
   by finite numbers of identical copies of state vectors, see
   Jones, K.R.W., [1994].}
are data smearing
owing to nonperfect detection, finite number of measurement events,
and discretization of continuous parameters, such as the
quadrature-component phase in balanced homodyning. In particular, the
latter is an example of leaving out observables $\hat{A}_{i}$ in
the expansion (\ref{3.0}) of the density operator $\hat{\varrho}$.
Whereas in balanced homodyning the distance between neighboring
phases can be diminished such that the systematic error is reduced,
in principle, below any desired level, there are also cases in which
a principally incomplete set of observables is available. Then
either additional knowledge of the quantum state is necessary to
compensate for the lack of observables or other principles must
be used to reconstruct the density operator according to the
actual observation level.


\subsubsection{Experimental inaccuracies}
\label{sec3.9.1}

Since in a realistic experiment the quantum efficiency $\eta$ is always
less than unity, the probability distributions of the measured
quantities are always more or less smeared, because of losses.
The problem of compensating for losses in direct sampling of the
quantum state from the quadrature-component distributions measurable
in balanced homodyning (\S~\ref{sec2.1.2}) has been studied widely.
As mentioned in \S~\ref{sec2.3}, active loss compensation may
be realized, in principle, by means of a squeezer, such as a degenerate
parametric amplifier (Leonhardt and Paul, H., [1994a]; see also
Leonhardt and Paul, H., [1995]). When the signal mode is preamplified before
detection by means of a squeezer [eq.~(\ref{2.52a}) with
$e^{i\phi}\hat{a}_{S}^{\dagger}$ in place of $\hat{a}_{I}^{\dagger}$],
then it can be shown that for appropriately phase matching the (scaled)
quadrature-component distribution of the preamplified signal measured
with quantum efficiency $\eta$ reads as\footnote{Note that squeezing
   the local oscillator has the same effect as antisqueezing the
   signal field under the assumption
   that the coherent component is large (Kim, M.S., and Sanders [1996]; see
   also Kim, M.S., [1997b]).}
\begin{equation}
p'(\sqrt{g}x,\varphi;\eta)
= \int {\rm d}x' \, p(x',\varphi) \,
p[\sqrt{g}(x\!-\!x');\eta]
\label{3.80a}
\end{equation}
($g$, amplification factor), with $p(x;\eta)$ being given by
eq.~(\ref{2.24}). From eq.~(\ref{3.80a}) together with eq.~(\ref{2.24})
it is seen that $\eta$ is effectively replaced with $g\eta/($
$\!1$ $\!-$ $\!\eta$ $\!+$ $\!g\eta)$, and hence
$\sqrt{g}$ $\!p'(\sqrt{g}x,\varphi;\eta)$ tends to
$p(x,\varphi)$ as the amplification becomes sufficiently
strong.\footnote{For details, see also Vogel, W., and Welsch [1994],
   and for a discussion of eqs.~(\ref{3.80a}) and (\ref{2.24}) in terms
   of moments of the measured quadrature components and those of
   $\hat{x}(\varphi)$, see Marchiolli, Mizrahi and Dodonov [1997]).}
Note that the degree of improving the quantum efficiency is limited
by the realizable mode-matching in degenerate parametric amplification,
which deteriorates with increasing pump strength.

Without active manipulation of the signal, it has been well
established that above a lower bound for $\eta$, it is
always possible to compensate for detection losses, introducing
a modified sampling function that depends on $\eta$ such that
performing the sampling algorithm on the really measurable (i.e., the
smeared) quadrature-component distributions $p(x,\varphi;\eta)$
yields the correct quantum state (D'Ariano, Leonhardt and
Paul, H., [1995]). This bound depends on the chosen state representation in
general. In particular, reconstruction of the density matrix
in the photon-number basis is always possible if $\eta$ $\!>$ $\!1/2$.
In this case,
\begin{equation}
\label{3.80}
\varrho_{mn} = \int_{0}^{\pi} {\rm d}\varphi \int {d}x
\, K_{mn}(x,\varphi;\eta) \, p(x,\varphi;\eta),
\end{equation}
where the now $\eta$-dependent sampling function
$K_{mn}(x,\varphi;\eta)$ is defined by eq.~(\ref{3.16})
[together with eq.~(\ref{3.16a})].

Another approach to the problem of loss compensation is
that the sampling function is
left unchanged [i.e., $K_{mn}(x,\varphi;\eta)$ $\!\to$
$K_{mn}(x,\varphi)$ $\!=$ $K_{mn}(x,\varphi;\eta$ $\!=$ $\!1)$ in
eq.~(\ref{3.80})] and, at the first stage, the density matrix
$\varrho_{mn}(\eta)$ of the quantum state that corresponds to the
smeared quadrature-component distributions $p(x,\varphi;\eta)$ is
reconstructed. After that, at the second stage, the true density matrix
$\varrho_{mn}$ $\!=$ $\!\varrho(\eta$ $\!=$ $\!1)$
is calculated from $\varrho_{mn}(\eta)$ (Kiss, Herzog and Leonhardt
[1995]; Herzog [1996a]). Extending the Bernoulli transformation
(\ref{A3.3}) and the inverse Bernoulli transformation
(\ref{A3.3a}) to off-diagonal
density-matrix elements, $\varrho_{mn}$ can be calculated from
$\varrho_{mn}(\eta)$ as\footnote{Replacing in eq.~(\ref{3.81}) $\eta^{-1}$
   with $\eta$ yields $\varrho_{mn}(\eta)$ in terms of $\varrho_{mn}$, which
   corresponds to a generalization of the Bernoulli transformation
   (\ref{A3.3}) to off-diagonal density-matrix elements. Since
   an imperfect detector can be regarded as a perfect detector
   with a beam splitter in front of it, $\hat{\varrho}(\eta)$ and $\hat{\varrho}$
   can be related to each other applying the beam-splitter transformation
   (\ref{2.11}), $\hat{\varrho}(\eta)$
   $\!=$ $\!_2\langle 0|e^{i\beta\hat{L}_{2}}\hat{\varrho}
   e^{-i\beta\hat{L}_{2}}|0\rangle_{2}$, with $\hat{L}_{2}$ according to
   eq.~(\ref{2.7}) and $\cos^{2}(\beta/2)$ $\!=$ $\!\eta$ (the mode indices $1$
   and $2$, respectively, refer to the signal and vacuum inputs). It can then
   be proved easily that $\hat{\varrho}(\eta)$ $\!=$
   $\eta^{\hat{L}}\hat{\varrho}$, where
   $\hat{L}$ $\!=$ $\frac{1}{2}(\hat{a}^{\dagger}\hat{a}\hat{\varrho}$
   $\!+$ $\!\hat{\varrho}\hat{a}^{\dagger}\hat{a}$
   $\!-$ $\!2\hat{a}\hat{\varrho}\hat{a}^{\dagger})$, which
   corresponds to the time evolution of a damped harmonic oscillator
   at zero temperature ($\eta$ $\!\to$ $\!e^{-t})$.}
\begin{equation}
\label{3.81}
\varrho_{mn} =
\eta^{-(m+n)/2}
\sum_{k=0}^{\infty}
{n\!+\!k \choose n}^{1/2} {m\!+\!k \choose m}^{1/2}
\left(1 - \frac{1}{\eta}\right)^{k} \varrho_{m+k\,n+k}(\eta).
\end{equation}
It is worth noting that eq.~(\ref{3.81}) is exact. In other words,
when (for precisely given overall detection efficiency) $\varrho_{mn}(\eta)$
is exactly known, then $\varrho_{mn}$ can always be calculated
precisely, irrespective of the value of $\eta$.

In practice however,
$\eta$ is not known precisely in general, and the experimentally
determined values of $\varrho_{mn}(\eta)$ always differ from the exact
ones. In particular, when \mbox{$\eta$ $\!\le$ $\!1/2$} and the error of
$\varrho_{mn}(\eta)$ does not vanish with increasing $m$ and $n$,
and when there is no {\em a priori} information about the quantum
state, such as the photon number at which it can be
truncated, then $\varrho_{mn}$ cannot be obtained -- for chosen number
of measurements -- from the measured $\varrho_{kl}(\eta)$,
because of error explosion. A typical example of such an error is
the statistical error with which $\varrho_{mn}(\eta)$ is sampled
from the quadrature-component distributions measured in balanced
homodyning. In this case loss compensation (for arbitrary
quantum states) is only possible if \mbox{$\eta$ $\!>$ $\!1/2$}
(D'Ariano and Macchiavello [1998]).\footnote{When the density matrix
   at initial time $t$ $\!=$ $\!0$ of a signal mode that undergoes
   phase-insensitive damping or amplification is tried to be reconstructed
   from the quadrature-component distributions $p(x,\varphi,t;\eta)$
   measured at time $t$ $\!>$ $\!0$, then the additional
   (phase-insensitive) noise gives rise to a modified overall
   quantum efficiency $\eta_{\ast}$, so that $\eta_{\ast}$ $\!>$
   $\!1/2$ must be valid in order to compensate for the losses
   (D'Ariano [1997b]; D'Ariano and Sterpi [1997]).}
Clearly when the quantum state to be reconstructed can be truncated
at some maximum photon number $n_{\max}$, then the sum in
eq.~(\ref{3.81}) is finite and the density-matrix elements
$\varrho_{mn}$ can also be obtained for $\eta$ $\!<$ $\!1/2$,
the accuracy being determined by that of $\varrho_{mn}(\eta)$.

Limitations on variables in real experiments always give rise to
systematic errors. This is the case, e.g., when the
quadrature components $\hat{x}(\varphi)$ are measured
at discrete phases $\varphi_{k}$ or when only some part of the
$\pi$ interval can be scanned experimentally
(for an example, see \S~\ref{sec4.5.1}).
In particular, in balanced
homodyning the quadrature-component distribution $p(x,\varphi)$
is measured at a finite number of phases $\varphi_{k}$ within
a $\pi$ interval and finite $x$-resolution.
When $p(x,\varphi)$ is measured precisely at $N$ equidistant
phases $\varphi_{k}$ $\!=$ $\!(\pi/N)k$, then the reconstructed
density-matrix elements are given by
\begin{equation}
\label{3.82}
\varrho_{mn}(N) = \frac{\pi}{N} \sum_{k=0}^{N-1} \int {d}x
\, K_{mn}(x,\varphi_{k}) \, p(x,\varphi_{k}),
\end{equation}
in place of eq.~(\ref{3.13}). A measure of the systematic error
is the difference $\Delta\varrho_{mn}$ $\!=$ $\varrho_{mn}(N)$
$\!-$ $\!\varrho_{mn}$, which reads
as\footnote{Equation (\ref{3.83}) can be derived by substituting in
   eq.~(\ref{3.82}) for $p(x,\varphi)$ the result of eq.~(\ref{3.17})
   and recalling eqs.~(\ref{3.14}) and (\ref{3.20}). Note that
   $G^{mn}_{kl}$ $\!=$ $\!\delta_{mk}\delta_{nl}$ only holds for
   $k$ $\!-$ $\!l$ $\!=$ $\!m$ $\!-$ $\!n$.}
\begin{equation}
\label{3.83}
\Delta\varrho_{mn}
= \sum_{j=1}^{\infty}
\sum_{k,l} G^{mn}_{kl}\,\varrho_{kl},
\quad {\rm with} \quad k\!-\!l = m\!-\!n \pm 2jN
\end{equation}
[$G^{mn}_{kl}$ $\!=$ $\pi\int{\rm d}x$ $\!f_{mn}(x)$ $\!g_{kl}(x)$]
(Leonhardt and Munroe [1996]; Leonhardt [1997b]). Equation (\ref{3.83})
reveals that when the quantum state can be truncated such that
\begin{equation}
\label{3.84}
\varrho_{mn} = 0 \quad {\rm for} \quad |m\!-\!n| \ge N,
\end{equation}
then the density matrix-elements $\varrho_{mn}$ for
$|m$ $\!-$ $\!n|$ $\!<$ $\!N$ can be reconstructed precisely
from $p(x,\varphi)$ at $N$ phases $\varphi_{k}$ according to
eq.~(\ref{3.82}). In particular,
one phase is required to reconstruct a completely dephased quantum
state that contains only diagonal density-matrix elements.
Since any quantum state can be approximated to any desired degree
of accuracy by setting $\varrho_{mn}$ $\!\approx$ $\!0$ for $m$($n$) $\!>$
$\!n_{\rm max}$, there is always an $N$ $\!\approx$ $\!n_{\rm max}$
for which the condition (\ref{3.84}) can be assumed to be satisfied,
so that all (essentially nonvanishing) density-matrix elements
can be reconstructed precisely. If $N$ is not known {\em a priori},
the quantum state can be reconstructed in an iterative way, increasing
$N$ in a sequence of (ensemble) measurements until $|\Delta\varrho_{mn}|$
is sufficiently small. Note that any normally ordered
moment $\langle\hat{a}^{\dagger n}\hat{a}^{m}\rangle$ can
exactly be reconstructed from $p(x,\varphi)$ at \mbox{$N$ $\!=$
$\!n$ $\!+$ $\!m$ $\!+$ $\!1$} phases (W\"{u}nsche [1996b];
see \S~\ref{sec3.8.1}).

The methods for quantum-state reconstruction are based on ensemble
measurements; i.e., on a sequence of individual measurements
carried out on identically prepared systems.\footnote{For the
   problem of measuring the state of single quantum systems, see
   Ueda and Kitagawa [1992]; Aharonov and Vaidman [1993];
   Aharonov, Anandan and Vaidman [1993]; Imamo\u{g}lu [1993]; Royer
   [1994, 1995]; Alter and Yamamoto [1995]; D'Ariano and Yuen [1996].}
Since the number of individual measurements is principally finite,
the measured quantities are always estimates of the true ones.
Hence all the quantities that can be inferred from the
measured quantities are also estimates, and the statistical error
with which the original quantities are measured
propagates to the quantities inferred from them. In particular,
when the quantities which are desired to be determined can
be sampled directly from the measured data, then the
statistical error can be estimated straightforwardly by also
using the sampling method. The problem of statistical error
in quantum-state reconstruction has been studied in a number of
papers, with special emphasis on balanced four-port homodyne detection
(Leonhardt, Munroe, Kiss, Raymer and Richter, Th., [1996]; D'Ariano [1997b];
D'Ariano and Paris [1997a]; D'Ariano, Macchiavello and Sterpi [1997];
Leonhardt [1997c]). Let us consider a quantity $\hat{F}$ and assume
that it can be sampled directly from the quadrature-component
statistics according to eq.~(\ref{3.62b}). When in an experiment
$n(\varphi_{k})$ individual measurements
are performed at phase $\varphi_{k}$, $k$ $\!=$
$\!0,1,\ldots,N$ $\!-$ $\!1$, then $\langle\hat{F}\rangle$ can be
estimated as
\begin{equation}
\label{3.85}
\langle\hat{F}^{\rm (est)}(N)\rangle
= \frac{\pi}{N}\sum_{k=0}^{N-1}
\frac{1}{n(\varphi_{k})} \sum_{n=1}^{n(\varphi_{k})}
K_{F}[x_{n}(\varphi_{k}),\varphi_{k}] ,
\end{equation}
where $x_{n}(\varphi_{k})$ is the result of the $n$th individual
measurement at phase $\varphi_{k}$. Taking the average over
all estimates $\langle\hat{F}^{\rm (est)}(N)\rangle$ yields
$\langle\hat{F}(N)\rangle$; i.e., the desired quantity within
the systematic error owing to phase discretization,
\begin{equation}
\label{3.86}
\overline{\langle\hat{F}^{\rm (est)}(N)\rangle}
=\langle\hat{F}(N)\rangle = \frac{\pi}{N}\sum_{k=0}^{N-1}
\overline{K_{F}[x(\varphi_{k}),\varphi_{k}]}
\end{equation}
where
\begin{equation}
\label{3.87}
\overline{K_{F}[x(\varphi_{k}),\varphi_{k}]}
= \int {\rm d}x \, p(x,\varphi_{k}) \, K_{F}(x,\varphi_{k}).
\end{equation}
Accordingly, the statistical fluctuation of
$\langle\hat{F}^{\rm (est)}(N)\rangle$ in terms of the averaged
estimates of the variance of $\langle\hat{F}^{\rm (est)}(N)\rangle$
can be given by, on taking into account that the individual
measuremenst are statistically independent of each other,\footnote{For
   simplicity, in eq.~(\ref{3.88}) it is assumed that $F$ is a real
   quantity.}
\begin{equation}
\label{3.88}
\sigma^{2}_{\langle\hat{F}\rangle}
= \frac{\pi}{N} \sum_{k=1}^{N-1}
N^{-1}(\varphi_{k})
\overline{\{\Delta K_{F}[x(\varphi_{k}),\varphi_{k}]\}^{2}} ,
\end{equation}
where
\begin{eqnarray}
\label{3.88-t}
\overline{\{\Delta K_{F}[x(\varphi_{k}),\varphi_{k}]\}^{2}}
= \int {\rm d}x \, p(x,\varphi_{k}) \, K_{F}^{2}(x,\varphi_{k})
- \left[ \int {\rm d}x \, p(x,\varphi_{k}) \, K_{F}(x,\varphi_{k})
\right]^{2}
\end{eqnarray}
is the variance of the sampling function, and $N(\varphi_{k})$ is the
number of measurements per phase interval,
$N(\varphi_{k})$ $\!=$ $n(\varphi_{k})N/\pi$.

Let us mention that when the kernel $K_{F}(x,\varphi)$  is a strongly
varying function of $x$ in regions where $p(x,\varphi)$ is
non-negligible, then the first term in eq.~(\ref{3.88-t}) is much larger
than the second one. In this case the second term can be neglected
and the statistical error can be approximated by averaging the square of
the kernel. Note that the same result is obtained if one assumes that
the numbers of events yielding the values of $x(\varphi_{k})$ in given
intervals of $x$ (bins) are independent Poissonian variables
(Leonhardt, Munroe, Kiss, Raymer and Richter, Th., [1996];
Leonhardt [1997c]; McAlister and Raymer [1997b]). However, these
variables are neither strictly independent [their sum is always
$N(\varphi_{k})$] nor Poissonian [probability that there is more than
$N(\varphi_{k})$ events in a bin is zero].
Therefore, care must be taken before one decides to use the simplified
error estimation. Whereas for the Fock-basis density matrix elements the
simplified estimation is very good (the kernels are strongly
oscillating), for the exponential moments of canonical phase one has to
take into account both terms in eq.~(\ref{3.88-t}), otherwise the error
would be overestimated (the kernels are slowly varying functions).

   From eq.~(\ref{3.88}) it can be expected that the statistical error
depends sensitively on the analytical form of the sampling function.
To give an example, let us consider the integral kernels $K_{mn}(x,\varphi)$,
eq.~(\ref{3.14}), needed for sampling the density-matrix elements in
the photon-number basis, $\varrho_{mn}$. For chosen $\varphi$,
$K_{mn}(x,\varphi)$ is an oscillating function of $x$, and with
increasing distance $d$ $\!=$ $|m$ $\!-$ $\!n|$ from the diagonal the
oscillations become faster and the oscillation range slowly increases.
It turns out that with increasing $m$ the diagonal-element variance
$\sigma^{2}_{\varrho_{mm}}$ becomes independent of $m$; i.e., saturation
of the statistical error for sufficiently large $m$ is observed,
$\sigma^{2}_{\varrho_{mm}}$ $\!\leq$ $\!2/\bar{N}$ ($\bar{N}$, mean
number of measurements per phase interval). On the contrary, the
off-diagonal statistical error increases with $d$, without saturation.
The influence of the quantum efficiency $\eta$ on the statistical
error of the density-matrix elements is very strong in general. For
chosen $m$ and $n$ the oscillation range of $K_{mn}(x,\varphi;\eta)$,
eq.~(\ref{3.16}), increases very rapidly as $\eta$ approaches the
lower bound $\eta$ $\!=$ $\!1/2$, and the statistical error increases
rapidly as well (for numerical examples, see D'Ariano, Macchiavello and
Sterpi [1997]).\footnote{For the error of the
   exponential phase moments sampled from the quadrature-component
   distributions in balanced homodyning, see Dakna, Opatrn\'{y} and
   Welsch [1998], and for the error in quantum-state measurement via
   unbalanced homodyning
   and direct photocounting, see Wallentowitz and Vogel, W., [1996a];
   Banaszek and W\'{o}dkiewicz [1997b]; Opatrn\'{y}, Welsch, Wallentowitz
   and Vogel, W., [1997].}
Using the experimentally
sampled density-matrix elements for calculating the expectation
values of other quantities, such as $\langle\hat{F}\rangle$
according to eq.~(\ref{3.62a}), the resulting statistical error is
determined by the law of error propagation. It is worth noting that
error propagation can lead to additional noise which is not observed
if the quantities are also directly sampled from the measured data
(provided that the sampling method applies).


\subsubsection{Least-squares method}
\label{sec3.9.2}

In the quantum-state reconstruction problems outlined in the foregoing
sections a set of measurable quantities (in the following also
referred to as data vector) is related linearly to a set of quantities
(state vector) that can be used to characterize the quantum state of
the system under study.  Both sets of quantities can be discrete or
continuous or of mixed type. Typical examples are the relations
(\ref{3.1}) and (\ref{3.17}), respectively, between the Wigner
function and the density-matrix elements in the Fock basis, and the
relations (\ref{3.24}) and (\ref{3.47}) between the Fourier components
of the displaced Fock-state distributions and the atomic-state
inversion, respectively, and the density-matrix elements in the Fock
basis. A powerful method for inversion of such relations has been
least-squares inversion.\footnote{The method of least squares was
  discovered by Legendre [1805] and Gauss [1809,1821] for solving the
  problem of reconstruction of orbits of planetoids from measured
  data.}
The method has been used for quantum-state reconstruction
in balanced optical homodyning (Tan [1997]), unbalanced homodyning
(Opatrn\'{y} and Welsch [1997]; Opatrn\'{y}, Welsch, Wallentowitz and
Vogel, W., [1997]), cavity QED (Bardroff, Mayr, Schleich, Domokos,
Brune, Raimond and Haroche [1996]; Bodendorf, Antesberger, Kim, M.S.
and Walther [1998]) and for orbital electronic motion (Cline, van der
Burgt, Westerveld and Risley [1994]). It has been used further to
reconstruct the quantum state of the center-of-mass motion of trapped
ions (Leibfried, Meekhof, King, Monroe, Itano and Wineland [1996]),
the quantum state of a particle in an anharmonic potential and the
quantum state of a particle that undergoes a damped motion in a
harmonic potential (Opatrn\'{y}, Welsch and Vogel, W., [1997c]).

An advantage of least-squares inversion is that it is a linear
method -- the density matrix elements can be reconstructed in
{\em real time} together with an estimation of the statistical
error. Moreover, it allows for an easy incorporation in the reconstruction
of various experimental peculiarities, such as non-unity quantum efficiency,
finite resolution or discretization of the data, finite observation time,
dissipative decay of the system, etc..
These aspects can hardly be treated on the basis of analytically determined
(and existing) sampling functions. On the other hand, the method does
not guarantee (similarly as any other linear method) that a
reconstructed density matrix is exactly positive-definite
(cf. \S~\ref{sec3.9.3}).

To illustrate the method, let us assume that a distribution
$p(x,\varphi)$ of the type of a quadrature-component distribution
is measured, and that $p(x,\varphi)$ can be given by a
linear combination of all density-matrix elements $\varrho_{nn'}$
of the quantum state to be reconstructed, with linearly
independent coefficient functions $S_{nn'}(x,\varphi)$,
\begin{equation}
\label{3.89}
p(x,\varphi) = \sum_{n,n'} S_{nn'}(x,\varphi) \, \varrho_{nn'},
\end{equation}
where $S_{nn'}(x,\varphi)$ need not be of the form used
in eq.~(\ref{3.17}).\footnote{In particular, when $x$ is the position
   of a moving particle and $\varphi$ corresponds to the time $t$,
   and the particle undergoes damping, then the quantum state evolves
   according to a master equation whose solution then determines
   $S_{nn'}(x,t)$.}
Since the density matrix of any physical state
can be truncated at some value $n_{\rm max}$, the sum in eq.~(\ref{3.89})
is effectively finite. Direct application of least-squares inversion
(Appendix \ref{app4}) yields the reconstructed density-matrix
elements $\varrho^{(M)}_{nn'}$ as
\begin{equation}
\label{3.90}
\varrho^{(M)}_{nn'} = \int_{X} \! {\rm d}x \int_{\Phi}\! {\rm d}\varphi \,
K_{nn'}(x,\varphi) \, p^{(M)}(x,\varphi) ,
\end{equation}
where $p^{(M)}(x,\varphi)$ is the experimentally measured distribution,
$X$ and $\Phi$ being the intervals accessible to measurement.
The integral kernel $K_{nn'}(x,\varphi)$ is given by
\begin{equation}
\label{3.91}
K_{nn'}(x,\varphi)
= \sum_{m,m' \le n_{\rm max}} F_{nn';mm'}
S_{mm'}^{\ast}(x,\varphi),
\end{equation}
and ${\bf F}$ $\!=$ $\!{\bf G}^{-1}$,
with the matrix ${\bf G}$ being defined by
\begin{equation}
\label{3.92}
G_{mm';nn'}
=\int_{X} \! {\rm d}x  \int_{\Phi} \! {\rm d}\varphi \,
S_{mm'}^{\ast}(x,\varphi)  S_{nn'}(x,\varphi).
\end{equation}
It can be proved by direct substitution that if the data correspond to the
exact quantities $p(x,\varphi)$, i.e., $p^{(M)}(x,\varphi)$ $\!=$
$\!p(x,\varphi)$, then the reconstructed density matrix
equals the correct one, i.e., $\varrho^{(M)}_{nn'}$ $\!=$
$\!\varrho_{nn'}$. On the other hand, if the experimental data suffer
from some inaccuracies, then the reconstructed density matrix has the
property that it reproduces the data as truly as possible
(in the sense of least squares). In the above given formulas we
have assumed that $x$ and $\varphi$ are continuos variables,
and $n$ is discrete. The formulas for other combinations of
discrete and/or continuous arguments can be obtained in a quite
similar way.

An essential point of the method is the inversion of the matrix
$\!{\bf G}$, which requires the matrix to be sufficiently far from
singularity; i.e., the data must carry enough information about all
the density-matrix elements that are desired to be reconstructed.
Otherwise regularized inversion must be applied (Appendix \ref{app4}).
Regularized inversion usually decreases the statistical error of the
reconstructed density-matrix elements, but on the other hand
they are biased. Therefore, in practice such a degree of
regularization should be used for which the introduced bias is
just below the statistical noise.


\subsubsection{Maximum-entropy principle}
\label{sec3.9.3}

As already mentioned, it is principally impossible to
measure the exact expectation values of an infinite number of operators
$\hat A_{i}$ in an expansion of the density operator of the type
given in eq.~(\ref{3.0}), because any realistic experiment can only
run for a finite time. So far, the exact formulas have been applied
to the analysis of the incomplete measurements including an
estimation of the error made. However the question may arise of
how to obtain an optimum result of reconstruction of a quantum state
when only a finite number of $\langle \hat{A}_{i} \rangle$ has been
measured in the experiment. An answer can be given using the Jaynes
principle of {\em maximum entropy\/}\footnote{See Jaynes
   [1957a,b]. Probability distributions (or density operators)
   describe our stage of knowledge about physical systems. If we do not
   know anything, we usually assign uniform distributions to the quantities
   (or a multiple of the unity operator to the density operators).
   If we have partial knowledge, we choose such distributions
   which are as broad as possible and still
   reflect our stage of knowledge. A suitable measure of the breadth is
   the entropy, one therefore seeks for such distributions (density operators)
   which maximize the entropy under the condition that known quantities are
   reproduced.}
(Bu\v{z}ek, Adam, G., and Drobn\'{y} [1996a,b];
Bu\v{z}ek, Drobn\'{y}, Adam, G., Derka and Knight [1997]).

Let us assume that the expectation values $\langle \hat A_{i} \rangle$
of $n$ quantities $\hat A_{i}$, $i$ $\!=$ $\!1,\dots n$ are
experimentally determined.\footnote{Note that the determination
   of $\langle \hat A_{i} \rangle$ already requires an infinite number
   of individual measurements which cannot be realized during
   a finite measurement time.}
The set of measured quantities
can be regarded as a measure of the realized observation level.
Certainly, there are a number of potential density operators
$\hat{\varrho}$, ${\rm Tr}$ $\!\hat{\varrho}$ $\!=$ $\!1$,
which are compatible with the experimental results, i.e.,
\begin{equation}
\label{3.93}
{\rm Tr}\! \left( \hat{\varrho} \hat{A}_{i} \right)
= \langle \hat{A}_{i}\rangle ,
\quad i = 1,\dots n .
\end{equation}
Among them that density operator is chosen
that maximizes the von Neumann entropy\footnote{See
   von Neumann [1932].}
\begin{equation}
\label{3.94}
S[\hat{\varrho}] =
- {\rm Tr}\!\left( \hat{\varrho}
\ln \hat{\varrho} \right).
\end{equation}
Introducing Lagrange multipliers, the resulting density operator
$\hat{\varrho}$ $\!=$ $\!\hat{\varrho}_{S}$ takes the familiar
(grand-canonical ensemble) form
\begin{equation}
\label{3.95}
\hat{\varrho}_{S} = Z^{-1}
\exp\!\left(-\sum_{i=1}^{n} \lambda_{i} \hat{A}_{i} \right) ,
\end{equation}
where
\begin{equation}
\label{3.96}
Z = {\rm Tr}\!\left[
\exp\!\left( - \sum_{i=1}^{n} \lambda_{i} \hat{A}_{i} \right)
\right],
\end{equation}
and represents a partially reconstructed (estimated) density
operator on the chosen observation level.
Substituting in eq.~(\ref{3.93}) for $\hat{\varrho}$ the
density operator $\hat{\varrho}_{S}$ from eq.~(\ref{3.95}),
a set of $n$ nonlinear equations is obtained for the calculation of
the $n$ Lagrange multipliers $\lambda_{i}$ from the measured
expectation values $\langle\hat{A}_{i}\rangle$. Any incomplete
observational level can be extended to a more complete
observational level, in principle, by including additional
observables in the scheme, which is usually associated with
a decrease of the entropy.
However, since rather involved calculations are required to be
performed, the method has been studied for
reconstructing the quantum state of a radiation-field
mode on particular (not very high) observational levels
(Bu\v{z}ek, Adam, G., and Drobn\'{y} [1996a,b];
Bu\v{z}ek, Drobn\'{y}, Adam, G., Derka and Knight [1997])
and/or low-dimensional systems, such as spin states
(Bu\v{z}ek, Drobn\'{y}, Adam, G., Derka and Knight [1997]).

Since the expectation values $\langle\hat{A}_{i}\rangle$
cannot be measured with infinite precision, only estimates
can be inserted into eq.~(\ref{3.93}),
and therefore it can happen that the solution does not exist;
i.e., the non-precisely measured averages are not compatible with
any density operator. To overcome this problem, the method can be
combined with least-squares minimization (Wiedemann [1996]\footnote{In
   this paper, which is unfortunately unpublished,
   the operators $\hat{A}_{i}$ are identified with the photon
   number $\hat{n}$ and  quadrature-component projectors
   $\!|x_{l},\varphi_{k}\rangle\langle x_{l},\varphi_{k}|$, where
   the subscript $l$ labels a finite subset of the continuous
   quadrature-components at chosen phase $\varphi_{k}$.
   Due to computational limits $4$ phases and $13$ values of
   $x$ at each phase are considered. The partial reconstruction
   of the quantum state (in phase space) from computer-simulated
   homodyne data is performed for various states and yields
   results that reflect typical properties of the states
   sufficiently well.});
i.e., the sum of squares of differences
\begin{equation}
\label{3.97}
C(\{\lambda_{i}\})
=\sum_{j=1}^{n}
\left\{
\langle\hat{A}_{j}\rangle^{(M)}
- {\rm Tr}\!\left[\hat{\varrho}_{S}(\{\lambda_{i}\})\hat{A}_{j}\right]
\right\}^{2}
\end{equation}
as a function of the parameters $\lambda_{i}$ is tried to
be minimized, $\langle\hat{A}_{j}\rangle^{(M)}$ being
the non-precisely measured averages.

\subsubsection{Bayesian inference}
\label{sec3.9.4}

The statistical fluctuations of the data are taken into account in the
{\em Bayesian inference} scheme (Helstrom [1976], Holevo [1982],
Jones, K.R.W., [1991, 1994]; Derka, Bu\v{z}ek, Adam, G., [1996]; Derka,
Bu\v{z}ek, Adam, G., and Knight [1996]). Let us assume that the system
under consideration is prepared in a pure state that belongs to a
continuous manifold of states in a {\em state space} $\Omega$ and
consider a repeated $N$-trial measurement of observables
$\hat{A_{i}}$, $i$ $\!=$ $\!1,\ldots,n$, with eigenvalues $A_{j_{i}}$.
The determination of the quantum state of the measured system is then
performed in a repeated three-step procedure:
\begin{itemize}
\item[(i)]
As a result of a
(single) measurement of $\hat{A}_{i}$ a conditional probabilty
$p(A_{j_{i}}|\hat{\varrho})$ is defined which specifies the result
$A_{j_{i}}$ if the measured system is in state $\hat{\varrho}$ $\!=$
$|\Psi\rangle\langle\Psi|$,
\begin{equation}
\label{3.98}
p(A_{j_{i}}|\hat{\varrho})
= {\rm Tr}\!\left(\hat{P}_{A_{j_{i}}} \hat{\varrho} \right),
\end{equation}
where $\hat{P}_{A_{j_{i}}}$ $\!=$ $|A_{j_{i}}\rangle\langle A_{j_{i}}|$.
\item[(ii)]
A probability distribution $p_{0}(\hat{\varrho})$
defined on the space $\Omega$ is specified such that it describes
the {\em a priori} knowledge of the state to be reconstructed.
The joint probability distribution $p(A_{j_{i}},\hat{\varrho})$
is then given by
\begin{equation}
\label{3.99}
p(A_{j_{i}},\hat{\varrho}) = p(A_{j_{i}}|\hat{\varrho})
\, p_{0}(\hat{\varrho}) .
\end{equation}
When no initial information about the measured syste is available, then
the prior probability distribution $p_{0}(\hat{\varrho})$ is chosen
to be constant.
\item[(iii)]
Finally, the Bayes rule\footnote{The Bayes rule expresses
   the relation between the conditional probabilities $p(x|y)$ and
   $p(y|x)$ as $p(x|y)$ $\!p(y)$ $\!=$ $\!p(x,y)$ $\!=$
   $\!p(y|x)$ $\!p(x)$, and hence
   $p(y|x)$ $\!=$ $\!p(x,y)/p(x)$ $\!=$ $\!p(x,y)/\int{\rm d}y$ $\!p(x,y)$.}
is used to obtain the probability
$p(\hat{\varrho}|A_{j_{i}})$ of the system being in state
$\hat{\varrho}$ under the condition that $A_{j_{i}}$ is measured,
\begin{equation}
\label{3.100}
p(\hat{\varrho}|A_{j_{i}})
= \frac{ p(A_{j_{i}},\hat{\varrho}) }
{ \int_{\Omega} {\rm d}\Omega \, p(A_{j_{i}},\hat{\varrho})  } \,.
\end{equation}
\end{itemize}
Now the procedure can be repeated, making a second measurement
of some observable $\hat{A_{k}}$ ($\hat{A_{k}}$ $\!=$ $\!\hat{A_{i}}$
or $\hat{A_{k}}$ $\!\neq$ $\!\hat{A_{i}}$) and using
$p(\hat{\varrho}|A_{j_{i}})$ obtained from the first measurement
as the prior probability distribution for the second measurement.
Proceeding in the way outlined, the $N$-trial conditional probability
distribution $p(\hat{\varrho}|\{ A_{j_{i}} \})$ is given by
\begin{equation}
\label{3.101}
p(\hat{\varrho} | \{ A_{j_{i}} \} ) =
\frac{ {\cal L} (\hat{\varrho}) \, p_{0}(\hat{\varrho})}
{\int_{\Omega} {\rm d}\Omega \,
{\cal L}(\hat{\varrho}) \, p_{0}(\hat{\varrho}) } \, ,
\end{equation}
where
\begin{equation}
\label{3.102}
{\cal L} (\hat \varrho) = \prod_{i=1}^{N}
p(A_{j_{i}}|\hat \varrho)
\end{equation}
is the {\em likelihood function} (regarded as a function of $\hat{\varrho}$;
the measured values $\{A_{j_{i}}\}$ playing the role of parameters).
Note that ${\cal L}(\hat \varrho)$
is the probability distribution of finding the
result $\{A_{j_{i}}\}$ in the sequence of $N$ measurements under the
condition that the state of the system is $\hat{\varrho}$.
The partially reconstructed density operator $\hat{\varrho}_{B}$
is then taken as the average over all the possible states
$\hat{\varrho}$,\footnote{Note that $\protect\hat{\varrho}_{B}$
   can correspond to a mixed state even though it is assumed
   that the system is prepared in a pure state.}
\begin{equation}
\label{3.103}
\hat{\varrho}_{B}
= \int_{\Omega} {\rm d} \Omega \,
p(\hat{\varrho} | \{ A_{j_{i}} \} )  \hat \varrho .
\end{equation}

To apply the method, the state space $\Omega$ of the measured system
and the corresponding integration measure ${\rm d}\Omega$ must be
defined and the prior probability $p_{0}(\hat \varrho)$ must be
specified.  In particular, the integration measure has to be invariant
under unitary transformations in the space $\Omega$. This requirement
uniquely determines the form of the measure. It should be pointed out
that this is no longer valid when $\Omega$ is tried to be extended to
mixed states. Since a system which is in a mixed state can always be
considered as a subsystem of a composite system that is in a pure
state, the Bayesian reconstruction can be applied to the composite
system and tracing the resulting density operator over the degrees of
freedom of the other subsystem (Derka, Bu\v{z}ek and Adam, G., [1996],
Derka, Bu\v{z}ek, Adam, G., and Knight [1996]).  As already mentioned, the
prior probability $p_{0}(\hat \varrho)$ can be chosen constant if
there is no {\em a priori} information about the state to be
reconstructed.\footnote{Note that when the integration measure ${\rm
  d}\Omega$ is not uniquely defined, then a prior probability that
  is constant with respect to a chosen integration measure need not be
  constant with respect to another one.}
It turns out that with
increasing number of measurements the method becomes rather
insensitive to the prior probability. In particular, when the number of
measurements approaches infinity, \mbox{$N$ $\!\to$ $\!\infty$},
eq.~(\ref{3.103}) corresponds, on chosen observation level, to the
principle of maximum entropy on a microcanonical
ensemble.\footnote{Note that on a chosen (incomplete) observational
  level the two methods yield different fluctuations of the
  observables in general.}
Contrary to least-squares inversion, Bayesian inference always yields
a (partially) reconstructed density operator which is (formally)
positive definite and normalized to unity.  However the price to pay
is a rather involved procedure that has to be carried out in practice.
For this reason, the method has been mostly considered -- similar to
the method of maximum entropy -- for spin systems; i.e., for systems
with small dimension of the state space (Jones, K.R.W., [1991], Derka,
Bu\v{z}ek and Adam, G., [1996], Derka, Bu\v{z}ek, Adam, G., and Knight [1996]).

Another statistical method closely related to the Bayesian
reconstruction method is the quantum state estimation
based on the maximization of the likelihood function
${\cal L} (\hat{\varrho})$,
eq.~(\ref{3.102}). Here, that state $\hat{\varrho}$ in the state
space is selected for which ${\cal L} (\hat{\varrho})$ attains
its maximum. Because of the difficulty of finding the maximum
of ${\cal L} (\hat \varrho)$ in higher-dimensional state spaces,
a procedure was proposed which is based on a sequence of general
inequalities satisfied by the likelihood function (Hradil [1997]).
In this way, the problem can be tranformed to that of the
diagonalization of an operator given by a linear combination of
the projectors $\hat{P}_{A_{j_{i}}}$,
where the expansion coefficients
must solve a set of nonlinear algebraic equations.
The method simplifies the Bayesian treatment but still guarantees the
positive definiteness. Significantly, all the solutions based on the
deterministic relation (\ref{3.0}) between counted data (frequencies)
and the desired density matrix are involved. Whenever such solution
exists as a positively defined density matrix, then it should maximize
the likelihood function as well.
The method was applied successfully to the reconstruction of
(low-dimensional) density matrices in the photon-number basis
of a radiation-field mode from
computer-simulated homodyne data, and a comparison with direct
inversion of the linear basic relation between the measurable quantities
and the density-matrix elements was given (Mogilevtsev, Hradil and
Pe\v{r}ina [1997]).


\section{Quantum states of matter systems}
\label{sec4}

In the preceding sections we have considered phase sensitive
measurements of radiation fields and methods for reconstructing
the quantum state of the fields from the measured data.
The problem of quantum-state measurement and reconstruction
has also been studied for various matter systems.
Different matter systems require, in general, different detection schemes
for measuring specific quantities that carry the full information
on the quantum state of the system. Although these methods may be,
at first glance, quite different from the methods outlined in
\S~\ref{sec2} for phase-sensitive measurements of light,
there have been a number of analogies between the reconstruction
concepts for radiation and matter.


\subsection{Molecular vibrations}
\label{sec4.1}

It was shown and demonstrated experimentally that the quantum state of
molecular vibrations can be determined using a tomographic method
(Dunn, Walmsley and Mukamel [1995]) which resembles the one for a light
mode outlined in \S~\ref{sec3.1}. The method, called {\em molecular
emission tomography}, is based on the fact that the time
resolved emission spectrum of a molecule allows one to visualise the
time dependence of a vibrational wave packet within the excited
electronic state from which the emission originates (Kowalczyk,
Radzewicz, Mostowski and Walmsley [1990]). Alternatively, the desired
information on the wave-packet dynamics can be obtained by
photoelectron spectroscopy (Assion, Geisler, Helbing, Seyfried and
Baumert [1996]).

Let us assume that the molecule is prepared in a given vibrational
quantum state in the excited electronic state. As can be
seen from Fig.~\ref{fig4.1}, for appropriately displaced potential
energy surfaces of the molecule the position of the vibrational wave
packet can be effectively mapped onto the frequency of the emitted
light. This fact is used for the tomographic reconstruction of the
vibrational wave packet by measuring the time-resolved emission
spectrum with a time resolution that is fast compared with the
characteristic time period of the molecular state to be studied.
\begin{figure}[htb]
 \unitlength=1cm
 \begin{center}
 \begin{picture}(6,6)
 \put(0,0){
\includegraphics{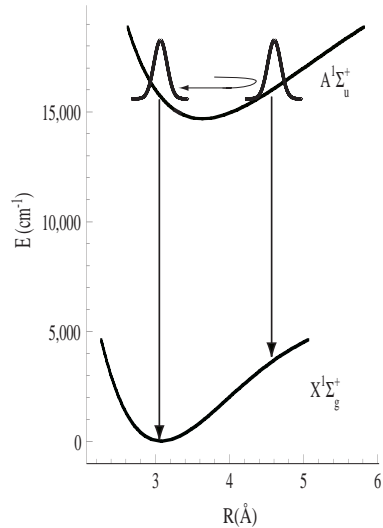}}
 \end{picture}
 \end{center}
\begin{center}
\protect\parbox{.9\textwidth}{
\caption{
The vibronic energies for the $A^1\Sigma^+_u \to X^1\Sigma^+_g$
transition of Na$_2$ clearly show the possibility to display
the vibrational motion (in the excited state) in the
time-resolved emission spectrum.
(After Kowalczyk, Radzewicz, Mostowski and Walmsley [1990].)
\label{fig4.1}
}
}
\end{center}
\end{figure}%
The experimental realization has been performed as follows (Dunn,
Walmsley and Mukamel [1995]). A sample of Na$_2$ molecules is illuminated
by a $4$~kHz train of laser pulses of $60$~fs duration and mean wavelength
of $630$~nm. The laser pulses generate vibrational wave packets in the
A$^1\Sigma_u^+$ state of the sodium dimer, which evolve with a time
period of $310$~fs. A fraction of the pulses is split off and plays the
role of a time-gate shutter. The light emitted from the molecular
sample is collected and focused synchronously with the split-off part
of the exciting pulse onto a nonlinear crystal. A prism monochromator
is used to filter the resulting sum-frequency and the field is
recorded by a photon-counting photomultiplier. The resulting temporal
resolution of the device is about $65$~fs.


\subsubsection{Harmonic regime}
\label{sec4.1.1}

The first reconstruction of the quantum state of molecular vibrations
from a time-resolved emission spectrum was based on the
assumption that only low vibrational quantum states are excited such that
the relevant potentials can be approximated by harmonic ones. Furthermore,
it was assumed that the vibrational frequencies in the two
electronic quantum states, which contribute to the emission spectrum,
are nearly equal. In this case, the vibronic coupling is caused solely by the
displacement of the equilibrium positions of the potentials in the two
electronic states.  When these approximations are justified, then the
time-gated spectrum $S(\Omega,T)$ can be related to the $s$-parametrized
phase-space distribution
$P(q,p;s)$ $\!\equiv$ $\!2^{-1}P[\alpha$ $\!=$ $\!2^{-1/2}(q$
$\!+$ $\!ip);s]$ as (Dunn, Walmsley and Mukamel [1995])
\begin{equation}
\label{eq4.1}
S(\Omega,T) = \int {\rm d}y \,
P[x(\Omega) \cos(\nu T) \!+\! y \sin(\nu T),
y \cos(\nu T) \!-\! x(\Omega) \sin(\nu T);s].
\end{equation}
Here, $\Omega$ and $T$ are the setting frequency of the spectral
filter and the setting time of the time gate, respectively ($\nu$,
vibrational frequency). The
function $x(\Omega)$ $\!=$ $\!(\Omega$ $\!-$ $\!\kappa^2\nu)/
\sqrt{2} \kappa \nu$ describes the mapping of the position of the
wave packet onto the emitted frequency, $\kappa$ being the ratio of
the displacement of the potentials to the width of the vibrational
ground state. The ordering parameter $s$ in eq.~(\ref{eq4.1}) is related to
the temporal width $\Gamma^{-1}$ of the time gate as
$s$ $\!=$ $\!-(\Gamma/\kappa \nu)^2$.

Obviously, the relation~(\ref{eq4.1}) between the phase-space distribution
$P(q,p;s)$ of the molecular vibration and the measured spectrum
$S(\Omega,T)$ is very close to the basic relation~(\ref{3.1}) of
optical homodyne tomography. Thus $P(q,p;s)$ can be obtained from
$S(\Omega,T)$ by means of inverse Radon transform (see
\S~\ref{sec3.1}). A typical experimental result
is shown in Fig.~\ref{fig4.2}.
\begin{figure}[htb]
 \unitlength=1cm
 \begin{center}
 \begin{picture}(6,5.3)
 \put(-1,-2.5){
\psfig{file=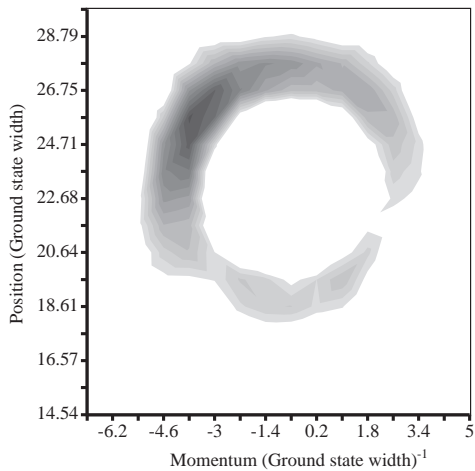,bbllx=29pt,bblly=30pt,bburx=583pt,bbury=762pt,clip=,%
width=8cm}
}
\end{picture}
\end{center}
\begin{center}
\protect\parbox{.9\textwidth}{
\caption{
Tomographic reconstruction of an $s$-parametrized phase-space
function for a vibrational wave packet in sodium from a measured set of
emission spectra. The time gate duration was $65$~fs, which implies
$s$ $\!=$ $-0.8$.
(After Dunn, Walmsley and Mukamel [1995].)
\label{fig4.2}
}
}
\end{center}
\end{figure}%


\subsubsection{Anharmonic vibrations}
\label{sec4.1.2}

In general, molecular vibrations are known to be significantly
anharmonic when their excitations are not restricted to very
small numbers of vibrational quanta. In such cases one cannot apply the
approximate reconstruction procedure based on eq.~(\ref{eq4.1}).
Due to the anharmonicity effects it is no longer possible to reconstruct
the quantum state only from one half of a vibrational period.

It was proposed (Shapiro, M., [1995]) to reconstruct the vibrational
quantum state from the time-resolved (spectrally integrated) intensity
$I(T)$ $\!=$ $\!\int {\rm d} \Omega \, S(\Omega,T) $ of the light
emitted by the molecular sample,
which can be related to the density-matrix elements $\varrho_{mn}$ of
the vibrational mode in the excited electronic state as
\begin{equation}
\label{eq4.2}
I(T) \sim \sum_{m,n} \varrho_{mn}
\sum_k \lambda_{km} \lambda^\ast_{kn}\,\exp(-i\nu_{mn}T) .
\end{equation}
Here, $\nu_{mn}$ are the vibrational transition frequencies in the
excited electronic state, and $\lambda_{km}$ $\!\sim$
$\!(\omega_{\,2\,1}^{mk})^2$ $\!_1\langle k|m\rangle_2$ is determined
by the Franck-Condon overlap $_1\langle k|m\rangle_2$ of the
vibrational wave functions in the two electronic states and the
vibronic transition frequency $\omega_{\,2\,1}^{mk}$. Equation
(\ref{eq4.2}) reveals that, as long as the transition frequencies
$\nu_{mn}$ are nondegenerate, the corresponding density-matrix
elements $\varrho_{mn}$ can be obtained, in principle, from
an analysis of $I(T)$ as a function of $T$.
However, the separation of the density-matrix elements from each
other may require a rather long time series.
Further, the dimension of the set of equations to be inverted can
be large, because of the large number of density-matrix elements
that may contribute to the intensity of the emitted light.
In the degenerate case, which is observed for the
diagonal elements of the density matrix and for some off-diagonal
elements due to the anharmonicities, additional information is needed.
It was proposed to use the stationary spectrum of the light, whose
determination requires an additional measurement.
Alternatively, the two measurements can be combined such that the
density-matrix elements are reconstructed from the time-resolved
spectrum (Trippenbach and Band [1996]).\footnote{Trippenbach and
  Band [1996] also discussed the inclusion of molecular rotations
  in the reconstruction.}

It is worth noting that the dimension
of the sets of equations that must be inverted numerically
can be reduced substantially by employing the full information
inherent in the time-resolved spectrum:
\begin{equation}
\label{eq4.3}
S(\Omega,T) \sim \sum_{m,n} \varrho_{mn}
\sum_k \lambda_{km} \lambda^\ast_{kn} \,
\exp(-i\nu_{mn}T) \,
g(\Omega\!-\!\omega_{21}^{mk}) \, g(\Omega\!-\! \omega_{21}^{nk})
\end{equation}
(Waxer, Walmsley and Vogel, W., [1997]).
In eq.~(\ref{eq4.3}), the blurring function $g(\omega)$ $\!=$
$\!\exp(-{\omega^2}/{4\Gamma^2})$ is determined by the resolution time
$\Gamma^{-1}$ of the time gate (see \S~\ref{sec4.1.1}).
In practice, the
time-dependent spectrum is available only in a finite time interval of
size $\tau$, which can be taken into account by multiplying
$S(\Omega,T)$ by the corresponding sampling-window function $G(T,\tau)$
in order to obtain
$S'(\Omega,T)$ $\!=$ $\!S(\Omega,T) G(T,\tau)$. The Fourier transform
of $S'(\Omega,T)$ with respect to $T$ reads
\begin{equation}
\label{eq4.4}
\underline{S}'(\Omega,\nu) \sim \sum_{m,m} \varrho_{mn}
\sum_k \lambda_{km} \lambda^\ast_{kn} \,
\underline{G}(\nu \! -\! \nu_{mn},\tau)
\,g(\Omega\!-\!\omega_{21}^{mk}) \, g(\Omega\!-\! \omega_{21}^{nk}),
\end{equation}
with $\underline{G}(\nu$ $\!-$ $\!\nu_{mn},\tau)$ being the Fourier
transform of the sampling window. Fixing the frequency $\nu$ of the
time-series spectrum, the structure of the window function ensures that
only some of the density matrix elements contribute at this frequency,
depending on the chosen size $\tau$ of the sampling window.  Let us
assume that the density matrix elements $\varrho_{m_i n_i}$ $(i$ $\!=$
$\!1,2,3,\dots,N)$ contribute to the spectrum at chosen frequency $\nu$.
By choosing $N$ frequencies $\Omega$ $\!=$ $\!\Omega_i$ of the emission
spectrum, one gets from eq.~(\ref{eq4.4}) a linear set of $N$ equations
of rather low dimension that can be inverted numerically.\footnote{Note
  that the emission frequencies $\Omega_i$ are determined by the
  two vibrational potentials involved in the vibronic emission, whereas the
  degenerate values of the vibrational frequencies $\nu_{mn}$ are determined
  solely by the vibrational potential in the excited electronic state.
  The (within the resolution of the sampling window) degenerate
  density-matrix elements usually contribute to the emission spectrum
  at distinct frequencies $\Omega_i$.}
In this way, the method allows one to reconstruct rather complicated
quantum states on a time scale that can  be shorter than the
fractional revival time,
as was demonstrated in a computer simulation of measurements
(Waxer, Walmsley and Vogel, W., [1997]).
Note that a reduction of the dimension of the set of equations to be
inverted improves the robustness of the method with respect to the
noise in the experimental data.

It was also proposed to reconstruct the density matrix from the
time-dependent position distribution according to eq.~(\ref{eq4.5a})
(Leonhardt and Raymer [1996]; Richter, Th., and W\"{u}nsche [1996a,b];
Leonhardt [1997a]; Leonhardt and Schneider [1997]).
This requires a scheme suitable for measuring either the position
distribution of molecular vibrations or another set of quantities that
can be mapped onto the position at different times. Note that this is
hardly possible in molecular emission tomography in general.
Finally it was proposed to use a wave-packet interference
technique (Chen and Yeazell [1997]; \S~\ref{sec4.3.0}) for
reconstructing pure vibrational states in the excited electronic
state (Leichtle, Schleich, Averbukh and Shapiro, M., [1998]).


\subsection{Trapped-atom motion}
\label{sec4.2}

Since the first observation of a single ion in a
{\em Paul trap\/}\footnote{For the trap, see Paul, W.,
   Osberghaus and Fischer [1958].}
(Neuhauser, Hohenstatt, Toschek and Dehmelt [1980])
much progress has been achie\-ved with respect to laser manipulation
of the quantized motion of single atoms in trap potentials.
Such systems are of particular interest since the quantized
low-frequency ($\sim$~MHz) motion is very stable, and laser
manipulations allow one to prepare very interesting nonclassical states.
Until now, motional Fock states and squeezed states (Meekhof, Monroe,
King, Itano and Wineland [1996]) as well as Schr\"odinger-cat type
superposition states (Monroe, Meekhof, King and Wineland [1996]) have
been realized experimentally.

One might expect that the reconstruction of the quantum state of the
center-of-mass motion of a trapped atom may be very similar to the
reconstruction of molecular vibrations. However, the vibronic
couplings in the two systems are basically different. In the case of
a molecule, the vibrating atoms are close together within atomic
dimensions and the change of the electronic state substantially alters
the potential of the nuclear motion. In the case of an atom in a
Paul trap the potential of the center-of-mass motion is externally
given by the trap. In this case, electronic transitions can hardly affect
the potential. The vibronic interaction in this system is induced by
the interaction with radiation. Therefore one may expect that appropriate
interactions of a trapped atom with laser fields may open various
possibilities for measuring the motional quantum state.


\subsubsection{Quadrature measurement}
\label{sec4.2.1}

The first proposals to reconstruct
the motional quantum state of a trapped atom were based on measuring
the quadrature components of the atomic center-of-mass motion in
the (harmonic) trap potential.
In the scheme in Fig.~\ref{fig4.4} a
weak electronic transition $|g\rangle$ $\!\equiv$ $|1\rangle$
$\!\leftrightarrow$ $|e\rangle$ $\!\equiv$ $\!|2\rangle$ of the atom
is proposed to be simultaneously driven by two (classical) laser beams
whose frequencies $\omega_{r}$ and $\omega_{b}$, respectively, are
tuned to the first motional sidebands, $\omega_{r}$ $\!=$
$\!\omega_{21}$ $\!-$ $\!\nu$ and $\omega_{b}$ $\!=$
$\!\omega_{21}+\nu$ of the electronic transition of frequency
$\omega_{21}$ (Wallentowitz and Vogel, W., [1995,1996b]). Since the
linewidth of the transition is very small, the motional sidebands can
be well resolved. For a long-living transition and in the {\em
  Lamb--Dicke regime\/}\footnote{In this regime, the (one-dimensional)
  spread of the motional wave packet, $\Delta x$, is small compared with
  the laser wavelength $\lambda_{L}$ over $2\pi$, $\Delta x\ll
  \lambda_{L}/2\pi$. The Lamb-Dicke parameter, $\eta_{LD}$ $\!=$
  $\!2\pi(\Delta x)_{0}/\lambda_{L}$, is a measure of the spread of
  the motional wave packet in the ground state of the trap potential
  relative to the wavelength.}
these interactions are well described
by Hamiltonians of the Jaynes-Cummings (and anti Jaynes-Cummings)
type.\footnote{For the description of the dynamics in the Lamb--Dicke
  regime by a Jaynes--Cummings type Hamiltonian, see Blockley, Walls
  and Risken [1992]; Wineland, Bollinger, Itano, Moore, and Heinzen
  [1992], Cirac, Blatt, Parkins and Zoller [1994]. Note that the fact
  that the method is based on the Lamb--Dicke regime is not a serious
  limitation. In recent experiments (Meekhof, Monroe, King, Itano and
  Wineland [1996]) the frequency of the weak transition is in the GHz
  range and it is driven by two lasers in a Raman configuration. In
  this case the Lamb--Dicke parameter can be changed easily by the
  laser beam geometry.}
Assuming that the Rabi frequencies of the two (classically treated)
laser fields are equal, the two Jaynes--Cummings interactions
can be combined to an electron-vibration coupling that in the
interaction picture reads as
\begin{equation}
\label{eq4.7}
  \hat{H}'
  = {\textstyle\frac{1}{2}}
\hbar \Omega_{L} \sqrt{2}  \left( \hat{\sigma}_{-} +
                     \hat{\sigma}_{+} \right) \hat{x}(\varphi),
\end{equation}
where $\Omega_{L}$ is the vibronic Rabi frequency of the driven
transitions, and $\varphi$ is the phase difference of the two lasers
which can be controlled precisely.
\begin{figure}[htb]
 \unitlength=1cm
 \begin{center}
 \begin{picture}(6,5)
 \put(0,0){
 \includegraphics{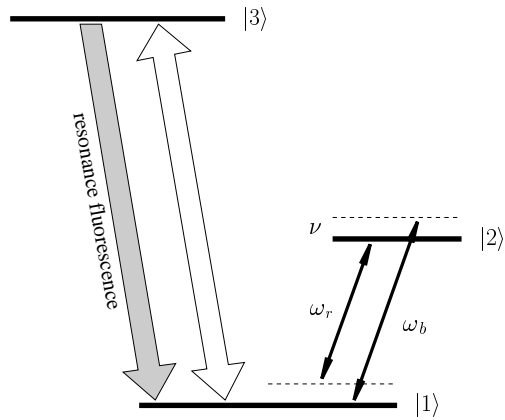}}
\end{picture}
\end{center}
\begin{center}
\protect\parbox{.9\textwidth}{
\caption{
Scheme of a trapped ion with a weak transition $|1\rangle$
$\!\leftrightarrow$ $\!|2\rangle$ and a strong transition
$|1\rangle$ $\!\leftrightarrow$ $\!|3 \rangle$.
Two incident lasers of frequencies $\omega_r$ $\!=$
$\!\omega_{21}$ $\!-$ $\!\nu$ and $\omega_b$ $\!=$
$\!\omega_{21}$ $\!+$ $\!\nu$ are detuned from the
electronic transition by the vibrational frequency
$\nu$ to the red and blue respectively. The laser driving
the strong transition is used for testing the ground state
occupation probability by means of resonance fluorescence.
(After Wallentowitz and Vogel, W., [1995].)
\label{fig4.4}
}
}
\end{center}
\end{figure}%
Assuming that the atom is initially prepared in the electronic
ground state, $P_{g}(t)|_{t=0}$ $\!=$ $\!1$ (incoherent preparation),
it can be shown that the atomic-state inversion at time $t$ is nothing
but the real part of the (scaled) characteristic
function $\Psi(x,\varphi)$ of the quadrature-component distribution
$p(x,\varphi)$ at phase $\varphi$ of the center-of-mass motion,
\begin{equation}
\label{eq4.8}
\Delta P^{({\rm inc})}(t)
= - {\rm Re}\,\Psi\!\left(\sqrt{2}\Omega_{L}t,\varphi\right)
\end{equation}
[$\Delta P$ $\!=$ $\!P_{e}$ $\!-$ $\!P_{g}$]. Similarly,
the imaginary part of the characteristic function can be obtained
by preparing the atom initially in an appropriate coherent
superposition of the two electronic quantum states such that
\begin{equation}
\label{eq4.9}
\Psi(\sqrt{2}\Omega_{L}t,\varphi)
= - \Delta P^{({\rm inc})}(t) - \Delta P^{({\rm coh})}(t).
\end{equation}

Hence, varying the phase difference between the two
driving laser beams, the complete information about the
motional quantum state can be determined from the time
evolution of the atomic-state inversion of the weak transition.
This occupation probability can be probed by testing an
auxiliary, strong transition $|1\rangle$ $\! \leftrightarrow$
$\!|3\rangle$ for
the appearance of fluorescence (cf. Fig.~\ref{fig4.4}).
The existence and the absence of fluorescence detects the weak transition
in state $|1\rangle$ and $|2\rangle$, respectively. This method has an
almost ideal quantum efficiency which is a great advantage for the present
purpose.\footnote{For the experimental detection of a weak transition
  via fluorescence from another, strong transition, see Nagourney,
  Sandberg and Dehmelt [1986]; Sauter, Neuhauser, Blatt and Toschek
  [1986]; Bergquist, Hulet, Itano and Wineland [1986].}
Applying eqs.~(\ref{3.9}) -- (\ref{3.10}),
the reconstruction of density matrix of the center-of-mass motion
in a quadrature-component basis from the measured signal can be
accomplished with a single Fourier integral, and application of
eqs.~(\ref{eq4.13}) and (\ref{eq4.14}) offers the possibility of
direct sampling of the density matrix in the Fock basis from
the fluorescence signal.
Further, using eqs.~(\ref{A2.19}) and (\ref{A2.22}), $s$-parametrized
phase-space functions can be inferred from the signal by performing
a two-fold integral.

The latter was demonstrated in another proposal suitable for measuring
$\Psi(z,\varphi)$ (D'Helon and Milburn [1996]). Here, the
$|1\rangle$ $\!\leftrightarrow$ $\!|2\rangle$ transition (Fig.~\ref{fig4.4})
is driven by a standing-wave laser pulse tuned to $\omega_{21}$ whose
duration $\tau_{P}$ is much shorter than the vibrational period,
$\nu\tau_{P}$ $\! \ll$ $\!1$, and it is assumed that
the center of the trap potential coincides with a node of the standing wave.
In the interaction picture, the interaction Hamiltonian
in the Lamb-Dicke regime then reads as\footnote{Note that the Hamiltonian
  (\ref{eq4.15}) is not based on a vibrational rotating wave approximation
  as it is the case for the Hamiltonian~(\ref{eq4.7}). In eq.~(\ref{eq4.15})
  the position operator results from the Lamb-Dicke approximation of the
  (operator-valued) mode function of the standing wave, $\sin(k_{L} \hat{x})$
  $\!\approx$ $\!k_{\rm L} \hat{x}$.}
\begin{equation}
\label{eq4.15}
\hat{H}'
= {\textstyle\frac{1}{2}}
\hbar \Omega_{L}(t) \sqrt{2}  \left( \hat{\sigma}_{-} +
                     \hat{\sigma}_{+} \right) \hat{x}(t).
\end{equation}
Here, $\hat{x}(t)$ corresponds to the freely evolving
position operator of the center-of-mass motion, which
for the harmonic motion considered is simply given by the
quadrature-component operator at phase $\varphi$ $\!=$ $\!\nu t$,
and the time-dependent (slowly varying) Rabi frequency $\Omega_{L}(t)$
includes the shape of the pulse. Owing to the shortness of the driving
pulse it can be assumed that its action at time $t$ gives rise to
a unitary kick described by the operator
\begin{equation}
\label{eq4.16}
\hat{K}
= T \exp\!\left[-i\hbar^{-1}\int {\rm d}t'\, H'(t') \right]
\approx \exp\!\left[- i {\textstyle\frac{1}{2}}
\theta ( \hat{\sigma}_{-} + \hat{\sigma}_{+}) \hat{x}(t)\right],
\end{equation}
where $\theta$ $\!=$ $\!\int {\rm d}t'$ $\!\Omega(t')$ is
the pulse area. Obviously, its action on the state is formally
the same as the action of the unitary operator that corresponds
to the interaction Hamiltonian (\ref{eq4.7}). Hence,
$\Delta^{(\rm inc)} P(t)$ and $\Delta^{(\rm coh)} P(t)$ measured
immediately after application of the pulse at time $t$
are again related to $\Psi(z,\varphi)$ according to eq.~(\ref{eq4.9}),
but now with $\theta$ and $\nu t$ in place of $\Omega_L t$ and
$\varphi$, respectively.\footnote{Whereas in the scheme of
   Wallentowitz and Vogel, W., [1995] the phase difference between
   the two laser beams must be controlled, the scheme of
   D'Helon and Milburn [1996] requires control of short-pulse
   areas.}

It is well known that a squeezed coherent state approaches
a quadrature-compo\-nent eigenstate as the strength of squeezing
goes to infinity; i.e.,
\begin{equation}
\label{eq4.19}
|x,\varphi\rangle
\propto \lim_{|\xi|\to\infty}
\hat{D}\big(xe^{i\varphi}/\sqrt{2}\big) \,
\hat{S}\big(|\xi| e^{i2\varphi}\big)|0\rangle ,
\end{equation}
with $\hat{D}(\alpha)$ the coherent displacement operator,
and $\hat{S}(\xi)$ $\!=$ $\!\exp[(\xi^\ast \hat{a}^2$ $\!-$
$\!\xi \hat{a}^{\dagger\,2})/2 ]$ the squeeze
operator.\footnote{For a proof, see, e.g.,
   Vogel, W., and Welsch [1994].}
Hence, the quadrature-component distribution $p(x,\varphi)$ can be
asymptotically given by
\begin{equation}
\label{eq4.20}
p(x,\varphi)
\propto \lim_{|\xi|\to\infty} \langle 0 | \hat{\varrho}'|0\rangle,
\end{equation}
where $\hat{\varrho}'$ is a coherently displaced and squeezed version of
the density operator $\hat{\varrho}$ to be determined,
\begin{equation}
\label{eq4.21}
 \hat{\varrho}'= \hat{S}^\dagger\big(|\xi| e^{i2\varphi}\big) \,
 \hat{D}^\dagger\big(x e^{i\varphi}/\sqrt{2}\big) \, \hat{\varrho} \,
 \hat{D}\big(x e^{i\varphi}/\sqrt{2}\big) \,
 \hat{S}\big(|\xi| e^{i2\varphi}\big).
\end{equation}
Equations (\ref{eq4.20}) and (\ref{eq4.21}) reveal that
when the quantum state to be measured can be squeezed strongly and
displaced coherently, then the quadrature-component distributions
can be obtained, in principle, from the occupation probability of the
motional ground state that is measured after these manipulations,
the phase being controlled by the free evolution of the vibrating
system (Poyatos, Walser, Cirac, Zoller and Blatt [1996]).
To realize the scheme, the following procedure was proposed:
\begin{itemize}
\item[(i)] Wait for a time $t$ such that $\varphi$ $\!=$ $\!\nu t$.
\item[(ii)] Perform a sudden displacement of the center of the trap to
the right for a distance $d$ such that $x$ $\!=$
$\!\sqrt{ m \nu/(2\hbar)}\,d$, $m$ being the mass of the atom.
\item[(iii)] Change the trap frequency instantaneously from $\nu$
to (lower) $\nu\,'$.\footnote{The sudden change of
  the trap frequency leads to squeezing with the squeeze parameter
  being $|\xi|= \ln(\nu/\nu')/2$ (Janszky and Yushin [1986]). Note
  that a significant change of the trap frequency is needed since the
  measurement scheme requires strong squeezing.}
\item[(iv)] Determine the population of the motional ground
state.\footnote{It was demonstrated experimentally that the motional
  number statistics can be determined from the measured Jaynes--Cummings
  revivals (Meekhof, Monroe, King, Itano and Wineland [1996];
  \S~\ref{sec4.2.2}).}
\end{itemize}
Knowing $p(x,\varphi)$ for all phases $\varphi$ in a $\pi$ interval,
the complete information on the motional quantum state is
available and the (tomographic) methods of quantum-state reconstruction
outlined in \S~\ref{sec3} applies.\footnote{The steps (i), (ii)
   and (iv) were proposed in order to measure the $Q$ function.}


\subsubsection{Measurement of the Jaynes--Cummings dynamics}
\label{sec4.2.2}

Let us consider the dynamics of a trapped ion that is driven
by a single (classical) laser beam in a resolved sideband
regime\footnote{In the
  practical realization the single laser may also be replaced with
  two lasers driving a dipole-forbidden transition in a Raman
  configuration, which essentially yields the same basic Hamiltonian.}
and assume that the laser frequency is tuned to the (red)
$k$th-order motional sideband of the weak electronic
transition $|1\rangle$ $\!\leftrightarrow$ $\!|2\rangle$,
$\omega_{21}$ $\!=$ $\! \omega_l$ $\! -$ $\! k\nu$. In this case
the resulting coupling (in the interaction picture) between the
electronic transition and the center-of-mass motion has the form
of a nonlinear multiquantum Jaynes-Cummings interaction
(Vogel, W., and de Matos Filho [1995]),
\begin{equation}
\label{eq4.22}
\hat{H}'
= {\textstyle\frac{1}{2}}
\hbar \Omega_{L}  \hat{\sigma}_{+}
\hat{f}^{(k)}(\hat{a}^\dagger \hat{a}) \,\hat{a}^k + {\rm H.c.},
\end{equation}
where the operator function $\hat{f}_k$ reads in normally
ordered form as
\begin{equation}
\label{eq4.23}
\hat{f}^{(k)}(\hat{a}^\dagger \hat{a})
= e^{-\eta_{LD}^{2}/2} \sum_{l=0}^\infty
\frac{(i\eta_{LD})^{2l+k}}{l!(l+k)!} \,\hat{a}^{\dagger\,l} \hat{a}^l
\end{equation}
($\eta_{LD}$, Lamb--Dicke parameter).
It describes the influence of the interference between the driving
laser wave and the (extended) atomic wave function on the dynamics of
the system. The nonlinear Jaynes--Cummings model can be solved exactly.
In particular, when the atom is initially prepared in the
excited electronic state, then the atomic-state inversion
is given by an equation of the type of
eq.~(\ref{3.44}),\footnote{The dynamics were realized experimentally
and used to determine the (initial) excitation statistics
  $\varrho_{nn}$ of the motional state by inverting eq.~(\ref{eq4.24})
  (Meekhof, Monroe, King, Itano and Wineland [1996]).
  The same dynamics can be
  realized by tuning the laser onto the corresponding blue sideband,
  $\omega_{21}$ $\!=$ $\! \omega_l$ $\! +$ $\! k\nu$, and preparing
  the electronic subsystem initially in the ground state.}
\begin{equation}
\label{eq4.24}
\Delta P(t) = \sum_{n=0}^{\infty}
\varrho_{nn} \cos\!\left(\Omega_{n\,n+k} t \right) ,
\end{equation}
where now $\Omega_{n\,n+k}$ $\!=$
$\!\Omega_{L}\langle n| \hat{f}^{(k)}\hat{a}^{k}|n\!+\!k\rangle$,
and $\varrho_{nn}$ are the motional density-matrix elements
at the initial time $t$ $\!=$ $\!0$.

The motional quantum state can be reconstructed from the measured
(nonlinear) Jaynes--Cummings dynamics by applying typical methods
outlined in \S~\ref{sec3}. When the motional quantum state
which is desired to be reconstructed is coherently shifted
[$\hat{\varrho}$ $\!\to$ $\!\hat{D}^{\dagger}(\alpha)
\hat{\varrho}\hat{D}(\alpha)$] before
$\Delta P(t)$ is measured, then the Fock-state probability
distribution $p_{n}$ $\!=$ $\!\varrho_{nn}$ in eq.~(\ref{eq4.24})
must be replaced with the displaced Fock-state probability distribution
$p_{n}(\alpha)$ $\!=$ $\!\varrho_{nn}(\alpha)$
$\!=$ $\langle n,\alpha|\hat{\varrho}|n,\alpha\rangle$
[cf. eq.~(\ref{3.45})], and inversion of eq.~(\ref{eq4.24}) yields
$p_{n}(\alpha)$. Changing $\alpha$ in a succession of
(ensemble) measurements, the complete information about
the original (unshifted) motional quantum state can be inferred
from the measured data.
In particular, applying the method outlined in \S~\ref{sec3.5},
pointwise reconstruction of phase-space functions is feasible.
\begin{figure}[htb]
 \unitlength=1cm
 \begin{center}
 \begin{picture}(6,6)
 \put(0,0){
\includegraphics{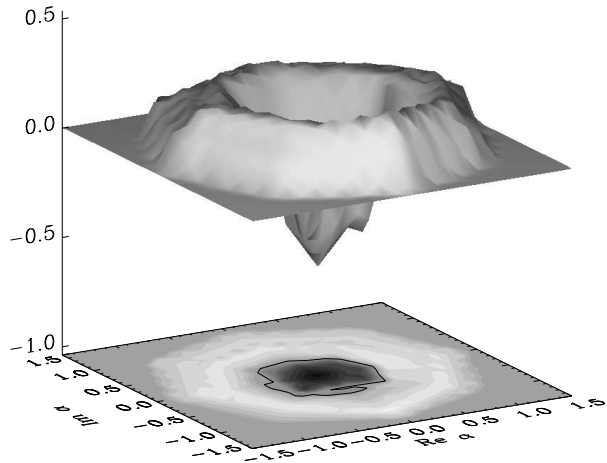}}
 \end{picture}
 \end{center}
\begin{center}
\protect\parbox{.9\textwidth}{
\caption{
Surface and contour plots of the reconstructed Wigner function
$W(\alpha)$ of the $|n$ $\!=$ $\!1\rangle$ motional number state.
The plotted points are the result of fitting a linear interpolation
between the actual data points to a $0.1$ by $0.1$ grid. The
octagonal shape is an artifact of the eight measured phases
per radius.
(After Leibfried, Meekhof, King, Monroe, Itano and Wineland [1996].)
\label{fig4.5}
}
}
\end{center}
\end{figure}%
This was demonstrated successfully by reconstructing the
Wigner function [eq.~(\ref{3.38})] of a single
$^9$Be$^+$ ion that is stored in a rf Paul trap
(Leibfried, Meekhof, King, Monroe, Itano and Wineland [1996]). In the
experiment, the relevant oscillation frequency in the trap potential is
$\nu/2\pi$ $\! \approx$ $11.2$~MHz, and the transition between the states
$|g\rangle$ and $|e\rangle$ is a stimulated Raman transition
between the hyperfine ground states $^2S_{1/2}$
($F$ $\!=$ $\!2$, $m_F$ $\!=$ $\!-2$) and $^2S_{1/2}$
($F$ $\!=$ $\!1$, $m_F$ $\!=$ $\!-1$), respectively, which are
separated by about $1.25$~GHz. The coherent displacement of
the initially prepared motional quantum state is realized by
applying a classical, spatially uniform rf field.
In Fig.~\ref{fig4.5} an example of a reconstructed Wigner
function of a motional number state is shown.

The density matrix in the Fock basis, $\varrho_{mn}$,
can be obtained from the displaced Fock-state probability
on a circle, applying the method outlined in \S~\ref{sec3.3.2}.
This was also demonstrated in the above mentioned experiment
(Leibfried, Meekhof, King, Monroe, Itano and Wineland [1996]).
In Fig.~\ref{fig4.6} an example of a reconstructed density
matrix of a superposition of two Fock states is shown.
\begin{figure}[htb]
 \unitlength=1cm
 \begin{center}
 \begin{picture}(6,4.5)
\put(-.5,-.5){
\psfig{file=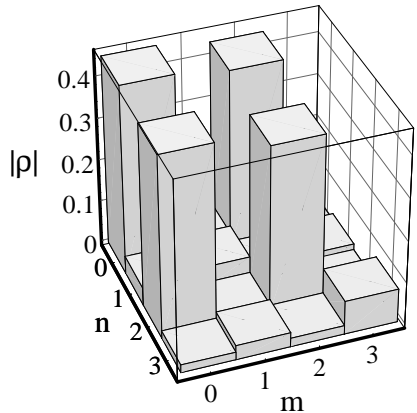,bbllx=148pt,bblly=374pt,bburx=470pt,bbury=700pt,clip=,%
width=5.5cm}
}
 \end{picture}
 \end{center}
\begin{center}
\protect\parbox{.9\textwidth}{
\caption{
Reconstructed density-matrix amplitudes $|\rho|$ $\!\equiv$
$\!|\varrho|_{mn}$ of an approximate
$1/\protect\sqrt{2}(|n$ $\!=$ $\!0\rangle$ $\!-$ $\!i|n$ $\!=$ $\!2\rangle)$
motional state. The state was displaced by $|\alpha|$ $\!=$ $\!0.79$
for eight phases on the circle.
(After Leibfried, Meekhof, King, Monroe, Itano and Wineland [1996].)
\label{fig4.6}
}
}
\end{center}
\end{figure}%

The method of coherently displacing the motional quantum state
to be detected can be extended in order to measure
the Wigner function rather than reconstructing it from the
measured data (Lutterbach and Davidovich [1997]).\footnote{For
   comparison, see the approach to direct measurement of the Wigner
   function in cavity QED as outlined in \S~\ref{sec2.4}.}
For this purpose the laser driving the Jaynes--Cummings dynamics is
tuned on resonance to the electronic transition, so that the
Hamiltonian (\ref{eq4.22}) for $k$ $\!=$ $\!0$ applies.
For sufficiently small Lamb--Dicke parameter, $\eta_{LD}$ $\!\ll$ $\!1$,
the nonlinear function $\hat{f}^{(0)}$ in eq.~(\ref{eq4.23}) can
be approximated by $\hat{f}^{(0)}$ $\! \approx$ $\!1$ $\!-$
$\!\eta_{LD}^{2}(\hat{a}^\dagger \hat{a}$ $\!+$ $\!1/2)$.
Consequently, the vibronic Rabi frequencies read as
$\Omega_{nn}$ $\!\approx$ $\Omega_{L}$ $\![1$ $\!-$ $\!\eta_{LD}^{2}$
$\!(n$ $\!+$ $\!1/2)]$. Inserting this result into
eq.~(\ref{eq4.24}), with $\varrho_{nn}(\alpha)$ in place
of $\varrho_{nn}$, and choosing the interaction time as
$t$ $\!=$ $\!t_{W}$ $\!=$ $\!\pi/(\Omega \eta_{LD}^{2})$ yields
\begin{equation}
\label{eq4.25}
\Delta P(t_{W})
\approx \sin(\pi/\eta_{LD}^{2}) \sum_{n=0}^{\infty} \varrho_{nn} (-1)^n
= {\textstyle\frac{1}{2}}\pi \sin(\pi/\eta_{LD}^{2})\,W(\alpha),
\end{equation}
which reveals that under the conditions given
the measured electronic-state occupation probability
directly gives the Wigner function at the phase-space
point $\alpha$.

Further it was proposed to reconstruct the density matrix in
the Fock basis following a line similar to that given by
eqs.~(\ref{3.40}) -- (\ref{3.42}) for cavity QED
(Bardroff, Leichtle, Schrade and Schleich [1996]). In
cavity QED the $k$-photon transitions needed
to calculate the off-diagonal density matrix elements
$\varrho_{n\,n+k}$ from the measured atomic-state inversion
(provided that initially a coherent superposition
of the two states is prepared) can hardly be realized for
larger values of $k$. In
the case of a trapped ion the situation is improved significantly. The
$k$-quantum interaction of Jaynes--Cummings type as given by
eq.~(\ref{eq4.22}) can be realized as outlined above, by tuning the
laser on resonance to the $k$th motional sideband. This enables one
to reconstruct the density matrix elements $\varrho_{n\,n+k}$ for
all $k$ in principle, by successively tuning the laser on all sidebands.
Needless to say, this requires sufficiently large values of the
Lamb--Dicke parameter, since otherwise the coupling of the laser
to high-order sidebands is very small. Formally, the
difference to cavity QED only consists in the frequencies
$\Omega_{n\,n+k}$, which are different for the two
systems.\footnote{At this point it should be noted that in
  a Paul trap the applied rf field may modulate the motion
  of a trapped atom, which gives rise to the so-called {\em micromotion}.
  The rf frequency is usually large compared to the (effective) motional
  frequency, and its effect may consist in a modification of the
  excitation-dependent function $\hat{f}^{(k)}$ in the interaction
  Hamiltonian (\ref{eq4.22}) (Bardroff, Leichtle, Schrade and Schleich
  [1996]). Apart from such modifications, the structure of the
  Hamiltonian is preserved.
  For details of the treatment of the micromotion, see
  the original article.}

It was also proposed to reconstruct the density matrix in the
coherent-state basis by combining coherent displacements with
motional ground state measurements (Freyberger [1997]).
The scheme allows a pure motional quantum state to
be reconstructed by coherent displacements on a circle.
Reconstruction of the density matrix of a mixed
state requires additional variation of the displacement amplitude.
A first displacement is introduced before a filtering measurement
is performed in order to decide whether or not the atom is in
the motional ground state. A second displacement is performed which
is followed by detecting the probability that the atom is in the
motional ground state, which may be obtained from the
Jaynes--Cummings dynamics as outlined above.


\subsubsection{Entangled vibronic states}
\label{sec4.2.3}

Let us consider the situation where the motional state is entangled
with the electronic state of the system
and ask for measuring the complete vibronic state.
Such a measurement may be realized by the interaction
Hamiltonian~(\ref{eq4.22}) for $k$ $\!=$ $\!0$
(Wallentowitz, de Matos Filho and Vogel, W., [1997]).
In the scheme the initially prepared vibronic state $\hat{\varrho}$
is first displaced coherently in the phase space of the motional
subsystem, $\hat{\varrho}$ $\!\to$ $\hat{\varrho}(\alpha)$ $\!=$
$\!\hat{D}^{\dagger}(\alpha)\hat{\varrho}\hat{D}(\alpha)$.
Next the driving laser causing the electron-motional coupling
is switched on for an interaction time $\tau$.
This procedure is followed by probing the atom for fluorescence on the
strong, auxiliary transition. Provided the atom is detected in the
excited electronic state $|e\rangle$ (no fluorescence), the density
operator $\hat{\varrho}(\tau)$ of the system reduces to
\begin{equation}
  \label{eq4.28}
\hat{\varrho}'(\tau) \sim |e\rangle\langle e| \otimes
\hat{\varrho}_{M}(\tau),
\end{equation}
where $\hat{\varrho}_{M}(\tau)$ $\!\sim$
$\!\langle e|\hat{\varrho}(\tau)|e\rangle$ is the corresponding
density operator of the motional subsystem. Its diagonal elements
in the Fock basis read
\begin{eqnarray}
\label{eq4.29}
\lefteqn{
\left[\varrho_{M}(\tau)\right]_{nn}
\sim {\rm Im}\!\left[ \varrho^{nn}_{ge}(\alpha)
\right]
\sin\!\left( \Omega_{nn} \tau \right)
}
\nonumber \\ && \hspace{10ex}
+ \, \varrho^{nn}_{ee}(\alpha)
\cos^2\!\left( {\textstyle\frac{1}{2}} \Omega_{nn} \tau \right)
+ \varrho^{nn}_{gg}(\alpha)
\sin^2\!\left( {\textstyle\frac{1}{2}} \Omega_{nn} \tau \right) .
\end{eqnarray}
Equation (\ref{eq4.29}) reveals that appropriately
chosen interaction times $\tau$ allow the displaced vibronic
density-matrix elements $\varrho_{ab}^{nn}(\alpha)$ to
be mapped onto the reduced motional number statistics
$[\varrho_{M}(\tau)]_{nn}$. It is worth noting that the
underlying vibronic coupling is not only suited for
this mapping but also for a QND measurement
of the reduced motional number statistics, since the Hamiltonian
commutes with the motional number operator.\footnote{For the
  application of this type of vibronic coupling to
  QND measurement of the motional
  excitation of a trapped atom, see de Matos Filho and
  Vogel, W., [1996]; Davidovich, Orszag and Zagury [1996].}
Note that the reduced motional number statistics can also
be determined using other methods, such as the methods
outlined in \S~\ref{sec4.2.2}.

Knowing the displaced vibronic density-matrix elements
$\varrho_{ab}^{nn}(\alpha)$ as functions of $\alpha$,
the density matrix $\varrho_{ab}^{mn}$ of the original
vibronic quantum state can be reconstructed, following
the lines given in \S~\ref{sec3}. In particular,
the elements $\varrho^{nn}_{ab}(\alpha)$ can be summed up to
obtain the vibronic quantum state in terms of the Wigner-function
matrix $W_{ab}(\alpha)$,
\begin{equation}
  \label{eq4.30}
  W_{ab}(\alpha) = \frac{2}{\pi} \sum_{n=0}^\infty (-1)^n
  \varrho_{ab}^{nn}(\alpha)
\end{equation}
(Wallentowitz, de Matos Filho and Vogel, W., [1997]). The Wigner-function
matrix $W_{ab}(\alpha)$ has the following properties. Its trace with
respect to the electronic subsystem is the (reduced) motional Wigner
function. Integrating $W_{ab}(\alpha)$ with respect
to the motional phase-space amplitude $\alpha$ yields the (reduced)
electronic density matrix. A typical quantum state for which such a
measurement scheme would be of interest is an entangled state of the
type $|\psi\rangle$ $\!\sim$ $\!(|2\rangle |\alpha\rangle$ $\!\pm$
$\!|1\rangle | -\alpha\rangle)$.\footnote{States of this type have been
  realized experimentally (Monroe, Meekhof, King and Wineland [1996]).
  For a reconstruction of $W_{ab}(\alpha)$ from a computer simulation
  of measurements, see Wallentowitz, de Matos Filho and Vogel, W., [1997]).}
Clearly, the nonclassical features of such a state are completely lost
when one measures only the reduced motional quantum state.


\subsection{Bose--Einstein condensates}
\label{sec4.4}

The recent progress in evaporative cooling of an atomic gas
has rendered it possible to realize Bose--Einstein condensation
experimentally (Anderson, Ensher,
Matthews, Wieman and Cornell [1995]; Bradley, Sackett, Tollett and Hulet
[1995]; Davies, Mewes, Andrews, van Druten, Durfee, Kurn and Ketterle
[1995]). The Bose--Einstein condensate (BEC) represents a macroscopic
occupation of the ground state of the gas, which is an important
signature of quantum-statistical mechanics. Moreover, coherence
effects, such as interference fringes between two condensates,
could be demonstrated
(Mewes, Andrews, Kurn, Durfee, Townsend and Ketterle [1997], Andrews,
Townsend, Miesner, Durfee, Kurn and Ketterle [1997];
for a theoretical interpretation, see Wallis, R\"{o}hrl, Naraschewski
and Schenzle [1997]). In view of these
feasibilities it is interesting to get more insight in the exact
nature of the quantum state of BEC.

Let us consider a two-mode BEC consisting of $N$ atoms,\footnote{The
  two-mode BEC is considered in view of the lack of a coherent
  reference state within a condensate of a fixed number of atoms.}
where the two modes may correspond to different hyperfine states. To
reconstruct the quantum state, it was proposed to introduce a
controllable phase shift $\phi$ between the modes and to mix them by
applying a (lossless) beam-splitter transformation. The
numbers of atoms in the two modes are counted, and the two-mode
quantum state is inferred from joint counting statistics (Bolda,
Tan and Walls [1997], Mancini and Tombesi [1997b], Walser
[1997]).\footnote{The three proposals are basically similar.
  Concerning the practical realization of such a measurement, we
  briefly outline the rather detailed scheme proposed by Bolda, Tan,
  and Walls [1997]. For the interaction with radiation of
  an atomic Bose gas in an isotropic harmonic oscillator potential,
  see also Javanainen [1994].}
Neglecting collisions between the atoms, the two-mode density operator
$\hat{\varrho}$ is transformed as
\begin{equation}
  \label{eq4.34}
  \hat{\varrho}'
  = \hat{U}^\dagger(\beta,\phi) \hat{\varrho} \hat{U}(\beta,\phi),
\end{equation}
with
\begin{equation}
  \label{eq4.35}
 \hat{U}(\beta,\phi)
 = \exp\!\left[{\textstyle\frac{1}{2}}i\beta
 \left(\hat{a}_1^\dagger \hat{a}_2
 e^{i\phi} +\hat{a}_2^\dagger \hat{a}_1 e^{-i\phi}\right)
 \right].
\end{equation}
[cf. eq.~(\ref{2.11})], where the mode subscripts $1$ and $2$ label the
corresponding hyperfine states. Raman transitions between these states
can be made by optical pulses that are off-resonance from an excited
state $3$. An rf field may be used to couple state $1$ to another state $4$,
and atoms in state $4$ are repelled from the trap. The beam splitter
transformation can be realized by applying two copropagating light pulses
that are detuned from the $1-3$ and $2-3$ transitions. The phase $\phi$ is
controlled by the phase difference of the lasers. The pulses are
followed by an rf $\pi$-pulse that transfers the state $1$ to $4$.
Consequently, the atoms in state $4$ fall freely and those in state $2$
remain trapped. Eventually, the seperate groups of trapped and
untrapped atoms are counted (for the experimental setup,
see also Mewes, Andrews, Kurn, Durfee, Townsend and Ketterle [1997],
Andrews, Townsend, Miesner, Durfee, Kurn and Ketterle [1997]).
The probability $P_{m}(\phi)$ $\!\equiv$ $P_{n,N-n}(\phi)$
of counting $n$ and $N$ $\!-$ $\!n$ atoms, respectively, in
the modes $1$ and $2$ for a phase shift setting of $\phi$ can be
given by\footnote{For simplicity, the abbreviating
  notation $|m\rangle$ for the two-mode states $|n\rangle_1$
  $\! \otimes$ $\!|N-n\rangle_2$ is used.}
\begin{equation}
  \label{eq4.36}
P_{m}(\phi)
=
\langle m|
\hat{U}^\dagger(\beta,\phi) \hat{\varrho} \hat{U}(\beta,\phi)
|m\rangle
=
\sum_{l,l'}
T_{m}^{ll'}(\phi)  \varrho_{ll'},
\end{equation}
where $T_{m}^{ll'}(\phi)$ is definded by the corresponding matrix
elements of the beam splitter transformation in the basis of the
states $|m\rangle$.

The density matrix of the two-mode state can be obtained by numerically
inverting eq.~(\ref{eq4.36}) (Bolda, Tan and Walls [1997],
Mancini and Tombesi [1997b]) by means of the
least-squares method (Appendix \ref{app4}; see also \S~\ref{sec3.9.2}).
A further simplification can be achieved by taking
the Fourier transform of $P_m(\phi)$ with respect to $\phi$ (Bolda,
Tan and Walls [1997]). Since the
$s$th Fourier component $P_{m}^{s}$ $\!=$ $\!(2\pi)^{-1}\int_0^{2\pi}
{\rm d}\phi$ $\!e^{is\phi} P_m (\phi)$ is related only to the matrix
elements whose row and column indices differ by $s$, the corresponding
blocks of equations can be inverted separately (cf.
\S~\ref{sec3.3.2}). Alternatively, the inversion can also be
performed analytically by applying orthogonality relations for
Clebsch-Gordan coefficients (Walser [1997]).

The complex amplitude of a BEC field, $\psi({\bf r},t)$
$\!=$ $\langle\hat{\psi}({\bf r},t)\rangle$, is frequently assumed
to satisfy a nonlinear Schr\"{o}dinger equation
\begin{eqnarray}
\label{bose-1}
i\hbar\frac{\partial \psi}{\partial t} = \left(
-\frac{\hbar^2}{2m} \Delta  + U + g|\psi|^{2} \right) \psi
\end{eqnarray}
($m$, particle mass; $U$, potential; $g$, atom--atom interaction
constant). The density of particles $|\psi|^{2}$ can be measured
by phase-contrast imaging (Andrews, Mewes, van Druten, Durfee, Kurn
and Ketterle [1996]). In order to determine $\psi({\bf r},t)$
for given $|\psi({\bf r},t)|^{2}$, it was proposed to
use a nonlinear propagation for a trial function $\varphi$,
so that the correct function $\psi$ can be found by
adaptive modification of the trial function
(Leonhardt and Bardroff [1997]).


\subsection{Atomic matter waves}
\label{sec4.5}

Stimulated by the
successful experimental demonstration of atom interferometry
(Carnal and Mlynek [1991], Keith, Ekstrom, Turchette and Pritchard
[1991]) the study of coherence properties of atomic matter waves
has been of increasing interest. In particular, several methods
have been considered for determining the (single-particle)
quantum state of both the transverse and the longitudinal center-of-mass
motion of the atoms, and first experiments have been performed.

\subsubsection{Transverse motion}
\label{sec4.5.1}

The density matrix of the (with respect to the direction of
propagation) transverse motion of a single particle
can be determined by methods of refractive (atom) optics (Raymer,
Beck and McAlister [1994], Janicke and Wilkens [1995]). Combining
the effect of lenses and propagation, it is possible to
reconstruct the quantum state from measured position distributions
by tomographic methods (\S~\ref{sec3.1}).
The desired rotation of the
quadrature components can be realized by appropriately changing the
parameters of the experimental set-up, such as the focal lengths of
the lenses, their positions and the positions of the detectors
(Raymer, Beck and McAlister [1994]).\footnote{For
  possible realizations of optical elements in atom optics,
  see, e.g., Adams, Sigel and Mlynek [1994].}

\begin{figure}[htb]
 \unitlength=1cm
 \begin{center}
 \begin{picture}(6,6)
 \put(0,0){
\includegraphics{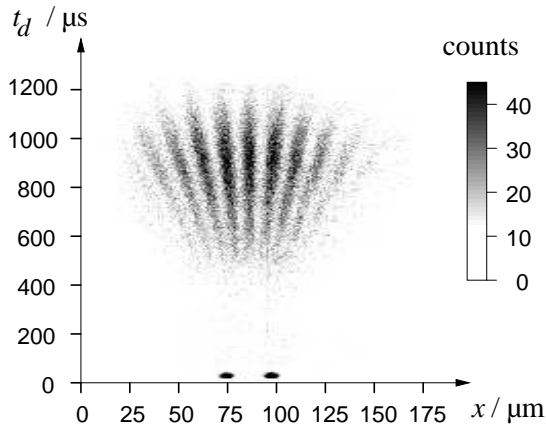}}
 \end{picture}
 \end{center}
\begin{center}
\protect\parbox{.9\textwidth}{
\caption{
Measured time-resolved spatial atomic distribution in the Fraunhofer
diffraction limit. The two wave packets emerging from the individual
slits overlap almost completely, and a maximal visibility of the
interference fringes is observed.
(After Pfau and Kurtsiefer [1997].)
\label{fig4.7a}
}
}
\end{center}
\end{figure}%
In the first experimental reconstruction of the
transverse quantum state of an atomic matter wave (Kurtsiefer, Pfau
and Mlynek [1997], Pfau and Kurtsiefer [1997]) the scheme is
simplified by avoiding the use of lenses. Metastable helium atoms are
used whose quantum state of the transverse motion is prepared by
a combination of an
entrance slit and a double slit, which realizes a
nonclassical two-peaked atom-wave state. The atoms then propagate
for some distance to a time- and space-resolving detector,
which allows one to measure the transverse atomic distribution of
the atoms with a spatial resolution down to $500$~nm, and the
arrival time of the atoms with an accuracy of $100$~ns.
This renders it possible to record spatial atomic distributions
for different free-evolution times of the atom-wave packet after the
double slit (Fig.~\ref{fig4.7a}), which are suited for reconstructing
the quantum state. The measured distributions are
position distributions of the sheared Wigner function
\begin{equation}
  \label{eq4.38}
W(x,p,t) = W\!\left(x-\frac{p}{m}\, t, p,0\right)
\end{equation}
($m$, mass of the atom) of the freely evolving atom-wave packet
after the double slit. Knowing the position distributions for a
sufficiently large set of evolution times, the original quantum
state can be obtained (\S~\ref{sec3.3.1}; for reconstruction of
an original Wigner function from  marginals of the sheared Wigner
function, see also Lohmann [1993]).
In particular, the measured position distributions correspond to the
quadrature-component distributions whose phase parameter is given by
\begin{equation}
  \label{eq4.39}
\varphi  = \arctan\!\left (\frac{t\,\hbar}{m x_0^2}\right ),
\end{equation}
($x_0$, scaling length).
Obviously, the phase $\varphi$ can only be scanned
between $0$ and $\pi/2$, and in a real experiment this interval is
further reduced in general. Since a precise reconstruction
of the quantum state requires a
phase interval of size $\pi$, some {\em a priori} information about
the state should be available (cf. \S~\ref{sec3.9.1}).
In the present case it may be possible to assume symmetry in position
space for the Wigner function, because of the preparation of the
quantum state by a double slit. The
Wigner function reconstructed (with a limited
range of phases $\varphi$) from the measured diffraction pattern in
Fig.~\ref{fig4.7a} by applying inverse Radon transform
(\S~\ref{sec3.1}) is shown in Fig.~\ref{fig4.7b}.
\begin{figure}[htb]
 \unitlength=1cm
 \begin{center}
 \begin{picture}(6,6)
 \put(0,0){
\includegraphics{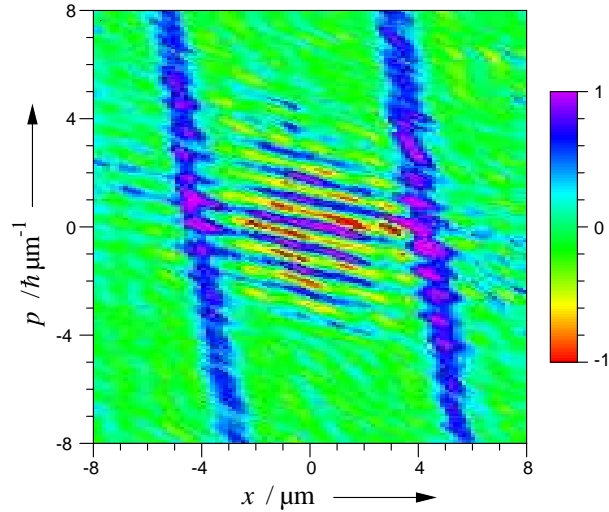}}
 \end{picture}
 \end{center}
\begin{center}
\protect\parbox{.9\textwidth}{
\caption{
Wigner function of the atomic motional state immediately behind
the double slit, reconstructed from the measured time-resolved
spatial atomic distribution in Fig.~\protect\ref{fig4.7a}.
(After Pfau and Kurtsiefer [1997].)
\label{fig4.7b}
}
}
\end{center}
\end{figure}%

Although the shape of the Wigner function and especially the
oscillating interference part are well reproduced, there is a
systematic error owing to the
limited range of phases accessible in the experiment.
For example, the reconstructed Wigner function appears sheared.
This shear could be avoided by forced symmetrization of
the Wigner function in position.
Using the {\em Wittaker-Shannon sampling theorem} (see, e.g.,
Marks [1991]), it can be proved
which features of the reconstructed Wigner function of an
atom-wave packet are obtained
correctly when the set of data is not tomographically complete
(Raymer [1997]). The analysis shows that the density matrix
in the momentum basis
is correct everywhere except in an excluded region of
width $(\delta p)_{\rm min}$ around the diagonal, with
$(\delta p)_{\rm min}$ being inversely proportional to the time
interval over which the diffraction pattern is recorded.
Transferring this result to the
reconstructed Wigner function, it turns out that there is a
low-frequency error in $W(x,p)$ as a function of $x$, whereas the
high-frequency behaviour is essentially correct.

In another tomographic scheme for reconstructing the
Wigner function of an initially prepared transverse motional
state, a set-up is considered, in which a classical standing
light field that is strongly detuned from the atomic transition
serves as a thick gradient-index lens for two-level atoms
which are prepared in the ground state and
cross the light field close to a node (Kienle, Fischer,
Schleich, Yakovlev and Freyberger [1997]; for a set-up of
this type, see also Fig.~\ref{fig2.8}).
It was found that
here the phase $\varphi$ of the quadrature-component
distribution associated with the measured time-resolved position
distribution can be tuned over a complete $\pi$ interval.

In a modified (interferometric) scheme it is assumed that the phase
$\varphi_{L}$ of the mode function of the standing-wave laser field
can be controlled via a movable mirror, which is used to create
the standing wave (Freyberger, Kienle and Yakovlev [1997];
Kienle, Fischer, Schleich, Yakovlev and Freyberger [1997]).
It is shown that when a pure motional state is realized, then
the phase of the free-evolving wave function can be inferred
from the atomic position distributions for phases of the laser field
$\varphi_{L}$ $\!=$ $\!-\pi/4$, $\!0$, $\!\pi/4$ and $\pi/2$
at chosen time. Knowing from a pre-measurement the absolute
value of the wave function for this time, the initial wave function
can then be reconstructed by inverting the free time evolution.

\subsubsection{Longitudinal motion}
\label{sec4.5.2}

In order to determine the quantum state of the longitudinal
center-of-mass motion of atoms in an atomic beam,
it was proposed to reconstruct the Wigner
function tomographically via state-selective time-dependent
measurement of longitudinal position distributions
(Kokorowski and Pritchard [1997]).
In the scheme, two-level atoms that are initially prepared in
the ground state are considered. Located at an
adjustable position in the apparatus is an electromagnetic
excitation region that is assumed to be effectively a
$\pi$ pulse; i.e., any particle exposed to the radiation
in the region is completely excited.
Downstream, a state-selective
detector counts the number of particles which have made a
transition into the excited state owing to the $\pi$ pulse.
Time dependence may be introduced either by operating the excitation
region in short pulses at definite times, or by using an excited
state whose lifetime is much shorter than the desired time
resolution. In the latter case, detection consists of observing
spontaneously emitted radiation, and the time is simply the
time of detection.
Measurement of the longitudinal position distribution as a
function of time then yields, similar to eq.~(\ref{eq4.39}),
the quadrature-component distributions for phase parameters in
a $\pi/2$ interval. Again, symmetry assumptions may be made
in order to compensate for the lack of accessible phases.

It was also proposed to reconstruct the quantum state in terms
of the density matrix in the energy basis by using
an interferometric method that is based on a generalization
of Ramsey's classic separated oscillatory fields
technique\footnote{For
   details, see Ramsey [1956].}
and the observation that the longitudinal momentum of an atom can
be shifted coherently via interaction with off-resonant radiation
(Dhirani, Kokorowski, Rubenstein, Hammond, Rohwedder, Smith,
Roberts and Pritchard [1997]). In the interferometer
an atom-wave in the internal (electronic) ground state is split into
ground-state and excited-state components by an electromagnetic
field (detuning $\delta_{1}$) at position $z_{1}$. The excited-state
component receives a momentum shift $\hbar\Delta k_{1}$
($\!\approx$ $\!\delta_{1}/v$ if the kinetic energy
$\!mv^{2}/2$ is much larger than $\hbar\delta_{1}$).
The remaining ground-state component is split again by a second field
(detuning $\delta_{2}$) at $x_{2}$, where the excited-state component
is momentum-shifted by $\hbar\Delta k_{2}$. Both field regions are
assumed to have different detunings ($\delta_{1}$ $\!\neq$
$\!\delta_{2}$). The excited state detected at the interferometer
output contains a coherent superposition of two distinct energy
components, each with a different corresponding longitudinal
momentum.
It can be shown that the probability of detecting an atom in the
excited state at location $z$ (output port) and time $t$ is
  given by\footnote{Here it is
  assumed that $\Omega'$ $\!-$ $\!\Omega''$
  $\! \ll$ $\! \Omega', \Omega''$; for the relation without
  this restriction, see Dhirani, Kokorowski, Rubenstein, Hammond,
  Rohwedder, Smith, Roberts and Pritchard [1997].}
\begin{eqnarray}
\label{eq4.40}
\lefteqn{
P_{e}(z,t)  =  \int {\rm d}\Omega' \int {\rm d}\Omega''
\, \varrho(\Omega',\Omega'')
\exp\!\left[-i(\Omega'-\Omega'')\left(t-z/v'\right)
\right]
}
\nonumber \\  && \hspace{6ex} \times
\cos^2 \!\left \{
{\textstyle\frac{1}{2}} \left[ (\delta_1-\delta_2)t -
    \Delta k'_1 (z-z_1) +\Delta k'_2 (z-z_2)\right ] \right \},
\end{eqnarray}
where $\varrho(\Omega',\Omega'')$ is the motional density matrix in
the energy basis (strictly speaking, $\Omega$ $\!=$ $\!E/\hbar$).
Equation~(\ref{eq4.40}) reveals that the measured
excited-state probability $P_{e}(z,t)$ is determined by a convolution
of the density-matrix elements $\varrho(\Omega',\Omega'')$ with
two factors. The exponential represents
the free evolution of the density matrix, and the cosine squared
contains the phase difference accumulated by the two paths of the
interferometer. Eventually, eq.~(\ref{eq4.40}) can be inverted in order
to obtain $\varrho(\Omega',\Omega'')$ in terms of $P_{e}(z,t)$
by Fourier transforming twice after appropriate change of the
variables (for details, see
Dhirani, Kokorowski, Rubenstein, Hammond, Rohwedder, Smith, Roberts,
and Pritchard [1997]).


\subsection{Electron motion}
\label{sec4.3}

So far knowledge of electronic quantum states
has been typically required in order to reconstruct motional
and vibrational quantum states of atoms and molecules. This may
have brought the reconstruction of the motional quantum state of
electrons into question (for the spin, see \S~\ref{sec4.6}).


\subsubsection{Electronic Rydberg wave packets}
\label{sec4.3.0}

Highly excited Rydberg states have offered novel possibilities
of preparing electron-wave packets. Since the early experiments
at the end of the $1980$s (Yeazell and Stroud, Jr., [1988];
ten Wolde, Noordam, Lagendijk and van Linden van den Heuvell [1988])
various schemes have been used in order to produce Rydberg wave
packets (for a brief review of generation and
detection of Rydberg wave packets, see Noordam and Jones, R.R., [1997]).
A Rydberg wave packet is produced whenever several Rydberg
states are excited coherently. In particular,
it has also been possible to prepare an atomic electron in a
coherent superposition of two wave packets which are localized spatially
at the opposite extremes of a Kepler orbit (Noel and Stroud, Jr.
[1996]). On a sufficiently short time scale these nonclassical states
are close to the even/odd coherent states of the harmonic
oscillator. To demonstrate important features of these states, two
kinds of measurements have been used. State selective ionization
allows one to verify that only every other atomic level is populated,
and a Ramsey fringe measurement verifies the coherence of the
superposition.

A wave-packet interference technique was proposed to measure
engineered atomic Rydberg wave functions (Chen and Yeazell
[1997]).\footnote{The method was also proposed for reconstructing
   pure vibrational states in a molecule (Leichtle, Schleich, Averbukh
   and Shapiro, M., [1998]).}
It is assumed that the Rydberg wave packet is produced by (a sequence of)
short optical (Gaussian) pulses whose interaction with the atomic system
can be treated in lowest-order perturbation theory. Describing the atomic
system by the radial wave function,
\begin{equation}
\Psi(r,t) = a_{g}(t)e^{-i\omega_{g}t}u_{g}(r)
+ \sum_{n}a_{n}(t)e^{-i\omega_{n}t}u_{n}(r),
\label{IRW1}
\end{equation}
where $a_{n}$ and $\omega_{n}$ are the Schr\"{o}dinger amplitudes
and eigenenergies for the Rydberg states and $a_{g}$ and $\omega_{g}$
are these for the ground state, the excited-state amplitudes
after the last excitation pulse, $a_{n}$, are given by
\begin{equation}
a_{n} = -{\textstyle\frac{1}{2}} i\Omega_{n} e^{-\Delta_{n}^{2}\sigma^{2}/2}
z_{n},
\label{IRW2}
\end{equation}
where $z_{n}$ $\!=$ $\!1$ $\!+$
$\!\sum_{i}\exp[i(\omega$ $\!+$ $\!\Delta_{n})\tau_{i}]$
($\Omega_{n}$, Rabi frequency of the excitation
between the ground state and a given Rydberg state $n$; $\Delta_{n}$
$\!=$ $\!\omega_{n}$ $\!-$ $\!\omega$, detuning; $\sigma$, pulse
width; $\tau_{i}$, time delay for the $i$th individual pulse).
When a probe pulse (identical to the previous pulses) interacts
with the atom, then the new excited-state amplitudes
\begin{equation}
a'_{n} = -{\textstyle\frac{1}{2}} i\Omega_{n} e^{-\Delta_{n}^{2}\sigma^{2}/2}
\left\{z_{n} + \exp[i(\omega\!+\!\Delta_{n})\tau] \right\}
\label{IRW3}
\end{equation}
are produced ($\tau$, delay time for the probe pulse).
By measuring the population in each eigenstate, $|a'_{n}|^{2}$,
for different phases between excitation and
probe pulses (on applying high-resolution technique, such as
state-selective field ionization),
the complex quantities $z_{n}$ can be obtained;
i.e., the (radial) wave function created by the excitation
pulse may be inferred. Note that the pump and probe pulses need
not be identical, but they must
be phase coherent with each other.

Using optical pulses,
Rydberg wave packets can usually be detected when they are near to an
ionic core. This limitation can be overcome by using subpicosecond,
unipolar electromagnetic field (half-cycle) pulses,\footnote{For these
   pulses, commonly referred to as HCP's, see also Greene,
   Federici, Dykaar, Jones, R.R., and Bucksbaum [1991];
   You, Jones, R.R., Bucksbaum and Dykaar [1993].}
which have been a powerful tool for studying dynamics in weakly
bound systems (Jones, R.R., You and Bucksbaum [1993]).
Such pulses can track the wave packet throughout
its orbit and detect wave-packet motion anywhere in the atom,
as was demonstrated experimentally (Raman, Conover, Sukenik
and Bucksbaum [1996]). In particular, it has been possible
to monitor the momentum-space probability of the
wave packet as a function of time (Jones, R.R., [1996];
for details, see also Noordam and Jones, R.R., [1997]).
In the experimental demonstration of the method also called
{\em impulsive momentum retrieval}
Na atoms are first excited by a tunable nanosecond laser to the
$25d$ Rydberg state. Subsequently, two half-cycle pulses are applied.
Kicking the electron with the first pulse, the
Rydberg population is redistributed and a complicated,
dynamically evolving wave packet is created. The second pulse
is used to measure the ionization as a function of its peak field
and delay relative to the first pulse.
If the duration of the probe pulse is negligible compared with the
time scale for variations in the position and momentum of the
electron, then the energy gained (or lost) by the electron
depends only on its initial momentum (i.e., the momentum immediately
before the application of the pulse) and the time-integrated field.
By measuring the ionization probability (at threshold) as a function
of field, the momentum distribution of the initial state along the
field axis can be obtained.


\subsubsection{Cyclotron state of a trapped electron}
\label{sec4.3.1}

Let us consider an electron in a {\em Penning trap}\footnote{For
  the trap, see Penning [1936].}
and assume a uniform magnetic field along the positive $z$ axis. The
cyclotron and the axial motions are well separated in their frequency
ranges (for a review, see Brown, L.S., and Gabrielse [1986]).
It was proposed to reconstruct the cyclotron state by mapping it
onto the axial degree of freedom (Mancini and Tombesi [1997a]).
By applying appropriate fields on the trapped electron, a
coupling between the two degrees of freedom may be realized such
that the Hamiltonian in the (cyclotron) interaction
picture is of the form
\begin{equation}
  \label{eq4.31}
  \hat{H}
  = \hbar \nu_{z}\left(\hat{a}^\dagger_{z}\hat{a}_{z}
  +{\textstyle\frac{1}{2}}\right)
  + \hbar g \hat{x}_{c}(\varphi) \hat{z}.
\end{equation}
Here, $\hat{a}^\dagger_{z}$ ($\hat{a}_{z}$) is the creation
(annihilation) operator for the axial motion of frequency $\nu_{z}$,
and $\hat{z}$ and $\hat{x}_{c}(\varphi)$, respectively,
are the axial positon operator and the quadrature operator of the
cyclotron motion, $g$ being the (field induced) coupling constant.
For an interaction time that is short compared with the axial period,
$\tau$ $ \!\ll$ $ \! 2\pi/\nu_{z}$, the Hamiltonian~(\ref{eq4.31}) gives
rise to a unitary kick and the time evolution of the mean axial momentum
yields the mean of the quadrature operator,
\begin{equation}
  \label{eq4.32}
  \langle\hat{p}_{z}(t\!+\!\tau)\rangle
  = \langle\hat{p}_{z}(t) \rangle
  + \hbar g \tau \langle \hat{x}_{c}(\varphi)\rangle \cos(\nu_{z} t).
\end{equation}
By measuring the current due to the induced charge variation
on the cap electrodes of the trap, one may obtain the axial
momentum and then the cyclotron quadrature component.
Knowing the cyclotron quadrature-component statistics,
the quantum state can be derived; e.g., by
applying the methods of \S\S~\ref{sec3.1}, \ref{sec3.2} and
\ref{sec3.3.1}.

To determine the cyclotron state from the displaced number
statistics (\S\S~\ref{sec3.3.2} and \ref{sec3.5}),
it was proposed to perform a QND  measurement of the cyclotron
excitation number (Mancini and Tombesi [1997a]). Using a magnetic
bottle configuration (Brown, L.S., and Gabrielse [1986]),
an interaction Hamiltonian of the type
\begin{equation}
  \label{eq4.33}
  \hat{H}'
  = \hbar \kappa \hat{a}^\dagger_{c} \hat{a}_{c}
  \hat{z}^2
\end{equation}
may be realized.
It gives rise to a dependence of the axial angular frequency on the
number of cyclotron excitations, which may be obtained by
probing the resonance frequency of the output electric signal.
The displacement of the cyclotron state may be realized
by a driving field acting immediately before the measurement process
induced by the Hamiltonian (\ref{eq4.33}).


\subsubsection{Electron beam}
\label{sec4.3.2}

A method was also proposed for determining the
quantum state of the (one-di\-mens\-ional) transverse motion
of a beam of identically prepared electrons (Tegmark [1996]).
Originating from the electron source, the beam crosses a
magnetic field and in conjunction with collimators a highly
monochromatic beam is prepared.
This beam enters a shielded box, wherein a harmonic potential acts in
the $x$-direction. The needed potential is obtained by subdividing the
walls of the box into a large number of metal plates which are
insulated from one another, and by fixing the potentials of the plates
appropriately. A slidable detector (along the $z$-direction of
longitudinal motion of the beam) allows one to measure the particle
density $p(x)$ along the $x$-direction. Performing these measurements
for a series of locations of the slidable detector, the density
matrix characterizing the transverse motion of the electrons
of the beam may be reconstructed.


\subsection{Spin and angular momentum systems}
\label{sec4.6}

In the history of quantum-state measurement spin (angular momentum) systems
have played a particular role, because of the finite-dimensional
Hilbert space (see, e.g., Fano [1957] and the references
in footnote \ref{general} for the Pauli problem).
Spin-quantum-state reconstruction is usually based on data
obtained from Stern--Gerlach experiments, in which
particles with magnetic momentum are deflected in a magnetic-field
gradient and the spin (or angular momentum) projection onto the direction
of the field gradient is measured (Stern [1921], Gerlach and Stern
[1921, 1922]; Feynman, Leighton and Sands [1965]; for
the optical Stern--Gerlach effect, i.e., state-dependent
deflection of particles in optical fields, see Kazantsev [1975, 1978],
Cook [1978], Tanguy, Reynaud and Cohen-Tannoudji [1984], and for its
experimental realization, see Sleator, Pfau, Balykin, Carnal and Mlynek
[1992]). Interacting the particles with fields of magnetic multipoles,
multipole moments of the particles can be measured (Bohn [1991]).

In a Stern--Gerlach analyzer as described by Feynman, Leighton and
Sands [1965] a beam of particles is split in an inhomogeneous magnetic
field into different paths provided with gates that may be opened or closed.
The diagonal density-matrix elements $\varrho_{mm}$ are then given by
the probability that the particle is deflected into the corresponding
$m$th path. Measurement of the off diagonal element $\varrho_{kn}$ requires
two analyzers. In the first analyzer, the $n$th and $k$th
paths are opened while the remaining paths are closed. In the second
analyzer the field gradient is in the plane that is
perpendicular to the quantization axis defined by the first analyzer.
Let the orientation of the field gradient of the second analyzer be
$\varphi$. The probability $p_{m}(\varphi)$ of the
particle being deflected into the $m$th path in the second analyzer
is then given by (Gale, Guth and Trammel [1968])
\begin{eqnarray}
\label{spin-6}
\lefteqn{
p_{m}(\varphi) = N \bigg\{
\left[ d^{(J)}_{mk} (\pi/2) \right]^{2} \varrho_{kk} +
\left[ d^{(J)}_{mn} (\pi/2) \right]^{2} \varrho_{nn}
}
\nonumber \\ && \hspace{5ex}
+ 2 |\varrho_{kn}| d^{(J)}_{mk} (\pi/2) d^{(J)}_{mn} (\pi/2)
\cos \left[ (k\!-\!n) \varphi \!+\! \varphi_{kn} \right]
\bigg\},
\end{eqnarray}
where $\varrho_{kn}$ $\!=$ $\!|\varrho_{kn}|$ exp$(i\varphi_{kn})$.
The matrix $d^{(J)}_{mk} (\vartheta)$
is the $\vartheta$-dependent part of the
Wigner rotation matrix (see Wigner [1959]), and
$N$ $\!=$ $\!(\varrho_{kk}$ $\!+$ $\!\varrho_{nn})^{-1}$.
Measuring (for known $\varrho_{kk}$ and $\varrho_{nn}$)
the probabilities for two different orientations $\varphi$,
the off diagonal-density matrix element $\varrho_{kn}$ can be
determined.

The quantum state can also be determined by using a single
Stern--Gerlach apparatus with variable orientation (Newton and
Young [1968]).
The probability $p_{m}(\vartheta,\varphi)$ that the projection of the
spin on the quantization axis $(\vartheta,\varphi)$ is $m$ can be
expressed in terms of the density matrix elements $\varrho_{kn}$ as
\begin{equation}
\label{spin-1}
p_{m}(\vartheta,\varphi) = \sum_{n,k=-J}^{J} e^{-i(n-k)\varphi}
d^{(J)}_{nm}(\vartheta)  d^{(J)}_{km}(\vartheta) \,
\varrho _{kn} ,
\end{equation}
which can be inverted to obtain
\begin{eqnarray}
\label{spin-2}
\varrho _{k\,k+w} = \sum_{m=-J}^{J}\sum_{j=0}^{2J}
C^{J,J,j}_{m,-m,0} C^{J,J,j}_{k+w,-k,w}
\,\frac{(-1)^{k\!-\!m}}{d^{(j)}_{w,0}(\vartheta)}\,
X(\vartheta)_{wm} ,
\end{eqnarray}
where
\begin{equation}
\label{spin-3}
X(\vartheta)_{wm}
= \frac{1}{4J\!+\!1} \sum_{s=-2J}^{2J} e^{iw\varphi_{s}}
p_{m}(\vartheta,\varphi_{s}),
\end{equation}
$\varphi_{s}$ $\!=$ $\!2\pi s/(4J$ $\!+$ $\!1)$, and
$C^{J,J,j}_{k,m,w}$ are Clebsch--Gordan coefficients.
It is seen that in order to reconstruct the density matrix,
it is sufficient to keep $\vartheta$ fixed and take
$\varphi$ at $4J\!+\!1$ values such that
$d^{(j)}_{w,0}(\vartheta)$
does not vanish for any combination $j$ and $w$.
It should be noted that the Stern--Gerlach apparatus can also be
allowed to take all possible orientations (Dodonov and Man'ko, V.I., [1997]).
In this case one need not search for an optimum choice of orientations,
since all possible orientations are considered on an equal footing.

As already outlined, the problem of reconstruction of spin density
matrices has also been treated by means of the principle of maximum
entropy (Bu\v{z}ek, Drobn\'{y}, Adam, G., Derka and Knight [1997]; see
\S~\ref{sec3.9.3}) and Bayesian inference (Jones, K.R.W., [1991];
Derka, Bu\v{z}ek and Adam, G., [1996]; Derka, Bu\v{z}ek, Adam, G., and Knight
[1996]; see \S~\ref{sec3.9.4}).  It was also proposed to
tomographically reconstruct
the discrete Wigner function of a spin system in
analogy to the inverse Radon transform for the continuous Wigner
function (Leonhardt [1995, 1996]). However, the scheme does not only
require measurement of probability distributions of spin (angular
momentum) components but also those of discrete phase states.

With regard to pure states, it can be shown that the
spin wave function can be reconstructed from the populations of the spin
projections on two directions deviated from each other by an
infinitesimal angle (Weigert [1992]). Particular attention was
paid to the determination of the class of pure states which are
eigenstates of rotated spin projection operators $J_{n}$ $\!=$
$\!n_{x}J_{x}$ $\!+$ $\!n_{y}J_{y}$ $\!+$ $\!n_{z}J_{z}$, where
${\bf n}$ $\!=$ $\!(n_x,n_y,n_z)$ is a direction vector of unit length.
As shown, measurement of spin projections on three directions is
sufficient to determine the unknown direction ${\bf n}$ and
the corresponding spin component eigenvalue (Ivanovi\'{c} [1993]).

Finally, experiments for reconstructing the angular momentum density
matrix of electrons in hydrogen atoms from measurements of the
Stark structure of the emitted light have been reported
(Havener, Rouze, Westerveld and Risley [1986];
Ashburn, Cline, Stone, van der Burgt, Westerveld and Risley [1989];
Ashburn, Cline, van der Burgt, Westerveld and Risley [1990];
Cline, van der Burgt, Westerveld and Risley [1990];
Renwick, Martell, Weaver and Risley [1993]; Seifert, Gibson and Risley [1995]).
In the experiments, the density matrix is considered for the
manifold of states of principal quantum number $n$ $\!=$ $\!3$
and $n$ $\!=$ $\!2$, the matrix elements being parametrized by
the orbital and magnetic quantum numbers $l$ and
$m$ respectively. The excited hydrogen atoms are produced by
collisions of protons with noble gas atoms.
Measuring the coherence parameters of the emitted light
[transitions $(n$ $\!=$ $\!3)$ $\to$ $(n$ $\!=$ $\!2)$ and
$(n$ $\!=$ $\!2)$ $\to$ $(n$ $\!=$ $\!1)$]
in dependence on the external electric field, one can infer
the density-matrix elements (for theoretical results,
see Jain, Lin and Fritsch [1987a,b, 1988]).


\subsection{Crystal lattices}
\label{sec4.7}

More than ten years before the first demonstration of reconstruction
of the quantum state of light experiments were already
performed for reconstructing off-diagonal elements of the
single-particle density matrix of crystal lattices (Golovchenko,
Kaplan, Kincaid, Levesque, Meixner, Robbins and Felsteiner [1981];
Sch\"{u}lke, Bonse and Mourikis [1981]). The methods are based on
Compton scattering or $X$-ray scattering. Before giving the basic idea
of these methods, let us outline briefly the information available in
standard $X$-ray and Compton scattering experiments.

In conventional $X$-ray diffraction the Bragg diffracted intensity
corresponding to a reciprocal lattice vector ${\bf g}$
is proportional to the squared modulus of the form factor
$F({\bf g})$. Provided that the phase of the form factor can be
determined, some limited insight into the one-particle
density matrix in momentum space, $\varrho({\bf p},{\bf p}+{\bf g})$,
can be gained. To be more specific, the form factor yields
the momentum average of the density matrix according to
\begin{equation}
  \label{eq4.41}
F({\bf g})  = \int {\rm d} {\bf p} \,
\varrho({\bf p},{\bf p}+{\bf g}).
\end{equation}
In conventional Compton scattering one may obtain
the position average of the density matrix in position space,
$\varrho({\bf r},{\bf r}')$. The one-dimensional Fourier transform of
the Compton profile, the so-called reciprocal form factor $B({\bf r})$,
can be given by
\begin{equation}
  \label{eq4.42}
B({\bf r})  = \int {\rm d}{\bf r}' \,
\varrho({\bf r}',{\bf r}'+{\bf r}).
\end{equation}
Obviously, conventional methods do not provide the full density
matrix, neither in position nor in momentum space.

The basic idea to overcome this problem consists in the use of coherent
scattering methods (Golovchenko, Kaplan, Kincaid, Levesque, Meixner,
Robbins and Felsteiner [1981]; Sch\"{u}lke, Bonse and Mourikis
[1981], Sch\"{u}lke and Mourikis [1986]; Sch\"{u}lke [1988]).
Let us consider the situation in
Compton scattering. In conventional experiments a propagating plane wave
is used, so that each point in position space is excited equally. This
explains the fact that one can only measure position-space averages of the
density matrix as given in eq.~(\ref{eq4.42}). To avoid this
limitation, one may introduce a weighting of certain positions within the
elementary cell. This can be done by using a standing-wave field with
a spatial periodicity of the nodes that is commensurable with the
lattice periodicity.  The spatial distribution of the corresponding
intensity is given by
\begin{equation}
  \label{eq4.43}
I({\bf r})  = I_1 + I_2 +2 \sqrt{I_1 I_2}
\cos({\bf g}\!\cdot\!{\bf r} + \varphi),
\end{equation}
where the phase $\varphi$ defines the position of the nodes with
respect to the atomic planes of the lattice. One may perform two
measurements of the Compton profiles by choosing the values of the
phase to be $\varphi$ $\!=$ $\!0, \pi$. The difference of the
corresponding form factors, $\Delta B({\bf r};{\bf g})$, is
then given by
\begin{equation}
  \label{eq4.44}
\Delta B({\bf r}; {\bf g})
\sim \int {\rm d}{\bf r}' \,
\varrho({\bf r}',{\bf r}'+{\bf r}) \,
\cos({\bf g}\!\cdot\!{\bf r}').
\end{equation}
That is, coherent Compton scattering allows to determine spatial
Fourier transforms of the one-particle density matrix.

%
%
%

\section*{Acknowledgments}

We thank all the colleagues who supported us during
the preparation of this article. We are indebted to them
for reading foregoing drafts and giving valuable comments
that improved the final version, for bringing particular
details of the subject to our attention, for providing us with
experimental data and for the help in preparing figures.
In particular,
we are grateful to E.L. Bolda, T. Coudreau, G.M. D'Ariano, E. Giacobino,
Z. Hradil, W.M. Itano, L. Kn\"{o}ll, D. Leibfried, U. Leonhardt, M.G.A. Paris,
J. Pe\v{r}ina, Th. Richter, S. Wallentowitz, D.F. Walls, I.A. Walmsley,
L. Waxer, D.J. Wineland and A. Zucchetti. Finally, we like to
emphasize that our research in the field of phase-sensitive measurements
and quantum-state reconstruction has been supported by the
Deutsche Forschungsgemeinschaft.


\begin{appendix}
\renewcommand{\thesection}{\Alph{section}.}
\renewcommand{\thesubsection}{\Alph{section}.\arabic{subsection}.}
\renewcommand{\theequation}{\Alph{section}.\arabic{equation}}

\section{Radiation field quantization}
\label{app1}
\setcounter{equation}{0}

   From Maxwell's theory it is well known that the vector potential of
the radiation field in free space can be given by an expansion
in transverse travelling waves. For dealing with radiation inside resonators,
such as cavities bounded by perfectly reflecting mirrors, a standing-wave
expansion is appropriate, where the spatial structure
of the waves sensitively depends on the resonator geometry through
the boundary conditions. An expansion of the vector potential
\begin{equation}
\label{A1.1}
{\bf A}({\bf r},t) = {\bf A}^{(+)}({\bf r},t)
+ {\bf A}^{(-)}({\bf r},t), \qquad
{\bf A}^{(-)}({\bf r},t)
= \left[{\bf A}^{(+)}({\bf r},t)\right]^\ast
\end{equation}
in orthogonal transverse waves,
\begin{equation}
\label{A1.2}
{\bf A}^{(+)}({\bf r},t)  =
\sum_{\lambda} {\bf A}_{\lambda}  ({\bf  r})
e^{-i \omega_{\lambda} t} \, a_{\lambda},
\end{equation}
is usually called mode expansion. The mode functions
${\bf A}_{\lambda}({\bf r})$ satisfy the Helmholtz equation
\begin{equation}
\label{A1.3}
\nabla\times\nabla\times{\bf A}_\lambda -
\frac{\omega_\lambda^2}{c^2} \, {\bf A}_\lambda = 0,
\end{equation}
where the condition of transversality,
$\nabla\!\cdot\!{\bf A}_{\lambda}$ $\!=$ $\!0$, implies
that $\nabla$ $\!\times$ $\!\nabla$ $\!\times$ $\!{\bf A}_\lambda$ $\!=$
$\!-\Delta {\bf A}_\lambda$. The mode amplitudes $a_{\lambda}(t)$ evolve
according to harmonic-oscillator equations of motion,
\begin{equation}
\label{A1.4}
\dot{a}_{\lambda} = -i \omega_{\lambda}\, a_{\lambda}.
\end{equation}

Since harmonic oscillators can be associated with the modes,
quantization of radiation reduces to quantization of harmonic oscillators.
In eq.~(\ref{A1.1}), the complex amplitues $a_{\lambda}$ and
$a_{\lambda}^{\ast}$ are regarded as non-Hermitian destruction and
creation operators
$\hat{a}_{\lambda}$ and $\hat{a}_{\lambda}^{\dagger}$, respectively,
which satisfy the well-known (equal-time) bosonic commutation relations:
\begin{equation}
\label{A1.5}
\big[\hat{a}_\lambda, \hat{a}_{\lambda '}^\dagger\big]
= \delta_{\lambda \lambda '},
\quad
\big[\hat{a}_\lambda, \hat{a}_{\lambda '}\big] = 0.
\end{equation}
The bosonic excitations introduced in this way are called photons.
It should be pointed out that the expansion in monochromatic modes of
the (Heisenberg) operator of the vector potential,
\begin{equation}
\label{A1.6}
\hat{\bf A}({\bf r},t) = \hat{\bf A}^{(+)}({\bf r},t)
+ \hat{\bf A}^{(-)}({\bf r},t),
\end{equation}
\begin{equation}
\label{A1.7}
\hat{\bf A}^{(+)}({\bf r},t)  =
\sum_{\lambda} {\bf A}_{\lambda}  ({\bf  r})
e^{-i \omega_{\lambda} t} \, \hat{a}_{\lambda},
\end{equation}
can be used to introduce photons that are associated with other than
monochromatic waves; viz.,
\begin{equation}
\label{A1.8}
\hat{a}'_{\mu} = \sum_{\lambda}
C_{\mu \lambda}^{\ast} \, \hat{a}_{\lambda},
\end{equation}
\begin{equation}
\label{A1.9}
\hat{\bf A}({\bf r},t)  =  \sum_{\mu}
\left [ {\bf A}'_{\mu}  ({\bf r},t) \, \hat{a}'_{\mu} +
{{\bf A}'}_{\mu}^{*} ({\bf r},t) \,
\hat{a}'\,\!_{\!\!\mu}^{\dagger}\right]\!,
\end{equation}
where $C_{\mu \lambda}$ are elements of a unitary matrix, so that the
$\hat{a}'_{\mu}$ and $\hat{a}'\,\!_{\!\!\mu}^{\dagger}$ are again photon
destruction and creation operators, respectively, and
\begin{equation}
\label{A1.10}
{\bf A}'_{\mu}  ({\bf r},t) = \sum_{\lambda} C_{\mu \lambda}
{\bf A}_{\lambda}({\bf r}) e^{-i \omega_{\lambda} t}
\end{equation}
(for details, see Titulaer and Glauber [1966]).
The spatial-temporal (nonmonochromatic) modes can be used
advantageously for describing and analysing pulse-like radiation.

The mirrors used in resonator experiments are of course not perfectly
reflecting. Moreover, fractionally transparent mirrors are deliberately
used to open input and output ports in resonator equipments.
Another typical example is the use of beam splitters in various interference
and correlation experiments. In many cases these and related (passive)
optical instruments can be regarded as macroscopic dielectric bodies
that respond linearly to radiation and whose action can be included
phenomenologically in the Maxwell theory through a space-dependent
permittivity $\varepsilon$ (Kn\"{o}ll, Vogel, W., and
Welsch [1987], Glauber and Lewenstein [1991]). When the radiation under
consideration is in a frequency intervall where the effect of dispersion
is sufficiently small, then the dependence of the permittivity on frequency
can be disregarded. Performing a mode expansion (\ref{A1.2}), the mode
functions are now solutions of a modified Helmholtz equation (\ref{A1.3}),
with $v$ $\!=$ $\!v({\bf r})$ $\!=$ $\!c/n({\bf r})$ in place of $c$, where
$n({\bf r})$ $\!=$ $\!\sqrt{\varepsilon({\bf r})}$ is the (real) refractive
index. The mode functions satisfy a generalized condition of transversality,
$\nabla\!\cdot\!(\varepsilon {\bf A}_{\lambda})\!=\!0$, and the
relation of orthogonality requires inclusion of $\varepsilon$
in the space integral (for details, see Vogel, W., and Welsch [1994]),
because of the dependence on ${\bf r}$ of the permittivity.\footnote{For
   extensions to dispersive and absorbing
   media, see Gruner and Welsch [1996] and references
   therein. In this case, the permittivity is a complex function of
   frequency, the real and imaginary parts being related to each
   other according the Kramers--Kronig relations, and a mode
   decomposition outlined here fails.}


\section{Quantum-state representations}
\label{app2}
\setcounter{equation}{0}

Among the wide variety of possible radiation-field states, there
are some fundamental states which, with regard to representation
and measurement, play a special role in quantum optics.
Let us consider a (free) radiation field whose vector potential is
given by a mode decomposition [eq.~(\ref{A1.7}) or eq.~(\ref{A1.9})]
and restrict our attention to single-mode states. The
single-mode states can then be used to build up multimode
states as (direct-)product states.


\subsection{Fock states}
\label{app2.1}

A mode (with photon destruction and creation operators $\hat{a}$
and $\hat{a}^{\dagger}$, respectively) is said to be in a Fock state
(photon-number state) $| n \rangle$, when the
state is an eigenstate of the photon-number operator
$\hat{n}\!=\!\hat{a}^{\dagger}\hat{a}$,
\begin{equation}
\label{A2.1}
       \hat{n}|n\rangle = n|n\rangle,  \qquad  n=0,1,2,\dots.
\end{equation}
Each number state $|n\rangle$ can be obtained from the
vacuum state $|0\rangle$ according to
\begin{equation}
\label{A2.2}
|n\rangle = \frac{1}{\sqrt{n!}}\,
\hat{a}^{\dagger n} |0\rangle.
\end{equation}
Since the photon-number states form an orthogonal and complete set of
states, the density operator of any (single-mode) quantum state
can be expanded in the photon-number basis,
\begin{equation}
\label{A2.3}
\hat{\varrho} = \sum_{n,n'} | n \rangle \langle n | \hat{\varrho} |n'\rangle
                \langle n' |,
\end{equation}
the diagonal elements $p_n$ $\!=$ $\!\langle n | \hat{\varrho} |n \rangle$
being the probabilities for observing $n$ photons.


\subsection{Quadrature-component states}
\label{app2.2}

Next let us consider a (single-mode) field-strength operator
of the type of $\hat{F}(\varphi)$ $\!=$
$\!|F| (\hat{a} e^{-i\varphi} + \hat{a}^{\dagger}
e^{i\varphi} )$.
Depending on the choice of the (scaling) factor $|F|$, the operator
$\hat{F}$ may represent the vector potential or the electric or
magnetic fields associated with the mode. In what follows we
set $\hat{F}$ $\!=$ $\!2^{1/2}|F|\hat{x}(\varphi)$, where
\begin{equation}
\label{A2.4}
\hat{x}(\varphi) = 2^{-1/2}
\left (\hat{a} e^{-i\varphi}+\hat{a}^{\dagger} e^{i\varphi} \right)
\end{equation}
is called quadrature-component operator. The quadrature components
$\hat{x}(\varphi)$ and $\hat{x}(\varphi')$ are observables that cannot
be measured simultaneously for $\varphi$ $\!\neq$ $\!\varphi'$
$\!+$ $\!k\pi$, $k$ $\!=$ $\!0,\pm 1,\ldots$,
because of $[\hat{x}(\varphi),$ $\!\hat{x}(\varphi')]$ $\!\neq$ $\!0$.
For chosen $\varphi$ quadrature-component states $|x,\varphi\rangle$
can be introduced (Schubert and Vogel, W., [1978a,b], see also
Vogel, W., and Welsch [1994]), solving the eigenvalue problem
\begin{equation}
\label{A2.5}
\hat{x}(\varphi) \,|x,\varphi\rangle = x \, |x,\varphi\rangle,
\end{equation}
$-\infty$ $\!\leq$ $\!x$ $\!\leq$ $\!\infty$.
Obviously, $|x,\varphi\rangle$ is related to $|x\rangle$ $\!=$
$\!|x,\varphi$ $\!=$ $\!0\rangle$ as $|x,\varphi\rangle$
$\!=$ $\!e^{i\hat{n}\varphi}$ $\!|x\rangle$, and can be given by
\begin{eqnarray}
\label{A2.6}
|x,\varphi\rangle
& = &
\pi^{-1/4} e^{-x^{2}/2}
\sum_{n}\frac{e^{in\varphi}}{\sqrt{2^{n}n!}}\,{\rm H}_{n}(x)\,|n\rangle
\nonumber \\[1ex]
& = &
\pi^{-1/4} e^{-x^{2}/2}
\exp\!\left[-{\textstyle\frac{1}{2}}
\left( e^{i\varphi} \hat{a}^\dagger \right)^2
+ 2^{1/2} x \,e^{i\varphi}\hat{a}^\dagger\right] |0\rangle
\end{eqnarray}
[${\rm H}_{n}(x)$, Hermite polynomial].
The states $|x,\varphi\rangle$ define an orthogonal Hilbert space basis
for each value of $\varphi$, so that the density operator can be expanded as
\begin{equation}
\label{A2.7}
\hat{\varrho} = \int {\rm d}x \int {\rm d}x' \,
|x,\varphi\rangle
\langle x,\varphi| \hat{\varrho} |x',\varphi\rangle
\langle x',\varphi|.
\end{equation}
In particular, $p(x,\varphi)$ $\!=$
$\!\langle x,\varphi| \hat{\varrho} |x,\varphi\rangle$ is the probability
density for observing the value $x$ of the quadrature component
$\hat{x}(\varphi)$. Note that the symmetry relation
$p(x,\varphi\!+\!\pi)$ $\!=$ $\!p(-x,\varphi)$ holds.


\subsection{Coherent states}
\label{app2.3}

The coherent states introduced by Schr\"{o}dinger [1926] to simulate
the motion of a (near)classical particle in a harmonic potential are
usually defined by the eigenvalue problem\footnote{The coherent states
   were recovered in the 1960's (Klauder [1960, 1963a,b]; Glauber [1963a];
   Sudarshan [1963]; Klauder and Sudarshan [1968]).}
\begin{equation}
\label{A2.8}
\hat{a}\, | \alpha \rangle = \alpha\, | \alpha \rangle
\end{equation}
($\alpha$, complex). Equivalently,
\begin{equation}
\label{A2.9}
 | \alpha \rangle = \hat{D} ( \alpha ) | 0 \rangle,
\end{equation}
where $\hat{D}(\alpha)$ is the coherent displacement operator,
\begin{equation}
\label{A2.10}
\hat{D} ( \alpha ) =
           \exp\!\left(\alpha \hat{a}^\dagger - \alpha^\ast \hat{a} \right)
\end{equation}
[note that $\hat{D}(\alpha)\hat{a}\hat{D}^{\dagger}(\alpha)$
$\!=$ $\!\hat{a}$ $\!-$ $\!\alpha$].
The coherent states are not orthogonal,
\begin{equation}
\label{A2.11}
\langle \alpha' | \alpha \rangle =
  \exp\!\left(- {\textstyle\frac{1}{2}} | \alpha - \alpha' |^2 \right)
        \exp\!\left[{\textstyle\frac{1}{2}}
            ( \alpha {\alpha'}^\ast - \alpha^\ast \alpha' ) \right],
\end{equation}
and they are overcomplete
(Cahill [1965]; Bacry, Grossmann and Zak [1975]), so that
there are various possibilities of representing the density
operator. In particular, it can be expanded as
\begin{equation}
\label{A2.12}
\hat{\varrho} = \frac{1}{\pi^2} \int {\rm d}^2 \alpha
\int {\rm d}^2 \alpha' \;
| \alpha \rangle \langle \alpha | \hat{\varrho} | \alpha' \rangle \langle \alpha' |
\end{equation}
(${\rm d}^2\alpha$ $\!=$  $\!{\rm d}\,{\rm Re}\,\alpha$
${\rm d}\,{\rm Im}\,\alpha$).

Equation (\ref{A2.9}) can be extented in order to define generalized
coherent states by the action of the displacement operator on arbitrary
states of the Hilbert space (see, e.g., Perelomov [1986]).
In particular, the action of $\hat{D}(\alpha)$
for all values of $\alpha$ on a set of states
$|\Psi_{\lambda}\rangle$ that resolve the unity
introduces a foliation of the
Hilbert space into orbits such that each state
belongs exactly to one orbit of generalized coherent states.
For example, the Fock states can be used to define
the generalized coherent states,
\begin{equation}
\label{A2.12a}
 |n,\alpha\rangle = \hat{D}(\alpha) |n\rangle,
\end{equation}
which are also called displaced Fock states.\footnote{Similar to the
   coherent states, these states correspond to wave packets of a particle
   in a harmonic potential which keep their shapes and follow the classical
   motion (Husimi [1953]; Senitzky [1954]; Plebanski [1954, 1955, 1956];
   Epstein [1959]).}
Their expansion in the ordinary Fock basis reads as
\begin{equation}
\label{A2.12b}
 |n,\alpha\rangle = \sum_{m} |m\rangle \langle m|n,\alpha\rangle,
\end{equation}
where the expansion coefficients $\langle m|n,\alpha\rangle$ can be given by
\begin{equation}
\label{A2.12c}
\langle m|n,\alpha\rangle
= \left(m!n!e^{-|\alpha|^{2}}\right)^{\frac{1}{2}}
\sum_{l=0}^{\{ m,n \}} (-1)^{m-l} \,
\frac{|\alpha|^{m+n-2l}e^{-i(m-n)\varphi}}
{l!(m-l)!(n-l)!}
\end{equation}
[$\{ m,n \}$ $\!=$ $\!{\rm min}\,(m,n)$, $\alpha$
$\!=$ $\!|\alpha|e^{i\varphi}$] or
\begin{equation}
\label{A2.12c-t}
\langle m|n,\alpha\rangle
= \exp \left( - {\textstyle\frac{1}{2}} |\alpha|^{2} \right)
\sqrt{\frac{n!}{m!}} \, \alpha ^{m-n}
{\rm L}_{n}^{m-n}\!\left( |\alpha |^{2}  \right)
\end{equation}
[${\rm L}_{n}^{m}(x)$, Laguerre polynomial].


\subsection{\protect\boldmath$s$-Parametrized phase-space functions}
\label{app2.4}

Representations of the density operator
in terms of phase-space functions which are
formally similar to classical probability distributions
are used frequently.
Defining an $s$-parameterized displacement operator by
\begin{equation}
\label{A2.13}
\hat{D}(\alpha;s)  =
e^{s|\alpha|^2/2} \hat{D}(\alpha),
\end{equation}
the following expansion of the density operator can be proved
correct:
\begin{equation}
\label{A2.14}
\hat{\varrho} = \frac{1}{\pi} \int {\rm d}^2 \alpha \,
{\rm Tr}\{\hat{\varrho}\,\hat{D}(-\alpha;s)\} \, \hat{D}(\alpha;-s)
\end{equation}
(for details, see, e.g., Cahill and Glauber [1969a,b];
Agarwal and Wolf [1970]; Pe\v{r}ina [1991]).\footnote{For
   $s$ $\!=$ $\!0$ the expansion (\ref{A2.14}) is known from Weyl's
   quantization method (Weyl [1927]).}
Introducing the Fourier transform of $\hat{D}(\alpha;s)$,
\begin{eqnarray}
\label{A2.15}
\hat{\delta}(\alpha-\hat{a};s) & = &
\frac{1}{\pi^2} \int {\rm d}^2 \beta \, \hat{D}(\beta;s)
\exp\!\left[\alpha \beta{^\ast} - \alpha^\ast \beta \right]
\nonumber \\
& = &\frac{2}{\pi(1\!-\!s)} \, \hat{D}(\alpha)
\left(\frac{s\!+\!1}{s\!-\!1}\right)^{\hat{a}^{\dagger} \hat{a}}
\hat{D}^{-1}(\alpha),
\end{eqnarray}
the expansion (\ref{A2.14}) can be rewritten as\footnote{Note that
   eqs.~(\ref{A2.16}) and (\ref{A2.17}) also apply to other
   than density operators. Let $\hat{F}$ be an operator function of
   $\hat{a}$ and $\hat{a}^{\dagger}$. Then it can be shown that
   $\hat{F}$ $\!=$ $\!\int {\rm d}^2\alpha$
   $\!F(\alpha;s)$ $\!\hat{\delta}(\alpha-\hat{a};s)$, where
   $F(\alpha;s)$ $\!=$ $\!\pi{\rm Tr}\{\hat{F}$ $\!\hat{\delta}(\alpha$
   $\!-$ $\!\hat{a};-s)\}$ is the associated $c$-number function in
   $s$ order.}
\begin{equation}
\label{A2.16}
\hat{\varrho} = \pi \int {\rm d}^2 \alpha \, P(\alpha;s) \,
\hat{\delta}(\alpha-\hat{a};-s),
\end{equation}
where
\begin{equation}
\label{A2.17}
P(\alpha;s) = {\rm Tr}\{\hat{\varrho}\,\hat{\delta}(\alpha-\hat{a};s)\}
\end{equation}
is the $s$-parametrized phase-space function, which is normalized
to unity.\footnote{Frequently the definition $P(q,p;s)$ $\!\equiv$
   $2^{-1}P[\alpha$ $\!=$ $\!2^{-1/2}(q$ $\!+$ $\!ip)]$ is used, so that
   $P(q,p;s)$ can be regarded as a function of ``position'' and ``momentum'',
   with $\int {\rm d}q$ $\!\int {\rm d}p$ $\!P(q,p;s)$ $\!=$ $\!1$.}
Cases of particular interest are the Glauber-Sudarshan $P$ function,
$P(\alpha$) $\!\equiv$ $\!P(\alpha;1)$
(Glauber [1963b, 1963c]; Sudarshan [1963]),
the Wigner function $W(\alpha)$ $\!\equiv$ $P(\alpha;0)$ (Wigner [1932])
and the $Q$ function
$\!Q(\alpha)$ $\!\equiv$ $\!P(\alpha;-1)$ (Husimi [1940]).
Note that $\hat{\delta}(\alpha$ $\!-$ $\!\hat{a}$,
$\!-1)$ $\!=$ $\!\pi^{-1}$ $\!|\alpha\rangle\langle\alpha|$, which implies
that $Q(\alpha)$ $\!=$ $\!\pi^{-1}$
$\!\langle\alpha|\hat{\varrho}|\alpha\rangle$.
The expectation value of an operator $\hat{F}$ can then be given by
\begin{equation}
\label{A2.17b}
\langle\hat{F}\rangle
= \int {\rm d}^{2}\alpha \,
P(\alpha;s) \, F(\alpha;s),
\end{equation}
where $F(\alpha;s)/\pi$ is defined according to eq.~(\ref{A2.17})
with $\hat{F}$ in place of $\hat{\varrho}$.

   From eqs.~(\ref{A2.15}) and (\ref{A2.17}) we see that
\begin{equation}
\label{A2.18}
\Phi(\alpha;s) =
{\rm Tr}\{\hat{\varrho}\,\hat{D}(\alpha;s)\},
\end{equation}
can be regarded as the characteristic function of the phase-space
function $P(\alpha;s)$,
\begin{equation}
\label{A2.19}
P(\alpha;s)=\frac{1}{\pi^2}\int {\rm d}^2\beta\,
    \exp\!\left(\alpha\beta^\ast-\alpha^\ast\beta\right)\!
    \Phi(\beta;s).
\end{equation}
Using eqs.~(\ref{A2.13}), (\ref{A2.18}), and (\ref{A2.19}), the
phase-space functions $P(\alpha;s)$ can be related to each other as
\begin{eqnarray}
\label{A2.20}
\lefteqn{   \hspace*{-5ex}
P(\alpha;s) = \frac{1}{\pi^2} \int {\rm d}^2\beta \,
   \exp\!\left[{\textstyle\frac{1}{2}}(s\!-\!s')|\beta|^2\right] }
                                                       \nonumber \\
&&  \hspace*{5ex} \times
  \int {\rm d}^2\gamma
   \exp\!\left[(\alpha\!-\!\gamma)\beta^\ast\!
   -\!(\alpha^\ast\!-\!\gamma^\ast)\beta\right]
      P(\gamma;s').
\end{eqnarray}

It should be noted that the formalism can be extended to finite
dimensional Hilbert spaces (Opatrn\'{y}, Welsch and Bu\v{z}ek [1996];
for the Wigner function, see Wootters [1987], Leonhardt [1995,1996];
for the $Q$ function, see Opatrn\'{y}, Bu\v{z}ek, Bajer and Drobn\'{y} [1995]
and Galetti and Marchiolli [1996]).\footnote{For an approach to
   phase-space functions for systems with finite-dimensional Hilbert spaces
   which uses a continuous phase space, see
   V\'{a}rilly and Gracia-Bond\'{\i}a [1989]; Dowling, Agarwal
   and Schleich [1994].}


\subsection{Quantum state and quadrature components}
\label{app2.5}

It is worth noting that there is a one-to-one correspondence between
the phase-space function $P(\alpha;s)$ and the set of quadrature-component
distributions $p(x,\varphi)$ for all values of $\varphi$ within a
$\pi$ interval (Vogel, K., and Risken [1989]).
Introducing the characteristic function $\Psi(z,\varphi)$ of the
quadrature-component distribution $p(x,\varphi)$,
\begin{equation}
\label{A2.21}
\Psi(z,\varphi) =
\int {\rm d} x \, e^{izx} \, p(x,\varphi),
\end{equation}
it can be shown that
\begin{equation}
\label{A2.22}
\Psi(z,\varphi) = e^{-sz^{2}/4} \,
  \Phi\!\left(iz2^{-1/2}e^{i\varphi};s\right)\!.
\end{equation}
Equations~(\ref{A2.19}), (\ref{A2.21}), and (\ref{A2.22}) reveal that
\begin{eqnarray}
\label{A2.23}
\lefteqn{ \hspace*{-5ex}
p(x,\varphi) =
  \frac{1}{2\pi} \int {\rm d} z \;
\exp\!\left(-iz\,x\!
 -\!{\textstyle\frac{1}{4}}sz^{2} \right)  }
\nonumber \\
&& \hspace*{5ex} \times
     \int {\rm d}^2\alpha\,
\exp\!\left[
iz\langle\alpha|\hat{x}(\varphi)|\alpha\rangle
\right] P(\alpha;s)
\end{eqnarray}
and
\begin{eqnarray}
\label{A2.24}
\lefteqn{
P(\alpha;s) = \frac{1}{2\pi^2}
\int_0^\pi {\rm d} \varphi \int {\rm d} z \,
\bigg\{
|z| \, e^{sz^{2}/4}
}
\nonumber \\ &&
\hspace{10ex} \times\,
\exp\!\left[
-iz\langle\alpha|\hat{x}(\varphi)|\alpha\rangle
\right]
\int {\rm d} x \, e^{izx} \, p(x,\varphi)
\bigg\}.
\end{eqnarray}
In other words, knowledge of $p(x,\varphi)$ for all values of $\varphi$
within a $\pi$ interval is equivalent to knowledge of the quantum
state.\footnote{Note that the $\varphi$ integral in eq.~(\ref{A2.24})
   can be performed over any $\pi$ interval.}
In particular, the expansion of the density operator as given
in eq.~(\ref{A2.14}) can be rewritten, on using
eqs.~(\ref{A2.18}) and (\ref{A2.22}), as
\begin{equation}
\label{A2.25}
\hat{\varrho}
= \frac{1}{2\pi} \int_0^\pi {\rm d}\varphi \int {\rm d}z
\,|z|\,e^{sz^{2}/4}
\Psi(-z,\varphi) \, \hat{D}\!\left(iz2^{-1/2}e^{i\varphi};-s\right),
\end{equation}
which offers the possibility of relating $\hat{\varrho}$ to $p(x,\varphi)$ as
\begin{equation}
\label{A2.26}
\hat{\varrho} =  \int_0^\pi {\rm d}\varphi \int {\rm d}x
\,\hat{K}(x,\varphi) \, p(x,\varphi),
\end{equation}
provided that the operator kernel
\begin{equation}
\label{A2.27}
\hat{K}(x,\varphi)
= \frac{1}{2\pi} \int {\rm d}z \, |z|\,
\exp\!\left\{iz\left[\hat{x}(\varphi)\!-\!x\right]\right\}\!,
\end{equation}
exists (D'Ariano [1995]; D'Ariano, Leonhardt and Paul, H., [1995]).


\section{Photodetection}
\label{app3}
\setcounter{equation}{0}

In order to give an introduction into the quantum theory of photoelectric
detection of light (Mandel [1958, 1963], Kelley and Kleiner [1964],
Glauber [1965]), let us consider, as a simple example of a photodetection
device, a large sample of atomic systems that are capable of absorbing light
through the photoemission of electrons in a certain time interval
$t$, $t$ $\!+$ $\!\Delta t$. Next suppose that the photoemission
of the actual number $m$ of photoelectrons is dominated by the process
in which exactly $m$ atomic systems are involved, so that each
emits just one electron. We further assume that the total number $N$ of
atomic systems is much larger compared with the mean number of emitted
electrons, so that for any (relevant) actual value $m$ the inequality
$m$ $\!<$ $\!N$ may be assumed. Under these assumptions, the main features
of the theory may be developed by applying Dirac's perturbation theory
to the basic process of light absorption and combining the corresponding
results with methods of classical statistics
with respect to the ensemble of photoelectrons generated
by the absorption processes (see, e.g., Vogel, W., and Welsch [1994]).

Here, we will give a more intuitive analysis rather than an exact
derivation. When each photon that falls -- in the chosen time interval --
on the detector gives rise to exactly one emitted electron, then the
number of counted electrons agrees exactly with the number of photons,
and the statistics of the counted electrons reflect exactly
the photon-number statistics; i.e., the probability $P_{m}$ of detecting
$m$ photoelectrons is equal to the probability $p_{m}$ of
$m$ photons being in the field,
\begin{equation}
\label{A3.1}
P_{m} = p_{m} =\langle m|\hat{\varrho}|m\rangle.
\end{equation}
However owing to losses, the probability $\eta$ of converting a photon
to an electron is less than unity in general ($0$ $\!\leq$ $\!\eta$
$\!\leq$ $\!1$). This probability is also called detection (or
quantum) efficiency. Since (under the assumptions made) the individual
events of emission of a photoelectron can be regarded as being independent
of each other, the probability $P_{m|n}(\eta)$ of observing $m$
photoelectrons under the condition that $n$ photons are present
corresponds to a Bernoulli process,
\begin{equation}
\label{A3.2}
P_{m|n}(\eta) = {n \choose m}
\eta^{m} (1-\eta )^{n-m}
\quad {\rm if} \quad
m \le n,
\end{equation}
and $P_{m|n}(\eta)$ $\!=$ $\!0$ if $m$ $\!>$ $\!n$.
The joint probability that $n$ photons are present and $m$ photoelectrons
are counted is then $P_{m|n}(\eta)p_n$, and hence the prior probability
of detecting $m$ photoelectrons is given by
\begin{equation}
\label{A3.3}
P_{m} = \sum_{n} P_{m|n}(\eta)\,p_{n}
= \sum_{n=m}^{\infty} {n \choose m}
\eta^{m} (1-\eta )^{n-m} p_{n}
\end{equation}
which for $\eta$ $\!\to$ $\!1$ (perfect detection) reduces
to eq.~(\ref{A3.1}). Note that eq.~(\ref{A3.3}) can be inverted
in order to obtain $p_{n}$ from $P_{m}$ by simply replacing $\eta$
with $\eta^{-1}$,
\begin{equation}
\label{A3.3a}
p_{n} = \sum_{m=n}^{\infty} {m \choose n}
\left( \frac{1}{\eta}\right)^{n}
\left(1-\frac{1}{\eta} \right)^{m-n} P_{m} .
\end{equation}
Equation (\ref{A3.3}) can be rewritten
as\footnote{Note that the idendity
   $|n\rangle\langle n|$ $\!=$ $\!:(\hat{n}^{n}/n!)e^{-\hat{n}}:$ is
   valid, which can easily be proved correct
   from the associated $c-$number function $|\langle\alpha|n\rangle|^{2}$
   of $|n\rangle\langle n|$ in normal order.}
\begin{equation}
\label{A3.4}
P_{m} =
\left\langle
:\frac{(\eta\hat{n})^{m}}{m!}\,e^{-\eta\hat{n}}:
\right\rangle,
\end{equation}
and the characteristic function
\begin{equation}
\label{A3.5}
\Omega(y) = \sum_{m} P_{m}\,e^{imy}
\end{equation}
can be given by
\begin{equation}
\label{A3.6}
\Omega(y) = \left\langle
:\exp\!\left[\eta\hat{n}(e^{iy}-1)\right]:
\right\rangle,
\end{equation}
where the symbol $:\,:$ indicates normal ordering, with the
operator $\hat{a}^{\dagger}$ to the left of the operator $\hat{a}$.
So far, we have assumed that the photons are effectively associated with
a single (nonmonochromatic) mode. The extension of the above given formulae
to multimode fields is straightforward. In particular, the multi-mode
version of the single-mode photocounting formula (\ref{A3.4}) reads as
\begin{equation}
\label{A3.7}
P_{m_{1},m_{2},\ldots} =
\left\langle
\prod_{k} :\frac{(\eta_{k}\hat{n}_{k})^{m_{k}}}{m_{k}!}
\,e^{-\eta_{k}\hat{n}_{k}}:
\right\rangle,
\end{equation}
where $\hat{n}_{k}$ is the photon-number operator of the $k$th mode,
and $\eta_{k}$ is the quantum efficiency which with the photons of
this mode are detected.


\section[Elements of least-squares inversion]{Elements of least-squares
inversion\protect\footnote{For details,
   see, e.g., Golub and van Loan [1989];
   Robinson [1991]; Press, Teukolsky, Vetterling and Flannery [1995].}
}
\label{app4}
\setcounter{equation}{0}

Let ${\bf f}$ be a (possibly unknown) $n_{0}$ dimensional ``state''
vector and consider a stochastic linear transform,
\begin{equation}
\label{A4.1}
{\bf y} = \mbox{\boldmath $A$}\,{\bf f} + {\bf n},
\end{equation}
yielding an $m_{0}$ dimensional ($m_{0}$ $\! \ge $ $\! n_{0}$) ``data''
vector $\bf y$ available from measurements. Here,
{\boldmath $A$} is a given $m_{0}$ $\!\times$ $\!n_{0}$ matrix, and
${\bf n}$ is an $m_{0}$ dimensional vector whose random elements
with zero means and covariance matrix $\mbox{\boldmath $W$}^{-1}$
describe the noise associated with realistic measurements.
The probability of a state vector ${\bf f}$ being realized under
the condition that there is a data vector ${\bf y}$ can be given by
\begin{equation}
\label{A4.2}
P({\bf f}|{\bf y}) \sim P({\bf y}|{\bf f}) \,P({\bf f}),
\end{equation}
where for Gaussian noise the conditional probability $P({\bf y}|{\bf f})$
of observing the data vector ${\bf y}$ is
\begin{equation}
\label{A4.3}
P({\bf y}|{\bf f}) \sim \exp\!\left[ -{\textstyle\frac{1}{2}}
({\bf y}-\mbox{\boldmath $A$}{\bf f})^{\dagger} \mbox{\boldmath $W$}
({\bf y}- \mbox{\boldmath $A$}{\bf f})
\right] .
\end{equation}
The probability $P({\bf f})$ is a measure of {\em a priori}
knowledge of the state and can be set constant if no state is
preferred. The most probable state $\tilde {\bf f}$ can then be
found by minimization of\footnote{If $\mbox{\boldmath $W$}$ is diagonal
   (i.e., the noise is uncorrelated), then eq.~(\protect\ref{A4.4})
   represents a sum of weighted squares of the differences between the
   components of the data vector ${\bf y}$ and the components of the
   transformed vector $\mbox{\boldmath $A$}{\bf f}$, each
   term of the sum being multiplied by a weight given by the corresponding
   element of $\mbox{\boldmath $W$}$.}
\begin{equation}
\label{A4.4}
C({\bf f}) =
({\bf y}-\mbox{\boldmath $A$}{\bf f})^{\dagger} \mbox{\boldmath $W$}
({\bf y}- \mbox{\boldmath $A$}{\bf f}),
\end{equation}
from which
\begin{equation}
\label{A4.5}
\mbox{\boldmath $A$}^{\dagger}\mbox{\boldmath $W$}
\mbox{\boldmath $A$}\tilde {\bf f}
=\mbox{\boldmath $A$}^{\dagger}\mbox{\boldmath $W$}{\bf y},
\end{equation}
and when $\mbox{\boldmath $A$}^{\dagger}\mbox{\boldmath $W$}
\mbox{\boldmath $A$}$ is not singular, then
\begin{equation}
\label{A4.6}
\tilde {\bf f}=(\mbox{\boldmath $A$}^{\dagger}\mbox{\boldmath $W$}
\mbox{\boldmath $A$})^{-1}
\mbox{\boldmath $A$}^{\dagger}\mbox{\boldmath $W$}{\bf y}.
\end{equation}
Otherwise, the inversion of
eq.~(\ref{A4.5}) is not unique and further criteria must be used to
select a solution.
When the matrix $\mbox{\boldmath $W$}$ is not known, then
it may be set a multiple of a unity matrix, so that
eq.~(\ref{A4.6}) reduces to
\begin{equation}
\label{A4.6a}
\tilde {\bf f}=(\mbox{\boldmath $A$}^{\dagger}
\mbox{\boldmath $A$})^{-1}
\mbox{\boldmath $A$}^{\dagger}{\bf y},
\end{equation}
provided that $\mbox{\boldmath $A$}^{\dagger}\mbox{\boldmath $A$}$
is not singular. Equation
(\ref{A4.6a}) still gives the correct averaged inversion, but the
statistical fluctuation of the result may be (slightly) enhanced.

If the data are not sensitive enough to some state-vector components,
then these components can hardly be determined with reasonable
accuracy. Mathematically, $\mbox{\boldmath $A$}^{\dagger}
\mbox{\boldmath $W$}\mbox{\boldmath $A$}$ becomes
(quasi-)singular and regularizations, such as Tikhonov
regularization and singular-value decomposition, are required
to solve approximately eq.~(\ref{A4.5}). For simplicity let us set
$\mbox{\boldmath $W$}$ $\!=$ $\!\mbox{\boldmath $I$}$
($\mbox{\boldmath $I$}$, unity matrix).
Using Tikhonov regularization, it is assumed that
some components of the state vector can be preferred by a
properly chosen {\em a priori} probability $P({\bf f})$, such as
\begin{equation}
\label{A4.7}
P({\bf f}) \sim \exp\!\left( - {\textstyle\frac{1}{2}} \lambda ^2
{\bf f}^{\dagger} {\bf f}\right),
\end{equation}
the parameter $\lambda$ ($\lambda$ $\!\ge$ $\!0$) being a measure of the
strength of regularization. Maximization of $P({\bf f}|{\bf y})$
then yields, on recalling eqs.~(\ref{A4.2}) and (\ref{A4.3}),
\begin{equation}
\label{A4.8}
\tilde {\bf f}=(\lambda ^{2} {\bf I}
+ \mbox{\boldmath $A$}^{\dagger}\mbox{\boldmath $A$})^{-1}
\mbox{\boldmath $A$}^{\dagger}{\bf y}.
\end{equation}
Note that $\lambda ^{2} {\bf I}$ $\!+$ $\!\mbox{\boldmath $A$}^{\dagger}
\mbox{\boldmath $A$}$ has only positive eigenvalues and is thus
always invertible. A possible choice of $\lambda$ is based on the
so-called $L$ curve, which is a log-log plot of $||{\bf f} ||$
versus $||\Delta {\bf y} ||$, $\Delta {\bf y}$ $\!=$ $\!{\bf y}$ $\!-$
$\!\mbox{\boldmath $A$}{\bf f}$,
for different values of $\lambda$.
The points on the horizontal branch correspond to large noise,
whereas the points on the vertical branch correspond to
large data misfit. Optimum choice of $\lambda$ corresponds to points
near the corner of the $L$ curve.

Applying singular-value decomposition, the inversion of the matrix
$\mbox{\boldmath $A$}^{\dagger}\mbox{\boldmath $A$}$ is
performed such that their eigenvalues
whose absolute values are smaller than the (positive) parameter
of regularization $\sigma_{0}$ are treated as zeros, but the inversions
are set zero (instead to infinity). This operation is called
``pseudoinverse'' of a matrix,
\begin{equation}
\label{A4.9}
\tilde {\bf f}= {\rm Pseudoinverse}
(\mbox{\boldmath $A$}^{\dagger}\mbox{\boldmath $A$}; \sigma_{0})
\mbox{\boldmath $A$}^{\dagger}{\bf y}.
\end{equation}
For $\sigma _{0}$ close to zero the result of eq.~(\ref{A4.9}) is
similar to that of eq.~(\ref{A4.6a}). With increasing $\sigma _{0}$,
smaller absolute values of components of $\tilde {\bf f}$ are preferred.

The effect of the regularization parameters $\lambda$ and $\sigma _{0}$
is similar. The statistical error of the reconstructed
state vector $\tilde {\bf f}$ is decreased, but bias
towards zero is produced simultaneously.
Hence, optimum parameters are those for which
the bias is just below the statistical fluctuation.
The bias can be estimated, e.g., by Monte Carlo generating new sets of
``synthetic'' data from the reconstructed state. From these sets one can
again reconstruct new sets of $\tilde {\bf f}$.
The difference between the mean value of
the states reconstructed from the synthetic data
and the originally reconstructed state estimates the bias.

\end{appendix}

\newpage

\addcontentsline{toc}{section}{References}


\pagestyle{myheadings}
\markboth{\protect\sl REFERENCES}{\protect\sl REFERENCES}
\section*{References}

Abbas, G.L., V.W.S. Chan and T.K. Yee, 1983,
Opt. Lett. {\bf 8}, 419. \\
Agarwal, G.S., and S. Chaturvedi, 1994,
Phys. Rev. A{\bf 49}, R665. \\
Adams, C.S., M. Sigel, and J. Mlynek, 1994,
Phys. Rep. {\bf 240}, 143. \\
Agarwal, G.S., and E. Wolf, 1970,
Phys. Rev. D{\bf 2}, 2161. \\
Aharonov, Y., and L. Vaidman, 1993,
Phys. Lett. A{\bf 178}, 38. \\
Aharonov, Y., D.Z. Albert and C.K. Au, 1981,
Phys. Rev. Lett. {\bf 47}, 1029. \\
Aharonov, Y., J. Anandan and L. Vaidman, 1993,
Phys. Rev. A{\bf 47}, 4616. \\
Akulin, V.M., Fam Le Kien and W.P. Schleich, 1991,
Phys. Rev. A{\bf 44}, R1462. \\
Alter, O., and Y. Yamamoto, 1995,
Phys. Rev. Lett. {\bf 74}, 4106. \\
Anderson, M.H., J.R. Ensher, M.R. Matthews,
C.E. Wieman and E.A. Cornell, 1995,
Science {\bf 269}, 198. \\
Andrews, M.R., M.-O. Mewes, N.J. van Druten, D.S. Durfee,
D.M. Kurn and W. Ketterle, 1996,
Science {\bf 273}, 84. \\
Andrews, M.R., C.G. Townsend, H.-J. Miesner,
D.S. Durfee, D.M. Kurn and W. Ketterle, 1997,
Science {\bf 275}, 637.\\
Arthurs, E., and J.L. Kelly Jr., 1965,
Bell. Syst. Tech. J. {\bf 44}, 725.\\
Ashburn, J.R., R.A. Cline, P.J.M. van der Burgt,
W.B. Westerveld and J.S. Risley, 1990,
Phys. Rev. A{\bf 41}, 2407.\\
Ashburn, J.R., R.A. Cline, C.D. Stone, P.J.M. van der Burgt,
W.B. Westerveld and J.S. Risley, 1989,
Phys. Rev. A{\bf 40}, 4885.\\
Assion, A., M. Geisler, J. Helbing, V. Seyfried and T. Baumert, 1996,
Phys. Rev. A{\bf 54}, R4605.\\
Bacry, H., A. Grossmann and J. Zak, 1975,
Phys. Rev. B{\bf 12}, 1118.\\
Ban, M., 1991a,
Phys. Lett. A{\bf 152}, 223.\\
Ban, M., 1991b,
Phys. Lett. A{\bf 155}, 397.\\
Ban, M., 1991c,
J. Math. Phys. {\bf 32}, 3077.\\
Ban, M., 1991d,
Physica A{\bf 179}, 103.\\
Ban, M., 1992,
J. Opt. Soc. Am. B{\bf 9}, 1189.\\
Ban, M., 1993,
Phys. Lett. A{\bf 176}, 47.\\
Banaszek, K., and K. W\'{o}dkiewicz, 1996,
Phys. Rev. Lett. {\bf 76}, 4344.\\
Banaszek, K., and K. W\'{o}dkiewicz, 1997a,
Phys. Rev. A{\bf 55}, 3117.\\
Banaszek, K., and K. W\'{o}dkiewicz, 1997b,
J. Mod. Opt. {\bf 44}, 2441.\\
Banaszek, K., and K. W\'{o}dkiewicz, 1998,
Optics Express {\bf 3}, 141.\\
Band, W., and J.L. Park, 1970,
Found. Phys. {\bf 1}, 133.\\
Band, W., and J.L. Park, 1971,
Found. Phys. {\bf 1}, 339.\\
Band, W., and J.L. Park, 1979,
Am. J. Phys. {\bf 47}, 188.\\
Bandilla, A., and H. Paul, 1969,
Ann. Phys. (Leipzig) {\bf 23}, 323.\\
Bandilla, A., and H. Paul, 1970,
Ann. Phys. (Leipzig) {\bf 24}, 119.\\
Bardroff, P.J., E. Mayr and W.P. Schleich, 1995,
Phys. Rev. A{\bf 51}, 4963.\\
Bardroff, P.J., C. Leichtle, G. Schrade and W.P. Schleich, 1996,
Phys. Rev. Lett. {\bf 77}, 2198.\\
Bardroff, P.J., E. Mayr, W.P. Schleich,
P. Domokos, M. Brune, J.M. Raimond and S. Haroche, 1996,
Phys. Rev. A{\bf 53}, 2736.\\
Baseia, B., M.H.Y. Moussa and V.S. Bagnato, 1997,
Phys. Lett. A{\bf 231}, 331.\\
Beck, M., D.T. Smithey and M.G. Raymer, 1993,
Phys. Rev. A{\bf 48}, R890.\\
Bergquist, J.C., R.G. Hulet, W.M. Itano and D.J. Wineland, 1986,
Phys. Rev. Lett. {\bf 57}, 1699.\\
Barnett, S.M., and D.T. Pegg, 1996,
Phys. Rev. Lett. {\bf 76}, 4148.\\
Bergou, J., and B.-G. Englert, 1991,
Ann. Phys. (NY) {\bf 209}, 479.\\
Bertrand, J., and P. Bertrand, 1987,
Found. Phys. {\bf 17}, 397.\\
Bia{\l}ynicka-Birula, Z., and I. Bia{\l}ynicki-Birula, 1994,
J. Mod. Opt. {\bf 41}, 2203.\\
Bia{\l}ynicka-Birula, Z., and I. Bia{\l}ynicki-Birula, 1995,
Appl. Phys. B{\bf 60}, 275.\\
Blockley, C.A., D.F. Walls and H. Risken, 1992,
Europhys. Lett. {\bf 17}, 509.\\
Blow, K.J., R. Loudon, S.J.D. Phoenix and T.J. Shepherd, 1990,
Phys. Rev. A{\bf 42}, 4102.\\
Bodendorf, C.T., G. Antesberger, M.S. Kim and H. Walther, 1998,
Phys. Rev. A{\bf 57}, 1371.\\
Bohn, J., 1991,
Phys. Rev. Lett. {\bf 66}, 1547.\\
Bolda, E.L., S.M. Tan and D.F. Walls, 1997,
Phys. Rev. Lett. {\bf 79}, 4719.\\
de Boor, C., 1987,
{\em A Practical Guide to Spines} (Springer, New York).\\
Bradley, C.C., C.A. Sackett, J.J. Tollett and R.G. Hulet, 1995,
Phys. Rev. Lett. {\bf 75}, 1687.\\
Braginsky, V.B., and F.Y. Khalili, 1992,
{\em Quantum Measurement}, (Cambridge University Press, Cambridge).\\
Braginsky, V.B., Y.I Vorontsov and F.Y. Khalili, 1977,
Sov. Phys.-JETP {\bf 46}, 705.\\
Braunstein, S.L., 1990,
Phys. Rev. A{\bf 42}, 474.\\
Braunstein, S.L., C.M. Caves and G.J. Milburn, 1991,
Phys. Rev. A{\bf 43}, 1153.\\
Breitenbach, G., and S. Schiller , 1997,
J. Mod. Optics {\bf 44}, 2207.\\
Breitenbach, G., S. Schiller and J. Mlynek, 1997,
Nature {\bf 387}, 471.\\
Breitenbach, G., T. M\"{u}ller, S.F. Pereira, J.-Ph. Poizat,
S. Schiller and J. Mlynek, 1995,
J. Opt. Soc. Am. B{\bf 12}, 2304.\\
Hanbury Brown, R., and R.Q. Twiss, 1956a,
Nature {\bf 177}, 27.\\
Hanbury Brown, R., and R.Q. Twiss, 1956b,
Nature {\bf 178}, 1046.\\
Hanbury Brown, R., and R.Q. Twiss, 1957a,
Proc. Roy. Soc. (London) A{\bf 242}, 300.\\
Hanbury Brown, R., and R.Q. Twiss, 1957b,
Proc. Roy. Soc. (London) A{\bf 243}, 291.\\
Brown, L.S., and G. Gabrielse, 1986,
Rev. Mod. Phys. {\bf 58} 233.\\
Brune, M., and S. Haroche, 1994,
in: {\em Quantum Dynamics of Simple Systems},
{\em The Forty Forth Scottish Universities Summer School in
Physics}, Stirling, eds G.-L. Oppo, S.M. Barnett, E. Riis and M. Wilkinson
(Institute of Physics Publishing, Bristol and Philadelphia, 1996) p. 49.\\
Brune, M., S. Haroche, V. Lefevre, J.M. Raimond and N. Zagury, 1990,
Phys. Rev. Lett. {\bf 65}, 976.\\
Brune, M., S. Haroche,  J.M. Raimond, L. Davidovich and N. Zagury, 1992,
Phys. Rev. A{\bf 45}, 5193.\\
Brune, M, J.M. Raimond, P. Goy, L. Davidovich and S. Haroche, 1987,
Phys. Rev. Lett. {\bf 59} 1899.\\
Brune, M., F. Schmidt-Kaler, A. Maali,
J. Dreyer, E. Hagley, J.M. Raimond and S. Haroche, 1996,
Phys. Rev. Lett. {\bf 76}, 1800.\\
Brunner, W., H. Paul and G. Richter, 1965,
Ann. Phys. (Leipzig) {\bf 15}, 17.\\
Busch, P., and P.J. Lahti, 1989,
Found. Phys. {\bf 19}, 633.\\
Bu\v{z}ek, V., and M. Hillery, 1996,
J. Mod. Opt. {\bf 43}, 1633.\\
Bu\v{z}ek, V., and P.L. Knight, 1995,
in: {\em Progress in Optics XXXIV\/}, ed. E. Wolf
(North-Holland, Amsterdam).\\
Bu\v{z}ek, V., G. Adam and G. Drobn\'{y}, 1996a,
Ann. Phys. (NY) {\bf 245}, 37.\\
Bu\v{z}ek, V., G. Adam and G. Drobn\'{y}, 1996b,
Phys. Rev. A{\bf 54}, 804.\\
Bu\v{z}ek, V., C.H. Keitel and P.L. Knight, 1995a,
Phys. Rev. A{\bf 51}, 2575.\\
Bu\v{z}ek, V., C.H. Keitel and P.L. Knight, 1995b,
Phys. Rev. A{\bf 51}, 2594.\\
Bu\v{z}ek, V.,  G. Drobn\'{y}, G. Adam, R. Derka and P.L. Knight, 1997,
J. Mod. Opt. {\bf 44}, 2607.\\
Cahill, K.E., 1965,
Phys. Rev. {\bf 138}B, 1566.\\
Cahill, K.E., and R.J. Glauber, 1969a,
Phys. Rev. {\bf 177}, 1857.\\
Cahill, K.E., and R.J. Glauber, 1969b,
Phys. Rev. {\bf 177}, 1882.\\
Campos, R.A., B.E.A. Saleh and M.C. Teich, 1989,
Phys. Rev. A{\bf 40}, 1371.\\
Carmichael, H.J., 1987,
J. Opt. Soc. Am. B{\bf 4}, 1588.\\
Carnal, O., and J. Mlynek, 1991,
Phys. Rev. Lett. {\bf 66}, 2689.\\
Caves, C.M., 1982,
Phys. Rev. D {\bf 26}, 1817.\\
Caves, C.M., and P.D. Drummond, 1994,
Rev. Mod. Phys. {\bf 66}, 481.\\
Caves, C.M., and B.L. Schumaker, 1985,
Phys. Rev. A{\bf 31}, 3068.\\
Chaturvedi, S., G.S. Agarwal and V. Srinivasan, 1994,
J. Phys. A{\bf 27}, L39.\\
Chen, X., and J.A. Yeazell, 1997,
Phys. Rev. A{\bf 56}, 2316.\\
Cirac, J.I., R. Blatt, A.S. Parkins and P. Zoller, 1994,
Phys. Rev. A{\bf 49}, 1202.\\
Cline, R., P.J.M. van der Burgt, W.B. Westerveld and
J.S. Risley, 1994,
Phys. Rev. A{\bf 49}, 2613.\\
Collett, M.J., R. Loudon and C.W. Gardiner, 1987,
J. Mod. Opt. {\bf 34}, 881.\\
Cook, R.J., 1978,
Phys. Rev. Lett. {\bf 41}, 1788.\\
Corbett, J.V., and C.A. Hurst, 1978,
J. Austral. Math. Soc. B{\bf 20}, 182.\\
Dakna, M., L. Kn\"{o}ll and D.-G. Welsch, 1997a,
Quantum Semiclass. Opt. {\bf 9}, 331.\\
Dakna, M., L. Kn\"{o}ll and D.-G. Welsch, 1997b,
Phys. Rev. A{\bf 55}, 2360.\\
Dakna, M., T. Opatrn\'{y} and D.-G. Welsch, 1998,
Opt. Commun. {\bf 148}, 355.\\
D'Ariano, G.M., 1995,
Quantum. Semiclass. Opt. {\bf 7}, 693.\\
D'Ariano, G.M., 1997a,
in: {\em Quantum Optics and Spectroscopy of
Solids}, eds T. Hakio\v{g}lu and A.S. Shumovsky
(Kluwer Academic Publisher, Amsterdam) p. 175.\\
D'Ariano, G.M., 1997b,
in: {\em Quantum Communication, Computing and Measurement,}
eds O. Hirota, A.S. Holevo and C.M. Caves (Plenum, New York) p. 253.\\
D'Ariano, G.M., 1997c,
Private communication.\\
D'Ariano, G.M., and C. Macchiavello, 1998,
Phys. Rev. A{\bf 57}, 3131.\\
D'Ariano, G.M., and M.G.A. Paris, 1997a,
Phys. Lett. A{\bf 233}, 49.\\
D'Ariano, G.M., and M.G.A. Paris, 1997b,
in: 5{\em th Central-European Workshop on Quantum Optics},
Prague, eds I. Jex and G. Drobn\'{y}
(Acta Physica Slovaca {\bf 47}, 281).\\
D'Ariano, G.M., and L. Maccone, 1997,
in: {\em Fifth International Conference on Squeezed States and
Uncertainty Relations}, Balatonf\"{u}red,
eds D. Han, J, Janszky, Y.S. Kim and V.I. Man'ko
(NASA/CP-1998-206855) p. 529.\\
D'Ariano, G.M., and N. Sterpi, 1997,
J. Mod. Opt. {\bf 44}, 2227.\\
D'Ariano, G.M., and H.P. Yuen, 1996,
Phys. Rev. Lett. {\bf 76}, 2832.\\
D'Ariano, G.M., U. Leonhardt and H. Paul, 1995,
Phys. Rev. A{\bf 52}, R1801.\\
D'Ariano, G.M., C. Macchiavello and M.G.A. Paris, 1994a,
Phys. Rev. A{\bf 50}, 4298.\\
D'Ariano, G.M., C. Macchiavello and M.G.A. Paris, 1994b,
Phys. Lett. A{\bf 195}, 31.\\
D'Ariano, G.M., C. Macchiavello and M.G.A. Paris, 1995,
Phys. Lett. A{\bf 198}, 286.\\
D'Ariano, G.M., C. Macchiavello and N. Sterpi, 1997,
Quantum Semiclass. Opt.  {\bf 9}, 929.\\
D'Ariano, G.M., S. Mancini, V.I. Man'ko and P. Tombesi, 1996,
Quantum Semiclass. Opt. {\bf 8}, 1017.\\
Davidovi\'{c}, and D.M., D. Lalovi\'{c}, 1993,
J. Phys. A{\bf 26}, 5099.\\
Davidovich, L., M. Orszag and N. Zagury, 1996,
Phys. Rev. A{\bf 54}, 5118.\\
Davies, E.B., 1976,
{\em Quantum theory of open systems} (Academic Press, New York).\\
Davies, E.B., and J.T. Lewis, 1970,
Commun. Math. Phys. {\bf 17}, 239.\\
Davies, K.B., M.-O. Mewes, M.R. Andrews, N.J. van Druten,
D.S. Durfee, D.M. Kurn and W. Ketterle, 1995,
Phys. Rev. Lett. {\bf 75}, 3969.\\
Derka, R., V. Bu\v{z}ek and G. Adam, 1996,
in: 4{\em th Central-European Workshop on Quantum Optics},
Budmerice, ed. V. Bu\v{z}ek
(Acta Physica Slovaca {\bf 46}, 355).\\
Derka, R., V. Bu\v{z}ek, G. Adam and P.L. Knight, 1996,
J. Fine Mechanics Optics {\bf 11-12}, 341.\\
D'Helon, C., and G.J. Milburn, 1996,
Phys. Rev. A{\bf 54}, R25.\\
Dhirani, Al-A., D.A. Kokorowski, R.A. Rubenstein, T.D. Hammond,
B. Rohwedder, E.T. Smith, A.D. Roberts, and D.E. Pritchard, 1997,
J. Mod. Opt. {\bf 44}, 2583.\\
Dodonov, V.V., I.A. Malkin, and V.I. Man'ko, 1974,
Physica {\bf 72}, 597.\\
Dodonov, V.V., and V.I. Man'ko, 1997,
Phys. Lett. A{\bf 229}, 335.\\
Dowling, J.P., G.S. Agarwal and W.P. Schleich, 1994,
Phys. Rev. A{\bf 49}, 4101.\\
Drummond, P.D., 1989,
in: {\em Fifth New Zealand Symposium on Quantum Optics},
eds J.D. Harvey and D.F. Walls
(Springer-Verlag, Berlin) p. 57. \\
Drummond, P.D., and C.W. Gardiner, 1980,
J. Phys. A{\bf 13}, 2353.\\
Dunn, T.J., I.A. Walmsley and S. Mukamel, 1995,
Phys. Rev. Lett. {\bf 74}, 884.\\
Dutra, S.M., P.L. Knight and H. Moya-Cessa, 1993,
Phys. Rev. A{\bf 48}, 3168.\\
Dutra, S.M., and P.L. Knight, 1994,
Phys. Rev. A{\bf 49}, 1506.\\
Epstein, S.T., 1959,
Am. J. Phys. {\bf 27}, 291.\\
d'Espagnat, B., 1976,
{\em Conceptual Foundations of Quantum Mechanics},
(Benjamin, Reading, MA).\\
Fano, U., 1957,
Rev. Mod. Phys. {\bf 29}, 74.\\
Fearn, H., and R. Loudon, 1987,
Opt. Commun. {\bf 64}, 485.\\
Feenberg, E., 1933,
Ph.D. thesis, Harvard University.\\
Feynman, R.P., R.B. Leighton and M. Sands, 1965,
{\em The Feynman Lectures on Physics} (Addison-Wesley, Reading, MA).\\
Freyberger, M., 1997,
Phys. Rev. A{\bf 55}, 4120.\\
Freyberger, M., S.H. Kienle, and V.P. Yakovlev, 1997,
Phys. Rev. A {\bf 56}, 195.\\
Freyberger, M., and W.P. Schleich, 1993,
Phys. Rev.  A{\bf 47}, R30.\\
Freyberger, M., K. Vogel and W.P. Schleich, 1993a,
Quantum Opt. {\bf 5}, 65.\\
Freyberger, M., and A.M. Herkommer, 1994,
Phys. Rev. Lett. {\bf 72}, 1952.\\
Freyberger, M., K. Vogel and W.P. Schleich, 1993b,
Phys. Lett. A{\bf 176}, 41.\\
Friedman, C.N., 1987,
J. Austral. Math. Soc. B{\bf 30}, 289.\\
Gale, W., E. Guth and G.T. Trammel, 1968,
Phys. Rev. {\bf 165}, 1434.\\
Galetti, D., and M.A. Marchiolli, 1996,
Ann. Phys. (NY) {\bf 249}, 454.\\
Gardiner C.W., 1983,
{\em Handbook of Stochastic Methods}
(Springer, Berlin).\\    
Gardiner, C.W., 1991,
{\em Quantum Noise} (Springer, Berlin).\\
Gauss, C.F., 1809,
{\em Theoria motus corporum coelestium in sectionibus
conicis solem ambientum} (Perthes and Besser, Hamburg). \\
Gauss, C.F., 1821,
{\em Theoria combinationis observationum erroribus minimis obnoxiae}
(Societati regiae scientiarum exhibita, G\"{o}ttingen).\\
Gerlach, W., and O. Stern, 1921,
Z. Phys. {\bf 8}, 110.\\
Gerlach, W., and O. Stern, 1922,
Z. Phys. {\bf 9}, 349.\\
Glauber, R.J., 1963a,
Phys. Rev. {\bf 130}, 2529.\\
Glauber, R.J., 1963b,
Phys. Rev. Lett. {\bf 10}, 84.\\
Glauber, R.J., 1963c,
Phys. Rev. {\bf 131}, 2766.\\
Glauber, R.J., 1965,
in: {\em Quantum Optics and
Electronics}, eds C. DeWitt, A. Blandin and C. Cohen--Tannoudji
(Gordon and Breach, New York)  p. 144. \\
Glauber, R.J., and M. Lewenstein, 1991,
Phys. Rev. A{\bf 43}, 467.\\
Golovchenko, J.A., D.R. Kaplan, B. Kincaid, R. Levesque, A. Meixner,
M.P. Robbins, and J. Felsteiner, 1981,
Phys. Rev. Lett. {\bf 46}, 1454.\\
Golub, G.H., and C.F. van Loan, 1989,
{\em Matrix Computations}
(J. Hopkins University Press, Baltimore and London). \\
Greene, B.I., J.F. Federici, D.R. Dykaar, R.R Jones
and P.H. Bucksbaum, 1991,
Appl. Phys. Lett. {\bf 59}, 893.\\
Gruner, T., and D.--G. Welsch, 1996,
Phys. Rev. A{\bf 53}, 1818.\\
Havener, C.C., N. Rouze, W.B. Westerveld and J.S. Risley, 1986,
Phys. Rev. A{\bf 33}, 276.\\
Helstrom, C.W., 1976,
{\em Quantum Detection and Estimation Theory}
(Academic, New York).\\
Herkommer, A.M., V.M. Akulin and W.P. Schleich, 1992,
Phys. Rev. Lett. {\bf 69}, 3298.\\
Herzog, U., 1996a,
Phys. Rev. A{\bf 53}, 1245.\\
Herzog, U., 1996b,
Phys. Rev. A{\bf 53}, 2889.\\
Ho, S.T., A.S. Lane, A. La Porta, R.E. Slusher and B. Yurke, 1990,
in: {\em International Quantum Electronics Conference, Technical
Digest Series Vol. 7} (Optical Society of America, Washington).\\
Holevo, A.S., 1973,
J. Multivar. Anal. {\bf 3}, 337.\\
Holevo, A.S., 1982,
{\em Probabilistic and Statistical Aspects of Quantum Theory}
(North-Holland, Amsterdam).\\
Hradil, Z., 1992,
Quantum Opt. {\bf 4}, 93.\\
Hradil, Z., 1993,
Phys. Rev. A{\bf 47}, 2376.\\
Hradil, Z., 1997,
Phys. Rev. A{\bf 55}, R1561.\\
Husimi, K., 1940,
Proc. Phys. Math. Soc. Jpn. {\bf 22}, 264.\\
Husimi, K.,  1953,
Proc. Theor. Phys. {\bf 9}, 381.\\
Huttner, B., J.J. Baumberg, J.F. Ryan and S.M. Barnett, 1992,
Opt. Commun. {\bf 90}, 128.\\
Imamo\u{g}lu, A., 1993,
Phys. Rev. A{\bf 47}, R4577.\\
Ivanovi\'{c}, I.D., 1981,
J. Phys. A{\bf 14}, 3241.\\
Ivanovi\'{c}, I.D., 1983,
J. Math. Phys. {\bf 24}, 1199.\\
Ivanovi\'{c}, I.D., 1993,
J. Phys. A{\bf 26}, L579.\\
Jain, A., C.D. Lin and W. Fritsch, 1987a,
Phys. Rev. A{\bf 35}, 3180.\\
Jain, A., C.D. Lin and W. Fritsch, 1987b,
Phys. Rev. A{\bf 36}, 2041.\\
Jain, A., C.D. Lin and W. Fritsch, 1988,
Phys. Rev. A{\bf 37}, 3611(E).\\
Janicke, U., and M. Wilkens, 1995,
J. Mod. Opt. {\bf 42}, 2183.\\
Janszky, J., and Y.Y. Yushin, 1986,
Opt. Commun. {\bf 59}, 151.\\
Janszky, J, P. Adam and Y. Yushin, 1992,
Opt. Commun. {\bf 93}, 191.\\
Javanainen, J., 1994,
Phys. Rev. Lett. {\bf 72}, 2375.\\
Jaynes, E.T., 1957a,
Phys. Rev. {\bf 106}, 620.\\
Jaynes, E.T., 1957b,
Phys. Rev. {\bf 108}, 171.\\
Jaynes, E.T., and F.W. Cummings, 1963,
Proc. IEEE {\bf 51}, 89.\\
Jex, I., S. Stenholm and A. Zeilinger, 1995,
Opt. Commun. {\bf 117}, 95.\\
Jones, K.R.W., 1991,
Ann. Phys. (NY) {\bf 207}, 140.\\
Jones, K.R.W., 1994,
Phys. Rev A{\bf 50}, 3682.\\
Jones, R.R., 1996,
Phys. Rev. Lett. {\bf 76}, 3927.\\
Jones, R.R., D. You and P.H. Bucksbaum, 1993,
Phys. Rev. Lett. {\bf 70}, 1236.\\
Kano, Y., 1965,
J. Math. Phys. {\bf 6}, 1913.\\
Kazantsev, A.P., 1975,
Zh. Eksp. Teor. Fiz. {\bf 67}, 1660;
(Sov. Phys. JETP {\bf 40}, 825).\\
Kazantsev, A.P., 1978,
Usp. Fiz. Nauk {\bf 124}, 113;
(Sov. Phys. Usp. {\bf 21}, 58).\\
Keith, D.W., C.R. Ekstrom, Q.A. Turchette, and D.E. Pritchard (1991),
Phys. Rev. Lett. {\bf 66}, 2693.\\
Kelley, P.L., and W.H. Kleiner, 1964,
Phys. Rev. {\bf 136}, A316.\\
Kemble, E.C., 1937,
{\em Fundamental Principles of Quantum Mechanics},
(McGraw-Hill, New York).\\
Kienle, S.H., D. Fischer, W.P. Schleich, V.P. Yakovlev
and M. Freyberger, 1997,
Appl. Phys. B{\bf 65}, 735.\\
Kim, M.S., 1997a,
Phys. Rev. A{\bf 56}, 3175.\\
Kim, M.S., 1997b,
J. Mod. Opt. {\bf 44}, 1437.\\
Kim, M.S., and N. Imoto, 1995,
Phys. Rev. A{\bf 52}, 2401. \\
Kim, M.S., and B.C. Sanders, 1996,
Phys. Rev. A{\bf 53}, 3694. \\
Kim, C., and  P. Kumar, 1994,
Phys. Rev. Lett. {\bf 73}, 1605. \\
Kiss, T., U. Herzog and U. Leonhardt, 1995,
Phys. Rev. A{\bf 52}, 2433.\\
Klauder, J.R., 1960,
Ann. Phys. (NY) {\bf 11}, 123.\\
Klauder, J.R., 1963a,
J. Math. Phys. {\bf 4}, 1055.\\
Klauder, J.R., 1963b,
J. Math. Phys. {\bf 4}, 1058.\\
Klauder, J.R., and E.C.G. Sudarshan, 1968,
{\em Fundamentals of Quantum Optics} (Benjamin, New York).\\
Kn\"oll, L., W. Vogel and D.--G. Welsch, 1987,
Phys. Rev. A{\bf 36}, 3803.\\
Kocha\'{n}ski, P., and K. W\'{o}dkiewicz, 1997,
J. Mod. Opt. {\bf 44}, 2343.\\
Kokorowski, D.A., and D.E. Pritchard, 1997,
J. Mod. Opt. {\bf 44}, 2575.\\
Kowalczyk, P., C. Radzewicz, J. Mostowski and I.A. Walmsley, 1990,
Phys. Rev. A{\bf 42}, 5622.\\
Kr\"{a}hmer, D.S., and U. Leonhardt, 1997a,
Phys. Rev. A{\bf 55}, 3275.\\
Kr\"{a}hmer, D.S., and U. Leonhardt, 1997b,
J. Phys. A{\bf 30}, 4783.\\
Kr\"{a}hmer, D.S., and U. Leonhardt, 1997c,
Appl. Phys. B{\bf 65}, 725.\\
Kreinovitch, V.Ja., 1977,
Theor. Math. Phys. {\bf 28}, 623.\\
K\"{u}hn, H., D.--G. Welsch and W. Vogel, 1994,
J. Mod. Opt. {\bf 41}, 1607.\\
K\"{u}hn, H., W. Vogel and D.--G. Welsch, 1995,
Phys. Rev. A{\bf 51}, 4240.\\
Kurtsiefer, Ch., T. Pfau and J. Mlynek, 1997,
Nature {\bf 386}, 150.\\
Kwiat, P.G., A.M. Steinberg, R.Y. Chiao, P.H. Eberhard and M.D. Petroff, 1993,
Phys. Rev. A{\bf 48}, R867.\\
Lai, Y., and H.A. Haus, 1989,
Quantum Opt. {\bf 1}, 99.\\
Lalovi\'{c}, D., D.M. Davidovi\'{c} and N. Bijedi\'{c}, 1992]
Phys. Rev. A{\bf 46}, 1206.\\
Lambrecht, A., T. Coudreau, A.M. Steinberg and E. Giacobino, 1996,
Europhys. Lett. {\bf 36}, 93.\\
Legendre, A.M., 1805,
{\em Nou\-vel\-les m\'{e}tho\-des pour la d\'{e}ter\-mi\-na\-tion
des orbites des com\`{e}tes} (Firmin--Didot, Paris).\\
Lee, C.T., 1992,
Phys. Rev. A{\bf 46}, 6097.\\
Lehner, J., U. Leonhardt and H. Paul, 1996]
Phys. Rev. A{\bf 53}, 2727.\\
Leibfried, D., D.M. Meekhof, B.E. King, C. Monroe,
W.M. Itano and D. Wineland, 1996,
Phys. Rev. Lett. {\bf 77}, 4281 (1996).\\
Leichtle, C., W.P. Schleich, I.Sh. Averbukh and M. Shapiro, 1998,
Phys. Rev. Lett. {\bf 80}, 1418.\\
Leonhardt, U., 1993,
Phys. Rev. A{\bf 48}, 3265.\\
Leonhardt, U., 1994,
Phys. Rev. A{\bf 49}, 1231.\\
Leonhardt, U., 1995,
Phys. Rev. Lett. {\bf 74}, 4101.\\
Leonhardt, U., 1996,
Phys. Rev. A{\bf 53}, 2998.\\
Leonhardt, U., 1997a,
Phys. Rev. A{\bf 55}, 3164.\\
Leonhardt, U., 1997b,
J. Mod. Opt. {\bf 44}, 2271.\\
Leonhardt, U., 1997c,
{\em Measuring the Quantum State of Light}
(Cambridge University Press, Cambridge).\\
Leonhardt, U., and P.J. Bardroff, 1997,
in: {\em 5th Central-European Workshop on Quantum Optics},
Prague, eds I. Jex and G. Drobn\`{y}
(Acta Physica Slovaca {\bf 47}, 225).\\
Leonhardt, U., and I. Jex, 1994,
Phys. Rev. A{\bf 49}, R1555.\\
Leonhardt, U., and M. Munroe, 1996,
Phys. Rev. A{\bf 54}, 3682.\\
Leonhardt, U., and S. Schneider, 1997,
Phys. Rev. A{\bf 56}, 2549.\\
Leonhardt, U., and H. Paul, 1993a,
Phys. Rev. A{\bf 47}, R2460.\\
Leonhardt, U., and H. Paul, 1993b,
Phys. Rev. A{\bf 48}, 4598.\\
Leonhardt, U., and H. Paul, 1994a,
Phys. Rev. Lett. {\bf 72}, 4086.\\
Leonhardt, U., and H. Paul, 1994b,
J. Mod. Opt. {\bf 41}, 1427.\\
Leonhardt, U., and H. Paul, 1995,
Progr. Quantum  Electron. {\bf 19}, 89.\\
Leonhardt, U., and M.G. Raymer, 1996,
Phys. Rev. Lett. {\bf 76}, 1985.\\
Leonhardt, U., H. Paul and G.M. D'Ariano, 1995,
Phys. Rev. A{\bf 52}, 4899.\\
Leonhardt, U., J.A. Vaccaro, B. B\"{o}hmer and H. Paul, 1995,
Phys. Rev. A{\bf 51}, 84.\\
Leonhardt, U., M. Munroe, T. Kiss, Th. Richter and M.G. Raymer, 1996,
Opt. Commun. {\bf 127}, 144.\\
Lohmann, A.W., 1993,
J. Opt. Soc. Am. A{\bf 10}, 2181.\\
London, F., 1926,
Z. Phys. {\bf 37}, 915.\\
London, F., 1927,
Z. Phys. {\bf 40}, 193.\\
Loudon, R., 1986,
in: {\em Coherence, Cooperation and Fluctuations},
eds F. Haake, L.M. Narducci and D.F. Walls
(Cambridge University Press, Cambridge) p. 240. \\
Luis, A., and J. Pe\v{r}ina, 1996a,
Quantum Semiclass. Opt. {\bf 8}, 873.\\
Luis, A., and J. Pe\v{r}ina, 1996b,
Quantum Semiclass. Opt. {\bf 8}, 887.\\
Luis, A., and L.L. S\'{a}nchez-Soto, 1995,
Quantum Semiclass. Opt. {\bf 7}, 153.\\
Luk\v{s}, A., and V. Pe\v{r}inov\'{a}, 1994,
Quantum Opt. {\bf 6}, 125.\\
Lutterbach, L.G., and L. Davidovich, 1997,
Phys. Rev. Lett. {\bf 78}, 2547.\\
Lynch, R., 1995,
Phys. Rep. {\bf 256}, 367.\\
Mancini, S., and P. Tombesi, 1997a,
Phys. Rev. A{\bf 56}, 3060.\\
Mancini, S., and P. Tombesi, 1997b,
Europhys. Lett. {\bf 40}, 351.\\
Mancini, S., V.I. Man'ko and P. Tombesi, 1995,
Quantum Semiclass. Opt. {\bf 7}, 615.\\
Mancini, S., V.I. Man'ko and P. Tombesi, 1996,
Phys. Lett. A{\bf 213}, 1.\\
Mancini, S., V.I. Man'ko and P. Tombesi, 1997,
J. Mod. Opt. {\bf 44}, 2281.\\
Mancini, S., P. Tombesi and V.I. Man'ko, 1997,
Europhys. Lett. {\bf 37}, 79.\\
Mandel, L., 1958,
Proc. Phys. Soc. Lond. {\bf 72}, 1037.\\
Mandel, L., 1963,
in: {\em Progress in Optics}, Vol. 2, ed. E. Wolf
(North-Holland, Amsterdam) p. 181. \\
Mandel, L., 1982,
Phys. Rev. Lett. {\bf 49}, 136.\\
Mandel, L., and E. Wolf, 1995,
{\em Optical Coherence and Quantum Optics}
(Cambridge University Press).\\
Man'ko, O.V., 1996,
J. Russ. Laser Research {\bf 17}, 579.\\
Man'ko, O.V., 1997, Phys. Lett. A{\bf 228}, 29.\\
Marchiolli, M.A., S.S. Mizrahi and V.V. Dodonov, 1997,
Phys. Rev. A{\bf 56}, 4278.\\
Marks, J.R., 1991,
{\em Introduction to Shannon Sampling and Interpolation Theory}
(Springer, New York).\\
de Matos Filho, R.L., and W. Vogel, 1996,
Phys. Rev. Lett. {\bf 76}, 4520.\\
Mattle, K., M. Michler, H. Weinfurter, A. Zeilinger and M. Zukowski, 1995]
Appl. Phys. B{\bf 60}, S111.\\
McAlister, D.F., and M.G. Raymer, 1997a,
Phys. Rev. A{\bf 55}, R1609.\\
McAlister, D.F., and M.G. Raymer, 1997b,
J. Mod. Opt. {\bf 44}, 2359.\\
Meekhof, D.M., C. Monroe, B.E. King, W.M. Itano and D.J. Wineland, 1996,
Phys. Rev. Lett. {\bf 76}, 1796.\\
Meschede, D., H. Walther and G. M\"uller, 1985,
Phys. Rev. Lett. {\bf 54}, 551.\\
Mewes, M.-O., M.R. Andrews, D.M. Kurn,
D.S. Durfee, C.G. Townsend and W. Ketterle, 1997,
Phys. Rev. Lett. {\bf 78}, 582.\\
Mogilevtsev, D., Z. Hradil and J. Pe\v{r}ina, 1997,
J. Mod. Opt. {\bf 44}, 2261.\\
Mollow, B.R., and R.J. Glauber, 1967a,
Phys. Rev. {\bf 160}, 1076.\\
Mollow, B.R., and R.J. Glauber, 1967b,
Phys. Rev. {\bf 160}, 1097.\\
Monroe, C., D.M. Meekhof, B.E. King and D.J. Wineland, 1996,
Science {\bf 272}, 1131.\\
Moroz, B.Z., 1983,
Int. J. Theor. Phys. {22}, 329.\\
Moroz, B.Z., 1984,
Int. J. Theor. Phys. {23}, 497(E).\\
Moya-Cessa, H., and P.L. Knight, 1993,
Phys. Rev. A {\bf 48}, 2479.\\
Munroe, M., D. Boggavarapu, M.E. Anderson and M.G. Raymer, 1995]
Phys. Rev. A{\bf 52}, R924.\\
Munroe, M., D. Boggavarapu, M.E. Anderson,
U. Leonhardt and M.G. Raymer, 1996,
in: {\em Coherence and Quantum Optics VII}, eds J.H. Eberly, L. Mandel
and E. Wolf (Plenum, New York) p. 53. \\
Nagourney, W., J. Sandberg, and H. Dehmelt, 1986,
Phys. Rev. Lett. {\bf 56}, 2797.\\
Natterer, F., 1986,
{\em The Mathematics of Computerized Tomography}
(Wiley, Chichester).\\
Neuhauser, W., M. Hohenstatt, P.E. Toschek and H.G. Dehmelt, 1980,
Phys. Rev. A{\bf 22}, 1137.\\
von Neumann, J., 1932,
{\em Mathematische Grundlagen der Quantenmechanik}
(Springer, Berlin).\\
Newton, R.G., and B. Young, 1968,
Ann. Phys. (NY) {\bf 49}, 393.\\
Noel, M.W., and C.R. Stroud, Jr., 1996,
Phys. Rev. Lett. {\bf 77}, 1913.\\
Noh, J.W., A. Foug\`{e}res and L. Mandel, 1991,
Phys. Rev. Lett. {\bf 67}, 1426.\\
Noh, J.W., A. Foug\`{e}res and L. Mandel, 1992a,
Phys. Rev. A{\bf 45}, 424.\\
Noh, J.W., A. Foug\`{e}res and L. Mandel, 1992b,
Phys. Rev. A{\bf 46}, 2840.\\
Noh, J.W., A. Foug\`{e}res and L. Mandel, 1993a,
Phys. Rev. Lett. {\bf 71}, 2579.\\
Noh, J.W., A. Foug\`{e}res and L. Mandel, 1993b,
Phys. Rev. A{\bf 47}, 4535.\\
Noh, J.W., A. Foug\`{e}res and L. Mandel, 1993c,
Phys. Rev. A{\bf 47}, 4541.\\
Noordam, L.D., and R.R. Jones, 1997,
J. Mod. Opt. {\bf 44}, 2515.\\
O'Connell, R.F., and A.K. Rajagopal, 1982,
Phys. Rev. Lett. {\bf 48}, 525. \\
Opatrn\'{y}, T., and D.-G. Welsch, 1997,
Phys. Rev. A{\bf 55}, 1462.\\
Opatrn\'{y}, T., M. Dakna and D.-G. Welsch, 1997,
in: {\em Fifth International Conference on Squeezed States and
Uncertainty Relations}, Balatonf\"{u}red,
eds D. Han, J, Janszky,
Y.S. Kim and V.I. Man'ko (NASA/CP-1998-206855) p. 621.\\
Opatrn\'{y}, T., M. Dakna and D.-G. Welsch, 1998,
Phys. Rev. A{\bf 57}, 2129.\\
Opatrn\'{y}, T., D.-G. Welsch and V. Bu\v{z}ek, 1996,
Phys. Rev. A{\bf 53}, 3822.\\
Opatrn\'{y}, T., D.-G. Welsch and  W. Vogel, 1996,
in: 4{\em th Central-European Workshop on Quantum Optics},
Budmerice, ed. V. Bu\v{zek}
(Acta Physica Slovaca {\bf 46}, 469).\\
Opatrn\'{y}, T., D.-G. Welsch and  W. Vogel, 1997a,
Phys. Rev. A{\bf 55}, 1416.\\
Opatrn\'{y}, T., D.--G. Welsch and W. Vogel, 1997b,
Optics Commun. {\bf 134}, 112.\\
Opatrn\'{y}, T., D.-G. Welsch and  W. Vogel, 1997c,
Phys. Rev. A{\bf 56}, 1788.\\
Opatrn\'{y}, T., V. Bu\v{z}ek, J. Bajer and G. Drobn\'{y}, 1995,
Phys. Rev. A{\bf 52}, 2419.\\
Opatrn\'{y}, T., D.--G. Welsch, S. Wallentowitz and W. Vogel, 1997,
J. Mod. Opt. {\bf 44}, 2405.\\
Or{\l}owski, A., and H. Paul, 1994,
Phys. Rev. A{\bf 50}, R921.\\
Ou, Z.Y., C.K. Hong and L. Mandel, 1987a,
Opt Commun. {\bf 63}, 118.\\
Ou, Z.Y., C.K. Hong and L. Mandel, 1987b,
Phys. Rev. A{\bf 36}, 192.\\
Ou, Z.Y., and H.J. Kimble, 1995,
Phys. Rev.  A{\bf 52}, 3126.\\
Paris, M.G.A., 1996a,
Phys. Lett. A{\bf 217}, 78.\\
Paris, M.G.A., 1996b,
Phys. Rev. A{\bf 53}, 2658.\\
Paris, M.G.A., 1996c,
Opt. Commun. {\bf 124}, 277.\\
Paris, M.G.A., A.V. Chizhov and O. Steuernagel, 1997,
Opt. Commun. {\bf 134}, 117.\\
Park, J.L., and W. Band, 1971,
Found. Phys. {\bf 1}, 211.\\
Park, J.L., W. Band and W. Yourgrau, 1980,
Ann. Phys. (Leipzig) {\bf 37}, 189.\\
Parkins, A.S., P. Marte, P. Zoller, O. Carnal and H.J. Kimble, 1995,
Phys. Rev. A{\bf 51}, 1578.\\
Paul, H., 1963,
Ann. Phys. (Leipzig) {\bf 11}, 411.\\
Paul, H., 1974,
Fortschr. Phys. {\bf 22}, 657.\\
Paul, H., W. Brunner and G. Richter, 1966,
Ann. Phys. (Leipzig) {\bf 17}, 262.\\
Paul, H., P. T\"{o}rm\"{a}, T. Kiss and I. Jex, 1996a,
Phys. Rev. Lett. {\bf 76}, 2464.\\
Paul, H., P. T\"{o}rm\"{a}, T. Kiss and I. Jex, 1996b,
J. Opt. Fine Mechanics {\bf 41}, 338.\\
Paul, H., P. T\"{o}rm\"{a}, T. Kiss and I. Jex, 1997a,
Phys. Rev. A{\bf 56}, 4076.\\
Paul, H., P. T\"{o}rm\"{a}, T. Kiss and I. Jex, 1997b,
J. Mod. Opt. {\bf 44}, 2395.\\
Paul, W., O. Osberghaus and E. Fischer, 1958,
{\em Forschungsberichte des Wirtschafts- und Verkehrsministeriums
Nordrhein-Westfalen}, Nr. 415 (Westdeutscher Verlag, K\"{o}ln und
Opladen).\\
Pauli, W., 1933,
in: {\em Handbuch der Physik}, Vol. 24, Part 1,
eds H. Geiger and K. Scheel (Springer, Berlin);
for an English translation, see Pauli, W., {\em General
Principles of Quantum Mechanics} (Springer, Berlin, 1980) p. 17. \\
Pavi\v{c}i\'{c}, M., 1987,
Phys. Lett. A{\bf 122}, 280.\\
Pegg, D.T., and S.M. Barnett, 1988,
Europhys. Lett. {\bf 6}, 483.\\
Pegg, D.T., and S.M. Barnett, 1997,
J. Mod. Opt. {\bf 44}, 225.\\
Pegg, D.T., S.M. Barnett and L.S. Phillips, 1997,
J. Mod. Opt. {\bf 44}, 2135.\\
Penning, F.M., 1936,
Physica {\bf 3}, 873.\\
Perelomov, A., 1986,
{\em Generalized Coherent States and Their Applications}
(Springer, Berlin).\\
Pe\v{r}ina, J., 1985,
{\em Coherence of Light}
(Reidel, Dordrecht).\\
Pe\v{r}ina, J., 1991,
{\em Quantum Statistics of Linear and Nonlinear Optical Phenomena}
(Reidel, Dordrecht).\\
Pfau, T., and Ch. Kurtsiefer, 1997,
J. Mod. Opt. {\bf 44}, 2551.\\
Plebanski, J., 1954,
Bull. Acad. Polon. {\bf 11}, 213.\\
Plebanski, J., 1955,
Bull. Acad. Polon. {\bf 14}, 275.\\
Plebanski, J., 1956,
Phys. Rev. {\bf 101}, 1825.\\
Polzik, E.S., J. Carri and H.J. Kimble, 1992,
Phys. Rev. Lett. {\bf 68}, 3020.\\
Popper, K.R., 1982,
{\em Quantum Theory and the Schism in Physics}
(Hutchinson, London).\\
Poyatos, J.F., R. Walser, J.I. Cirac, P. Zoller and R. Blatt, 1996,
Phys. Rev. A{\bf 53}, R1966.\\
Prasad, S., M.O. Scully and W. Martienssen, 1987,
Opt. Commun. {\bf 62}, 139.\\
Press, W.H., S.A. Teukolsky, W.T. Vetterling and B.P. Flannery, 1995,
{\em Numerical Recipes in C} (Cambridge University Press, New York).\\
Prugove\v{c}ki, E., 1977,
Int. J. Theor. Phys. {16}, 321.\\
Radon, J., 1917,
Berichte \"{u}ber die Verhandlungen der K\"{o}niglich-S\"{a}chsischen
Gesellschaft der Wissenschaften zu Leipzig, Mathematisch-Physikalische
Klasse {\bf 69}, 262.\\
Rajagopal, A.K., 1983,
Phys. Rev. A{\bf 27}, 558. \\
Raman, C., C.W.S. Conover, C.I. Sukenik and P.H. Bucksbaum, 1996,
Phys. Rev. Lett. {\bf 76}, 2436.\\
Ramsey, N.F., 1956,
{\em Molecular Beams} (Clarendon Press, Oxford).\\
Raymer, M.G., 1994,
Am. J. Phys. {\bf 62}, 986.\\
Raymer, M.G., 1997,
J. Mod. Opt. {\bf 44}, 2565.\\
Raymer, M.G., M. Beck, and D.F. McAlister, 1994,
Phys. Rev. Lett. {\bf 72}, 1137.\\
Raymer, M.G., D.F. McAlister and U. Leonhardt, 1996,
Phys. Rev. A{\bf 54}, 2397.\\
Raymer, M.G., J. Cooper, H.J. Carmichael, M. Beck and D.T. Smithey, 1995,
J. Opt. Soc. Am. B{\bf 12}, 1801.\\
Raymer, M.G., D.T. Smithey, M. Beck, M.E. Anderson and D.F.McAlister, 1993,
{\em Third International Wigner Symposium}, Oxford.\\
Reck, M., A. Zeilinger, H.J. Bernstein and P. Bertani, 1994,
Phys. Rev. Lett. {\bf 73}, 58.\\
Reichenbach, H., 1946,
{\em Philosophical Foundations of Quantum Mechanics}
(University of California Press, Berkeley).\\
Rempe, G., H. Walther and N. Klein, 1987,
Phys. Rev. Lett. {\bf 58}, 353.\\
Renwick, S.P., E.C. Martell, W.D. Weaver and J.S. Risley, 1993,
Phys. Rev. A{\bf 48}, 2910.\\
Richter, Th., 1996a,
Phys. Lett. A{\bf 211}, 327.\\
Richter, Th., 1996b,
Phys. Rev. A{\bf 53}, 1197.\\
Richter, Th., 1996c,
Phys. Rev. A{\bf 54}, 2499.\\
Richter, Th., 1997a,
J. Mod. Opt. {\bf 44}, 2385.\\
Richter, Th., 1997b,
Phys. Rev. A{\bf 55}, 4629.\\
Richter, Th., and A. W\"{u}nsche, 1996a,
Phys. Rev. A{\bf 53}, R1974.\\
Richter, Th., and A. W\"{u}nsche, 1996b,
in: 4{\em th Central-European Workshop on Quantum Optics},
Budmerice, ed. V. Bu\v{zek}
(Acta Physica Slovaca {\bf 46}, 487).\\
Richter, G., W.H Brunner and H. Paul, 1964,
Ann. Phys. (Leipzig) {\bf 14}, 239.\\
Robinson, D.J.S., 1991,
{\em A Course in Linear Algebra with Applications}
(World Scientific, Singapore).\\
Royer, A., 1977,
Phys. Rev. A{\bf 15}, 449.\\
Royer, A., 1985,
Phys. Rev. Lett. {\bf 55}, 2745.\\
Royer, A., 1989,
Found. Phys. {\bf 19}, 3.\\
Royer, A., 1994,
Phys. Rev. Lett. {\bf 73}, 913.\\
Royer, A., 1995,
Phys. Rev. Lett. {\bf 74}, 1040(E).\\
Royer, A., 1996,
Phys. Rev. A{\bf 53}, 70.\\
Sauter, Th., W. Neuhauser, R. Blatt and P.E. Toschek, 1986,
Phys. Rev. Lett. {\bf 57}, 1696.\\
Schiller, S., G. Breitenbach, S.F. Pereira,
T. M\"{u}ller and J. Mlynek, 1996,
Phys. Rev. Lett. {\bf 77}, 2933.\\
Schneider, S., A.M. Herkommer, U. Leonhardt and W.P. Schleich, 1997,
J. Mod. Opt. {\bf 44}, 2333.\\
Schr\"{o}dinger, E., 1926,
Naturwiss. {\bf 14}, 664.\\
Schubert, M., and W. Vogel, 1978a,
Wiss. Zeitschr. FSU Jena, Math.--Nat. R. {\bf 27}, 179.\\
Schubert, M., and W. Vogel, 1978b,
Phys. Lett. A{\bf 68}, 321.\\
Sch\"{u}lke, W., 1988,
Portgal. Phys. {\bf 19}, 421. \\
Sch\"{u}lke, W., and S. Mourikis, 1986,
Acta Cryst. A{\bf 42}, 86.\\
Sch\"{u}lke, W., U. Bonse, and S. Mourikis, 1981,
Phys. Rev. Lett. {\bf 47}, 1209.\\
Schumaker, B.L., 1984,
Opt. Lett. {\bf 9}, 189.\\
Seifert, N., N.D. Gibson and J.S. Risley, 1995,
Phys. Rev. A{\bf 52}, 3816.\\
Senitzky, I.R., 1954,
Phys. Rev. {\bf 95}, 1115.\\
Shapiro, J.H., 1985,
IEEE J. Quant. Electron. {\bf 21}, 237.\\
Shapiro, J.H., and S.S. Wagner, 1984,
IEEE J. Quant. Electron. {\bf 20}, 803.\\
Shapiro, J.H., and H.P. Yuen, 1979,
IEEE Trans. Inform. Theor. {\bf 25}, 179.\\
Shapiro, J.H., H.P. Yuen and J.A. Machado Mata, 1979,
IEEE Trans. Inform. Theor. {\bf 24}, 657.\\
Shapiro, M., 1995,
J. Chem. Phys. {\bf 103}, 1748.\\
Shore, B.W., and P.L. Knight, 1993,
J. Mod. Opt. {\bf 40}, 1195.\\
Sleator, T., T. Pfau, V. Balykin, O. Carnal and J. Mlynek, 1992,
Phys. Rev. Lett. {\bf 68}, 1996.\\
Smithey D.T., M. Beck, J. Cooper and M.G. Raymer, 1993,
Phys. Rev. A{\bf 48}, 3159.\\
Smithey, D.T., M. Beck, M.G. Raymer and A. Faridani, 1993,
Phys. Rev. Lett. {\bf 70}, 1244.\\
Smithey, D.T., M. Beck, J. Cooper, M.G. Raymer and A. Faridani, 1993,
Phys. Scr. T{\bf 48}, 35.\\
Song, S., C.M. Caves and B. Yurke, 1990,
Phys. Rev. A{\bf 41}, 5261.\\
Stenholm, S., 1992,
Ann. Phys. (N.Y.) {\bf 218}, 233.\\
Stenholm, S., 1995,
Appl. Phys. B{\bf 60}, 243.\\
Stern, O., 1921,
Z. Phys. {\bf 7}, 249.\\
Steuernagel, O., and J.A. Vaccaro, 1995,
Phys. Rev. Lett. {\bf 75}, 3201.\\
Stulpe, W., and M. Singer, 1990,
Found. Phys. Lett. {\bf 3}, 153.\\
Sudarshan, E.C.G.,  1963,
Phys. Rev. Lett. {\bf 10}, 277.\\
Susskind, L., and J. Glogower, 1964,
Physics {\bf 1}, 49.\\
Takahashi, K., and N. Sait\^{o}, 1985,
Phys. Rev. Lett. {\bf 55}, 645.\\
Tan, S.M., 1997,
J. Mod. Opt. {\bf 44}, 2233.\\
Tanguy, C., S. Reynaud and C. Cohen-Tannoudji, 1984,
J. Phys. B{\bf 17}, 4623.\\
Tegmark, M., 1996,
Phys. Rev. A {\bf 54}, 2703. \\
Titulaer, U.M., and R.J. Glauber, 1966,
Phys. Rev. {\bf 145}, 1041.\\
T\"{o}rm\"{a}, P., S. Stenholm and I. Jex, 1995,
Phys. Rev. A{\bf 52}, 4853.\\
T\"{o}rm\"{a}, P., I. Jex and S. Stenholm, 1996,
J. Mod. Opt. {\bf 43}, 245.\\
Trippenbach, M., and Y.B. Band, 1996,
J. Chem. Phys. {\bf 105}, 8463.\\
Turski, L.A., 1972,
Physica {\bf 57}, 432.\\
Ueda, M., and M. Kitagawa, 1992,
Phys. Rev. Lett. {\bf 68}, 3424.\\
Vaccaro, J.A., and S.M. Barnett, 1995,
J. Mod. Opt. {\bf 42}, 2165.\\
V\'{a}rilly, J.C., and J.M. Gracia-Bond\'{\i}a, 1989,
Ann. Phys. (NY) {\bf 190}, 107.\\
Vogel, K., and H. Risken, 1989,
Phys. Rev. A{\bf 40}, 2847.\\
Vogel, W., 1991,
Phys. Rev. Lett. {\bf 67}, 2450.\\
Vogel, W., 1995,
Phys. Rev. A{\bf 51}, 4160 (1995).\\
Vogel, W., and J. Grabow, 1993,
Phys. Rev. A{\bf 47}, 4227.\\
Vogel, W., and R.L. de Matos Filho, 1995,
Phys. Rev. A{\bf 52}, 4214.\\
Vogel, W., and W. Schleich, 1991,
Phys. Rev. A{\bf 44}, 7642.\\
Vogel, W., and D.--G. Welsch, 1994,
{\em Lectures on Quantum Optics} (Akademie Verlag, Berlin).\\
Vogel, W., and D.-G. Welsch, 1995,
in: 3{\em th Central-European Workshop on Quantum Optics},
Budmerice, ed. V. Bu\v{z}ek
(Acta Physica Slovaca {\bf 45}, 313).\\
Vogel, W., D.-G. Welsch and L. Leine, 1987,
J. Opt. Soc. Am. B{\bf 4}, 1633).\\
Vogt, A., 1978,
in: {\em Mathematical Foundations of Quantum Theory},
ed. A.R. Marlow (Academic, New York) p. 365.\\
Walker, N.G., 1987,
J. Mod. Opt. {\bf 34}, 15.\\
Walker, N.G., and J.E. Carroll, 1984,
Electron. Lett. {\bf 20}, 981.\\
Wallentowitz S., and W. Vogel, 1995,
Phys. Rev. Lett. {\bf 75}, 2932.\\
Wallentowitz, S., and W. Vogel, 1996a,
Phys. Rev. A{\bf 53}, 4528.\\
Wallentowitz S., and W. Vogel, 1996b,
Phys. Rev. A{\bf 54}, 3322.\\
Wallentowitz, S., R.L. de Matos Filho, and W. Vogel, 1997,
Phys. Rev. A{\bf 56}, 1205.\\
Wallis, H., A. R\"{o}hrl, M. Naraschewski and A. Schenzle, 1997,
Phys. Rev. A{\bf 55}, 2109.\\
Walser, R., 1997,
Phys. Rev. Lett. {\bf 79}, 4724.\\
Walser, R., J.I. Cirac and P. Zoller, 1996,
Phys. Rev. Lett. {\bf 77}, 2658.\\
Waxer, L.J., I.A. Walmsley and W. Vogel, 1997,
Phys. Rev. A{\bf 56}, R2491.\\
Weigert, S., 1992,
Phys. Rev. A{\bf 45}, 7688.\\
Weyl, H., 1927,
Z. Phys. {\bf 46}, 1.\\
Wiedemann, H., 1996,
unplished.\\
Wiesbrock, H.-W., 1987,
Int. J. Theor. Phys. {\bf 26}, 1175.\\
Wigner, E.P., 1932,
Phys. Rev. {\bf 40}, 749.\\
Wigner, E.P., 1959,
{\em Group Theory} (Academic Press, New York).\\
Wilkens, M., and P. Meystre, 1991,
Phys. Rev. A{\bf 43}, 3832.\\
Wineland, D.J., J.J. Bollinger, W.M. Itano, F.L. Moore, and
D.J. Heinzen, 1992,
Phys. Rev. A {\bf 46}, R6797.\\
W\'{o}dkiewicz, K., 1984,
Phys. Rev. Lett. {\bf 52}, 1064.\\
W\'{o}dkiewicz, K., 1986,
Phys. Lett. A{\bf 115}, 304.\\
W\'{o}dkiewicz, K., 1988,
Phys. Lett. A{\bf 129}, 1.\\
ten Wolde, A., L.D. Noordam, A. Lagendijk and H.B. van Linden
van den Heuvell, 1988,
Phys. Rev. Lett. {\bf 61}, 2099.\\
Wootters, W.K., 1987,
Ann. Phys. (NY) {\bf 176}, 1.\\
Wootters, W.K., and B.D. Fields, 1989,
Ann. Phys. (NY) {\bf 191}, 363.\\
Wootters, W.K., and W.H. Zurek, 1979,
Phys. Rev. D{\bf 19}, 473.\\
W\"{u}nsche, A., 1990,
Quantum Opt. {\bf 2}, 453.\\
W\"{u}nsche, A., 1991,
Quantum Opt. {\bf 3}, 359.\\
W\"{u}nsche, A., 1996a,
Quantum Semiclass. Opt. {\bf 8}, 343.\\
W\"{u}nsche, A., 1996b,
Phys. Rev. A{\bf 54}, 5291.\\
W\"{u}nsche, A., 1997,
J. Mod. Opt. {\bf 44}, 2293.\\
W\"{u}nsche, A., and V. Bu\v{z}ek, 1997,
Quantum Semiclass. Opt. {\bf 9}, 631.\\
Yeazell, J.A., and C.R. Stroud, Jr., 1988,
Phys. Rev. Lett. {\bf 60}, 1494.\\
Yuen, H.P., 1982,
Phys. Lett. A{\bf 91}, 101.\\
Yuen, H.P., and V.W.S. Chan, 1983,
Opt. Lett. {\bf 8}, 177.\\
Yuen, H.P., and J.H. Shapiro, 1978b,
IEEE Trans. Inf. Theory {\bf 24}, 657.\\
Yuen, H.P., and J.H. Shapiro, 1978b,
in: {\em Coherence and Quantum Optics IV},
eds L. Mandel and E. Wolf (Plenum, New York) p. 719. \\
Yuen, H.P., and J.H. Shapiro, 1980,
IEEE Trans. Inf. Theory {\bf 26}, 78.\\
You, D., R.R. Jones, P.H. Bucksbaum and D.R. Dykaar, 1993,
Opt. Lett. {\bf 18}, 290.\\
Yurke, B., 1985,
Phys. Rev. A{\bf 32}, 300, 311.\\
Yurke, B., and D. Stoler, 1987,
Phys. Rev. A{\bf 36}, 1955.\\
Yurke, B., S.L. McCall and J.R. Klauder, 1986,
Phys. Rev. A{\bf 33}, 4033.\\
Yurke, B., P. Grangier, R.E. Slusher and M.J. Potasek, 1987,
Phys. Rev. A{\bf 35}, 3586.\\
Zaugg, T., M. Wilkens and P. Meystre, 1993,
Found. Phys. 23, 857.\\
Zucchetti, A., W. Vogel and D.--G. Welsch, 1996,
Phys. Rev. A{\bf 54}, 856.\\
Zucchetti, A., W. Vogel, M. Tasche and D.--G. Welsch, 1996,
Phys. Rev. A{\bf 54}, 1678.


\end{document}